\documentclass[11pt,a4paper]{amsart}
\usepackage{enumitem}
\usepackage{color}
\usepackage[numbers,sort&compress]{natbib}
\usepackage{pifont,latexsym,ifthen,amsthm,rotating,calc,textcase,booktabs}
\usepackage{amsfonts,amssymb,amsbsy,amsmath}
\usepackage{amsthm}
\numberwithin{equation}{section}
\usepackage[dvipsnames, svgnames, x11names]{xcolor}
\usepackage{graphicx}
\usepackage{tikz}
\usepackage{stmaryrd}
\usepackage{graphicx}
\usepackage[latin1]{inputenc}
\usepackage[T1]{fontenc}
\usepackage[english]{babel}
\usepackage{listings}
\usepackage{xcolor,mathrsfs,url}
\usepackage{ifthen}
\usepackage{fancyhdr}
\usepackage{verbatim}
\usepackage{todonotes}
\usepackage{float}
\usepackage{subfigure}

\usepackage{amsmath,amssymb,tikz}
\usetikzlibrary{arrows.meta,decorations.markings}

\usepackage{hyperref}
\usepackage{appendix}
\usepackage[percent]{overpic}
\renewcommand{\Re}{\mathrm{Re}\,}
\renewcommand{\Im}{\mathrm{Im}\,}

\newtheorem{theorem}{Theorem}[section]
\newtheorem{lemma}[theorem]{Lemma}

\newtheorem{remark}[theorem]{Remark}
\newtheorem{proposition}[theorem]{Proposition}
\newtheorem{definition}[theorem]{Definition}

\newtheorem{RHP}[theorem]{RH Problem}
\newtheorem{assumption}[theorem]{Assumption}
\usepackage {caption}

\hypersetup{hypertex=true,
            colorlinks=true,
            linkcolor=blue,
            anchorcolor=blue,
            citecolor=red}

\newcommand*{\dif}{\mathop{}\!\mathrm{d}}

\allowdisplaybreaks[4]

\begin{document}
%\title[Painlev\'{e}  XXXIV asymptotics for defocusing mKdV  equation]{Painlev\'{e}  XXXIV  asymptotics for the defocusing modified Korteweg-de	Vries equation}
	
	\title[Long-time asymptotic of the defocusing mKdV equation]
	{Long-time asymptotic behavior of the defocusing mKdV equation with a finite-genus algebro-geometric background}

%\author[Engui Fan,\ Yuan Li \  and  \ Yiling Yang]{Engui Fan$^{1,2}$,\ Yuan Li$^{1}$ \  and  \ Yiling Yang$^{3}$   }
%\footnotetext[1]{\ School of Mathematical Sciences  and Key Laboratory   for Nonlinear Science, Fudan   University, Shanghai 200433, P. R. China. E-mail: \texttt{faneg@fudan.edu.cn, liy\_fd@fudan.edu.cn}}
%\footnotetext[2]{\  College of Mathematics and Systems Science, Shandong University of Science and Technology, Qingdao, 266590,  China.}

%\footnotetext[3]{\ College of Mathematics and Statistics,	Chongqing University,	Chongqing 401331,	P. R. China. E-mail: \texttt{ylyang19@fudan.edu.cn}}
%\date{  }

\author[Engui Fan, Yuan Li and Yiling Yang]{%
	Engui Fan$^{1,2}$, Yuan Li$^{1}$ and Yiling Yang$^{3}$%
}

\footnotetext[1]{\ School of Mathematical Sciences and Key Laboratory
	for Nonlinear Science, Fudan University, Shanghai 200433, P. R. China.}

\footnotetext[2]{\ College of Mathematics and Systems Science,
	Shandong University of Science and Technology, Qingdao 266590, P. R. China.}

\footnotetext[3]{\ College of Mathematics and Statistics,
	Chongqing University, Chongqing 401331, P. R. China.}

\begingroup
\renewcommand{\thefootnote}{}
\footnotetext[4]{\textit{E-mail addresses:}
	\texttt{faneg@fudan.edu.cn} (E.~Fan),
	\texttt{liy\_fd@fudan.edu.cn} (Y.~Li),
	\texttt{ylyang19@fudan.edu.cn} (Y.~Yang).}
\endgroup

\maketitle
\begin{abstract}
In this paper, we  investigate long-time asymptotics   of the defocusing modified Korteweg--de Vries equation with a finite-genus algebro-geometric background by utilizing 
   the nonlinear steepest descent analysis.  Using the critical lines arising
from  the collision between  phase points and branch points,  we divide
the $(x,t)$-half  plane into
four types of asymptotic regions $\mathcal{T}_{\mathrm{I}}$--$\mathcal{T}_{\mathrm{IV}}$  for both even and odd genus, together with an additional central transition region
$\mathcal{T}_{\mathrm{0}}$ for odd genus. Among them, the regions $\mathcal{T}_{\mathrm{III}}$ and $\mathcal{T}_{\mathrm{IV}}$ correspond to  the  Zakharov--Manakov region and the  fast-decaying region, respectively.
While in   two  transition regions $\mathcal{T}_{\mathrm{I}}$ and $\mathcal{T}_{\mathrm{II}}$ where  the two symmetric saddle points approach their respective branch points,    the  long-time  asymptotics can be expressed in terms of
the solution of  a Painlev\'e XXXIV model.  In
 the   transition region $\mathcal{T}_{\mathrm{0}}$ where  the two real saddle points coalesce at the origin
 for  odd genus,  the long-time  asymptotics can be  described by the  Painlev\'e II transcendent.
 \\[4pt]

\end{abstract}

\setcounter{tocdepth}{2}
\tableofcontents

\quad

\section{Introduction}
%\hspace*{\parindent}
%\baselineskip=15pt

%Defocusing modified Korteweg-de Vries equation
%\begin{equation}q_{t}+6q^{2}q_{x}+q_{xxx}=0,\end{equation}
%where $q=q(x,t)$ is pure imaginary function.

	In this paper, we study the long-time asymptotics of solutions to the defocusing modified Korteweg-de Vries (mKdV) equation
\begin{equation}\label{mkdv}
	q_{t}+6q^{2}q_{x}+q_{xxx}=0,\qquad x\in\mathbb R,\quad t>0,
\end{equation}
with a finite-genus algebro-geometric background
\begin{equation}\label{boundray}
	q_0(x)\sim q^{(AG)}(x,t)\big|_{t=0},\qquad x\to\pm\infty.
\end{equation}
Here \(q=q(x,t)\) is a purely imaginary function, $q_0(x)=q(x,0)$ and
\(q^{(\mathrm{AG})}(x,t)\) denotes an algebro-geometric solution of the mKdV equation~\eqref{mkdv}.
In fact, the usual form of the defocusing mKdV equation is
\begin{equation}
	u_t-6u^2u_x+u_{xxx}=0,  \qquad x\in\mathbb R,\quad t>0,
\end{equation}
where $u$ is  real-valued. The substitution $q=iu$ gives the equivalent form \eqref{mkdv}, which is more directly amenable to the 
Riemann--Hilbert (RH) analysis used here for an algebro-geometric background.

The mKdV equation  is a modification of the classical Korteweg--de Vries
(KdV) equation with cubic nonlinearity and serves as a fundamental
model for nonlinear wave propagation, describing the interplay
between cubic nonlinearity and dispersion
\cite{O92,KO69,KEC98,GPT97,T81}.
It  arises in the study of
long-wavelength disturbances in anharmonic lattices \cite{O92},
Alfv\'en waves in plasmas \cite{KO69,KEC98},
internal waves in stratified fluids \cite{GPT97},
and propagating double layers in plasmas \cite{T81}.
In particular, for internal waves, the sign of the cubic nonlinear
coefficient depends on the density stratification and leads to
qualitatively different wave dynamics \cite{GPT97}.

The well-posedness of the mKdV equation \eqref{mkdv} has been extensively studied.
For initial data in $H^s(\mathbb{R})$, Kenig, Ponce, and Vega
\cite{KPV93} established local well-posedness for $s\geq \frac14$.
Global well-posedness was subsequently proved for $s>\frac14$
by Colliander, Keel, Staffilani, Takaoka, and Tao~\cite{CKSTT03},
with the endpoint $s=\frac14$ treated by Guo \cite{G09}
and Kishimoto \cite{K09}. Recently, Harrop-Griffiths, Killip, and Vi\c{s}an~\cite{HKV24} combined the method of commuting flows with local smoothing estimates to extend global well-posedness to $s>-\frac12$.

%	For non-decaying initial data, Egorova and Teschl \cite{ET10}	established the existence and uniqueness of global classical	solutions to the defocusing mKdV equation with steplike	quasi-periodic finite-gap backgrounds, under suitable smoothness	and weighted integrability assumptions on the initial perturbations.  Egorova and Teschl~\cite{ET10} proved global existence and uniqueness of classical solutions to the defocusing mKdV equation for a class of steplike finite-gap initial data. 	

Of particular interest are the long-time asymptotics of solutions to the defocusing mKdV equation~\eqref{mkdv} under different boundary conditions.
For zero boundary conditions, 
%where the initial data satisfy
%\[q_0(x)\to 0,\qquad x\to\pm\infty,\]
Segur and Ablowitz~\cite{SA81} studied the asymptotic behavior of mKdV solutions in the absence of solitons. A foundational contribution was later made by Deift and Zhou~\cite{DZ93}, who introduced the nonlinear steepest descent method for oscillatory RH problems and established rigorous long-time asymptotics for the defocusing mKdV equation  in different space-time regions.  
Subsequent work relaxed the regularity and decay assumptions on the initial data \cite{L17,CL22}. 
%Long-time asymptotics have also been studied for initial data with nonzero  boundary conditions
%\[q_0(x)\to q_\pm,\qquad x\to\pm\infty,\]where \(q_+\) and \(q_-\) are constants.
For initial data with nonzero boundary conditions, asymptotic formulas for solutions of the defocusing mKdV equation have been obtained in both the solitonless regions~\cite{XZF23} and the solitonic
region~\cite{ZXF25}, while Wang, Xu, and Fan~\cite{WXF25} derived the corresponding asymptotics in the transition region.  
In the step-like case, Xu and Zhang~\cite{XZ26} studied the defocusing mKdV equation  in the rarefaction wave regime, and Wang and Wen~\cite{WW26} analyzed the dispersive shock wave regime.
Related results for the focusing mKdV equation with step-like data include the asymptotic description of a modulated elliptic wave by Kotlyarov and Minakov~\cite{KM10} and the analysis of dispersive shock waves, solitons, and breathers by Grava and Minakov~\cite{GM20}. In all the studies reviewed above, the initial data satisfy constant boundary conditions.

A natural extension is to study the long-time asymptotics of solutions whose initial data approach an algebro-geometric background, as in~\eqref{boundray}. The theory of algebro-geometric solutions developed from
the study of periodic spectral problems for integrable equations~\cite{DMN76,K77}.
Through the construction of Baker--Akhiezer functions on algebraic curves, these solutions admit representations in terms of Riemann theta functions and describe periodic or quasi-periodic nonlinear waves~\cite{AGbook,AGbook2}.  Once algebro-geometric solutions are constructed, a further question concerns the long-time behavior of solutions arising from localized perturbations of these backgrounds. An early answer to this question came from Kamvissis and
Teschl~\cite{KT07,KT2012,K11}, who extended the Deift-Zhou nonlinear steepest descent method to Riemann surfaces of general genus and used it to derive the long-time asymptotics of the periodic Toda lattice
under short-range perturbations.  Mikikits-Leitner and Teschl~\cite{MT} later applied this extension to the corresponding problem for the KdV equation on a finite-gap background. More recently, Fan, Li, Yang, and Zhang~\cite{NLSAG} obtained asymptotic formulas for the defocusing NLS equation with a finite-genus algebro-geometric background.

To our knowledge, previous studies of long-time asymptotics for the defocusing mKdV equation~\eqref{mkdv} have focused primarily on constant
boundary conditions~\cite{SA81,DZ93,L17,CL22,XZF23,ZXF25,WXF25,XZ26,WW26}, while the corresponding asymptotics  for the algebro-geometric background~\eqref{boundray} remain unknown. In this paper, we address this problem by deriving an asymptotic formula for  solutions of~\eqref{mkdv}  in each of the space-time regions specified in Definition~\ref{regions} below.

\begin{figure}
	\centering
	\begingroup
	
	\definecolor{mkdvIIIcolor}{HTML}{EAF7D9}
	\definecolor{mkdvIVcolor}{HTML}{C8DEF8}
	\definecolor{mkdvTransition}{HTML}{FFF8BD}
	\definecolor{mkdvCentral}{HTML}{E5E5E5}
	
	\def\mkdvW{10.8}
	\def\mkdvH{6.2}
	\def\mkdvEvenOrigin{6.2}
	\def\mkdvOddOrigin{7.5}
	
	\pgfmathsetmacro{\mkdvEvenThree}{6.2-6.2/2.2*6.2}
	\pgfmathsetmacro{\mkdvEvenTwo}{6.2-6.2/3.7*6.2}
	\def\mkdvEvenOne{4.25}
	
	\pgfmathsetmacro{\mkdvOddFive}{7.5-7.5/2.0*6.2}
	\pgfmathsetmacro{\mkdvOddFour}{7.5-7.5/3.5*6.2}
	\def\mkdvOddThree{1.55}
	\def\mkdvOddTwo{4.25}
	\def\mkdvOddZero{6.65}
	
	% Arguments: origin, top intercept, top half-width, white padding, color.
	\def\mkdvBand#1#2#3#4#5{%
		\fill[white] (#1,0)
		-- ({#2-#3-#4},\mkdvH)
		-- ({#2+#3+#4},\mkdvH) -- cycle;
		\fill[#5] (#1,0)
		-- ({#2-#3},\mkdvH)
		-- ({#2+#3},\mkdvH) -- cycle;
	}
	
	\def\mkdvAxes#1{%
		\begin{scope}[font=\fontsize{14}{16}\selectfont]
			\draw[
			black!85,
			line width=0.8pt,
			dash pattern=on 5pt off 3pt,
			-{Stealth[length=5pt]}
			] (0,0) -- ({\mkdvW+0.16},0);
			
			\draw[
			black!85,
			line width=0.8pt,
			dash pattern=on 5pt off 3pt,
			-{Stealth[length=5pt]}
			] (#1,0) -- (#1,{\mkdvH+0.16});
			
			\node[below,inner sep=2pt] at (#1,0) {$0$};
			\node[right,inner sep=2pt] at ({\mkdvW+0.16},0) {$x$};
			\node[above,inner sep=2pt] at (#1,{\mkdvH+0.16}) {$t$};
		\end{scope}
	}
	
	\resizebox{\linewidth}{!}{%
		\begin{tikzpicture}[
			x=1cm,
			y=1cm,
			% All region labels use the original font size of T_0.
			font=\fontsize{12}{14}\selectfont,
			mkdv band label/.style={
				fill=mkdvTransition,
				inner sep=0.4pt
			},
			mkdv speed label/.style={
				font=\fontsize{14}{16}\selectfont
			},
			mkdv panel label/.style={
				font=\fontsize{14}{16}\selectfont,
				anchor=north
			}
			]
			
			% (a) Even genus
			\begin{scope}
				\begin{scope}
					\clip (0,0) rectangle (\mkdvW,\mkdvH);
					
					\fill[mkdvIVcolor]
					(0,0) rectangle (\mkdvW,\mkdvH);
					
					\fill[mkdvIIIcolor]
					(0,0) -- (\mkdvEvenOrigin,0) -- (0,2.2) -- cycle;
					
					\fill[mkdvIIIcolor]
					(\mkdvEvenOrigin,0)
					-- (\mkdvEvenTwo,\mkdvH)
					-- (\mkdvEvenOne,\mkdvH) -- cycle;
					
					% Yellow transition bands
					\mkdvBand{\mkdvEvenOrigin}{\mkdvEvenThree}
					{2.05}{0.12}{mkdvTransition}
					\mkdvBand{\mkdvEvenOrigin}{\mkdvEvenTwo}
					{1.20}{0.08}{mkdvTransition}
					\mkdvBand{\mkdvEvenOrigin}{\mkdvEvenOne}
					{0.38}{0.045}{mkdvTransition}
					
					% Critical rays
					\foreach \a in {\mkdvEvenThree,\mkdvEvenTwo,\mkdvEvenOne}
					\draw[black!85,line width=0.8pt]
					(\mkdvEvenOrigin,0) -- (\a,\mkdvH);
					
					% Bulk-region labels
					\node at (1.42,0.72) {$\mathcal T_{III}$};
					\node at (0.60,2.64) {$\mathcal T_{IV}$};
					\node at (2.10,4.60) {$\mathcal T_{III}$};
					\node at (8.20,3.55) {$\mathcal T_{IV}$};
					
					% Transition-region labels
					\node[mkdv band label]
					at ({\mkdvEvenOrigin
						+(\mkdvEvenThree-\mkdvEvenOrigin)*2.08/\mkdvH},2.08)
					{$\mathcal T_{I}$};
					
					\node[mkdv band label]
					at ({\mkdvEvenOrigin
						+(\mkdvEvenTwo-\mkdvEvenOrigin)*3.30/\mkdvH},3.30)
					{$\mathcal T_{II}$};
					
					\node[mkdv band label]
					at ({\mkdvEvenOrigin
						+(\mkdvEvenOne-\mkdvEvenOrigin)*5.60/\mkdvH},5.60)
					{$\mathcal T_{I}$};
				\end{scope}
				
				\node[mkdv speed label,left,inner sep=3pt]
				at (0,2.2) {$\xi_3$};
				\node[mkdv speed label,left,inner sep=3pt]
				at (0,3.7) {$\xi_2$};
				\node[mkdv speed label,above,inner sep=3pt]
				at (\mkdvEvenOne,\mkdvH) {$\xi_1$};
				
				\mkdvAxes{\mkdvEvenOrigin}
				
				\node[mkdv panel label] at (5.4,-0.60)
				{(a) Even genus 2  ($n=1$)};
			\end{scope}
			
			% (b) Odd genus
			\begin{scope}[xshift=12.4cm]
				\begin{scope}
					\clip (0,0) rectangle (\mkdvW,\mkdvH);
					
					\fill[mkdvIVcolor]
					(0,0) rectangle (\mkdvW,\mkdvH);
					
					\fill[mkdvIIIcolor]
					(0,0) -- (\mkdvOddOrigin,0) -- (0,2.0) -- cycle;
					
					\fill[mkdvIIIcolor]
					(\mkdvOddOrigin,0)
					-- (\mkdvOddFour,\mkdvH)
					-- (\mkdvOddThree,\mkdvH) -- cycle;
					
					\fill[mkdvIIIcolor]
					(\mkdvOddOrigin,0)
					-- (\mkdvOddTwo,\mkdvH)
					-- (\mkdvOddZero,\mkdvH) -- cycle;
					
					% Yellow transition bands and gray central band
					\mkdvBand{\mkdvOddOrigin}{\mkdvOddFive}
					{2.85}{0.15}{mkdvTransition}
					\mkdvBand{\mkdvOddOrigin}{\mkdvOddFour}
					{1.30}{0.09}{mkdvTransition}
					\mkdvBand{\mkdvOddOrigin}{\mkdvOddThree}
					{0.50}{0.045}{mkdvTransition}
					\mkdvBand{\mkdvOddOrigin}{\mkdvOddTwo}
					{0.45}{0.045}{mkdvTransition}
					\mkdvBand{\mkdvOddOrigin}{\mkdvOddZero}
					{0.28}{0.045}{mkdvCentral}
					
					% Critical rays
					\foreach \a in {\mkdvOddFive,\mkdvOddFour,\mkdvOddThree,\mkdvOddTwo}
					\draw[black!85,line width=0.8pt]
					(\mkdvOddOrigin,0) -- (\a,\mkdvH);
					
					\draw[
					black!65,
					line width=0.8pt,
					dash pattern=on 3pt off 2pt
					] (\mkdvOddOrigin,0) -- (\mkdvOddZero,\mkdvH);
					
					% Bulk-region labels
					\node at (1.30,0.65) {$\mathcal T_{III}$};
					\node at (0.60,2.52) {$\mathcal T_{IV}$};
					\node at (1.70,4.60) {$\mathcal T_{III}$};
					\node at (3.60,5.10) {$\mathcal T_{IV}$};
					\node at (5.95,4.65) {$\mathcal T_{III}$};
					\node at (8.90,3.50) {$\mathcal T_{IV}$};
					
					% Transition-region labels
					\node[mkdv band label]
					at ({\mkdvOddOrigin
						+(\mkdvOddFive-\mkdvOddOrigin)*1.90/\mkdvH},1.90)
					{$\mathcal T_{I}$};
					
					\node[mkdv band label]
					at ({\mkdvOddOrigin
						+(\mkdvOddFour-\mkdvOddOrigin)*3.20/\mkdvH},3.20)
					{$\mathcal T_{II}$};
					
					\node[mkdv band label]
					at ({\mkdvOddOrigin
						+(\mkdvOddThree-\mkdvOddOrigin)*5.60/\mkdvH},5.60)
					{$\mathcal T_{I}$};
					
					\node[mkdv band label]
					at ({\mkdvOddOrigin
						+(\mkdvOddTwo-\mkdvOddOrigin)*5.60/\mkdvH},5.60)
					{$\mathcal T_{II}$};
					
					\node[fill=mkdvCentral,inner sep=0.4pt]
					at ({\mkdvOddOrigin
						+(\mkdvOddZero-\mkdvOddOrigin)*5.10/\mkdvH},5.10)
					{$\mathcal T_0$};
				\end{scope}
				
				\node[mkdv speed label,left,inner sep=3pt]
				at (0,2.0) {$\xi_5$};
				\node[mkdv speed label,left,inner sep=3pt]
				at (0,3.5) {$\xi_4$};
				\node[mkdv speed label,above,inner sep=3pt]
				at (\mkdvOddThree,\mkdvH) {$\xi_3$};
				\node[mkdv speed label,above,inner sep=3pt]
				at (\mkdvOddTwo,\mkdvH) {$\xi_2$};
				\node[mkdv speed label,above,inner sep=3pt]
				at (\mkdvOddZero,\mkdvH) {$\xi_0$};
				
				\mkdvAxes{\mkdvOddOrigin}
				
				\node[mkdv panel label] at (5.4,-0.60)
				{(b) Odd genus 3 ($n=2$)};
			\end{scope}
			
		\end{tikzpicture}%
	}
	\endgroup
	
	\caption{Asymptotic regions for the defocusing mKdV equation:	(a) Even genus 2; (b) Odd genus 3.
		The gray strip denotes the central transition region~$\mathcal T_0$.}
	\label{fig:mkdv-asymptotic-regions}
\end{figure}
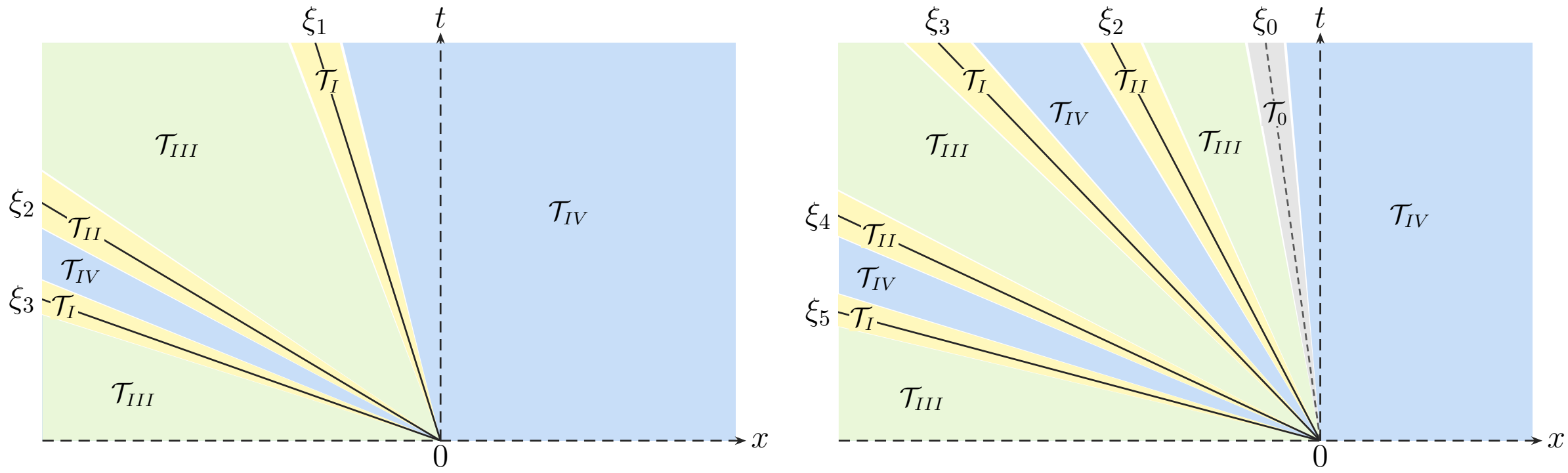

\begin{definition}\label{regions}
	Let $n$ be a positive integer, fix $C>0$, and set $\xi=\frac{x}{t}$. We define the following space-time  regions.
	\begin{itemize}
		\item  Transition regions for Painlev\'e XXXIV: 	
		\begin{align}
		&	\mathcal{T}_{\mathrm{I}}
			:=\left\{
			\begin{aligned}
		&	\bigcup_{k=0}^{n}
			\left\{
			\xi:\left|\xi-\xi_{2k+1}\right|t^{\frac23}\leq C
			\right\},\quad \text{genus}\   2n,\\
		&	\bigcup_{k=1}^{n}
			\left\{
			\xi:\left|\xi-\xi_{2k+1}\right|t^{\frac23}\leq C
			\right\},\quad \text{genus}\   2n-1,\\
		\end{aligned}
		\right.\\
		&\mathcal{T}_{\mathrm{II}}
			:=
			\bigcup_{k=1}^{n}
			\left\{
			\xi:\left|\xi-\xi_{2k}\right|t^{\frac23}\leq C
			\right\}.
		\end{align}
				\item  Zakharov--Manakov region: 	
	\begin{equation}
			\mathcal{T}_{\mathrm{III}}
			:=\left\{
			\begin{aligned}
		&	(-\infty,\xi_{2n+1})
			\cup
			\bigcup_{k=1}^{n}
			(\xi_{2k},\xi_{2k-1}),\quad \text{genus}\   2n,\\
			&(-\infty,\xi_{2n+1})\cup
			\bigcup_{k=2}^{n}(\xi_{2k},\xi_{2k-1})\cup(\xi_2,\xi_0), \quad \text{genus}\   2n-1.
		\end{aligned}
		\right.
		\end{equation}
		\item Fast-decaying region: 	
	\begin{equation}
				\mathcal{T}_{\mathrm{IV}}
				:=\left\{	\begin{aligned}
					&\bigcup_{k=1}^{n}
				(\xi_{2k+1},\xi_{2k})\cup(\xi_1,+\infty),\quad \text{genus}\   2n,\\
				&\bigcup_{k=1}^{n}(\xi_{2k+1},\xi_{2k})
				\cup(\xi_0,+\infty), \quad \text{genus}\   2n-1.
			\end{aligned}
			\right.
		\end{equation}
		\item Central transition region for Painlev\'e II:
		\[
		\mathcal T_0
		:=\left\{\xi:|\xi-\xi_0|t^{\frac{2}{3}}\le C\right\},\quad \text{genus}\   2n-1.
		\]
	\end{itemize}
Here   $\xi_k$ in the regions  $\mathcal{T}_{\mathrm{I}}$--$\mathcal{T}_{\mathrm{IV}}$  are   the critical points   defined by (\ref{critical-value}).     $\xi_0$ in the region   $\mathcal{T}_{\mathrm{0}}$   is a critical point
 defined by  (\ref{xi0}).  With  $\xi_k$ and $\xi_0$, we divide the $(x,t)$-half plane
 into the above  different  space-time regions as shown  in Figure \ref{fig:mkdv-asymptotic-regions}.
\end{definition}

We analyze the long-time asymptotics of the mKdV equation~\eqref{mkdv} separately for even genus~$2n$ and odd genus~$2n-1$. The reason is that the associated RH problem~\ref{RH1} possesses the additional spectral symmetry $z\mapsto -z$ given in \eqref{symmetry}. Consequently, the branch points occur in pairs $\pm E_j$, and the cuts on the real axis are arranged symmetrically about the
origin (see Figure~\ref{fig:mkdv-cuts}).
For even genus, the origin lies on the central cut $(-E_1,E_1)$, whereas for odd genus it lies in the central gap $(-E_2,E_2)$. 
%This  marks a fundamental difference from the defocusing NLS equation studied in \cite{NLSAG}.
This additional spectral symmetry marks a fundamental difference between the mKdV equation \eqref{mkdv} and the defocusing NLS equation studied in~\cite{NLSAG}, and it leads to different critical behavior for even and odd genus.

According to where the saddle points of the phase function $\theta(z;\xi)$ lie relative to the cuts as $\xi$ varies, we distinguish four types of asymptotic regions $\mathcal{T}_{\mathrm{I}}$--$\mathcal{T}_{\mathrm{IV}}$ for both even and odd genus, together with a central transition region $\mathcal{T}_{\mathrm{0}}$ for odd genus (see Definition~\ref{regions} and Figure~\ref{fig:mkdv-asymptotic-regions}). Regions $\mathcal{T}_{\mathrm{III}}$ and $\mathcal{T}_{\mathrm{IV}}$ describe the asymptotic behavior away from the critical speeds, while regions
$\mathcal{T}_{\mathrm{I}}$ and $\mathcal{T}_{\mathrm{II}}$ describe the two types of transition between them. In particular, for odd genus $2n-1$, the final transition from region $\mathcal{T}_{\mathrm{III}}$ to region $\mathcal{T}_{\mathrm{IV}}$
gives rise to an additional central transition region $\mathcal{T}_{\mathrm{0}}$ (see Figure~\ref{fig:mkdv-asymptotic-regions}(b)).  The key to the analysis is the treatment of the transition regions $\mathcal{T}_{\mathrm{I}}$--$\mathcal{T}_{\mathrm{II}}$ and $\mathcal{T}_{\mathrm{0}}$.

The transient asymptotics of integrable equations were investigated in early work by Segur and Ablowitz~\cite{SA81}, who connected the asymptotic behavior of mKdV solutions with a Painlev\'e~II transcendent.  Deift and Zhou~\cite{DZ93} subsequently derived the corresponding asymptotic formula with rigorous error bounds.  In the self-similar region, Charlier and Lenells~\cite{Lenellsmkdv} refined this result by obtaining an expansion to all orders, including Airy and Painlev\'e~II asymptotics, and Huang and Zhang~\cite{HZ22} extended this analysis to the mKdV hierarchy.   Painlev\'e~II transcendents also arise in transition regions for the defocusing NLS and mKdV equations with nonzero boundary conditions~\cite{WF23,WXF25}. For the Camassa--Holm equation, Boutet de Monvel, Its, and Shepelsky~\cite{BIS10} showed that Painlev\'e~II governs the asymptotics in two distinct transition regions, a result with a counterpart for the modified Camassa--Holm equation due to Xu, Yang, and Zhang~\cite{XYZ24}.  For the defocusing NLS equation with an algebro-geometric background, Fan, Li, Yang, and Zhang~\cite{NLSAG} were the first to show that the long-time asymptotics in the transition regions can be expressed in terms of the Painlev\'e~XXXIV transcendent.  This model had previously appeared in the study of critical phenomena in orthogonal polynomials and random matrix theory~\cite{DXZ19,ikj2008,XZ2011}.

In this work, we use the  Painlev\'e~XXXIV model to describe  the long-time asymptotics in the  transition regions $\mathcal{T}_{\mathrm{I}}$ and $\mathcal{T}_{\mathrm{II}}$, where the two
symmetric saddle points approach their respective branch points.  Notably, for odd genus, it is shown that the long-time asymptotics in the central transition region $\mathcal{T}_{\mathrm{0}}$ can be expressed by the Painlev\'e~II transcendent. Here the two symmetric saddle points coalesce at the origin. The remaining regions $\mathcal{T}_{\mathrm{III}}$ and
$\mathcal{T}_{\mathrm{IV}}$ correspond to the Zakharov--Manakov region and the fast-decaying region, respectively.

\begin{assumption}\label{assum}
		Throughout the paper, we work under the following assumption.
\begin{enumerate}
	\item[\textup{(i)}] There exists $R>0$ such that $q_0(x)=q^{(AG)}(x,0)$ for all $|x|>R$.
	\item[\textup{(ii)}]
	Let $\boldsymbol{E}$ denote the set of branch points given by~\eqref{E} and $\Gamma$ the union of the open cuts given by~\eqref{gamma}.
	The reflection coefficient $r_j$, $j=1,\ldots,4$, defined in \eqref{r} is bounded on $\Gamma$ and satisfies $|r_j(p)|=1$ for  $p\in\boldsymbol{E}$.
\end{enumerate}
\end{assumption}
Condition (i) requires the initial data $q_0$ to agree with the background outside a compact set. This allows us to extend the scattering matrix analytically off the cuts and simplifies the error estimates.
In (ii), the endpoint condition $|r_j(p)|=1$  holds in the generic case described in Proposition~\ref{propsca}(vi) (see \cite{DT79} for a similar case). Boundedness of the reflection coefficients on $\Gamma$ excludes spectral singularities arising from zeros of the scattering data $S_{11}$ or $S_{22}$ in the interiors of the cuts.

\subsection{Main results}
We now state our main result on the long-time asymptotics of the mKdV equation \eqref{mkdv} in the regions introduced in Definition~\ref{regions}.

\begin{theorem}\label{theorem}
	Let $q(x,t)$ be the solution of the Cauchy problem
	\eqref{mkdv}--\eqref{boundray} with the finite-genus algebro-geometric background
	$q^{(\mathrm{AG})}(x,t;\mathbf E,\boldsymbol\phi)$ given by \eqref{qAG}.
	Under Assumption~\ref{assum}, the following asymptotics hold as
	$t\to+\infty$ 	in the regions   specified
	in Definition~\ref{regions}.

	\begin{itemize}
		\item \textbf{\rm Transition region $\mathcal{T}_{\mathrm{I}}.$ }
		For $|\xi-\xi_{2k_0+1}|t^{\frac{2}{3}}\le C$, with fixed
	$k_0\in\{0,\ldots,n\}$ for genus $2n$ and
	$k_0\in\{1,\ldots,n\}$ for genus $2n-1$,
	\begin{align}\label{eq:main-I}
		q(x,t)
		=
		q^{(\mathrm{AG})}(x,t;\mathbf E,\widetilde{\boldsymbol\phi})
		+
		H_{E_{2k_0+1}}a(s)t^{-\frac{1}{3}}
		+
		\mathcal{O}(t^{-\epsilon}),
		\qquad
		\epsilon\in\left(\frac13,\frac23\right).
	\end{align}
	%Here $q^{(\mathrm{AG})}$ is given by \eqref{qAG}, with the shifted
    Here
	the shifted phase vector $\widetilde{\boldsymbol\phi}$ of the algebro-geometric solution in the leading term is  defined by
	\eqref{phi}, and $H_{E_{2k_0+1}}$ is  a pure imaginary function of $x$ and $t$ constructed by the Riemann theta function given by \eqref{eq:coefficient-I}.
	The real function $a(s)=a(s;\frac14,0)$ is specified by \begin{equation}\label{def:a}
 a(s)=\int_{-\infty}^s\left(u(\zeta)+\frac{\zeta}{2}\right)\dif\zeta, \ \  s=\frac{\theta^{(1)}(\xi)(\xi-\xi_{2k_0+1})t^{\frac{2}{3}} }{|\theta^{(3)}(\xi)|^{\frac{1}{3}}},
\end{equation}
 with real nonzero constants $\theta^{(1)}(\xi)$ and $\theta^{(3)}(\xi)$ given in \eqref{theta1} and \eqref{theta3}, respectively. 
	Here, $u(s)$ is the unique  solution of the Painlev\'e XXXIV equation
\begin{align*}
		 	u''(s)=4u(s)^2+2su(s)+\frac{u'(s)^2- \frac{1}{4}} {2u(s)}
\end{align*}
satisfying
\begin{equation*}
    u(s)=\left\{ \begin{array}{ll}
				\frac{1}{4\sqrt{s}}+\mathcal{O}(s^{-2}), &\qquad s\to +\infty,
				\\
				-\frac{s}{2}+\mathcal{O}(s^{-2}), &\qquad s\to -\infty.
			\end{array}
			\right .
\end{equation*}

		\item \textbf{\rm Transition region  $\mathcal{T}_{\mathrm{II}}.$}
			For $|\xi-\xi_{2k_0}|t^{\frac{2}{3}}\le C$, with fixed
		$k_0\in\{1,\ldots,n\}$,
		\begin{align}\label{eq:main-II}
			q(x,t)
			=
			q^{(\mathrm{AG})}(x,t;\mathbf E,\widetilde{\boldsymbol\phi})
			-
			 H_{E_{2k_0}}a(s)t^{-\frac{1}{3}}
			+
			\mathcal{O}(t^{-\epsilon}),
			\qquad
			\epsilon\in\left(\frac13,\frac23\right).
		\end{align}
		Here the shifted phase vector  $\widetilde{\boldsymbol\phi}$ is given by \eqref{eq:II-shifted-phases}, $H_{E_{2k_0}}$  is  a  pure imaginary function of $x$ and $t$ constructed by the Riemann theta function given by \eqref{eq:coefficient-II}, and now
        \begin{align*}
            s
	=\frac{\theta^{(1)}(\xi)(\xi-\xi_{2k_0})t^{\frac{2}{3}}}
	{|\theta^{(3)}(\xi)|^{\frac{1}{3}}},
        \end{align*}
with real nonzero constants $\theta^{(1)}(\xi)$ and $\theta^{(3)}(\xi)$ given in \eqref{II-phase-coefficients}.

		\item \textbf{\rm Zakharov--Manakov region  $\mathcal{T}_{\mathrm{III}}.$}
		For
		\[
		\xi\in
		\begin{cases}
			(-\infty,\xi_{2n+1})
			\displaystyle\cup\bigcup_{k=1}^{n}(\xi_{2k},\xi_{2k-1}),
			& \text{genus }2n,\\[2mm]
			(-\infty,\xi_{2n+1})
			\displaystyle\cup\bigcup_{k=2}^{n}(\xi_{2k},\xi_{2k-1})
			\cup(\xi_2,\xi_0),
			& \text{genus }2n-1,
		\end{cases}
		\]
		we have
		\begin{equation}\label{eq:asymptotics-region-III}
			\begin{aligned}
				q(x,t)
				={}&q^{(AG)}(x,t;\mathbf E,\widetilde{\boldsymbol\phi})
				\\
                &+\frac{2it^{-\frac12}}{\sqrt{\theta^{(2)}}}
   \Re \left(e^{-\frac{i\delta_{2n}}{2}}\bigl(
    \gamma_{21}
    M^{(\mathrm{glo})}_{11}(z_0;\mathbf E,\widetilde{\boldsymbol\phi})^2
    -\gamma_{12}
    M^{(\mathrm{glo})}_{12}(z_0;\mathbf E,\widetilde{\boldsymbol\phi})^2
    \bigr)
    \right)
				\\
                &+\mathcal O(t^{-1}),
			\end{aligned}
		\end{equation}
where $\widetilde{\boldsymbol\phi}$ is defined in \eqref{phi-region-III} with $\delta_j$ now determined by \eqref{deltaj-region-III}; $M^{(\mathrm{glo})}$ (a function of $x$ and $t$ constructed by the Riemann theta function) and $\gamma_{12},\gamma_{21}$ can be found in \eqref{def:Mglo} and \eqref{eq:III-beta}, respectively; and $\theta^{(2)}>0$ is a constant specified in \eqref{eq:III-phase-coefficient}.

		\item \textbf{\rm Fast decaying region $\mathcal{T}_{\mathrm{IV}}.$}
For
		\[
		\xi\in
		\begin{cases}
			\displaystyle
			\bigcup_{k=1}^{n}(\xi_{2k+1},\xi_{2k})\cup(\xi_1,+\infty),
			& \text{genus }2n,\\[2mm]
			\displaystyle
			\bigcup_{k=1}^{n}(\xi_{2k+1},\xi_{2k})\cup(\xi_0,+\infty),
			& \text{genus }2n-1,
		\end{cases}
		\]
		we have
		\begin{equation}\label{eq:asymptotics-region-IV}
			q(x,t)
			=
			q^{(\mathrm{AG})}(x,t;\mathbf E,\widetilde{\boldsymbol\phi})
			+\mathcal O(t^{-1}),
		\end{equation}
		where $\widetilde{\boldsymbol\phi}$ has the same definition as in \eqref{phi-region-III} but  $\delta_j$ is
now determined by the equations in~\eqref{deltaj-IV}.
	
	\item
	\textbf{\rm Central transition region $\mathcal{T}_{\mathrm{0}}.$}
	For genus $2n-1$ and $|\xi-\xi_0|t^{\frac{2}{3}}\leq C$, we have
	\begin{equation}\label{eq:central-asymptotics}
		q(x,t)
		=
		q^{(\mathrm{AG})}(x,t;\mathbf E,\boldsymbol\phi)
		+2it^{-\frac{1}{3}}(H_0)_{12}
		+\mathcal O\!\left(t^{-\frac{1}{3}-\epsilon}\right),
		\qquad
		0<\epsilon<\frac{1}{3},
	\end{equation}
	where
    \begin{align*}
  \begin{aligned}
(H_0)_{12}
=\frac{1}{|\theta^{(3)}(0;\xi_0)|^{\frac{1}{3}}}
\Bigg(
&v(s)
\left(M^{(\mathrm{glo})}_{11}(0;\mathbf E,\boldsymbol\phi)^2
-M^{(\mathrm{glo})}_{12}(0;\mathbf E,\boldsymbol\phi)^2\right)
\\
&\ \ +2iM^{(\mathrm{glo})}_{11}(0;\mathbf E,\boldsymbol\phi)
M^{(\mathrm{glo})}_{12}(0;\mathbf E,\boldsymbol\phi)
\int_s^\infty v(\eta)^2\,d\eta
\Bigg).
\end{aligned}
\end{align*}
	Here, $\theta^{(3)}(0;\xi_0)<0$ is a constant, $M^{(\mathrm{glo})}$ is a function of $x$ and $t$ constructed by the Riemann theta function  given in \eqref{def:Mglo} with $M^{(\mathrm{glo})}_{11}(0;\mathbf E,\boldsymbol\phi)\in\mathbb{R}$ and $M^{(\mathrm{glo})}_{12}(0;\mathbf E,\boldsymbol\phi)\in i\mathbb{R}$, and $v$ is the  unique  solution of
the Painlev\'e II equation
\begin{equation*}
	v''(s)=s v(s)+2v(s)^3,
\end{equation*}
characterized by
\begin{equation*}
	v(s;\kappa)
	=
	ir_1(0)\operatorname{Ai}(s)
	+\mathcal O\big(
	s^{-\frac{1}{4}}e^{-2s^{\frac{3}{2}}}
	\big),
	\qquad s\to+\infty,
\end{equation*}
where $\operatorname{Ai}$ denotes the Airy function and $r_1$ is the  reflection coefficient in
Proposition~\ref{propsca} with $i r_1(0)\in(-1,1)$.
	\end{itemize}
\end{theorem}

\begin{remark}
In the proof of Theorem~\ref{theorem}, we distinguish the asymptotic regions by analyzing the sign distribution of   $\operatorname{Im}\theta(z;\xi)$, where the phase function $\theta$ is defined in \eqref{alpha}.
Unlike the NLS equation studied in \cite{NLSAG}, the mKdV equation has an additional spectral symmetry. Therefore, we must not only treat the two saddle points simultaneously but also determine how the sign distribution of $\operatorname{Im}\theta$ changes when the two real saddle points coalesce at the origin in the odd genus case. To this end, we exploit the evenness of $F(z;\xi)$ in \eqref{F} to restrict attention to the positive real saddle point $z_0(\xi)$. For odd genus, we determine the coalescence speed from $F(0;\xi_0)=0$ and use the cubic expansion of $\theta$ at the origin to derive the local signature tables of $\operatorname{Im}\theta$ on either side of $\xi_0$.  See Figures~\ref{figim1} and~\ref{figim2} for an illustration.
\end{remark}

\begin{remark}
The main difficulty in proving Theorem~\ref{theorem} lies in constructing local parametrices in the transition regions. In regions $\mathcal{T}_{\mathrm{I}}$ and $\mathcal{T}_{\mathrm{II}}$, the two saddle points approach the positive and negative members of a symmetric pair of branch points, respectively. We construct a Painlev\'e XXXIV local parametrix near the positive branch point, match it to the finite-genus global parametrix, and use the spectral symmetry to obtain its counterpart near the negative branch point.
In the central transition region $\mathcal{T}_{\mathrm{0}}$ for odd genus, the parabolic cylinder approximation used in region $\mathcal{T}_{\mathrm{III}}$ loses uniformity as the two saddle points coalesce at the origin. We therefore construct a single Painlev\'e II local parametrix near the origin to obtain a uniform approximation as $\xi$ passes through $\xi_0$.

\end{remark}

This paper is organized as follows. In Section~\ref{sec2}, we construct the finite-genus algebro-geometric background solution and formulate the RH problem for the perturbed initial data. We then analyze the sign distribution of $\operatorname{Im}\theta$ to obtain the jump factorizations used in the steepest descent analysis. Sections~\ref{sec3} and~\ref{sec4} treat transition regions $\mathcal{T}_{\mathrm{I}}$ and $\mathcal{T}_{\mathrm{II}}$, respectively, using Painlev\'e XXXIV local parametrices near the corresponding pairs of branch points. In each case, we match the local parametrices to the global parametrix and compute the leading contribution from both branch points to the error matrix. In Section~\ref{sec5}, we analyze regions $\mathcal{T}_{\mathrm{III}}$ and $\mathcal{T}_{\mathrm{IV}}$ as well as the central transition region that arises for odd genus when the two real saddle points coalesce at the origin. To describe this transition, we use the cubic expansion of the phase to construct a single Painlev\'e II local parametrix for both points. Finally, in Section~\ref{sec6}, we use the reconstruction formula for $q(x,t)$ to derive the asymptotics in each region and complete the proof of Theorem~\ref{theorem}.

\subsection{Notation} Throughout this paper, the following notation will be used:
\begin{itemize}
\item The classical Pauli matrices $\{\sigma_j\}_{j=1,2,3}$ are defined by
\begin{equation}\label{def:PauliM}
	\sigma_1:=\begin{pmatrix}0 & 1 \\ 1 & 0\end{pmatrix}, \quad
	\sigma_2:=\begin{pmatrix}0 & -i \\ i & 0\end{pmatrix}, \quad
	\sigma_3:=\begin{pmatrix}1 & 0 \\ 0 & -1\end{pmatrix}.
\end{equation}
\item For a $2\times 2$ matrix $A$, we  define
\begin{equation*}%\label{def:hatsigma}
	e^{\hat{\sigma}_j}A:=e^{\sigma_j}Ae^{-\sigma_j}, \quad j=1,2,3
\end{equation*}
and
\begin{equation*}
|A|:=\max_{i,j=1,2}\{|A_{ij}|\},
\end{equation*}
where $A_{ij}$ denotes the $(i,j)$ entry of $A.$

\item For a given oriented contour $\Sigma$ and any function $f\in L^p(\Sigma)$, $1\leqslant p<\infty$, the Cauchy operator $\mathcal{C}_{\Sigma}$  is defined  by
\begin{align*}
	\mathcal{C}_{\Sigma}f(z)=\frac{1}{2\pi i}\int_{\Sigma}\frac{f(\zeta)}{\zeta-z}\dif\zeta, \quad  z\in\mathbb{C}\setminus \Sigma.
\end{align*}
 The projection operators are
given by
\begin{align}\label{def:opCpm}
	\mathcal{C}_\Sigma^\pm f(z):= \displaystyle\lim_{\substack{z'\to z\\\text{from the }\pm\text{ side of }\Sigma}}\frac{1}{2\pi i}\int_{\Sigma}\frac{f(\zeta)}{\zeta-z'}\dif\zeta.
\end{align}
Here $\mathcal{C}_{\Sigma}^{+}f$ and $\mathcal{C}_{\Sigma}^{-}f$ denote
the non-tangential boundary values of $\mathcal{C}_{\Sigma}f$ from the
left and right sides of the oriented contour $\Sigma$, respectively.

\end{itemize}

\section{Algebro-geometric solutions and a RH problem}\label{sec2}
\subsection{Lax pair and a basic RH problem}
The mKdV equation \eqref{mkdv} is integrable and   admits the Lax
pair representation in the form of an overdetermined system of linear ordinary equations for a $2\times2$ matrix-valued function \cite{AKNS,Wadati1973}
\begin{equation}\label{lax}
	\left\{
	\begin{aligned}
		&\psi_{x}=-iz\sigma_{3}\psi+U(q)\psi,\\
		&\psi_{t}=-4iz^{3}\sigma_{3}\psi+V(q)\psi,
	\end{aligned}
	\right.
\end{equation}
where  $z$ is the spectral variable and
\begin{align*}
	U(q)&=\begin{pmatrix}
		0&q\\
		-q&0
	\end{pmatrix},\\
	V(q)&=4z^{2}\begin{pmatrix}
		0&q\\
		-q&0
	\end{pmatrix}+2z\begin{pmatrix}
		iq^{2}&iq_{x}\\
		iq_{x}&-iq^{2}
	\end{pmatrix}-\begin{pmatrix}0&2q^{3}+q_{xx}\\-2q^{3}-q_{xx}&0\end{pmatrix}.
\end{align*}

For later use, we first introduce  a  basic RH problem for the mKdV equation \eqref{mkdv}.
\begin{RHP}\label{RH1}
	\hfill
	\begin{itemize}
		\item $m(z)=m(z;x,t)$ is analytic in $\mathbb{C}\setminus\overline{\Gamma}$, where $\Gamma$ represents the jump contour and the overline denotes closure. 
		\item $m(z)$ satisfies the jump condition
		$$ m_{+}(z)=m_{-}(z)e^{-i(zx+4z^{3}t)\hat{\sigma}_{3}}J, \quad z\in\Gamma,$$
		where \(J\) is piecewise constant with respect to \(z\) on \(\Gamma\) and is independent of \(x\) and \(t\).
		\item $m(z)$ satisfies the symmetry relations
		\begin{equation}\label{symmetry}
			m(z)=\sigma_1\overline{m(\bar{z})}\sigma_1=\sigma_2m(-z)\sigma_2.
		\end{equation}
		\item   $m(z)=I+\mathcal{O}(z^{-1})$, as $z\to\infty$.
	\end{itemize}
\end{RHP}
 Then we    have the following result.
\begin{proposition}
     Let $m$ satisfy the  RH problem \ref{RH1}.  Then
  $$\psi(z;x,t)=m(z;x,t)e^{-i(zx+4z^{3}t)\sigma_{3}} $$
is a solution of \eqref{lax} with
\begin{align*}
    q=2i\lim_{z\to\infty}zm_{12}(z),
\end{align*}
where the subscript ``$12$'' denotes the $(1,2)$ entry of the matrix function.  Moreover, $q$ also gives a purely imaginary solution for the mKdV  equation \eqref{mkdv}.
\end{proposition}
\begin{proof}
   By analyzing the asymptotic expansion of $m(z;x,t)$ as $z\to\infty$ and using the symmetry \eqref{symmetry}, we have
\begin{equation}\label{asmy}
\begin{aligned}
	m(z;x,t)=&\, I+\frac{m_1(x,t)}{z}+\frac{m_2(x,t)}{z^2}+\frac{m_3(x,t)}{z^3}+\mathcal{O}(z^{-4})\\
=&\, I+\frac{\sigma_1\overline{m_1(x,t)}\sigma_1}{z}+\frac{\sigma_1\overline{m_2(x,t)}\sigma_1}{z^2}+\frac{\sigma_1\overline{m_3(x,t)}\sigma_1}{z^3}+\mathcal{O}(z^{-4})\\
=&\,I-\frac{\sigma_2m_1(x,t)\sigma_2}{z}+\frac{\sigma_2m_2(x,t)\sigma_2}{z^2}-\frac{\sigma_2m_3(x,t)\sigma_2}{z^3}+\mathcal{O}(z^{-4}).
\end{aligned}
\end{equation}
From \eqref{asmy} it follows that
$$m_1=\begin{pmatrix}
	i\alpha_1&\beta_1\\
	\beta_1&	-i\alpha_1
\end{pmatrix},\qquad
m_{2}=\begin{pmatrix}\alpha_2&i\beta_2\\ -i\beta_2&\alpha_2\end{pmatrix},\qquad
m_{3}=\begin{pmatrix}	i\alpha_3&\beta_3\\
	\beta_3&	-i\alpha_3\end{pmatrix},
$$
with $\alpha_j, \beta_j\in\mathbb{R}$ for $j=1,2,3.$
It then follows that
$$\psi_{x}\psi^{-1}=m_{x}m^{-1}-izm\sigma_{3}m^{-1}.$$
Assuming that $\psi$  satisfies
$$\psi_{x}\psi^{-1}=-iz\sigma_{3}+\begin{pmatrix}
	0&2i\beta_1\\
	-2i\beta_1&0
\end{pmatrix},$$
we find that $m$ must fulfill
\begin{equation}\label{xlax}
	m_{x}=-iz[\sigma_{3},m]+\begin{pmatrix}
		0&2i\beta_1\\
		-2i\beta_1&0
	\end{pmatrix}m.
\end{equation}
Substituting the first asymptotic expansion given in \eqref{asmy} into equation \eqref{xlax} leads to
$$\begin{pmatrix}
	i\alpha_{1,x}&\beta_{1,x}\\\beta_{1,x}&-i\alpha_{1,x}
\end{pmatrix}=\begin{pmatrix}0&2\beta_{2}\\2\beta_{2}&0\end{pmatrix}+\begin{pmatrix}2i\beta_1^2&2\alpha_{1}\beta_1\\2\alpha_{1}\beta_1&-2i\beta_1^2\end{pmatrix},$$
$$\begin{pmatrix}	\alpha_{2,x}&i\beta_{2,x}\\-i\beta_{2,x}&\alpha_{2,x}
\end{pmatrix}=\begin{pmatrix}0&-2i\beta_{3}\\2i\beta_{3}&0\end{pmatrix}+\begin{pmatrix}2\beta_1\beta_{2}&2i\alpha_{2}\beta_{1}\\-2i\alpha_{2}\beta_{1}&2\beta_1\beta_{2}\end{pmatrix}.$$
Then,  as $z\to{\infty}$, we have
\begin{align*}
	m\sigma_{3}m^{-1}=&\,\sigma_{3}+\frac{1}{z}\begin{pmatrix}
		0&-2\beta_1\\2\beta_1&0
	\end{pmatrix}+\frac{1}{z^2}\begin{pmatrix}
		\alpha_1^2+\beta_1^2+2\alpha_2&-2i\alpha_{1}\beta_1-2i\beta_2\\-2i\alpha_{1}\beta_1-2i\beta_2&-\alpha_1^2-\beta_1^2-2\alpha_2
	\end{pmatrix}\\
	&+\frac{2}{z^3}\begin{pmatrix}
		0&-\alpha_{2}\beta_1+\alpha_{1}\beta_2-\beta_3\\\alpha_{2}\beta_1-\alpha_{1}\beta_2+\beta_3&0
	\end{pmatrix}+\mathcal{O}(z^{-4}),\\
	=&\,\sigma_{3}+\frac{1}{z}\begin{pmatrix}
		0&-2\beta_1\\2\beta_1&0
	\end{pmatrix}+\frac{1}{z^2}\begin{pmatrix}2\beta_{1}^{2}&-i\beta_{1,x}\\-i\beta_{1,x}&-2\beta_{1}^{2}\end{pmatrix}\\
	&+\frac{1}{z^3}\begin{pmatrix}0&-4\beta_{1}^{3}+\frac{1}{2}\beta_{1,xx}\\4\beta_{1}^{3}-\frac{1}{2}\beta_{1,xx}&0\end{pmatrix}+\mathcal{O}(z^{-4}).
\end{align*}
Thus,
\begin{align*}
	\psi_{t}\psi^{-1}=&\,m_{t}m^{-1}-4iz^{3}m\sigma_{3}m^{-1}\\
	=&-4iz^{3}\sigma_{3}-4iz^2\begin{pmatrix}0&-2\beta_{1}\\2\beta_{1}&0\end{pmatrix}-4iz\begin{pmatrix}2\beta_{1}^{2}&-i\beta_{1,x}\\-i\beta_{1,x}&-2\beta_{1}^{2}\end{pmatrix}\\
	&-4i\begin{pmatrix}0&-4\beta_{1}^{3}+\frac{1}{2}\beta_{1,xx}\\4\beta_{1}^{3}-\frac{1}{2}\beta_{1,xx}&0\end{pmatrix}.
\end{align*}
If we set $q=2i\beta_{1}$, one can easily obtain that  $q$ is purely imaginary. We also have that  $\psi$ solves the equations \eqref{lax}. In addition,
$$\psi_{tx}=\psi_{xt}\iff q_{t}+6q^{2}q_{x}+q_{xxx}=0.$$
Thus $q$ solves the mKdV equation \eqref{mkdv}.
%Consequently, the mKdV equation \eqref{mkdv} is associated with the Lax pair:
\end{proof}
%Based on the RH problem, we derive the Lax pair of the mKdV equation \eqref{mkdv}.

\subsection{The algebro-geometric solution}
This subsection is devoted to constructing the algebro-geometric solution of the mKdV equation \eqref{mkdv} via the standard procedure \cite{AGbook,AGbook2}.   Because the RH problem \ref{RH1} for the mKdV equation \eqref{mkdv} possesses the symmetry \eqref{symmetry}, we must treat the cases of even and odd genus separately.

We consider the hyperelliptic Riemann surface \(\mathcal R\) of genus \(2n\) or \(2n-1\) defined by
\begin{equation}\label{w}
	w^{2}(z):=\left\{
	\begin{aligned}
		&\prod_{j=1}^{2n+1}(z^{2}-E_{j}^{2}),\quad \text{genus}\   2n,\\
		&\prod_{j=2}^{2n+1}(z^{2}-E_{j}^{2}),\quad \text{genus}\ 2n-1,
	\end{aligned}
	\right.
\end{equation}
with the real branch points arranged in the order
\begin{align*}
	-E_{2n+1}<-E_{2n}<\cdots<-E_{1}<E_{1}<\cdots<E_{2n}<E_{2n+1}, \quad \text{genus} \ 2n,
\end{align*}
and
\begin{align*}
	-E_{2n+1}<-E_{2n}<\cdots<-E_{2}<E_{2}<\cdots<E_{2n}<E_{2n+1}, \quad \text{genus} \ 2n-1,
\end{align*}
as illustrated in Figure~\ref{fig:mkdv-cuts}. For convenience, we denote the set of points as
\begin{equation}\label{E}
	\boldsymbol{E}=\left\{
	\begin{aligned}
		&\bigcup_{j=1}^{2n+1}\{-E_j,E_j\},\quad \text{genus}\ 2n,\\
		&\bigcup_{j=2}^{2n+1}\{-E_j,E_j\},\quad \text{genus}\ 2n-1.
	\end{aligned}
	\right.
\end{equation}

\begin{figure}
	\centering
	\definecolor{mkdvAxisBlue}{RGB}{74,144,226}
	\resizebox{0.98\linewidth}{!}{%
		\begin{tikzpicture}[
			x=1cm,y=1cm,>=Latex,
			axis/.style={draw=mkdvAxisBlue,line width=0.72pt},
			band/.style={draw=black,line width=1.6pt},
			every node/.style={font=\footnotesize}
			]
			% Upper row: even genus. Lower row: odd genus.
			\foreach \yrow/\central in {0/1,-1.65/0}{
				\begin{scope}[shift={(0,\yrow)}]
					
					% Real axis.
					\draw[axis,->] (-6.35,0)--(6.45,0)
					node[right] {$\mathbb R$};
					
					% Cuts common to both panels.
					\foreach \a/\b in {
						-5.85/-4.55,-3.25/-1.95,
						1.95/3.25, 4.55/5.85
					}{
						\draw[band] (\a,0)--(\b,0);
						\fill[black] (\a,0) circle (2.2pt);
						\fill[black] (\b,0) circle (2.2pt);
					}
					
					% Branch-point labels.
					\foreach \x/\lab in {
						-5.85/{-E_{2n+1}},-4.55/{-E_{2n}},
						-3.25/{-E_3},     -1.95/{-E_2},
						1.95/{E_2},       3.25/{E_3},
						4.55/{E_{2n}},    5.85/{E_{2n+1}}
					}{
						\node[below=4pt] at (\x,0) {$\lab$};
					}
					
					\node[below=4pt] at (-3.90,0) {$\cdots$};
					\node[below=4pt] at ( 3.90,0) {$\cdots$};
					
					% Central cut and panel captions.
					\ifnum\central=1\relax
					\draw[band] (-0.65,0)--(0.65,0);
					\fill[black] (-0.65,0) circle (2.2pt);
					\fill[black] ( 0.65,0) circle (2.2pt);
					\node[below=4pt] at (-0.65,0) {$-E_1$};
					\node[below=4pt] at ( 0.65,0) {$E_1$};
					\node at (0,-0.82)
					{(a) Cuts on the real axis for genus $2n$.};
					\else
					\node at (0,-0.82)
					{(b) Cuts on the real axis for genus $2n-1$.};
					\fi
					
					% Origin.
					\fill[mkdvAxisBlue] (0,0) circle (1.6pt);
					\node[mkdvAxisBlue,above=3pt] at (0,0) {$0$};
					
				\end{scope}
			}
		\end{tikzpicture}%
	}
	\caption{Symmetry of the spectral cuts about the origin.}
	\label{fig:mkdv-cuts}
\end{figure}

Define two hyperelliptic integrals $f$ and $g$ by
\begin{equation}\label{fz}
	f(z):=\left\{
	\begin{aligned}
		&\int_{-E_{2n+1}}^{z}\frac{s\prod_{k=1}^{n}(s^{2}-(z_{k}^{f})^{2})}{w(s)}ds,\quad \text{genus}\ 2n,\\
		&\int_{-E_{2n+1}}^{z}\frac{\prod_{k=1}^{n}(s^{2}-(z_{k}^{f})^{2})}{w(s)}ds,\quad \text{genus}\ 2n-1,
	\end{aligned}
	\right.
\end{equation}
and
\begin{equation}\label{gz}
	g(z):=\left\{
	\begin{aligned}
		&\int_{-E_{2n+1}}^{z}\frac{12s\prod_{k=1}^{n+1}(s^{2}-(z_{k}^{g})^{2})}{w(s)}ds,\quad \text{genus}\ 2n,\\
		&\int_{-E_{2n+1}}^{z}\frac{12\prod_{k=1}^{n+1}(s^{2}-(z_{k}^{g})^{2})}{w(s)}ds,\quad \text{genus}\ 2n-1,
	\end{aligned}
	\right.
\end{equation}
where $w$ is given in \eqref{w}. The constants $z^f_k$ and  $z^g_k$ are uniquely determined by the  conditions
\begin{equation}\label{fzz}
	\left\{
	\begin{aligned}
		&f(z)=z+\mathcal{O}(1), \quad \hbox{$z\to\infty$,} \\
		&\int^{-E_{2k+1}}_{-E_{2k}}\dif f=\int^{E_{2k+1}}_{E_{2k}}\dif f=0, \quad \hbox{$k=1,2,\ldots,n$},
	\end{aligned}
	\right.
	\quad
\end{equation}
and
\begin{equation}\label{gzz}
	\left\{
	\begin{aligned}
		&g(z)=4z^3+\mathcal{O}(1), \quad \hbox{$z\to\infty$,} \\
		&\int^{-E_{2k+1}}_{-E_{2k}}\dif g=\int^{E_{2k+1}}_{E_{2k}}\dif g=0, \quad \hbox{$k=1,2,\ldots,n$},
	\end{aligned}
	\right.
	\quad
\end{equation}
respectively.

We denote
\begin{equation}\label{f0g0}
f_{0}:=\lim_{z\to\infty}(f(z)-z),\qquad g_{0}:=\lim_{z\to\infty}(g(z)-4z^{3}),
\end{equation}
and then we have
\begin{align*}
	f_{0}&=\int_{-E_{2n+1}}^{-\infty}(f'(s)-1)\,ds+(f(s)-s)\Big|_{-E_{2n+1}}\\
	&=\int_{-E_{2n+1}}^{-\infty}(f'(s)-1)\,ds+E_{2n+1}\\
	&=-\int_{E_{2n+1}}^{+\infty}(f'(s)-1)\,ds+E_{2n+1},
\end{align*}
where the last equality uses the fact that $f'(z)$ is an even function.  By the same analysis and using the result above, we have
\begin{align*}
f_{0}&=\int_{-E_{2n+1}}^{+\infty}(f'(s)-1)\,ds+E_{2n+1}\\
	&=\int_{-E_{2n+1}}^{E_{2n+1}}(f'(s)-1)\,ds+\int_{E_{2n+1}}^{+\infty}(f'(s)-1)\,ds+E_{2n+1}\\
	&=\int_{-E_{2n+1}}^{E_{2n+1}}(f'(s)-1)\,ds-f_0+2E_{2n+1},
\end{align*}
and therefore
\begin{equation}\label{afk}
2(f_{0}-E_{2n+1})=\int_{-E_{2n+1}}^{E_{2n+1}}(f'(s)-1)\,ds.
\end{equation}
More precisely, in the case of genus $2n$, we compute
 \begin{align*}
 	\int_{-E_{2n+1}}^{E_{2n+1}} (f'(s)-1)\,ds &= \sum_{k=1}^{n} \left( \int_{-E_{2k}}^{-E_{2k-1}} f'(s)\,ds + \int_{E_{2k-1}}^{E_{2k}} f'(s)\,ds \right) - 2E_{2n+1}\\
 	&= \sum_{k=1}^{n} 2 \int_{-E_{2k}}^{-E_{2k-1}} f'(s)\,ds - 2E_{2n+1}\\
 	& = 2f(-E_1) - 2E_{2n+1}.
 \end{align*}
For the case of genus  $2n-1$, a similar calculation gives
 \begin{align*}
 	\int_{-E_{2n+1}}^{E_{2n+1}} (f'(s)-1)\,ds= 2f(0) - 2E_{2n+1}.
 \end{align*}
Combining these two identities with \eqref{afk} yields
\begin{equation}
	f_{0}=\left\{
	\begin{aligned}
		&f(-E_{1})=f(E_{1}),\quad \text{genus}\ 2n,\\
		&f(0),\quad \text{genus}\ 2n-1.
	\end{aligned}
	\right.
\end{equation}
Similarly, from the definition of $g_0$ in \eqref{f0g0}, we can obtain
\begin{equation}
	g_{0}=\left\{
	\begin{aligned}
		&g(-E_{1}),\quad \text{genus}\ 2n,\\
		&g(0),\quad \text{genus}\ 2n-1.
	\end{aligned}
	\right.
\end{equation}
Clearly,  $f_0, g_0\in\mathbb{R}$. In addition,  for $k=1,\cdots,n$,  we set the constants
\begin{equation}\label{ckf}
	C_k^f:=\left\{
	\begin{aligned}
		& 2 \int_{E_{1}}^{E_{2k}} f'(s) \, ds,\quad \text{genus}\ 2n,\\
		&2 \int_0^{E_{2k}} f'(s) \, ds,\quad \text{genus}\ 2n-1,
	\end{aligned}
	\right.
	\qquad
		C_k^g:=\left\{
	\begin{aligned}
		& 2 \int_{E_{1}}^{E_{2k}} g'(s) \, ds,\quad \text{genus}\ 2n,\\
		&2 \int_0^{E_{2k}} g'(s) \, ds,\quad \text{genus}\ 2n-1.
	\end{aligned}
	\right.
\end{equation}

Following the basic  RH problem \ref{RH1}, we now consider the RH problem associated with the algebro-geometric solution of the mKdV equation \eqref{mkdv}.
\begin{RHP}
	\hfill
	\begin{itemize}
		\item $m^{(AG)}(z)=m^{(AG)}(z;x,t)$ is analytic in $\mathbb{C}\setminus\overline{\Gamma}$, where $\Gamma$ is given by
		\begin{equation}\label{gamma}
			\Gamma=\left\{
			\begin{aligned}
				&\bigcup_{k=1}^{n} \big( (-E_{2k+1}, -E_{2k}) \cup (E_{2k}, E_{2k+1}) \big)
				\cup (-E_1, E_1),
				&&\ \text{genus } 2n,\\
				&\bigcup_{k=1}^{n} \big( (-E_{2k+1}, -E_{2k}) \cup (E_{2k}, E_{2k+1}) \big),
				&&\  \text{genus } 2n-1,
			\end{aligned}
			\right.
		\end{equation}
		see Figure~\ref{fig:mkdv-cuts}.
		\item $m^{(AG)}(z)$ satisfies the jump condition
		\begin{equation}\label{jumpm}
		m^{(AG)}_{+}(z)=m^{(AG)}_{-}(z)e^{-i(zx+4z^{3}t)\hat{\sigma}_{3}}J(z), \quad z\in\Gamma,
		\end{equation}
		where $J(z)$ is given by
		\begin{equation*}
			J(z)=\left\{
			\begin{aligned}
				&\begin{pmatrix}
					0&-ie^{i\phi_{k}}\\-ie^{-i\phi_{k}}&0
				\end{pmatrix},\quad z\in(-E_{2k+1},-E_{2k}),\\
				&\begin{pmatrix}
					0&-ie^{-i\phi_{k}}\\-ie^{i\phi_{k}}&0
				\end{pmatrix},\quad z\in(E_{2k},E_{2k+1}).
			\end{aligned}
			\right.
		\end{equation*}
		Here $\boldsymbol{\phi}=(\phi_1,\ldots,\phi_n)\in \mathbb{R}^n$ is an arbitrary real vector. Specifically, for genus $2n$, $$J(z)=\begin{pmatrix}
			0&-i\\-i&0
		\end{pmatrix},\quad z\in(-E_{1},E_{1}).$$
		\item $m^{(AG)}(z)$ satisfies the symmetry relations
		\begin{equation}\label{sym}
			m^{(AG)}(z)=\sigma_1\overline{m^{(AG)}(\bar{z})}\sigma_1=\sigma_2m^{(AG)}(-z)\sigma_2.
		\end{equation}
		\item As $z\to\infty$, $m^{(AG)}(z)=I+\mathcal{O}(z^{-1})$.
			\item  As  $ z\to p\in\boldsymbol{E}$,
$m^{(AG)}(z)=\mathcal{O}\big((z-p)^{-\frac{1}{4}}\big).$
	\end{itemize}
\end{RHP}
The jump on $(-E_{1},E_{1})$ for genus $2n$ inspires us to define $\phi_0=0$ for convenience.
To obtain the expression for  $m^{(AG)}$,  we define
\begin{equation}
\widetilde{\mu}^{(AG)}(z;x,t)=m^{(AG)}(z;x,t)e^{i((f-f_{0})x+(g-g_{0})t-(zx+4z^{3}t))\sigma_{3}},
\end{equation}
with $f$ and $g$ defined in  \eqref{fz}--\eqref{gz}, and  $f_{0},g_{0}$ as in \eqref{f0g0}. From \eqref{jumpm}, we have
$$ \widetilde{\mu}^{(AG)}_{+}(z)=\widetilde{\mu}^{(AG)}_{-}(z)\widetilde{J}(z), \quad z\in\Gamma,$$
where $\Gamma$ is given by \eqref{gamma} and  $$\widetilde{J}(z)=e^{-i((f_{-}-f_{0})x+(g_{-}-g_{0})t)\sigma_{3}}J(z)e^{i((f_{+}-f_{0})x+(g_{+}-g_{0})t)\sigma_{3}}.$$
%For \(\tilde J\) to be independent of \(z\), we require that for $z\in\Gamma$,
It follows from \eqref{fz}--\eqref{gz} and  \eqref{f0g0} that
\begin{equation}\label{ckkf}
	f_{+}(z)+f_{-}(z)-2f_{0}=\left\{
	\begin{aligned}
		C_{k}^{f},\quad &z\in(E_{2k},E_{2k+1}),\\
		-C_{k}^{f},\quad &z\in(-E_{2k+1},-E_{2k}),
	\end{aligned}
	\right.
\end{equation}
and
\begin{equation}\label{ckkg}
	g_{+}(z)+g_{-}(z)-2g_{0}=\left\{
	\begin{aligned}
		C_{k}^{g},\quad &z\in(E_{2k},E_{2k+1}),\\
		-C_{k}^{g},\quad &z\in(-E_{2k+1},-E_{2k}),
	\end{aligned}
	\right.
\end{equation}
where $C_{k}^{f}$ and $C_{k}^{g}$ are defined by \eqref{ckf}.  Then $\widetilde{\mu}^{(AG)}$ satisfies the following RH problem.
\begin{RHP}\label{RHmu}
	\hfill
	\begin{itemize}
		\item $\widetilde{\mu}^{(AG)}(z)=\widetilde{\mu}^{(AG)}(z;x,t)$ is analytic in $\mathbb{C}\setminus\overline{\Gamma}$, where $\Gamma$ is given by \eqref{gamma}.
		\item $\widetilde{\mu}^{(AG)}(z)$ satisfies the jump condition
		$$ \widetilde{\mu}^{(AG)}_{+}(z)=\widetilde{\mu}^{(AG)}_{-}(z)\widetilde{J}(z), \qquad z\in\Gamma,$$
		where
		\begin{equation*}
			\widetilde{J}(z)=\left\{
			\begin{aligned}
				&\begin{pmatrix}
					0 & -ie^{i(\phi_k + C_k^f x + C_k^g t)} \\
					-ie^{-i(\phi_k + C_k^f x + C_k^g t)} & 0
				\end{pmatrix},\qquad z\in(-E_{2k+1},-E_{2k}),\\
				&\begin{pmatrix}
					0 & -ie^{-i(\phi_k + C_k^f x + C_k^g t)} \\
					-ie^{i(\phi_k + C_k^f x + C_k^g t)} & 0
				\end{pmatrix},\qquad z\in(E_{2k},E_{2k+1}),
			\end{aligned}
			\right. 
		\end{equation*}
		where $k=1,\cdots,n$.
		 Specifically, for genus $2n$, $$\widetilde{J}(z)=\begin{pmatrix}
			0&-i\\-i&0
		\end{pmatrix},\qquad z\in(-E_{1},E_{1}).$$
		\item $\widetilde{\mu}^{(AG)}(z)$ satisfies the symmetry relations
		\begin{equation}\label{sym}
			\widetilde{\mu}^{(AG)}(z)=\sigma_1\overline{\widetilde{\mu}^{(AG)}(\bar{z})}\sigma_1=\sigma_2\widetilde{\mu}^{(AG)}(-z)\sigma_2.
		\end{equation}
		\item   As $z\to\infty$, $\widetilde{\mu}^{(AG)}(z)=I+\mathcal{O}(z^{-1})$.
		\item  As  $ z\to p\in\boldsymbol{E}$,
	$\widetilde{\mu}^{(AG)}(z)=\mathcal{O}\big((z-p)^{-\frac{1}{4}}\big).$
	\end{itemize}
\end{RHP}

\begin{figure}
	\tikzset{every picture/.style={line width=0.75pt}} %set default line width to 0.75pt
	\begin{tikzpicture}[x=0.75pt,y=0.75pt,yscale=-0.7,xscale=0.7]
		%uncomment if require: \path (0,420); %set diagram left start at 0, and has height of 420	
		%Straight Lines [id:da09115869923316389]
		%Curve Lines [id:da2671035008602405]
		%bn-1
		\draw [color={rgb, 255:red, 74; green, 144; blue, 226 }  ,draw opacity=1 ]   (80,200) .. controls (90,160) and (190,160) .. (200,200) ;
		\draw [color={rgb, 255:red, 74; green, 144; blue, 226 }  ,draw opacity=1 ] [dash pattern={on 4.5pt off 4.5pt}]  (200,200) .. controls (190,240) and (90,240) .. (80,200) ;
		\draw [shift={(145,170)}, rotate = 182.35] [fill={rgb, 255:red, 74; green, 144; blue, 226 }  ,fill opacity=1 ][line width=0.08]  [draw opacity=0] (10.72,-5.15) -- (0,0) -- (10.72,5.15) -- (7.12,0) -- cycle  ;
		\draw [shift={(135,230)}, rotate = 2.35] [fill={rgb, 255:red, 74; green, 144; blue, 226 }  ,fill opacity=1 ][line width=0.08]  [draw opacity=0] (10.72,-5.15) -- (0,0) -- (10.72,5.15) -- (7.12,0) -- cycle    ;
		\draw (150,150) node [anchor=north west][inner sep=0.75pt]  [color={rgb, 255:red, 74; green, 144; blue, 226}  ,opacity=1 ,xscale=0.8,yscale=0.8]  {$b_{n-1}$};		
		%bn+1
		\draw [color={rgb, 255:red, 74; green, 144; blue, 226 }  ,draw opacity=1 ]   (50,200) .. controls (60,90) and (510,90) .. (520,200) ;
		\draw [color={rgb, 255:red, 74; green, 144; blue, 226 }  ,draw opacity=1 ] [dash pattern={on 4.5pt off 4.5pt}]  (520,200) .. controls (510,310) and (60,310) .. (50,200) ;
		\draw [shift={(300,118)}, rotate = 182.35] [fill={rgb, 255:red, 74; green, 144; blue, 226 }  ,fill opacity=1 ][line width=0.08]  [draw opacity=0] (10.72,-5.15) -- (0,0) -- (10.72,5.15) -- (7.12,0) -- cycle  ;
		\draw [shift={(300,282)}, rotate = 2.35] [fill={rgb, 255:red, 74; green, 144; blue, 226 }  ,fill opacity=1 ][line width=0.08]  [draw opacity=0] (10.72,-5.15) -- (0,0) -- (10.72,5.15) -- (7.12,0) -- cycle  ;
		\draw (385,103) node [anchor=north west][inner sep=0.75pt]  [color={rgb, 255:red, 74; green, 144; blue, 226}  ,opacity=1 ,xscale=0.8,yscale=0.8]  {$b_{n}$};		
		%b2n
		\draw [color={rgb, 255:red, 74; green, 144; blue, 226 }  ,draw opacity=1 ]   (35,200) .. controls (60,50) and (655,50) .. (680,200) ;
		\draw [color={rgb, 255:red, 74; green, 144; blue, 226 }  ,draw opacity=1 ] [dash pattern={on 4.5pt off 4.5pt}]  (680,200) .. controls (655,350) and (60,350) .. (35,200) ;
		\draw [shift={(380,87)}, rotate = 182.35] [fill={rgb, 255:red, 74; green, 144; blue, 226 }  ,fill opacity=1 ][line width=0.08]  [draw opacity=0] (10.72,-5.15) -- (0,0) -- (10.72,5.15) -- (7.12,0) -- cycle  ;
		\draw [shift={(380,311)}, rotate = 2.35] [fill={rgb, 255:red, 74; green, 144; blue, 226 }  ,fill opacity=1 ][line width=0.08]  [draw opacity=0] (10.72,-5.15) -- (0,0) -- (10.72,5.15) -- (7.12,0) -- cycle  ;
		\draw (500,75) node [anchor=north west][inner sep=0.75pt]  [color={rgb, 255:red, 74; green, 144; blue, 226}  ,opacity=1 ,xscale=0.8,yscale=0.8]  {$b_{2n-1}$};		
		%an-1
		\draw [color={rgb, 255:red, 65; green, 117; blue, 5 }  ,draw opacity=1 ]   (180,200) .. controls (190,175) and (250,175) .. (260,200) ;
		\draw [color={rgb, 255:red, 65; green, 117; blue, 5 }  ,draw opacity=1 ]   (260,200) .. controls (250,225) and (190,225) .. (180,200) ;
		\draw [shift={(227,181)}, rotate = 181.08] [fill={rgb, 255:red, 65; green, 117; blue, 5 }  ,fill opacity=1 ][line width=0.08]  [draw opacity=0] (10.72,-5.15) -- (0,0) -- (10.72,5.15) -- (7.12,0) -- cycle    ;
		\draw [shift={(215,219)}, rotate = 2.35] [fill={rgb, 255:red, 65; green, 117; blue, 5 }  ,fill opacity=1 ][line width=0.08]  [draw opacity=0] (10.72,-5.15) -- (0,0) -- (10.72,5.15) -- (7.12,0) -- cycle   ;
		\draw (230,165) node [anchor=north west][inner sep=0.75pt]  [color={rgb, 255:red, 65; green, 117; blue, 5 }  ,opacity=1 ,xscale=0.8,yscale=0.8]  {$a_{n-1}$};
		%an+1
		\draw [color={rgb, 255:red, 65; green, 117; blue, 5 }  ,draw opacity=1 ]   (500,200) .. controls (510,175) and (570,175) .. (580,200) ;
		\draw [color={rgb, 255:red, 65; green, 117; blue, 5 }  ,draw opacity=1 ]   (580,200) .. controls (570,225) and (510,225) .. (500,200) ;
		\draw [shift={(547,181)}, rotate = 181.08] [fill={rgb, 255:red, 65; green, 117; blue, 5 }  ,fill opacity=1 ][line width=0.08]  [draw opacity=0] (10.72,-5.15) -- (0,0) -- (10.72,5.15) -- (7.12,0) -- cycle    ;
		\draw [shift={(535,219)}, rotate = 2.35] [fill={rgb, 255:red, 65; green, 117; blue, 5 }  ,fill opacity=1 ][line width=0.08]  [draw opacity=0] (10.72,-5.15) -- (0,0) -- (10.72,5.15) -- (7.12,0) -- cycle   ;
		\draw (550,165) node [anchor=north west][inner sep=0.75pt]  [color={rgb, 255:red, 65; green, 117; blue, 5 }  ,opacity=1 ,xscale=0.8,yscale=0.8]  {$a_{n}$};	
		%an+1
		\draw [color={rgb, 255:red, 65; green, 117; blue, 5 }  ,draw opacity=1 ]   (660,200) .. controls (670,175) and (730,175) .. (740,200) ;
		\draw [color={rgb, 255:red, 65; green, 117; blue, 5 }  ,draw opacity=1 ]   (740,200) .. controls (730,225) and (670,225) .. (660,200) ;
		\draw [shift={(707,181)}, rotate = 181.08] [fill={rgb, 255:red, 65; green, 117; blue, 5 }  ,fill opacity=1 ][line width=0.08]  [draw opacity=0] (10.72,-5.15) -- (0,0) -- (10.72,5.15) -- (7.12,0) -- cycle    ;
		\draw [shift={(695,219)}, rotate = 2.35] [fill={rgb, 255:red, 65; green, 117; blue, 5 }  ,fill opacity=1 ][line width=0.08]  [draw opacity=0] (10.72,-5.15) -- (0,0) -- (10.72,5.15) -- (7.12,0) -- cycle   ;
		\draw (710,165) node [anchor=north west][inner sep=0.75pt]  [color={rgb, 255:red, 65; green, 117; blue, 5 }  ,opacity=1 ,xscale=0.8,yscale=0.8]  {$a_{2n-1}$};
		
		%blackline
		\draw [line width=1.5]    (20,200) -- (100,200) ;
		%blackdot
		\draw [shift={(20,200)}, rotate = 0.7] [color={rgb, 255:red, 0; green, 0; blue, 0 }  ][fill={rgb, 255:red, 0; green, 0; blue, 0 }  ][line width=1.5]      (0, 0) circle [x radius= 4.36, y radius= 4.36] ;
		\draw [shift={(100,200)}, rotate = 0.7] [color={rgb, 255:red, 0; green, 0; blue, 0 }  ][fill={rgb, 255:red, 0; green, 0; blue, 0 }  ][line width=1.5]      (0, 0) circle [x radius= 4.36, y radius= 4.36]   ;
		% Text Node
		\draw (-15,208) node [anchor=north west][inner sep=0.75pt]  [font=\footnotesize,xscale=0.8,yscale=0.8] [align=left] {$\displaystyle -E_{2n+1}$};
		\draw (93,208) node [anchor=north west][inner sep=0.75pt]  [font=\footnotesize,xscale=0.8,yscale=0.8] [align=left] {$\displaystyle -E_{2n}$};
		
		%blackline
		\draw [line width=1.5]    (180,200) -- (260,200) ;
		%blackdot
		\draw [shift={(180,200)}, rotate = 0.7] [color={rgb, 255:red, 0; green, 0; blue, 0 }  ][fill={rgb, 255:red, 0; green, 0; blue, 0 }  ][line width=1.5]      (0, 0) circle [x radius= 4.36, y radius= 4.36] ;
		\draw [shift={(260,200)}, rotate = 0.7] [color={rgb, 255:red, 0; green, 0; blue, 0 }  ][fill={rgb, 255:red, 0; green, 0; blue, 0 }  ][line width=1.5]      (0, 0) circle [x radius= 4.36, y radius= 4.36]   ;
		% Text Node
		\draw (153,208) node [anchor=north west][inner sep=0.75pt]  [font=\footnotesize,xscale=0.8,yscale=0.8] [align=left] {$\displaystyle -E_{3}$};
		\draw (248,208) node [anchor=north west][inner sep=0.75pt]  [font=\footnotesize,xscale=0.8,yscale=0.8] [align=left] {$\displaystyle -E_{2}$};
		
		%dash line
		\draw  [dash pattern={on 0.84pt off 2.51pt}]  (130,200) -- (160,200);
		
		%blackline
		\draw [line width=1.5]    (500,200) -- (580,200) ;
		%blackdot
		\draw [shift={(500,200)}, rotate = 0.7] [color={rgb, 255:red, 0; green, 0; blue, 0 }  ][fill={rgb, 255:red, 0; green, 0; blue, 0 }  ][line width=1.5]      (0, 0) circle [x radius= 4.36, y radius= 4.36] ;
		\draw [shift={(580,200)}, rotate = 0.7] [color={rgb, 255:red, 0; green, 0; blue, 0 }  ][fill={rgb, 255:red, 0; green, 0; blue, 0 }  ][line width=1.5]      (0, 0) circle [x radius= 4.36, y radius= 4.36]   ;
		% Text Node
		\draw (485,208) node [anchor=north west][inner sep=0.75pt]  [font=\footnotesize,xscale=0.8,yscale=0.8] [align=left] {$\displaystyle E_{2}$};
		\draw (575,208) node [anchor=north west][inner sep=0.75pt]  [font=\footnotesize,xscale=0.8,yscale=0.8] [align=left] {$\displaystyle E_{3}$};
		
		%blackline
		\draw [line width=1.5]    (660,200) -- (740,200) ;
		%blackdot
		\draw [shift={(660,200)}, rotate = 0.7] [color={rgb, 255:red, 0; green, 0; blue, 0 }  ][fill={rgb, 255:red, 0; green, 0; blue, 0 }  ][line width=1.5]      (0, 0) circle [x radius= 4.36, y radius= 4.36] ;
		\draw [shift={(740,200)}, rotate = 0.7] [color={rgb, 255:red, 0; green, 0; blue, 0 }  ][fill={rgb, 255:red, 0; green, 0; blue, 0 }  ][line width=1.5]      (0, 0) circle [x radius= 4.36, y radius= 4.36]   ;
		% Text Node
		\draw (645,208) node [anchor=north west][inner sep=0.75pt]  [font=\footnotesize,xscale=0.8,yscale=0.8] [align=left] {$\displaystyle E_{2n}$};
		\draw (735,208) node [anchor=north west][inner sep=0.75pt]  [font=\footnotesize,xscale=0.8,yscale=0.8] [align=left] {$\displaystyle E_{2n+1}$};
		
		%dash line
		\draw  [dash pattern={on 0.84pt off 2.51pt}]  (610,200) -- (640,200);
	\end{tikzpicture}
	\caption{\footnotesize The canonical homology basis $\{a_j,b_j\}_{j=1}^{2n-1}$ on the Riemann surface $\mathcal{R}$ of genus~$2n-1$.}\label{fig2n-1+}
\end{figure}
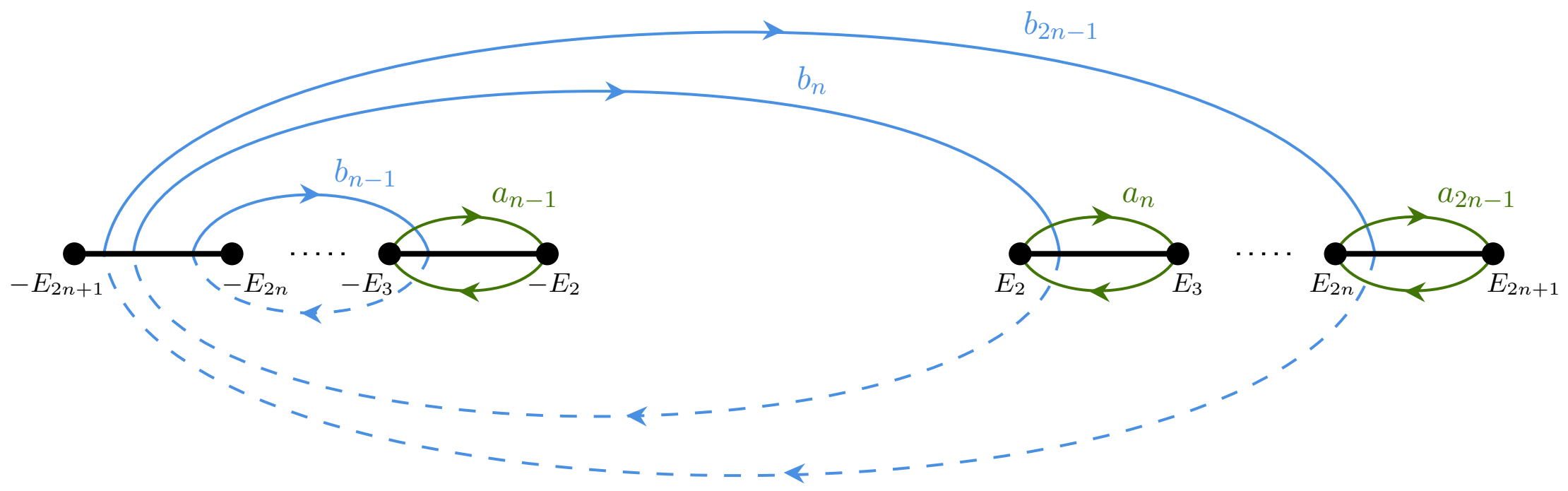

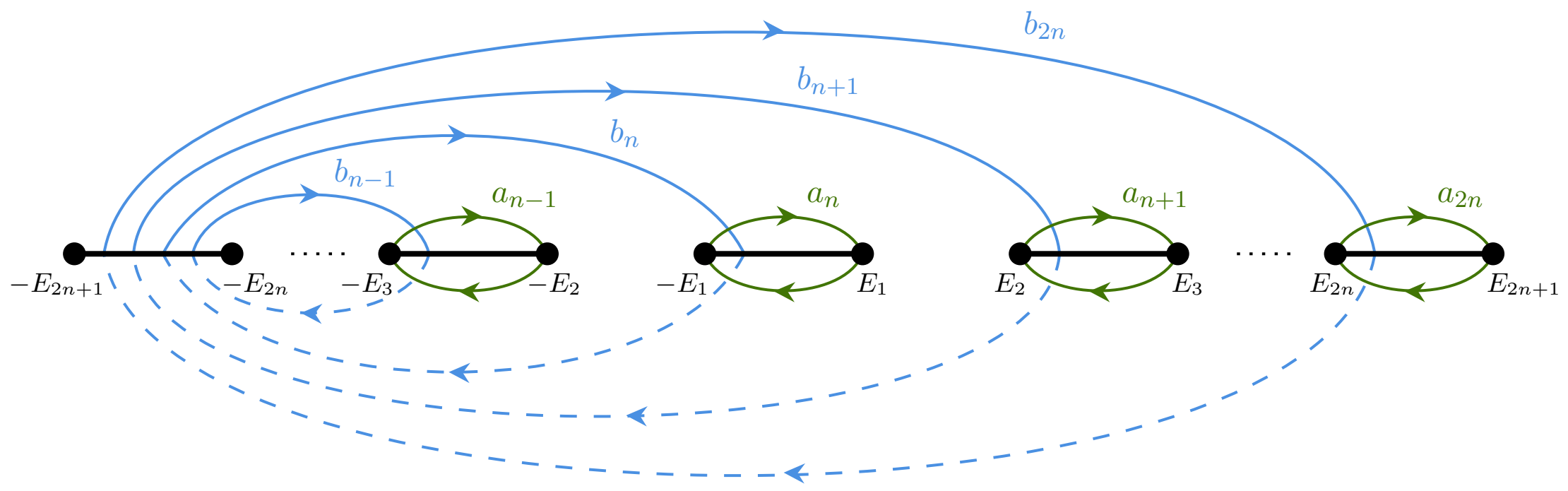
\begin{figure}	
 	\tikzset{every picture/.style={line width=0.75pt}} %set default line width to 0.75pt
 	\begin{tikzpicture}[x=0.75pt,y=0.75pt,yscale=-0.7,xscale=0.7]
 		%uncomment if require: \path (0,420); %set diagram left start at 0, and has height of 420	
 		%Straight Lines [id:da09115869923316389]
 		%Curve Lines [id:da2671035008602405]		
 		%bn-1
 		\draw [color={rgb, 255:red, 74; green, 144; blue, 226 }  ,draw opacity=1 ]   (80,200) .. controls (90,160) and (190,160) .. (200,200) ;
 		\draw [color={rgb, 255:red, 74; green, 144; blue, 226 }  ,draw opacity=1 ] [dash pattern={on 4.5pt off 4.5pt}]  (200,200) .. controls (190,240) and (90,240) .. (80,200) ;
 		\draw [shift={(145,170)}, rotate = 182.35] [fill={rgb, 255:red, 74; green, 144; blue, 226 }  ,fill opacity=1 ][line width=0.08]  [draw opacity=0] (10.72,-5.15) -- (0,0) -- (10.72,5.15) -- (7.12,0) -- cycle  ;
 		\draw [shift={(135,230)}, rotate = 2.35] [fill={rgb, 255:red, 74; green, 144; blue, 226 }  ,fill opacity=1 ][line width=0.08]  [draw opacity=0] (10.72,-5.15) -- (0,0) -- (10.72,5.15) -- (7.12,0) -- cycle    ;
 		\draw (150,150) node [anchor=north west][inner sep=0.75pt]  [color={rgb, 255:red, 74; green, 144; blue, 226}  ,opacity=1 ,xscale=0.8,yscale=0.8]  {$b_{n-1}$};		
 		%bn
 		\draw [color={rgb, 255:red, 74; green, 144; blue, 226 }  ,draw opacity=1 ]   (65,200) .. controls (105,120) and (320,120) .. (360,200) ;
 		\draw [color={rgb, 255:red, 74; green, 144; blue, 226 }  ,draw opacity=1 ] [dash pattern={on 4.5pt off 4.5pt}]  (360,200) .. controls (320,280) and (105,280) .. (65,200) ;
 		\draw [shift={(220,140)}, rotate = 182.35] [fill={rgb, 255:red, 74; green, 144; blue, 226 }  ,fill opacity=1 ][line width=0.08]  [draw opacity=0] (10.72,-5.15) -- (0,0) -- (10.72,5.15) -- (7.12,0) -- cycle  ;
 		\draw [shift={(210,260)}, rotate = 2.35] [fill={rgb, 255:red, 74; green, 144; blue, 226 }  ,fill opacity=1 ][line width=0.08]  [draw opacity=0] (10.72,-5.15) -- (0,0) -- (10.72,5.15) -- (7.12,0) -- cycle  ;
 		\draw (290,130) node [anchor=north west][inner sep=0.75pt]  [color={rgb, 255:red, 74; green, 144; blue, 226}  ,opacity=1 ,xscale=0.8,yscale=0.8]  {$b_{n}$};		
 		%bn+1
 		\draw [color={rgb, 255:red, 74; green, 144; blue, 226 }  ,draw opacity=1 ]   (50,200) .. controls (60,90) and (510,90) .. (520,200) ;
 		\draw [color={rgb, 255:red, 74; green, 144; blue, 226 }  ,draw opacity=1 ] [dash pattern={on 4.5pt off 4.5pt}]  (520,200) .. controls (510,310) and (60,310) .. (50,200) ;
 		\draw [shift={(300,118)}, rotate = 182.35] [fill={rgb, 255:red, 74; green, 144; blue, 226 }  ,fill opacity=1 ][line width=0.08]  [draw opacity=0] (10.72,-5.15) -- (0,0) -- (10.72,5.15) -- (7.12,0) -- cycle  ;
 		\draw [shift={(300,282)}, rotate = 2.35] [fill={rgb, 255:red, 74; green, 144; blue, 226 }  ,fill opacity=1 ][line width=0.08]  [draw opacity=0] (10.72,-5.15) -- (0,0) -- (10.72,5.15) -- (7.12,0) -- cycle  ;
 		\draw (385,103) node [anchor=north west][inner sep=0.75pt]  [color={rgb, 255:red, 74; green, 144; blue, 226}  ,opacity=1 ,xscale=0.8,yscale=0.8]  {$b_{n+1}$};		
 		%b2n
 		\draw [color={rgb, 255:red, 74; green, 144; blue, 226 }  ,draw opacity=1 ]   (35,200) .. controls (60,50) and (655,50) .. (680,200) ;
 		\draw [color={rgb, 255:red, 74; green, 144; blue, 226 }  ,draw opacity=1 ] [dash pattern={on 4.5pt off 4.5pt}]  (680,200) .. controls (655,350) and (60,350) .. (35,200) ;
 		\draw [shift={(380,87)}, rotate = 182.35] [fill={rgb, 255:red, 74; green, 144; blue, 226 }  ,fill opacity=1 ][line width=0.08]  [draw opacity=0] (10.72,-5.15) -- (0,0) -- (10.72,5.15) -- (7.12,0) -- cycle  ;
 		\draw [shift={(380,311)}, rotate = 2.35] [fill={rgb, 255:red, 74; green, 144; blue, 226 }  ,fill opacity=1 ][line width=0.08]  [draw opacity=0] (10.72,-5.15) -- (0,0) -- (10.72,5.15) -- (7.12,0) -- cycle  ;
 		\draw (500,75) node [anchor=north west][inner sep=0.75pt]  [color={rgb, 255:red, 74; green, 144; blue, 226}  ,opacity=1 ,xscale=0.8,yscale=0.8]  {$b_{2n}$};
 		%an-1
 		\draw [color={rgb, 255:red, 65; green, 117; blue, 5 }  ,draw opacity=1 ]   (180,200) .. controls (190,175) and (250,175) .. (260,200) ;
 		\draw [color={rgb, 255:red, 65; green, 117; blue, 5 }  ,draw opacity=1 ]   (260,200) .. controls (250,225) and (190,225) .. (180,200) ;
 		\draw [shift={(227,181)}, rotate = 181.08] [fill={rgb, 255:red, 65; green, 117; blue, 5 }  ,fill opacity=1 ][line width=0.08]  [draw opacity=0] (10.72,-5.15) -- (0,0) -- (10.72,5.15) -- (7.12,0) -- cycle    ;
 		\draw [shift={(215,219)}, rotate = 2.35] [fill={rgb, 255:red, 65; green, 117; blue, 5 }  ,fill opacity=1 ][line width=0.08]  [draw opacity=0] (10.72,-5.15) -- (0,0) -- (10.72,5.15) -- (7.12,0) -- cycle   ;
 		\draw (230,165) node [anchor=north west][inner sep=0.75pt]  [color={rgb, 255:red, 65; green, 117; blue, 5 }  ,opacity=1 ,xscale=0.8,yscale=0.8]  {$a_{n-1}$};	
 		%an
 		\draw [color={rgb, 255:red, 65; green, 117; blue, 5 }  ,draw opacity=1 ]   (340,200) .. controls (350,175) and (410,175) .. (420,200) ;
 		\draw [color={rgb, 255:red, 65; green, 117; blue, 5 }  ,draw opacity=1 ]   (420,200) .. controls (410,225) and (350,225) .. (340,200) ;
 		\draw [shift={(387,181)}, rotate = 181.08] [fill={rgb, 255:red, 65; green, 117; blue, 5 }  ,fill opacity=1 ][line width=0.08]  [draw opacity=0] (10.72,-5.15) -- (0,0) -- (10.72,5.15) -- (7.12,0) -- cycle    ;
 		\draw [shift={(375,219)}, rotate = 2.35] [fill={rgb, 255:red, 65; green, 117; blue, 5 }  ,fill opacity=1 ][line width=0.08]  [draw opacity=0] (10.72,-5.15) -- (0,0) -- (10.72,5.15) -- (7.12,0) -- cycle   ;
 		\draw (390,165) node [anchor=north west][inner sep=0.75pt]  [color={rgb, 255:red, 65; green, 117; blue, 5 }  ,opacity=1 ,xscale=0.8,yscale=0.8]  {$a_{n}$};
 		%an+1
 		\draw [color={rgb, 255:red, 65; green, 117; blue, 5 }  ,draw opacity=1 ]   (500,200) .. controls (510,175) and (570,175) .. (580,200) ;
 		\draw [color={rgb, 255:red, 65; green, 117; blue, 5 }  ,draw opacity=1 ]   (580,200) .. controls (570,225) and (510,225) .. (500,200) ;
 		\draw [shift={(547,181)}, rotate = 181.08] [fill={rgb, 255:red, 65; green, 117; blue, 5 }  ,fill opacity=1 ][line width=0.08]  [draw opacity=0] (10.72,-5.15) -- (0,0) -- (10.72,5.15) -- (7.12,0) -- cycle    ;
 		\draw [shift={(535,219)}, rotate = 2.35] [fill={rgb, 255:red, 65; green, 117; blue, 5 }  ,fill opacity=1 ][line width=0.08]  [draw opacity=0] (10.72,-5.15) -- (0,0) -- (10.72,5.15) -- (7.12,0) -- cycle   ;
 		\draw (550,165) node [anchor=north west][inner sep=0.75pt]  [color={rgb, 255:red, 65; green, 117; blue, 5 }  ,opacity=1 ,xscale=0.8,yscale=0.8]  {$a_{n+1}$};	
 		%an+1
 		\draw [color={rgb, 255:red, 65; green, 117; blue, 5 }  ,draw opacity=1 ]   (660,200) .. controls (670,175) and (730,175) .. (740,200) ;
 		\draw [color={rgb, 255:red, 65; green, 117; blue, 5 }  ,draw opacity=1 ]   (740,200) .. controls (730,225) and (670,225) .. (660,200) ;
 		\draw [shift={(707,181)}, rotate = 181.08] [fill={rgb, 255:red, 65; green, 117; blue, 5 }  ,fill opacity=1 ][line width=0.08]  [draw opacity=0] (10.72,-5.15) -- (0,0) -- (10.72,5.15) -- (7.12,0) -- cycle    ;
 		\draw [shift={(695,219)}, rotate = 2.35] [fill={rgb, 255:red, 65; green, 117; blue, 5 }  ,fill opacity=1 ][line width=0.08]  [draw opacity=0] (10.72,-5.15) -- (0,0) -- (10.72,5.15) -- (7.12,0) -- cycle   ;
 		\draw (710,165) node [anchor=north west][inner sep=0.75pt]  [color={rgb, 255:red, 65; green, 117; blue, 5 }  ,opacity=1 ,xscale=0.8,yscale=0.8]  {$a_{2n}$};	
 		%blackline
 		\draw [line width=1.5]    (20,200) -- (100,200) ;
 		%blackdot
 		\draw [shift={(20,200)}, rotate = 0.7] [color={rgb, 255:red, 0; green, 0; blue, 0 }  ][fill={rgb, 255:red, 0; green, 0; blue, 0 }  ][line width=1.5]      (0, 0) circle [x radius= 4.36, y radius= 4.36] ;
 		\draw [shift={(100,200)}, rotate = 0.7] [color={rgb, 255:red, 0; green, 0; blue, 0 }  ][fill={rgb, 255:red, 0; green, 0; blue, 0 }  ][line width=1.5]      (0, 0) circle [x radius= 4.36, y radius= 4.36]   ;
 		% Text Node
 		\draw (-15,208) node [anchor=north west][inner sep=0.75pt]  [font=\footnotesize,xscale=0.8,yscale=0.8] [align=left] {$\displaystyle -E_{2n+1}$};
 		\draw (93,208) node [anchor=north west][inner sep=0.75pt]  [font=\footnotesize,xscale=0.8,yscale=0.8] [align=left] {$\displaystyle -E_{2n}$};		
 		%blackline
 		\draw [line width=1.5]    (180,200) -- (260,200) ;
 		%blackdot
 		\draw [shift={(180,200)}, rotate = 0.7] [color={rgb, 255:red, 0; green, 0; blue, 0 }  ][fill={rgb, 255:red, 0; green, 0; blue, 0 }  ][line width=1.5]      (0, 0) circle [x radius= 4.36, y radius= 4.36] ;
 		\draw [shift={(260,200)}, rotate = 0.7] [color={rgb, 255:red, 0; green, 0; blue, 0 }  ][fill={rgb, 255:red, 0; green, 0; blue, 0 }  ][line width=1.5]      (0, 0) circle [x radius= 4.36, y radius= 4.36]   ;
 		% Text Node
 		\draw (153,208) node [anchor=north west][inner sep=0.75pt]  [font=\footnotesize,xscale=0.8,yscale=0.8] [align=left] {$\displaystyle -E_{3}$};
 		\draw (248,208) node [anchor=north west][inner sep=0.75pt]  [font=\footnotesize,xscale=0.8,yscale=0.8] [align=left] {$\displaystyle -E_{2}$};
 		
 		%dash line
 		\draw  [dash pattern={on 0.84pt off 2.51pt}]  (130,200) -- (160,200);
 		
 		%blackline
 		\draw [line width=1.5]    (340,200) -- (420,200) ;
 		%blackdot
 		\draw [shift={(340,200)}, rotate = 0.7] [color={rgb, 255:red, 0; green, 0; blue, 0 }  ][fill={rgb, 255:red, 0; green, 0; blue, 0 }  ][line width=1.5]      (0, 0) circle [x radius= 4.36, y radius= 4.36] ;
 		\draw [shift={(420,200)}, rotate = 0.7] [color={rgb, 255:red, 0; green, 0; blue, 0 }  ][fill={rgb, 255:red, 0; green, 0; blue, 0 }  ][line width=1.5]      (0, 0) circle [x radius= 4.36, y radius= 4.36]   ;
 		% Text Node
 		\draw (313,208) node [anchor=north west][inner sep=0.75pt]  [font=\footnotesize,xscale=0.8,yscale=0.8] [align=left] {$\displaystyle -E_{1}$};
 		\draw (415,208) node [anchor=north west][inner sep=0.75pt]  [font=\footnotesize,xscale=0.8,yscale=0.8] [align=left] {$\displaystyle E_{1}$};
 		
 		%blackline
 		\draw [line width=1.5]    (500,200) -- (580,200) ;
 		%blackdot
 		\draw [shift={(500,200)}, rotate = 0.7] [color={rgb, 255:red, 0; green, 0; blue, 0 }  ][fill={rgb, 255:red, 0; green, 0; blue, 0 }  ][line width=1.5]      (0, 0) circle [x radius= 4.36, y radius= 4.36] ;
 		\draw [shift={(580,200)}, rotate = 0.7] [color={rgb, 255:red, 0; green, 0; blue, 0 }  ][fill={rgb, 255:red, 0; green, 0; blue, 0 }  ][line width=1.5]      (0, 0) circle [x radius= 4.36, y radius= 4.36]   ;
 		% Text Node
 		\draw (485,208) node [anchor=north west][inner sep=0.75pt]  [font=\footnotesize,xscale=0.8,yscale=0.8] [align=left] {$\displaystyle E_{2}$};
 		\draw (575,208) node [anchor=north west][inner sep=0.75pt]  [font=\footnotesize,xscale=0.8,yscale=0.8] [align=left] {$\displaystyle E_{3}$};
 		
 		%blackline
 		\draw [line width=1.5]    (660,200) -- (740,200) ;
 		%blackdot
 		\draw [shift={(660,200)}, rotate = 0.7] [color={rgb, 255:red, 0; green, 0; blue, 0 }  ][fill={rgb, 255:red, 0; green, 0; blue, 0 }  ][line width=1.5]      (0, 0) circle [x radius= 4.36, y radius= 4.36] ;
 		\draw [shift={(740,200)}, rotate = 0.7] [color={rgb, 255:red, 0; green, 0; blue, 0 }  ][fill={rgb, 255:red, 0; green, 0; blue, 0 }  ][line width=1.5]      (0, 0) circle [x radius= 4.36, y radius= 4.36]   ;
 		% Text Node
 		\draw (645,208) node [anchor=north west][inner sep=0.75pt]  [font=\footnotesize,xscale=0.8,yscale=0.8] [align=left] {$\displaystyle E_{2n}$};
 		\draw (735,208) node [anchor=north west][inner sep=0.75pt]  [font=\footnotesize,xscale=0.8,yscale=0.8] [align=left] {$\displaystyle E_{2n+1}$};
 		
 		%dash line
 		\draw  [dash pattern={on 0.84pt off 2.51pt}]  (610,200) -- (640,200);
 	\end{tikzpicture}
 	\caption{\footnotesize The canonical homology basis $\{a_j,b_j\}_{j=1}^{2n}$ on the Riemann surface $\mathcal{R}$ of genus~$2n$. }\label{fig2n+}
 \end{figure}

To give the exact expression of $\widetilde{\mu}^{(AG)}$, we first introduce the following notations.
Let $P:=(w,z)\in \mathbb{C}^2$ be a point on the Riemann surface $\mathcal{R}$, where  the canonical projection $\pi$ is given by $\pi(P)=z$. We construct the canonical homology bases $\{a_j,b_j\}_{j=1}^{2n}$ and $\{a_j,b_j\}_{j=1}^{2n-1}$ for the genera $2n$ and $2n-1$, respectively, as shown in Figures  \ref{fig2n+} and  \ref{fig2n-1+}.

For the case of genus $2n$, the cycles in Figure \ref{fig2n+} are chosen such that their intersection matrix reads
\begin{equation*}
	a_j\circ b_l=\delta_{jl}=\left\{\begin{aligned}&1,\quad j=l,\\
	&0,\quad j\neq l,
	\end{aligned}
	\qquad j,l=1,\ldots,2n.
	\right.
\end{equation*}
   The normalized holomorphic differentials $\{\omega_j=\rho_j(z)\dif z\}_{j=1}^{2n}$  are  uniquely determined by the normalization relations  with respect to the cycles $a_{1},\ldots, a_{2n}$:
\begin{align*}
	\oint_{a_l} \omega_j=\delta_{jl},\qquad j,l=1,\ldots,2n.
\end{align*}
Then the $b$-period matrix $B=(B_{jl})_{j,l=1}^{2n}$  is given by
\begin{align*}
	B_{jl}:=\oint_{b_l}\omega_j.
\end{align*}
Associated with the matrix $B$ there is the Riemann theta function defined for $\boldsymbol{\varphi}=(\varphi_1,\ldots,\varphi_{2n})\in\mathbb{C}^{2n}$ by the Fourier series
\begin{equation}\label{theta}
	\Theta(\boldsymbol{\varphi}):=\sum_{\boldsymbol{m}\in \mathbb{Z}^{2n}}\exp\left(\pi i \boldsymbol{m}^TB\boldsymbol{m}+2\pi i\boldsymbol{m}^T\boldsymbol{\varphi} \right).
\end{equation}
It is an even function, i.e. $\Theta(\boldsymbol{\varphi})=\Theta(-\boldsymbol{\varphi})$,
and has the following periodicity properties:
\begin{align*}
	\Theta(\boldsymbol{\varphi}\pm \boldsymbol{e}_j)=\Theta(\boldsymbol{\varphi}),\qquad \Theta(\boldsymbol{\varphi}\pm B \boldsymbol{e}_j)=e^{\mp 2\pi i \varphi_j-\pi i B_{jj}}\Theta(\boldsymbol{\varphi}),
\end{align*}
where $\boldsymbol{e}_j = (0,\ldots, 0, 1, 0,\ldots, 0)$ is the $j$-th basis vector in $\mathbb{R}^{2n}$.  And the column vector of the Riemann constants is given by
\begin{equation}\label{def:K}
	\boldsymbol{K}:=(K_1,\ldots,K_{2n})^{T}, \qquad K_j=\frac{1}{2}\sum_{l=1}^{2n}B_{jl}+\frac{j}{2}, \qquad j=1,\ldots,2n.
\end{equation}
The Abel mapping $\mathcal{A}$ is defined by
\begin{equation}\label{abel}
	\mathcal{A}(P):=(\mathcal{A}_1(P),\ldots, \mathcal{A}_{2n}(P))^T,\qquad	\mathcal{A}_j(P)=\int_{P_0}^{P}\omega_j,
\end{equation}
where the point $P_0$ is fixed by the condition $\pi(P_{0})=-E_{2n}$.  Let  $\mathcal{D}=P_1+\cdots+P_{2n}$ be the unique divisor of degree $2n$ such that $P_j$ ($j=1,\ldots,2n$) are in the upper sheet of $\mathcal{R}$ with
\begin{equation*}
	\pi(P_j)\in\left\{\begin{aligned}&(-E_{2n-2j+2}, -E_{2n-2j+1}),\quad 1\leq j\leq n,\\
		&(E_{2j-2n-1},E_{2j-2n}),\quad \ \  n+1\leq j\leq 2n,
	\end{aligned}
	\right.
\end{equation*}
satisfying
\begin{align*}
\prod_{l=0}^{n} (\pi(P_j) + E_{2l+1}) \prod_{l=1}^{n} (\pi(P_j) - E_{2l}) = \prod_{l=1}^{n} (\pi(P_j) + E_{2l}) \prod_{l=0}^{n} (\pi(P_j) - E_{2l+1}).
\end{align*}
The $2n$-dimensional column vector  $\boldsymbol{C}(x,t;\boldsymbol{\phi})$ is given by
\begin{equation}\label{C}
\boldsymbol{C}(x,t;\boldsymbol{\phi})=
\frac{1}{2\pi}
\begin{pmatrix}
	\phi_{n-1}+C_{n-1}^{f}x+C_{n-1}^{g}t-\phi_{n}-C_{n}^{f}x-C_{n}^{g}t \\
	\vdots \\
	\phi_{1}+C_{1}^{f}x+C_{1}^{g}t-\phi_{n}-C_{n}^{f}x-C_{n}^{g}t \\	
	-\phi_{n}-C_{n}^{f}x-C_{n}^{g}t\\
	-\phi_{1}-C_{1}^{f}x-C_{1}^{g}t-\phi_{n}-C_{n}^{f}x-C_{n}^{g}t \\	
	\vdots \\
	-\phi_{n-1}-C_{n-1}^{f}x-C_{n-1}^{g}t-\phi_{n}-C_{n}^{f}x-C_{n}^{g}t \\
	-2\phi_{n}-2C_{n}^{f}x-2C_{n}^{g}t
\end{pmatrix},
\end{equation}
where $\boldsymbol{\phi}=(\phi_1,\ldots,\phi_n)\in \mathbb{R}^n$ is an arbitrary real vector and  $C_{k}^{f}, C_{k}^{g}$ ($k=1,\cdots,n$) are defined by \eqref{ckf}.

For the case of genus $2n-1$, the cycles in Figure \ref{fig2n-1+} are chosen such that their intersection matrix reads
\begin{equation*}
	a_j\circ b_l=\delta_{jl}=\left\{\begin{aligned}&1,\quad j=l,\\
		&0,\quad j\neq l,
	\end{aligned}
	\qquad j,l=1,\ldots,2n-1.
	\right.
\end{equation*}
The normalized holomorphic differentials $\{\omega_j=\rho_j(z)\dif z\}_{j=1}^{2n-1}$, the Riemann theta function $\Theta$ for $\boldsymbol{\varphi}\in\mathbb{C}^{2n-1}$, $\mathcal{D}=P_1+\cdots+P_{2n-1}$ and the Abel mapping $\mathcal{A}$ are similarly defined. In this setting, the $(2n-1)$-dimensional column vector $\boldsymbol{C}(x,t;\boldsymbol{\phi})$ is given by
\begin{equation}\label{C1}
	\boldsymbol{C}(x,t;\boldsymbol{\phi})=
\frac{1}{2\pi}
\begin{pmatrix}
	\phi_{n-1}+C_{n-1}^{f}x+C_{n-1}^{g}t-\phi_{n}-C_{n}^{f}x-C_{n}^{g}t \\
	\vdots \\
	\phi_{1}+C_{1}^{f}x+C_{1}^{g}t-\phi_{n}-C_{n}^{f}x-C_{n}^{g}t \\	
	-\phi_{1}-C_{1}^{f}x-C_{1}^{g}t-\phi_{n}-C_{n}^{f}x-C_{n}^{g}t \\	
	\vdots \\
	-\phi_{n-1}-C_{n-1}^{f}x-C_{n-1}^{g}t-\phi_{n}-C_{n}^{f}x-C_{n}^{g}t \\
	-2\phi_{n}-2C_{n}^{f}x-2C_{n}^{g}t
\end{pmatrix}.
\end{equation}

Following the standard procedure to obtain the algebro-geometric solution solution \cite{AGbook,AGbook2}, the solution to the RH problem \ref{RHmu} is given by
\begin{align}\label{muag}
	\widetilde{\mu}^{(AG)}(z;x,t)=&\frac{1}{2}\begin{pmatrix}
		\frac{(\varkappa(z)+\varkappa(z)^{-1})F_{11}(z;x,t,\boldsymbol{\phi})}{F_{11}(\infty;x,t,\boldsymbol{\phi})} & \frac{(\varkappa(z)-\varkappa(z)^{-1})F_{12}(z;x,t,\boldsymbol{\phi})}{F_{11}(\infty;x,t,\boldsymbol{\phi})}e^{i(\phi_{n}+2f_{0}x+2g_{0}t)}\\
		\frac{(\varkappa(z)-\varkappa(z)^{-1})F_{21}(z;x,t,\boldsymbol{\phi})}{F_{22}(\infty;x,t,\boldsymbol{\phi})}e^{-i(\phi_{n}+2f_{0}x+2g_{0}t)} & 	\frac{(\varkappa(z)+\varkappa(z)^{-1})F_{22}(z;x,t,\boldsymbol{\phi})}{F_{22}(\infty;x,t,\boldsymbol{\phi})}
	\end{pmatrix},
\end{align}
and the algebro-geometric solution $q^{(AG)}$ of the mKdV equation \eqref{mkdv} can be expressed as
\begin{equation}
	q^{(AG)}(x,t)=\lim_{z\to\infty}2iz\widetilde{\mu}^{(AG)}_{12}(z;x,t).
\end{equation}
Here, the function $\varkappa(z)$ is defined by
\begin{equation}\label{varkappa}
	\varkappa(z)=\left\{
	\begin{aligned}
		&\left(\frac{\prod_{k=0}^{n}(z+E_{2k+1})\prod_{k=1}^{n}(z-E_{2k})}{\prod_{k=1}^{n}(z+E_{2k})\prod_{k=0}^{n}(z-E_{2k+1})}\right)^{\frac{1}{4}},\quad \text{genus}\ 2n,\\
		&\left(\frac{\prod_{k=1}^{n}(z+E_{2k+1})\prod_{k=1}^{n}(z-E_{2k})}{\prod_{k=1}^{n}(z+E_{2k})\prod_{k=1}^{n}(z-E_{2k+1})}\right)^{\frac{1}{4}},\quad \text{genus}\ 2n-1,
	\end{aligned}
	\right.
\end{equation}
which is analytic in $\mathbb{C}\setminus\overline{\Gamma}$, satisfies the jump condition $\varkappa_{+}(z)=-i\varkappa_{-}(z)$ for $z\in \Gamma$, and is fixed by $\varkappa(\infty)=1$.   $F_{ij}(z;x,t,\boldsymbol{\phi})$ $(1\leq i,j\leq2)$ is expressed by
\begin{align}
	F_{11}(z;x,t,\boldsymbol{\phi})&
	=\dfrac{\Theta(\mathcal{A}(z)+\boldsymbol{C}(x,t;\boldsymbol{\phi})+\mathcal{A}(\mathcal{D})+\boldsymbol{K})}{\Theta(\mathcal{A}(z)+\mathcal{A}(\mathcal{D})+\boldsymbol{K})},\label{f11}\\ 	F_{12}(z;x,t,\boldsymbol{\phi})&
	=\dfrac{\Theta(-\mathcal{A}(z)+\boldsymbol{C}(x,t;\boldsymbol{\phi})+\mathcal{A}(\mathcal{D})+\boldsymbol{K})}{\Theta(-\mathcal{A}(z)+\mathcal{A}(\mathcal{D})+\boldsymbol{K})},\\ F_{21}(z;x,t,\boldsymbol{\phi})&=
	\dfrac{\Theta(\mathcal{A}(z)+\boldsymbol{C}(x,t;\boldsymbol{\phi})-\mathcal{A}(\mathcal{D})-\boldsymbol{K})}{\Theta(\mathcal{A}(z)-\mathcal{A}(\mathcal{D})-\boldsymbol{K})},
	\\	
	F_{22}(z;x,t,\boldsymbol{\phi})&=
	\dfrac{\Theta(\mathcal{A}(z)-\boldsymbol{C}(x,t;\boldsymbol{\phi})+\mathcal{A}(\mathcal{D})+\boldsymbol{K})}{\Theta(\mathcal{A}(z)+\mathcal{A}(\mathcal{D})+\boldsymbol{K})}.
	\label{f22}
\end{align}
%Where   $\Theta$ is Riemann theta function and $\mathcal{A}$ is Abel map
Consequently, the finite-genus algebro-geometric solution $q^{(AG)}$ of the mKdV equation
 \eqref{mkdv} is given by
\begin{align}\label{qAG}
	q^{(AG)}(x,t)=&	\,q^{(AG)}(x,t;\boldsymbol{E}, \boldsymbol{\phi})\nonumber\\
	=&\,i\beta e^{2i(f_0x+g_0t)+i\phi_{n}} \times
	\dfrac{\Theta(\mathcal{A}(\infty)+\mathcal{A}(\mathcal{D})+\boldsymbol{K})\Theta(-\mathcal{A}(\infty)+\boldsymbol{C}(x,t;\boldsymbol{\phi})
		+\mathcal{A}(\mathcal{D})+\boldsymbol{K})}{\Theta(-\mathcal{A}(\infty)+\mathcal{A}(\mathcal{D})+\boldsymbol{K})\Theta(\mathcal{A}(\infty)
		+\boldsymbol{C}(x,t;\boldsymbol{\phi})+\mathcal{A}(\mathcal{D})+\boldsymbol{K})},
\end{align}
where
\begin{equation*}
	\beta=\left\{
	\begin{aligned}
		&\sum_{k=0}^{n}E_{2k+1}-\sum_{k=1}^{n}E_{2k},\quad \text{genus}\ 2n,\\
		&\sum_{k=1}^{n}(E_{2k+1}-E_{2k}),\quad \text{genus}\ 2n-1,
	\end{aligned}
	\right.
\end{equation*}
$f_{0},g_{0}$ are defined by \eqref{f0g0} and $\phi_{n}\in\mathbb{R}$ is a constant.

%In the following, we explain in detail each  notation contained in expression \eqref{qAG}.

\subsection{The Jost functions}
Consider the Jost function
\begin{align}\label{Psiag}
	\Psi^{(AG)}(z;x,t)=\widetilde{\mu}^{(AG)}(z;x,t)e^{-i((f-f_0)x+(g-g_0)t)\sigma_3},
\end{align}
associated with the algebro-geometric solution $q^{(AG)}$ (defined in \eqref{qAG}), where $\widetilde{\mu}^{(AG)}$ is given by \eqref{muag}, $f$, $g$ and $f_0$, $g_0$ are defined in \eqref{fz}--\eqref{gz} and \eqref{f0g0}. This Jost function satisfies the  Lax pair \eqref{lax} when $q$ is replaced by  $q^{(AG)}$.

In order to obtain the Jost solution of the Lax pair \eqref{lax} with  $q$ being the solution of the mKdV equation \eqref{mkdv} with the finite-genus algebro-geometric background \eqref{boundray}, we consider $\mu^\pm$ satisfying the following Volterra  integral equation:
\begin{align}\label{mupm}
	\mu^\pm(z;x,t)= \widetilde{\mu}^{(AG)}(z;x,t)+\int_{\pm\infty}^x\Lambda(z;x,s,t)U(q-q^{(AG)})(s,t)\mu^\pm(z;s,t)e^{i(f-f_0)(x-s)\sigma_3}\dif s,
\end{align}
where
\begin{align}\label{lambda}
	\Lambda(z;x,s,t)&=\Psi^{(AG)}(z;x,t)\Psi^{(AG)}(z;s,t)^{-1}\nonumber\\
	&=\widetilde{\mu}^{(AG)}(z;x,t)e^{-i(f-f_0)(x-s)\sigma_3}\widetilde{\mu}^{(AG)}(z;s,t)^{-1}.
\end{align}
Set
\begin{align}\label{defmu12}
		\mu^\pm(z;x,t)=\begin{pmatrix}
	\mu_{1}^\pm(z;x,t) & \mu_{2}^\pm(z;x,t)
\end{pmatrix}, \end{align}
with the subscripts $1$ and $2$ indicating the first and second columns of the matrix, respectively.

\begin{proposition}\label{propmu}
	 $\mu_{1}^\pm$ and $\mu_{2}^\pm$ satisfy the following properties:
	\begin{itemize}
		\item  [$\mathrm{(i)}$]  $\mu^{-}_1 (z;x,t)$ and $\mu^{+}_2(z;x,t)$ exist uniquely for $z\in\overline{\mathbb{C}^{+}}\setminus\boldsymbol{E}$, which are
		analytic in $\mathbb{C}^+$ and admit continuous extensions to $\overline{\mathbb{C}^{+}}\setminus\boldsymbol{E}$; while $\mu_{1}^+(z;x,t)$ and $\mu_{2}^-(z;x,t)$
		exist uniquely for $z\in\overline{\mathbb{C}^{-}}\setminus\boldsymbol{E}$, which are analytic for $z$ in $\mathbb{C}^-$ and admit continuous extensions to $\overline{\mathbb{C}^{-}}\setminus\boldsymbol{E}$. Here $\boldsymbol{E}$ is defined by \eqref{E}.
	\item [$\mathrm{(ii)}$]
	As $z\to\infty$,
	\begin{align*}
	\begin{pmatrix}
		\mu^-_1(z;x,t)&\mu^+_2(z;x,t)
	\end{pmatrix}&=I+\mathcal{O}(z^{-1}),\qquad z\in {\mathbb{C}^{+}},
		\\
		\begin{pmatrix}
			\mu^+_1(z;x,t)&\mu^-_2(z;x,t)
		\end{pmatrix}&=I+\mathcal{O}(z^{-1}), \qquad z\in {\mathbb{C}^{-}}.
	\end{align*}	
		\item [$\mathrm{(iii)}$]
		 $\mu_{1}^{\pm}$ and $\mu_{2}^{\pm}$ satisfy the symmetry relations
		\begin{align}\label{symm}
			\mu^\pm_2(z;x,t)=\sigma_1\overline{\mu^\pm_1(\bar{z};x,t)},\qquad
			\mu^\pm_2(z;x,t)=-i\sigma_2\mu^\pm_1(-z;x,t).
		\end{align}
		For $z\in \Gamma$,  we have
		\begin{equation}\label{symm1}
			\mu^+_{2,+}(z;x,t)=\left\{
			\begin{aligned}
				&-ie^{i(C_k^fx+C_k^gt+\phi_k)}\mu^+_{1,-}(z;x,t),\quad z\in(-E_{2k+1},-E_{2k}),\\
				&-ie^{-i(C_k^fx+C_k^gt+\phi_k)}\mu^+_{1,-}(z;x,t),\quad z\in(E_{2k},E_{2k+1}),
			\end{aligned}
			\right.
		\end{equation}
		\begin{equation}\label{symm2}
			\mu^-_{2,-}(z;x,t)=\left\{
			\begin{aligned}
				&ie^{i(C_k^fx+C_k^gt+\phi_k)}\mu^-_{1,+}(z;x,t),\quad z\in(-E_{2k+1},-E_{2k}),\\
				&ie^{-i(C_k^fx+C_k^gt+\phi_k)}\mu^-_{1,+}(z;x,t),\quad z\in(E_{2k},E_{2k+1}),
			\end{aligned}
			\right.
		\end{equation}
		where $k=1,\cdots,n$.
		Specifically, for $z\in(-E_{1},E_{1})$,
		\begin{align*}
		 \mu^+_{2,+}(z;x,t)&=-i\mu^+_{1,-}(z;x,t),\\ \mu^-_{2,-}(z;x,t)&=i\mu^-_{1,+}(z;x,t).
	   \end{align*}
			\item [$\mathrm{(iv)}$]
		$\det \mu^\pm(z;x,t)=1$, $z\in\mathbb{R}\setminus \overline{\Gamma}$, where $\Gamma$ is given  by \eqref{gamma}.
	\end{itemize}
\end{proposition}
\begin{proof}
By the definitions of $\mu^\pm_1$ and $\mu^\pm_2$ in \eqref{defmu12} and the integral representation for $\mu^\pm$ provided in \eqref{mupm}, $\mu^\pm_1$ and $\mu^\pm_2$ satisfy the following Volterra integral equations respectively:
\begin{align}
	\mu^\pm_1(z;x,t)=&\,\widetilde{\mu}^{(AG)}_1(z;x,t)	+\int_{\pm\infty}^xe^{i(f-f_0)(x-s)}\Lambda(z;x,s,t)U(q-q^{(AG)})(s,t)\mu^\pm_1(z;s,t)\dif s,\label{mu1}\\
	\mu^\pm_2(z;x,t)=&\, \widetilde{\mu}^{(AG)}_2(z;x,t)+\int_{\pm\infty}^xe^{-i(f-f_0)(x-s)}\Lambda(z;x,s,t)U(q-q^{(AG)})(s,t)\mu^\pm_2(z;s,t)\dif s. \label{mu2}
\end{align}
In the following, we will use these two integral equations to prove items (i)--(iv) respectively.
	\begin{itemize}
	\item  [$\mathrm{(i)}$]
 We take  $\mu^-_1$ as an example to give the proof. From the definition of $\Lambda$ in \eqref{lambda}, the definition of $f$ in \eqref{fz}, and the fact that $\widetilde{\mu}^{(AG)}$ satisfies the  RH problem \ref{RHmu}, we have the function $\Lambda(z;x,s,t)$ is entire. Moreover, since
 \[
 \operatorname{Im}(f(z)-f_0)=\operatorname{Im}f(z)>0,
 \qquad z\in\mathbb C^+,
 \]
 and \(s<x\), we have
 \[
 \left|e^{i(f-f_0)(x-s)}\right|
 =
 e^{-\operatorname{Im}f(z)(x-s)}
 \leq 1.
 \]
 Therefore,
 \(
 e^{i(f-f_0)(x-s)}\Lambda(z;x,s,t)
 \)
 is uniformly bounded for \(s<x\). It follows that $$e^{i(f-f_0)(x-\cdot)}\Lambda(z;x,\cdot,t)U(q-q^{(AG)})(\cdot,t)\in L^1(-\infty,x).$$ Therefore, the property of $\mu^-_1$ is  obtained from  Picard-Lindel\"of iteration of the integral equation \eqref{mu1}.

	\item  [$\mathrm{(ii)}$] We prove only the asymptotic behavior for $\mu^-_1$ as  $z\to\infty$:
	\begin{equation}\label{asmu1}
		\mu^-_1(z;x,t)=\begin{pmatrix}
			1\\0\\
		\end{pmatrix}+\mathcal{O}(z^{-1}),\qquad z\in {\mathbb{C}^{+}},
	\end{equation}
	the remaining cases being analogous. From the RH problem \ref{RHmu}, we have $\widetilde{\mu}^{(AG)}(z)=I+\mathcal{O}(z^{-1})$, as $z\to\infty$. According to the definition of  $\Lambda$ in \eqref{lambda}, we obtain
	\begin{equation}\label{asylam}
		\begin{aligned}
		&e^{if(x-s)}\Lambda(z;x,s,t)U(q-q^{(AG)})(s,t)\\
		=&\begin{pmatrix}
			(q^{(AG)}-q)(s,t)\mathcal{O}(z^{-1})&(q-q^{(AG)})(s,t)\mathcal{O}(1)\\
			(q^{(AG)}-q)(s,t)\mathcal{O}(z^{-2})& (q^{(AG)}-q)(s,t)\mathcal{O}(z^{-1})\\
		\end{pmatrix}.
		\end{aligned}
	\end{equation}
	Then, applying Lebesgue's dominated convergence theorem to the Volterra equation \eqref{mu1} yields
	\begin{align*}
		\lim_{z\to\infty}\mu^-_1(z;x,t)=\begin{pmatrix}
			1\\0\\
		\end{pmatrix}.
	\end{align*}
	A further use of \eqref{mu1}, \eqref{asylam} and the limit above leads to
	\begin{align*}
	z\left(\mu^-_1(z;x,t)-\begin{pmatrix}
			1\\0\\
		\end{pmatrix}\right)=\mathcal{O}(1),\qquad z\to\infty,
	\end{align*}
	hence \eqref{asmu1} holds.
	\item  [$\mathrm{(iii)}$] The symmetry properties \eqref{symm}--\eqref{symm2} follow from the uniqueness of solutions to the Volterra integral equations \eqref{mu1} and \eqref{mu2}.
	\item  [$\mathrm{(iv)}$] This is a direct consequence of Abel's formula.
\end{itemize}
\end{proof}

Based on  $\mu^\pm$ in \eqref{mupm}, two linearly independent solutions $\Psi^-$ and $\Psi^+$ of the Lax pair \eqref{lax} are constructed as
\begin{align}\label{Psipm}
	\Psi^\pm(z;x,t):=\mu^\pm(z;x,t)e^{-i((f-f_0)x+(g-g_0)t)\sigma_3}.
\end{align}
These solutions are related by a scattering matrix $S(z)$ independent of $(x, t)$ satisfying
\begin{align}
	\Psi^+(z;x,t)=\Psi^-(z;x,t)S(z), \qquad z\in\mathbb{R}\setminus \overline{\Gamma},
\end{align}
that is,
\begin{align}\label{Sz}
	\mu^+(z;x,t)=\mu^-(z;x,t)e^{-i((f-f_0)x+(g-g_0)t)\hat{\sigma}_3}\begin{pmatrix}S_{11}(z)&S_{12}(z)\\S_{21}(z)&S_{22}(z)\end{pmatrix}.
\end{align}
Then the scattering data can be expressed as
\begin{equation}\label{data}
\left\{\begin{aligned}
	&S_{11}(z)=\det\left(\mu^+_1(z),\mu^-_2(z)\right),\\ &S_{12}(z)=e^{2i((f-f_0)x+(g-g_0)t)}\det\left(\mu^+_2(z),\mu^-_2(z)\right),\\
	&S_{21}(z)=e^{-2i((f-f_0)x+(g-g_0)t)}\det\left(\mu^-_1(z),\mu^+_1(z)\right), \\ &S_{22}(z)=\det\left(\mu^-_1(z),\mu^+_2(z)\right).
\end{aligned}\right.
\end{equation}
We subsequently define the reflection coefficients
\begin{align}\label{r}
	r_1(z)=\frac{S_{12}(z)}{S_{11}(z)},\quad r_2(z)=\frac{S_{21}(z)}{S_{22}(z)}, \quad r_3(z)=\frac{S_{21}(z)}{S_{11}(z)},\quad r_4(z)=\frac{S_{12}(z)}{S_{22}(z)},\quad z\in\mathbb{R}\setminus \overline{\Gamma}.
\end{align}
The following proposition can thereby be established.

\begin{proposition}\label{propsca}
	The scattering data given by \eqref{data} and the reflection coefficients defined in \eqref{r} satisfy the following properties.
	\begin{itemize}
		\item [$\mathrm{(i)}$]  $S_{11}(z)$ and $ S_{22}(z)$ are  analytic in $\mathbb{C}^-$ and $\mathbb{C}^+$, respectively;  moreover, they admit continuous extensions $\overline{\mathbb{C}^{-}}\setminus \boldsymbol{E}$ and $\overline{\mathbb{C}^{+}}\setminus \boldsymbol{E}$,  respectively.
		\item [$\mathrm{(ii)}$] $S_{12}(z)$ and $ S_{21}(z)$  can be extended to $\mathbb{R}\setminus\boldsymbol{E}$. Particularly, for $z\in \Gamma$,
		\begin{align*}
			&S_{12}(z)=\det\left(e^{i((f_+-f_0)x+(g_+-g_0)t)}\mu^+_2(z), \, e^{i((f_--f_0)x+(g_--g_0)t)}\mu^-_2(z)\right),\\ &S_{21}(z)=\det\left(e^{-i((f_+-f_0)x+(g_+-g_0)t)}\mu^-_1(z),\,e^{-i((f_--f_0)x+(g_--g_0)t)}\mu^+_1(z)\right).
		\end{align*}
		\item [$\mathrm{(iii)}$] For $z\in\mathbb{R}\setminus\overline{\Gamma}$, $\det S(z)=S_{11}(z)S_{22}(z)-S_{12}(z)S_{21}(z)=1$.
		\item [$\mathrm{(iv)}$] For $ z\in\mathbb{C}^-$,
		$$S_{11}(z)=\overline{S_{22}(\bar{z})},\qquad S_{11}(z)=S_{22}(-z).$$
		For $z\in \mathbb{R}\setminus\overline{\Gamma}$,
		\begin{align*}
			&S_{11}(z)=\overline{S_{22}(z)},\quad S_{11}(z)=S_{22}(-z),\quad S_{12}(z)=\overline{S_{21}(z)},\quad S_{12}(z)=-S_{21}(-z),\\
			&r_1(z)=\overline{r_2(z)},\qquad r_1(z)=-r_2(-z),\qquad
			r_3(z)=\overline{r_4(z)},\qquad 	r_3(z)=-r_4(-z).
		\end{align*}
	 Moreover, 	for $z\in\Gamma$, we have
		\[
		S_{11}(z)=S_{22}(z),
		\]
		\begin{equation}
			S_{12}(z)
			=
			\begin{cases}
				e^{2i\phi_k}S_{21}(z),
				& z\in(-E_{2k+1},-E_{2k}),\\[1mm]
				e^{-2i\phi_k}S_{21}(z),
				& z\in(E_{2k},E_{2k+1}),\\[1mm]
				S_{21}(z),
				& z\in(-E_1,E_1),
			\end{cases}
		\end{equation}
		and
		\begin{equation}\label{r12}
			r_1(z)
			=
			\begin{cases}
				e^{2i\phi_k}r_2(z),
				& z\in(-E_{2k+1},-E_{2k}),\\[1mm]
				e^{-2i\phi_k}r_2(z),
				& z\in(E_{2k},E_{2k+1}),\\[1mm]
				r_2(z),
				& z\in(-E_1,E_1).
			\end{cases}
		\end{equation}
		In particular, for genus $2n-1$, we have $r_{j}(0)\in i\mathbb{R}$ and $|r_{j}(0)|< 1$, $j=1,\cdots,4.$
		\item [$\mathrm{(v)}$] As $z\to\infty$, we have
		$$S_{11}(z),\ S_{22}(z)=1+\mathcal{O}(z^{-1}), \qquad  S_{12}(z),\ S_{21}(z)=\mathcal{O}(z^{-1}),$$ and  $$r_j(z)=\mathcal{O}(z^{-1}),\qquad j=1,\ldots,4. $$
		\item [$\mathrm{(vi)}$]
		For the generic case, as $z\to p\in \boldsymbol{E}$ from $\mathbb{R}\setminus\overline{\Gamma}$, we have
		\begin{align}\label{eq:Sloc}
			S_{11}(z), \ S_{12}(z)=\mathcal{O}((z-p)^{-\frac{1}{2}}),
			\end{align}
	and
	\begin{align}\label{eq:riloc}
		r_1(z)=
		\begin{cases}
			\pm  e^{\mp i\phi_k}+\mathcal{O}\bigl((z-p)^{\frac12}\bigr),
			& p=\pm E_{2k+1},\\[1mm]
			\mp  e^{\mp i\phi_k}+\mathcal{O}\bigl((z-p)^{\frac12}\bigr),
			& p=\pm E_{2k},\\[1mm]
			\pm 1+\mathcal{O}\bigl((z-p)^{\frac12}\bigr),
			& p=\pm E_1,
		\end{cases}
		\qquad k=1,\ldots,n,
	\end{align}
	where the last line is only for the case of genus $2n$ since $\phi_0=0$.
	\end{itemize}
\end{proposition}
\begin{proof}
	Items (i)--(v) can be directly obtained from Proposition \ref{propmu}. We now prove item (vi). Let $\nu^{\pm} (z;x,t)$ be the solution of the  integral equation
\begin{align*}
	\nu^\pm(z;x,t)
	={}&I+\int_{\pm\infty}^{x}
	\Lambda(z;x,s,t)\,
	U(q-q^{(AG)})(s,t)\,
	\nu^\pm(z;s,t)\,
	\Lambda(z;x,s,t)^{-1}\,ds,
\end{align*}
where  $\Lambda$  is  defined in \eqref{lambda}. It then follows that $\nu^{\pm} (z;x,t)$  is  continuous, bounded on $\mathbb{R}$, and satisfies the symmetry relation $$\nu^\pm(z;x,t)
=\sigma_1\overline{\nu^\pm(\bar z;x,t)}\sigma_1.$$ Moreover, as $z\to p\in\boldsymbol{E}  \text{ from }\mathbb{C}^\mp$, we have
\begin{align}\label{nu}
	\nu^{\pm} (z;x,t)= \nu^{\pm} (p;x,t)+\mathcal{O}((z-p)^{\frac12}).
\end{align}
According to the Volterra  integral equations \eqref{mu1}  and \eqref{mu2}, we get
\begin{align}
	\mu_1^\pm(z;x,t)
	&=\nu^\pm(z;x,t)\widetilde\mu_1^{(AG)}(z;x,t),\label{257}\\
	\mu_2^\pm(z;x,t)
	&=\nu^\pm(z;x,t)\widetilde\mu_2^{(AG)}(z;x,t).\label{258}
\end{align}
Consequently, the scattering data in  \eqref{data} can be represented using $\nu^\pm$ and $\widetilde{\mu}^{(AG)}$. As the scattering data does not depend on $x$ and $t$, it suffices to compute it at $(x,t)=(0,0)$. Combining the expression \eqref{muag} for $\widetilde{\mu}^{(AG)}$ with \eqref{nu} yields \eqref{eq:Sloc}.

We now provide the proof for $r_1$ in \eqref{eq:riloc}.  At the right endpoint $E_{2k+1}$,
the background jump and its fourth-root behavior imply
\[
\widetilde\mu_2^{(AG)}(z)
=e^{-i\phi_k}\widetilde\mu_1^{(AG)}(z)
+\mathcal O\bigl((z-E_{2k+1})^{\frac{1}{4}}\bigr).
\]
Using \eqref{257}--\eqref{258} and the boundedness of $\nu^\pm$,
we obtain the same relation for $\mu^\pm$. Hence,
by \eqref{data},
\[
S_{12}(z)=e^{-i\phi_k}S_{11}(z)+\mathcal O(1).
\]
Since the leading coefficient of the
$(z-E_{2k+1})^{-\frac{1}{2}}$ term in $S_{11}$ is nonzero
in the generic case, this yields
\[
r_1(z)=e^{-i\phi_k}
+\mathcal O\bigl((z-E_{2k+1})^{\frac{1}{2}}\bigr).
\]
At $E_{2k}$, the corresponding leading column relation
has coefficient $-e^{-i\phi_k}$.
The negative endpoints follow from the symmetry
$r_1(z)=-\overline{r_1(-z)}$ on $\mathbb R\setminus\Gamma$.
The central endpoints in the even-genus case are
treated in the same way.

\end{proof}

\subsection{Construction of the RH problem}
To derive the asymptotic  behavior  of the solution $q$ of Cauchy problem \eqref{mkdv}--\eqref{boundray}, we construct the RH  problem \ref{RHM}  below. From the scattering relation \eqref{Sz}, we define
\begin{align}\label{def M}	
M(z)&=M(z;x,t):=	\left\{ \begin{array}{ll}
	\left(  \frac{\mu^-_1  (z;x,t)}{S_{22}(z)}  , \  \mu^+_2(z;x,t)\right), \qquad   z\in \mathbb{C}^+,\\
	\left( \mu^+_1(z;x,t), \ \frac{\mu^-_2(z;x,t)}{S_{11}(z)}\right)  , \qquad z\in \mathbb{C}^-.\\
\end{array}\right.
\end{align}
It is then straightforward to verify that $M$ satisfies the following RH problem.
\begin{RHP}\label{RHM}
	\hfil
	\begin{itemize}
		\item $M(z)$ is analytic in $\mathbb{C}\setminus \mathbb{R}$.
		\item
		$M(z)$  satisfies the jump condition $M_{+}(z)=M_{-}(z)V(z)$ with
			\begin{equation}\label{V}
			V(z)=\left\{
			\begin{aligned}
				&
				\begin{pmatrix}
					0 & -ie^{i(\phi_k + C_k^f x + C_k^g t)} \\
					-ie^{-i(\phi_k + C_k^f x + C_k^g t)} & 0
				\end{pmatrix},\quad z\in(-E_{2k+1},-E_{2k}),\\
				&\begin{pmatrix}
					0 & -ie^{-i(\phi_k + C_k^f x + C_k^g t)} \\
					-ie^{i(\phi_k + C_k^f x + C_k^g t)} & 0
				\end{pmatrix},\quad z\in(E_{2k},E_{2k+1}),\\
				&\begin{pmatrix}
					1-r_1(z)r_2(z) & r_1(z)e^{2it\theta(z)}\\
					-r_2(z)e^{-2it\theta(z)} & 1
				\end{pmatrix}, \quad z\in \mathbb{R}\setminus \overline{\Gamma},	
			\end{aligned}
			\right.
		\end{equation}
		where  $k=1,\cdots,n$,  $\Gamma$ is given by \eqref{gamma} and
		\begin{align}\label{alpha}
			\theta(z)=\theta(z;\xi):=-(f(z)-f_0)\xi-\left(g(z)-g_0 \right),\quad\xi=\frac{x}{t}.
		\end{align}
		Specifically, for genus $2n$,
		$$V(z)=\begin{pmatrix}
			0&-i\\-i&0
		\end{pmatrix},\quad z\in(-E_{1},E_{1}). $$
			\item $M(z)$ satisfies the symmetry relations
		\begin{equation*}
			M(z)=\sigma_1\overline{M(\bar{z})}\sigma_1=\sigma_2M(-z)\sigma_2.
		\end{equation*}
		\item  As $z\to\infty$, we have $$M(z)=I+\mathcal{O}(z^{-1}).$$
		\item For $p\in\boldsymbol E$, we have
		\[
		M(z)=
		\bigl(\mathcal{O}((z-p)^{\frac{1}{4}}),\mathcal{O}((z-p)^{-\frac{1}{4}})\bigr),
		\qquad z\to p,\quad z\in\mathbb C^+,
		\]
		\[
		M(z)=
		\bigl(\mathcal{O}((z-p)^{-\frac{1}{4}}),\mathcal{O}((z-p)^{\frac{1}{4}})\bigr),
		\qquad z\to p,\quad z\in\mathbb C^-.
		\]
	\end{itemize}
\end{RHP}

The uniqueness of the solution of RH problem \ref{RHM}  is guarantee by Liouville's theorem, and the solution $q$ is recovered by
	\begin{align}\label{reco}
	q=2i\lim_{z\to\infty}zM_{12}(z).
\end{align}

\subsection{Phase analysis and jump factorizations}\label{sec25}
We now proceed to factorize the jump matrix given by \eqref{V}, based on the signature table of the phase function
 $\theta$ defined in \eqref{alpha}. For this purpose, we introduce the following definition:
\begin{definition}\label{def z0}
	Let  $z_0=z_0(\xi) \in \mathbb{R}$. Then $z_{0}$ is said to be a saddle point of  $\theta$ if it satisfies  one of the following conditions:
	\begin{itemize}
		\item [\rm(a) ] $\theta'(z_0)=0$ for $z_0\in \mathbb{R}\setminus \Gamma$;
		\item [\rm(b) ] $\Im \theta(z_0)=0$ for $z_0\in \Gamma$,
	\end{itemize}
	where $\Gamma$ is given by \eqref{gamma}.
\end{definition}

From the expression of $\theta$ in \eqref{alpha} together with the definitions of $f$ and $g$ in \eqref{fz}--\eqref{gz}, we derive
\begin{equation}
	\theta'(z)=\left\{
	\begin{aligned}
		&-\frac{z}{w(z)} \left[\xi\prod_{k=1}^{n}(z^{2}-(z_{k}^{f})^{2})+ 12\prod_{k=1}^{n+1} (z^2 - (z_k^{g})^2) \right],\quad \text{genus}\ 2n,\\
		&-\frac{1}{w(z)} \left[\xi\prod_{k=1}^{n}(z^{2}-(z_{k}^{f})^{2})+ 12\prod_{k=1}^{n+1} (z^2 - (z_k^{g})^2) \right],\quad \text{genus}\ 2n-1.
	\end{aligned}
	\right.
\end{equation}
By \eqref{fzz} and \eqref{gzz} we have $f(\pm E_{2k+1})=f(\pm E_{2k})$ and $g(\pm E_{2k+1})=g(\pm E_{2k})$, which implies $\theta(\pm E_{2k+1})=\theta(\pm E_{2k})$,  $k=1,\dots,n$. Therefore, $\theta'$ admits a pair of zeros in each interval $(-E_{2k+1}, -E_{2k}) \cup (E_{2k}, E_{2k+1})$, $k=1,\ldots,n$. Let us set
\begin{equation}\label{F}
F(z;\xi)=\xi\prod_{k=1}^{n}(z^{2}-(z_{k}^{f})^{2})+ 12\prod_{k=1}^{n+1} (z^2 - (z_k^{g})^2).
\end{equation}
Away from \(z=0\), the zeros of \(\theta'\) are the zeros of \(F\). For genus \(2n\), \(z=0\) is an additional stationary point on the central cut.

Since $F$ is a polynomial of degree  $2n+2$,  the fundamental theorem of algebra guarantees that $F$ has exactly  $2n+2$ zeros. The positions of   $2n$ zeros are already known, while the remaining pair of zeros is yet to be determined.  Since $F(z;\xi)$ is an even polynomial, the remaining two zeros
 form a symmetric pair with respect to the origin. This symmetric pair may be real or purely imaginary,
 and may coalesce at the origin. Their locations
 determine the topology of the signature table of $\operatorname{Im}\theta$. If the remaining pair lies in $\mathbb{R}\setminus\Gamma$, then
 they are zeros of $\theta'$ and give a pair of saddle points of
 type (a) in Definition \ref{def z0}, which we denote by $\pm z_0$ with
 $z_0>0$. On the other hand, suppose that the remaining pair enters a pair
 of symmetric components of $\Gamma$,
 \[
 (-E_{2k+1},-E_{2k})\cup(E_{2k},E_{2k+1}).
 \]
 Then each of these two intervals contains two zeros of $F$.
 Since $\theta'$ is purely imaginary on $\Gamma$ and
 $\operatorname{Im}\theta$ vanishes at the endpoints of each
 interval, the corresponding zero-level curve of
 $\operatorname{Im}\theta$ intersects the interval at an interior
 point. These two symmetric intersection points are the saddle
 points of type (b) in Definition \ref{def z0}, again denoted by $\pm z_0$.

 \textbf{For case (a)}, let $z_0=z_0(\xi)>0$ denote the positive saddle
 point, so that $-z_0$ is also a saddle point by symmetry.
 We note that
 \begin{equation}\label{critical-speed}
 	F(E_j;\xi_j)=F(- E_j;\xi_j)=0,
 \end{equation}
 where
 \begin{equation}
 \xi_j
 =
 -12\,
 \frac{\prod_{k=1}^{n+1}
 	\big(E_j^2-(z_k^g)^2\big)}
 {\prod_{k=1}^{n}
 	\big(E_j^2-(z_k^f)^2\big)}, \label{critical-value}
\end{equation}
 with $j=1,\ldots,2n+1$ for genus $2n$ and
 $j=2,\ldots,2n+1$ for genus $2n-1$.
Then,
 \[
 z_0(\xi_j)=E_j,
 \qquad
 -z_0(\xi_j)=-E_j.
 \]
Moreover, away from the critical values $\xi_j$, the saddle
point $z_0$ is a simple zero of $F$. By the implicit
function theorem, we have
\[
\partial_\xi z_0(\xi)
=
-\frac{\partial_\xi F(z_0;\xi)}
{\partial_zF(z_0;\xi)}
=
-\frac{\prod_{k=1}^{n}
	\big(z_0^2-(z_k^f)^2\big)}
{\partial_zF(z_0;\xi)}.
\]
On each component of $\mathbb{R}\setminus\Gamma$ containing
$z_0$, the quantities
\(
\partial_zF(z_0;\xi)
\) and \(
\prod_{k=1}^{n}\big(z_0^2-(z_k^f)^2\big)
\)
have the same sign. Therefore, $z_0(\xi)$ is continuous and monotonically decreasing
with respect to $\xi$ on each component of
$\mathbb{R}\setminus\Gamma$. More precisely,
$z_0$ maps $(-\infty,\xi_{2n+1})$ onto
$(E_{2n+1},+\infty)$, and maps
$(\xi_{2k},\xi_{2k-1})$ onto
$(E_{2k-1},E_{2k})$.
Here $k=1,\ldots,n$ for genus $2n$, whereas
$k=2,\ldots,n$ for genus $2n-1$.

For genus $2n-1$, the innermost component requires separate
consideration. In this case, $z_0$ decreases from $E_2$ to the
origin as $\xi$ increases from $\xi_2$ to the critical value
$\xi_0$, where
\(
F(0;\xi_0)=0.
\)
Consequently,
\begin{equation}\label{xi0}
\xi_0
=
12\,
\frac{\prod_{k=1}^{n+1}(z_k^g)^2}
{\prod_{k=1}^{n}(z_k^f)^2}.
\end{equation}
This is the critical speed associated with the central transition
region $\mathcal{T}_0$ in Definition~\ref{regions}. In this case, the central interval $(-E_2,E_2)$ lies in
$\mathbb R\setminus\Gamma$. At $\xi=\xi_0$, the two real
saddle points $\pm z_0$ coalesce at the origin.
For $\xi>\xi_0$, the remaining zeros of $F$ form a purely
imaginary pair, and there is no real saddle point in
$(-E_2,E_2)$. Thus, $(\xi_2,\xi_0)$ and $(\xi_0,+\infty)$
are the innermost parts of regions $\mathcal{T}_{\mathrm{III}}$ and $\mathcal{T}_{\mathrm{IV}}$, respectively,
separated by the central transition region $\mathcal T_0$.

\textbf{ For case (b)}, for each $k=1,\ldots,n$, the positive saddle point
$z_0(\xi)$ lies in $(E_{2k},E_{2k+1})$ and decreases continuously
from $E_{2k+1}$ to $E_{2k}$ as $\xi$ increases from
$\xi_{2k+1}$ to $\xi_{2k}$. By symmetry, the second saddle point
$-z_0$ moves symmetrically in $(-E_{2k+1},-E_{2k})$.
For genus $2n$, $(-E_1,E_1)$ is a central cut,
and the range $\xi>\xi_1$ corresponds to region $\mathcal{T}_{\mathrm{IV}}$
in Definition~\ref{regions}. The signature tables of $\operatorname{Im}\theta$ in the generic cases are displayed in Figure~\ref{figim1}, while those in the central transition region for genus $2n-1$ are shown in Figure~\ref{figim2}.

\begin{figure}
	\centering
	\begingroup
	\definecolor{mkdvAxisBlue}{RGB}{74,144,226}
	\definecolor{mkdvSigPink}{RGB}{255,240,244}
	
	% #1: positive crossing coordinate; #2: subfigure annotation.
	\newcommand{\mkdvSignaturePanel}[2]{%
		\begin{tikzpicture}[
			x=1cm,y=0.7cm,
			font=\small,
			line cap=round,
			line join=round,
			blueline/.style={draw=mkdvAxisBlue,line width=0.75pt}
			]
			\def\mkdvZ{#1}
			\def\mkdvW{6.05}
			
			% Positive region in the upper half-plane.
			%\path[fill=mkdvSigPink]
			(-\mkdvZ,0)
			.. controls (-\mkdvZ,0.55) and ({-\mkdvZ-0.28},1.10)
			.. ({-\mkdvZ-0.65},1.55)
			-- ({\mkdvZ+0.65},1.55)
			.. controls ({\mkdvZ+0.28},1.10) and (\mkdvZ,0.55)
			.. (\mkdvZ,0) -- cycle;
			
			% Positive regions in the lower half-plane.
			%\path[fill=mkdvSigPink]
			(-\mkdvW,0) -- (-\mkdvZ,0)
			.. controls (-\mkdvZ,-0.55) and ({-\mkdvZ-0.28},-1.10)
			.. ({-\mkdvZ-0.65},-1.55)
			-- (-\mkdvW,-1.55) -- cycle;
			
			%\path[fill=mkdvSigPink]
			(\mkdvZ,0) -- (\mkdvW,0) -- (\mkdvW,-1.55)
			-- ({\mkdvZ+0.65},-1.55)
			.. controls ({\mkdvZ+0.28},-1.10) and (\mkdvZ,-0.55)
			.. (\mkdvZ,0) -- cycle;
			
			% Real axis and the two symmetric zero-level curves.
			\draw[blueline,-{Latex}] (-\mkdvW,0)--(6.35,0);
			\node[anchor=west] at (6.35,0) {$\mathbb R$};
			
			\foreach \s in {-1,1}{
				\draw[blueline]
				({\s*(\mkdvZ+0.65)},-1.55)
				.. controls ({\s*(\mkdvZ+0.28)},-1.10)
				and ({\s*\mkdvZ},-0.55)
				.. ({\s*\mkdvZ},0)
				.. controls ({\s*\mkdvZ},0.55)
				and ({\s*(\mkdvZ+0.28)},1.10)
				.. ({\s*(\mkdvZ+0.65)},1.55);
			}
			
			% Spectral cuts and branch points.
			\foreach \a/\b in {-5.4/-3.9,-2.3/-0.8,0.8/2.3,3.9/5.4}{
				\draw[black,line width=2.4pt] (\a,0)--(\b,0);
				\fill[black] (\a,0) circle (1.8pt);
				\fill[black] (\b,0) circle (1.8pt);
			}
			
			\foreach \x/\lab in {
				-5.4/{-E_{2k+3}},-3.9/{-E_{2k+2}},
				-2.3/{-E_{2k+1}},-0.8/{-E_{2k}},
				0.8/{E_{2k}},2.3/{E_{2k+1}},
				3.9/{E_{2k+2}},5.4/{E_{2k+3}}
			}{
				\node[anchor=north,inner sep=2pt] at (\x,-0.08) {$\lab$};
			}
			\node[inner sep=1pt] at (0,-0.30) {$\cdots$};
			
			% Saddle points.
			\fill[mkdvAxisBlue] (-\mkdvZ,0) circle (2.3pt);
			\fill[mkdvAxisBlue] ( \mkdvZ,0) circle (2.3pt);
			\node[text=mkdvAxisBlue,anchor=south east,inner sep=3pt]
			at (-\mkdvZ,0.08) {$-z_0$};
			\node[text=mkdvAxisBlue,anchor=south west,inner sep=3pt]
			at (\mkdvZ,0.08) {$z_0$};
			
			% Signs of Im theta.
			\foreach \x/\y/\lab in {
				-4.8/0.9/{-},0/0.9/{+},4.8/0.9/{-},
				-4.8/-0.95/{+},0/-0.95/{-},4.8/-0.95/{+}
			}{
				\node[font=\Large] at (\x,\y) {$\lab$};
			}
			
			\node[anchor=north,inner sep=0pt] at (0,-1.83) {#2};
		\end{tikzpicture}%
	}
	
	\resizebox{0.845\linewidth}{!}{%
		\mkdvSignaturePanel{3.1}{%
			(a) $z_0\in(E_{2k+1},E_{2k+2})$,\quad
			$\xi\in(\xi_{2k+2},\xi_{2k+1})$.}%
	}
	\par\vspace{5mm}
	\resizebox{0.845\linewidth}{!}{%
		\mkdvSignaturePanel{1.55}{%
			(b) $z_0\in(E_{2k},E_{2k+1})$,\quad
			$\xi\in(\xi_{2k+1},\xi_{2k})$.}%
	}
	
	\caption{Two examples for the signature table of $\operatorname{Im}\theta$.
		Black segments denote cuts, and blue curves indicate
		$\operatorname{Im}\theta=0$ off the cuts.
		The signs $``+"$ and $``-"$ indicate $\operatorname{Im}\theta>0$ and $\operatorname{Im}\theta<0$, respectively. 
		These schematic diagrams apply to both genus $2n$ and genus $2n-1$.}
	\label{figim1}
	\endgroup
\end{figure}

The signature tables in Figures~\ref{figim1} and~\ref{figim2} motivate the following
factorizations of the jump matrices $V$ for the RH problem \ref{RHM}.

\begin{lemma}
	\label{lem:jump-factorization}
	Let $V$ be the jump matrix in \eqref{V}, and let $\theta$
	be defined by \eqref{alpha}.
	The following factorizations hold:
	
	For
	\(
	z\in\bigcup_{k=1}^{n}
	\bigl((-E_{2k+1},-E_{2k})\cup(E_{2k},E_{2k+1})\bigr),
\)
	we have
	\begin{equation}\label{fac1}
		\begin{aligned}
			V(z)
			&=
			\begin{pmatrix}
				1 & r_1(z)e^{2it\theta_-(z)}\\
				0 & 1
			\end{pmatrix}
			\begin{pmatrix}
				0 & -ie^{i\breve{\theta}(z)}\\
				-ie^{-i\breve{\theta}(z)} & 0
			\end{pmatrix}
			\begin{pmatrix}
				1 & 0\\
				-r_2(z)e^{-2it\theta_+(z)} & 1
			\end{pmatrix}
			\\[2mm]
			&=
			\begin{pmatrix}
				1 & 0\\
				\frac{-r_2(z)e^{-2it\theta_-(z)}}
				{1-r_1(z)r_2(z)} & 1
			\end{pmatrix}
			\begin{pmatrix}
				0 & -ie^{i\breve{\theta}(z)}\\
				-ie^{-i\breve{\theta}(z)} & 0
			\end{pmatrix}
			\begin{pmatrix}
				1 & \frac{r_1(z)e^{2it\theta_+(z)}}
				{1-r_1(z)r_2(z)}\\
				0 & 1
			\end{pmatrix},
		\end{aligned}
	\end{equation}
	where	\begin{equation}\label{breal}
		\breve{\theta}(z)
		=
		\begin{cases}
			t\bigl(\theta_+(z)+\theta_-(z)\bigr)+\phi_k,
			& z\in(-E_{2k+1},-E_{2k}),\\
			t\bigl(\theta_+(z)+\theta_-(z)\bigr)-\phi_k,
			& z\in(E_{2k},E_{2k+1}),
		\end{cases}
		\qquad k=1,\ldots,n,
	\end{equation}
    with
	\begin{equation}\label{gur}
		\theta_+(z)+\theta_-(z)
		=
		\begin{cases}
			C_k^f\xi+C_k^g,
			& z\in(-E_{2k+1},-E_{2k}),\\
			-C_k^f\xi-C_k^g,
			& z\in(E_{2k},E_{2k+1}),
		\end{cases}
		\qquad k=1,\ldots,n.
	\end{equation}

	For $z\in\mathbb R\setminus\overline{\Gamma}$, we have
	\begin{equation}\label{fac2}
		\begin{aligned}
			V(z)
			&=
			\begin{pmatrix}
				1 & r_1(z)e^{2it\theta(z)}\\
				0 & 1
			\end{pmatrix}
			\begin{pmatrix}
				1 & 0\\
				-r_2(z)e^{-2it\theta(z)} & 1
			\end{pmatrix}
			\\[2mm]
			&=
			\begin{pmatrix}
				1 & 0\\
				-\frac{r_2(z)e^{-2it\theta(z)}}
				{1-r_1(z)r_2(z)} & 1
			\end{pmatrix}
			\begin{pmatrix}
				1-r_1(z)r_2(z) & 0\\
				0 & \frac{1}{1-r_1(z)r_2(z)}
			\end{pmatrix}
			\begin{pmatrix}
				1 & \frac{r_1(z)e^{2it\theta(z)}}
				{1-r_1(z)r_2(z)}\\
				0 & 1
			\end{pmatrix}.
		\end{aligned}
	\end{equation}

\end{lemma}

\begin{proof}
	Identity \eqref{gur} follows from
	\eqref{ckkf}--\eqref{ckkg} and the definition of $\theta$ in \eqref{alpha}.
	Using the relations in \eqref{r12}, direct multiplication
	verifies \eqref{fac1}.
	Likewise, \eqref{fac2} follows by direct
	multiplication.
\end{proof}

\begin{figure}
	\centering
	\begingroup
	\definecolor{mkdvAxisBlue}{RGB}{74,144,226}
	\definecolor{mkdvCtrPink}{RGB}{255,240,244}
	
	% #1 = 1: before; 2: at; 3: after the coalescence at xi_0.
	\newcommand{\mkdvCenterPanel}[1]{%
		\begin{tikzpicture}[
			x=0.77cm,y=1.40cm,
			font=\scriptsize,
			line cap=round,
			line join=round,
			blueline/.style={draw=mkdvAxisBlue,line width=0.75pt}
			]
			\def\mkdvW{2.88}
			
			\ifcase#1\relax
			\or
			% Two real saddle points +/-z_0.
			\def\mkdvZ{0.58}
			
			\path[]
			plot[domain=0:1.35,samples=41,variable=\t]
			({-sqrt(\mkdvZ^2+(\t)^2/3)},\t)
			-- ({sqrt(\mkdvZ^2+1.35^2/3)},1.35)
			-- plot[domain=1.35:0,samples=41,variable=\t]
			({sqrt(\mkdvZ^2+(\t)^2/3)},\t)
			-- cycle;
			
			\foreach \s in {-1,1}{
				\path[ ]
				({\s*\mkdvW},0)--({\s*\mkdvZ},0)
				-- plot[domain=0:-1.35,samples=41,variable=\t]
				({\s*sqrt(\mkdvZ^2+(\t)^2/3)},\t)
				-- ({\s*\mkdvW},-1.35)--cycle;
				
				\draw[blueline]
				plot[domain=-1.35:1.35,samples=81,variable=\t]
				({\s*sqrt(\mkdvZ^2+(\t)^2/3)},\t);
			}
			
			\or
			% Coalescence at the origin.
			\path[]
			(0,0)
			.. controls (-0.26,0.45) and (-0.50,1.00)
			.. (-1.22,1.35)
			-- (1.22,1.35)
			.. controls (0.50,1.00) and (0.26,0.45)
			.. (0,0)
			-- cycle;
			
			\foreach \s in {-1,1}{
				\path[]
				(0,0)
				.. controls ({\s*0.26},-0.45) and ({\s*0.50},-1.00)
				.. ({\s*1.22},-1.35)
				-- ({\s*\mkdvW},-1.35)
				-- ({\s*\mkdvW},0)
				-- cycle;
			}
			
			\foreach \s in {-1,1}{
				\draw[blueline]
				({-\s*1.22},-1.35)
				.. controls ({-\s*0.50},-1.00) and ({-\s*0.26},-0.45)
				.. (0,0)
				.. controls ({\s*0.26},0.45) and ({\s*0.50},1.00)
				.. ({\s*1.22},1.35);
			}
			
			\or
			% No real saddle point in the central interval.
			\def\mkdvV{0.48}
			\pgfmathsetmacro{\mkdvEnd}{sqrt((1.35^2-\mkdvV^2)/3)}
			
			\path[]
			plot[domain=-\mkdvEnd:\mkdvEnd,samples=61,variable=\t]
			(\t,{sqrt(3*(\t)^2+\mkdvV^2)})
			-- cycle;
			
			\path[]
			(-\mkdvW,0)--(\mkdvW,0)--(\mkdvW,-1.35)
			-- (\mkdvEnd,-1.35)
			-- plot[domain=\mkdvEnd:-\mkdvEnd,samples=61,variable=\t]
			(\t,{-sqrt(3*(\t)^2+\mkdvV^2)})
			-- (-\mkdvW,-1.35)--cycle;
			
			\foreach \s in {-1,1}{
				\draw[blueline]
				plot[domain=-\mkdvEnd:\mkdvEnd,samples=61,variable=\t]
				(\t,{\s*sqrt(3*(\t)^2+\mkdvV^2)});
			}
			\fi
			
			% Real axis and spectral cuts.
			\draw[blueline,-{Latex}] (-\mkdvW,0)--(2.98,0);
			\node[anchor=west,inner sep=1pt] at (2.98,0) {$\mathbb R$};
			
			\foreach \a/\b in {-2.15/-1.30,1.30/2.15}{
				\draw[black,line width=2pt] (\a,0)--(\b,0);
				\fill[black] (\a,0) circle (1.5pt);
				\fill[black] (\b,0) circle (1.5pt);
			}
			
			\foreach \x/\lab in {
				-2.15/{-E_3},-1.30/{-E_2},
				1.30/{E_2},2.15/{E_3}
			}{
				\node[anchor=north,inner sep=2pt] at (\x,-0.07) {$\lab$};
			}
			
			% Saddle points and origin.
			\ifnum#1=1
			\foreach \s in {-1,1}{
				\fill[mkdvAxisBlue] ({\s*\mkdvZ},0) circle (2pt);
			}
			\node[text=mkdvAxisBlue,anchor=south east,inner sep=2pt]
			at (-\mkdvZ,0.05) {$-z_0$};
			\node[text=mkdvAxisBlue,anchor=south west,inner sep=2pt]
			at (\mkdvZ,0.05) {$z_0$};
			\fill[black] (0,0) circle (1.5pt);
			\node[anchor=north,inner sep=2pt] at (0,-0.1) {$0$};
			\else\ifnum#1=2
			\fill[mkdvAxisBlue] (0,0) circle (2pt);
			\fill[black] (0,0) circle (1.5pt);
			\node[anchor=north west,inner sep=2pt] at (0.18,-0.05) {$0$};
			\else
			\fill[black] (0,0) circle (1.5pt);
			\node[anchor=north,inner sep=2pt] at (0,-0.1) {$0$};
			\fi\fi
			
			% Signs of Im theta.
			\foreach \x/\y/\lab in {
				-2.0/0.83/{-},0/1.0/{+},2.0/0.83/{-},
				-2.0/-0.83/{+},0/-1.04/{-},2.0/-0.83/{+}
			}{
				\node[font=\normalsize] at (\x,\y) {$\lab$};
			}
		\end{tikzpicture}%
	}
	
	\begin{minipage}[t]{0.32\linewidth}
		\centering
		\mkdvCenterPanel{1}\par\smallskip
		{\scriptsize (a) $\xi_2<\xi<\xi_0$; $0<z_0<E_2$.}
	\end{minipage}\hfill
	\begin{minipage}[t]{0.32\linewidth}
		\centering
		\mkdvCenterPanel{2}\par\smallskip
		{\scriptsize (b) $\xi=\xi_0$; $z_0=0$.}
	\end{minipage}\hfill
	\begin{minipage}[t]{0.32\linewidth}
		\centering
		\mkdvCenterPanel{3}\par\smallskip
		{\scriptsize (c) $\xi>\xi_0$.}
	\end{minipage}
	
	\caption{The signature tables of $\operatorname{Im}\theta$
		near the central transition $\xi=\xi_0$ for genus $2n-1$.
		The real saddle points $\pm z_0$ in (a) coalesce at the origin in (b);
		no real saddle point remains in $(-E_2,E_2)$ in (c).
		Black segments denote cuts, and blue curves indicate
		$\operatorname{Im}\theta=0$ off the cuts. 
			The signs $``+"$ and $``-"$ indicate $\operatorname{Im}\theta>0$ and $\operatorname{Im}\theta<0$, respectively. }
	\label{figim2}
	\endgroup
\end{figure}

\section{Asymptotic analysis of transition  region  $\mathcal{T}_{\mathrm{I}}$}\label{sec3}
In this section, we analyze RH Problem~\ref{RHM} to derive the
long-time asymptotics of $q$ in transition region  $\mathcal{T}_{\mathrm{I}}$.
Throughout this section, we fix
\[
k_0\in
\begin{cases}
	\{0,\ldots,n\}, & \text{for genus }2n,\\
	\{1,\ldots,n\}, & \text{for genus }2n-1,
\end{cases}
\]
and assume that
\[
-C\leq (\xi-\xi_{2k_0+1})t^{\frac{2}{3}}\leq 0,
\qquad \xi=\frac{x}{t},
\]
since the analysis in the other half of the transition region is similar. For sufficiently large $t$, we have
\[
\xi_{2k_0+2}<\xi\leq\xi_{2k_0+1},
\]
and hence, by the discussion in Section~\ref{sec25},
\[
z_0\in[E_{2k_0+1},E_{2k_0+2}),
\qquad
-z_0\in(-E_{2k_0+2},-E_{2k_0+1}].
\]
 At $\xi=\xi_{2k_0+1}$, the saddle points
reach the branch points $\pm E_{2k_0+1}$.

\subsection{Construction of $M^{(1)}$ and $M^{(2)}$}

The first transformation involves an auxiliary scalar
function and the opening of global lenses.  For genus $2n$, we define
\begin{equation}\label{delta}
	\begin{aligned}
		\delta(z):=&\,\frac{w(z)}{2\pi i} \left(\sum_{j=1}^{n-1}\delta_j\int_{-E_{2n-2j+1}}^{-E_{2n-2j}}+\delta_{n}\int_{-E_1}^{E_1}+\sum_{j=n+1}^{2n}\delta_j\int_{E_{2j-2n}}^{E_{2j-2n+1}}\right)\frac{i}{w_+(s)(s-z)}\dif s\\
		&+\frac{w(z)}{2\pi i}\int_{I_0}\frac{\log (1-r_1(s)r_2(s))}{w(s)(s-z)}\dif s,
	\end{aligned}
\end{equation}
where
\begin{equation}\label{I}
	I_0:=(-E_{2k_0},E_{2k_0})\setminus \Gamma,
\end{equation}
 $w$, $r_1$ and $r_2$ are defined in \eqref{w} and \eqref{r}, respectively, and the constants $\delta_j$, $j=1,\ldots,2n$, are determined by  the following linear systems:
\begin{equation}\label{deltaj}
	\begin{aligned}
	&\left(\sum_{j=1}^{n-1}\delta_j\int_{-E_{2n-2j+1}}^{-E_{2n-2j}}+\delta_{n}\int_{-E_1}^{E_1}+\sum_{j=n+1}^{2n}\delta_j\int_{E_{2j-2n}}^{E_{2j-2n+1}}\right)\frac{is^m}{w_+(s)}\dif s\\
	=&-\int_{I_0}\frac{\log (1-r_1(s)r_2(s))s^m}{w(s)}\dif s,
	\end{aligned}
\end{equation}
with $m=0,\cdots,2n-1$. For genus $2n-1$, the treatment is analogous: one simply deletes the integral term $\delta_{n}\int_{-E_1}^{E_1}$ in \eqref{delta} and \eqref{deltaj}, and takes $m=0,\cdots,2n-2$. Namely, we have
\begin{equation}\label{dfa}
	\begin{aligned}
		\delta(z):=&\,\frac{w(z)}{2\pi i} \left(\sum_{j=1}^{n-1}\delta_j\int_{-E_{2n-2j+1}}^{-E_{2n-2j}}+\sum_{j=n+1}^{2n}\delta_j\int_{E_{2j-2n}}^{E_{2j-2n+1}}\right)\frac{i}{w_+(s)(s-z)}\dif s\\
		&+\frac{w(z)}{2\pi i}\int_{I_0}\frac{\log (1-r_1(s)r_2(s))}{w(s)(s-z)}\dif s,
	\end{aligned}
\end{equation}
where  the constants $\delta_j$, $j=1,\ldots,2n$, are determined by  the following linear systems:
\begin{equation}\label{deltaj1}
	\begin{aligned}
   & \delta_n=0,\\
	&\left(\sum_{j=1}^{n-1}\delta_j\int_{-E_{2n-2j+1}}^{-E_{2n-2j}}+\sum_{j=n+1}^{2n}\delta_j\int_{E_{2j-2n}}^{E_{2j-2n+1}}\right)\frac{is^m}{w_+(s)}\dif s\\
	=&-\int_{I_0}\frac{\log (1-r_1(s)r_2(s))s^m}{w(s)}\dif s,\qquad\qquad m=0,\cdots,2n-2.
	\end{aligned}
\end{equation}

From the definition of $\delta$ in \eqref{delta}, \eqref{dfa}, and Proposition \ref{propsca}, it follows that $\delta$ satisfies the following properties.
\begin{proposition}\label{Prodelta}
	\hfill
	\begin{itemize}
		\item  [$\mathrm{(i)}$]
		$\delta(z)$ is analytic in $\mathbb{C}\setminus\big(\overline{\Gamma}\cup(-E_{2k_0+1},E_{2k_0+1})\big)$ and satisfies the symmetry relation $$\delta(z)=-\overline{\delta(\bar{z})}, \qquad \delta(z)+\delta(-z)=2\delta(\infty).$$
			\item [$\mathrm{(ii)}$ ] $\delta(z)$  satisfies the jump conditions	
		\begin{equation*}
			\delta_+(z)+\delta_-(z)=i\delta_j,\quad z\in\left\{
			\begin{aligned}
				&(-E_{2n-2j+1},-E_{2n-2j}),& 1\leq j\leq n-1,\\
				&(-E_1,E_1),& j=n\text{ for genus } 2n,\\
				&(E_{2j-2n},E_{2j-2n+1}),& n+1\leq j\leq 2n.
			\end{aligned}\right.
		\end{equation*}
		\begin{equation*}
			\delta_+(z)=\delta_-(z)+\log (1-r_1(z)r_2(z)),\quad z\in I_0,
		\end{equation*}
		where $I_0$ is given by \eqref{I}.
		\item[(iii)]
		As $z\to\infty$, we have
		\begin{equation}\label{asde}
			\delta(z)
			=\delta(\infty)+\frac{\delta^{(1)}}{z}
			+\mathcal{O}(z^{-2}).
		\end{equation}
		For genus $2n$, the coefficients are given by
		\begin{align*}
			\delta(\infty)
			={}&-\frac{1}{2\pi i}
			\int_{I_0}
			\frac{\log(1-r_1(s)r_2(s))s^{2n}}{w(s)}\,ds
			\nonumber\\
			&-\frac{1}{2\pi i}
			\Bigg(
			\sum_{j=1}^{n-1}\delta_j
			\int_{-E_{2n-2j+1}}^{-E_{2n-2j}}
			+\delta_n\int_{-E_1}^{E_1}
			+\sum_{j=n+1}^{2n}\delta_j
			\int_{E_{2j-2n}}^{E_{2j-2n+1}}
			\Bigg)
			\frac{i s^{2n}}{w_+(s)}\,ds,
			\\
			\delta^{(1)}
			={}&-\frac{1}{2\pi i}
			\int_{I_0}
			\frac{\log(1-r_1(s)r_2(s))s^{2n+1}}{w(s)}\,ds
			\nonumber\\
			&-\frac{1}{2\pi i}
			\Bigg(
			\sum_{j=1}^{n-1}\delta_j
			\int_{-E_{2n-2j+1}}^{-E_{2n-2j}}
			+\delta_n\int_{-E_1}^{E_1}
			+\sum_{j=n+1}^{2n}\delta_j
			\int_{E_{2j-2n}}^{E_{2j-2n+1}}
			\Bigg)
			\frac{i s^{2n+1}}{w_+(s)}\,ds.
		\end{align*}
		For genus $2n-1$, the corresponding formulas are obtained
		by deleting the central-cut terms containing $\delta_n$
		and replacing the powers $s^{2n}$ and $s^{2n+1}$ in the
		formulas for $\delta(\infty)$ and $\delta^{(1)}$
		by $s^{2n-1}$ and $s^{2n}$, respectively.
		
	\item[(iv)]
		In the generic case, as $z\to p$ from $\mathbb{C}^+$,
		we have
		\begin{equation}
			e^{\delta(z)}=\mathcal{O}\bigl((z-p)^{\frac{1}{2}}\bigr),
		\end{equation}
		where
		\[
		p\in
		\begin{cases}
			\{\pm E_j:1\leq j\leq 2k_0\},
			& \text{for genus }2n,\\
			\{\pm E_j:2\leq j\leq 2k_0\},
			& \text{for genus }2n-1.
		\end{cases}
		\]
		At the two critical branch points, we have
		\begin{equation}	\label{delp}
			\left\{
		\begin{aligned}
			\delta(z)
			&=\frac{i}{2}\delta_{n-k_0}
			+\mathcal{O}\bigl((z+E_{2k_0+1})^{\frac{1}{2}}\bigr),
			&& z\to -E_{2k_0+1},
			\\
			\delta(z)
			&=\frac{i}{2}\delta_{n+k_0}
			+\mathcal{O}\bigl((z-E_{2k_0+1})^{\frac{1}{2}}\bigr),
			&& z\to E_{2k_0+1},
		\end{aligned}
		\right.
		\end{equation}
		where $\delta_0=0$ and the limits are taken off the cuts.
	\end{itemize}
\end{proposition}
From the relationship of $r_1$ and $r_2$ in Proposition \ref{propsca}, the equations of $\delta_j$ in \eqref{deltaj} and \eqref{deltaj1} also give that
\begin{align}\label{equ:deletajdelta2n}
\delta_{n-j}+\delta_{n+j}=\delta_{2n},
\qquad j=1,\ldots,n.
\end{align}
Combining this with  (i) and (ii) in the above proposition, we also obtain that
\begin{align}\label{equ:deltainfty}
    \delta(\infty)=\frac{i}{4}\delta_{2n}.
\end{align}

\begin{figure}
	\centering
	\definecolor{mkdvAxisBlue}{RGB}{74,144,226}
	\definecolor{mkdvLensGreen}{RGB}{60,130,70}
	\resizebox{0.90\linewidth}{!}{%
		\begin{tikzpicture}[
			x=1cm,
			y=1cm,
			>=Latex,
			axis/.style={draw=mkdvAxisBlue,line width=0.72pt},
			band/.style={draw=black,line width=2.15pt},
			outer/.style={
				draw=mkdvLensGreen,
				line width=1.25pt,
				postaction={decorate},
				decoration={markings,
					mark=at position 0.6 with
					{\arrow{Latex[length=3.0mm,width=1.9mm]}}
				}
			},
			inner/.style={draw=mkdvLensGreen,line width=1.15pt},
			innerarrow/.style={
				inner,
				postaction={decorate},
				decoration={markings,
					mark=at position 0.55 with
					{\arrow{Latex[length=3.0mm,width=1.9mm]}}
				}
			},
			point/.style={circle,fill=black,inner sep=1.45pt},
			every node/.style={font=\footnotesize}
			]
			
			% Schematic coordinates.
			\def\zouter{4.85}
			\def\Eodd{4.18}
			\def\zsaddle{3.40}
			\def\Eeven{1.65}
			\def\rho{1.68}
			\def\rayx{7.08}
			\def\rayy{1.58}
			
			% Real axis and spectral bands.
			\draw[axis,->] (-7.28,0)--(7.42,0) node[right] {$\mathbb R$};
			\draw[band] (-\Eodd,0)--(-\Eeven,0);
			\draw[band] ( \Eeven,0)--( \Eodd,0);
			\node[black!58] at (0,-0.19) {$\cdots$};
			
			% Outer contours.
			\draw[outer] (-\rayx, \rayy)--(-\zouter,0);
			\draw[outer] ( \zouter,0)--( \rayx, \rayy);
			\draw[outer] (-\rayx,-\rayy)--(-\zouter,0);
			\draw[outer] ( \zouter,0)--( \rayx,-\rayy);
			
			% Inner contours.
			\draw[innerarrow] (-\zsaddle,0)--(0, \rho);
			\draw[innerarrow] (0, \rho)--(\zsaddle,0);
			\draw[innerarrow] (-\zsaddle,0)--(0,-\rho);
			\draw[innerarrow] (0,-\rho)--(\zsaddle,0);
			
			% Branch points and contour vertices on the real axis.
			\foreach \x in {-\Eodd,-\Eeven,\Eeven,\Eodd}
			\node[point] at (\x,0) {};
			
			\foreach \x in {-\zouter,\zouter,-\zsaddle,\zsaddle}
			\node[point,fill=mkdvLensGreen] at (\x,0) {};
			
			% Point labels.
			\node[above=4pt] at (-\zouter,0)  {$-z_0$};
			\node[above=4pt] at ( \zouter,0)  {$z_0$};
			\node[below=4pt] at (-\Eodd,0)    {$-E_{2k_0+1}$};
			\node[below=4pt] at ( \Eodd,0)    {$E_{2k_0+1}$};
			\node[above=4pt] at (-\zsaddle,0) {$-z_{k_0}$};
			\node[above=4pt] at ( \zsaddle,0) {$z_{k_0}$};
			\node[below=4pt] at (-\Eeven,0)   {$-E_{2k_0}$};
			\node[below=4pt] at ( \Eeven,0)   {$E_{2k_0}$};
			
			% Contour labels.
			\node[mkdvLensGreen] at (-6, 1.5) {$\Sigma_1^{(l)}$};
			\node[mkdvLensGreen] at ( 6, 1.5) {$\Sigma_1^{(r)}$};
			\node[mkdvLensGreen] at (-6,-1.5) {$\Sigma_1^{(l)*}$};
			\node[mkdvLensGreen] at ( 6,-1.5) {$\Sigma_1^{(r)*}$};
			\node[mkdvLensGreen,above=5pt] at (0, \rho) {$\Sigma_2$};
			\node[mkdvLensGreen,below=5pt] at (0,-\rho) {$\Sigma_2^*$};
			
			% Region labels.
			\node[black] at (-6.34, 0.49) {$\Omega_1^{(l)}$};
			\node[black] at ( 6.34, 0.49) {$\Omega_1^{(r)}$};
			\node[black] at (-6.34,-0.49) {$\Omega_1^{(l)*}$};
			\node[black] at ( 6.34,-0.49) {$\Omega_1^{(r)*}$};
			\node[black] at (0, 0.61) {$\Omega_2$};
			\node[black] at (0,-0.61) {$\Omega_2^*$};
			
		\end{tikzpicture}%
	}
	\caption{The opened contours $\Sigma_j,\Sigma_j^*$ and the opened regions
		$\Omega_j,\Omega_j^*$, $j=1,2$, in transition region~$\mathcal{T}_{\mathrm{I}}$.}
	\label{fig:mkdv-region-I-contours}
\end{figure}
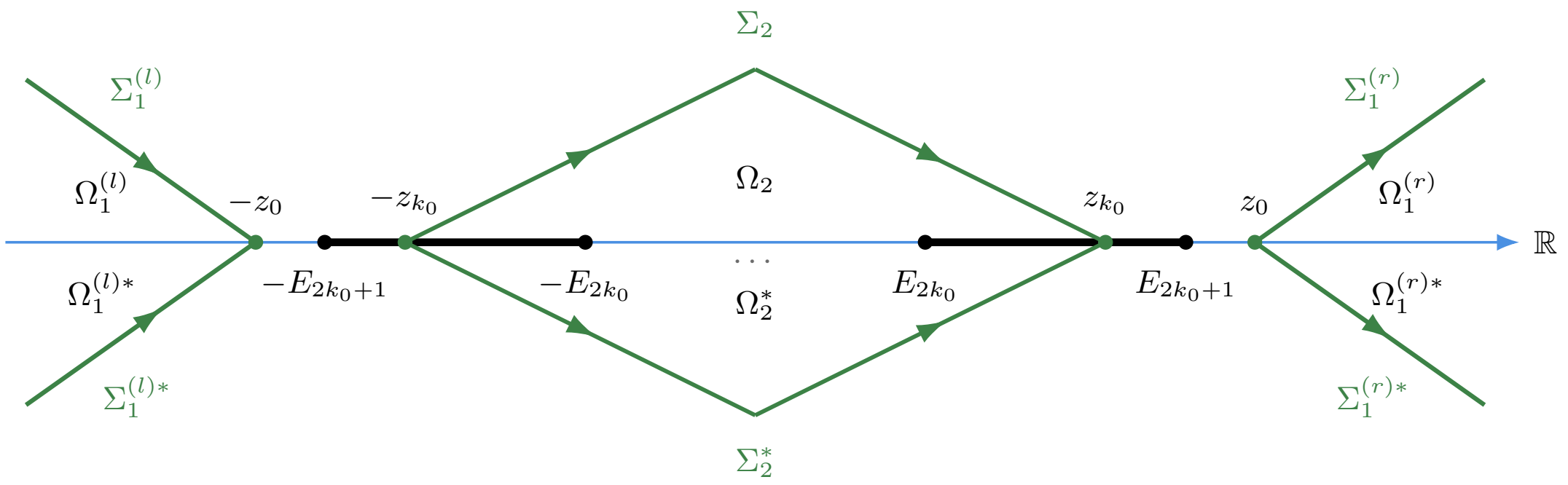

To improve the singularity in the RH problem \ref{RHM}, we  perform a transformation on $M$. We begin by introducing the contours and regions used
to open the lenses. Choose a fixed point $z_{k_0}$ such that $z_{k_0}\in(E_{2k_0},E_{2k_0+1})$. Let $z_0=z_0(\xi)\in [E_{2k_0+1},E_{2k_0+2})$ be the positive saddle point introduced
in Section~\ref{sec25}.
We set
\begin{equation}\label{eq:Sigma1}
	\begin{aligned}
		\Sigma_1
		:=&\,\Sigma_1^{(l)}\cup\Sigma_1^{(r)}\\
	=&\,\{-z_0+s e^{i(\pi-\varphi_0)}:s\geq0\}\cup \{z_0+s e^{i\varphi_0}:s\geq0\},
	\end{aligned}
\end{equation}
and the corresponding upper-half-plane regions are
\begin{equation}\label{eq:Omega1}
	\begin{aligned}
		\Omega_1
		:=&\,\Omega_1^{(l)}\cup\Omega_1^{(r)}\\
	=&\,\{z\in\mathbb C^+:
		\pi-\varphi_0<\arg(z+z_0)<\pi\}\cup\{z\in\mathbb C^+:
		0<\arg(z-z_0)<\varphi_0\},
	\end{aligned}
\end{equation}
where $\varphi_0$ is chosen such that
\begin{equation}\label{eq:inequality1}
		\Im \theta(z) <0, \quad z\in\Sigma_1\setminus\{\pm z_0\}.
\end{equation}
Define the inner contour by
\begin{equation}\label{eq:Sigma2}
	\begin{aligned}
		\Sigma_2
		:={}&\bigl\{-(1-s)z_{k_0}+is\rho_0:
		0\leq s\leq1\bigr\}\cup
		\bigl\{sz_{k_0}+i(1-s)\rho_0:
		0\leq s\leq1\bigr\},
	\end{aligned}
\end{equation}
and let
\begin{equation}\label{eq:Omega2}
	\Omega_2:=
	\left\{
	z\in\mathbb C:
	-z_{k_0}<\operatorname{Re}z<z_{k_0},\
	0<\operatorname{Im}z
	<\rho_0\left(1-\frac{|\operatorname{Re}z|}{z_{k_0}}\right)
	\right\},
\end{equation}
where $\rho_0$ is chosen such that
\begin{equation}\label{eq:inequality2}
	\Im \theta(z) >0, \quad  z\in\Sigma_2\setminus\{\pm z_{k_0}\}.
\end{equation}
For $j=1,2$, define the lower-half-plane contours and
regions by
\[
\Sigma_j^*:=\{\bar z:z\in\Sigma_j\},
\qquad
\Omega_j^*:=\{\bar z:z\in\Omega_j\}.
\]
See Figure \ref{fig:mkdv-region-I-contours} for an illustration.

  Next, we define
\begin{align}\label{M1}
	M^{(1)}(z;\xi,t):=e^{\delta(\infty)\sigma_3}M(z)G(z)e^{-\delta(z)\sigma_3},
\end{align}
where $\delta$ is given in \eqref{delta}, \eqref{dfa} and
\begin{equation}\label{def:G}
	G(z): = G(z;\xi) =\left\{
	\begin{array}{ll}
		\begin{pmatrix}
			1 & 0\\
			r_2(z)e^{-2it\theta(z)} & 1
		\end{pmatrix}, & \hbox{$z\in\Omega_1$,} \\[8pt]
		\begin{pmatrix}
			1 & r_1(z)e^{2it\theta(z)} \\
			0 & 1
		\end{pmatrix}, & \hbox{$z\in\Omega_1^*$,}
		\\[8pt]
		\begin{pmatrix}
			1 &	-\frac{ r_1(z)e^{2it\theta(z)}}{1-r_1(z)r_2(z)}\\
			0 & 1
		\end{pmatrix}, & \hbox{$z\in\Omega_2$,} \\[12pt]
		\begin{pmatrix}
			1 &	0\\
			-\frac{r_2(z)e^{-2it\theta(z)}}{1-r_1(z)r_2(z)} & 1
		\end{pmatrix}, & \hbox{$z\in\Omega_2^*$,}
		\\
		I, & \hbox{elsewhere.}
	\end{array}
	\right.
\end{equation}
Then, from the factorizations \eqref{fac1} and \eqref{fac2} of the RH problem \ref{RHM}, we deduce that $M^{(1)}$ satisfies the following RH problem.
\begin{RHP}	\label{RHPM1}
	\hfil
\begin{itemize}
	\item  $M^{(1)}(z)$ is analytic in $\mathbb{C}\setminus\overline{\Sigma^{(1)}}$, where  \begin{equation}\label{sigma1}
			\Sigma^{(1)}=\Gamma\cup\Sigma_1\cup\Sigma_1^*\cup\Sigma_2\cup\Sigma_2^*\cup (-z_0,-E_{2k_0+1})\cup (E_{2k_0+1},z_0),
		\end{equation}
		see Figure~\ref{fig:mkdv-region-I-contours}.
		\item   $M^{(1)}(z)$ satisfies the jump condition  $$M^{(1)}_{+}(z)=M^{(1)}_{-}(z)V^{(1)}(z),\quad z\in\Sigma^{(1)}.$$
		For $z\in\Gamma$, with $k=1,\dots,n$ and $\delta_0=0$, we have
		\begin{equation}\label{V11}
			V^{(1)}(z)=\left\{
			\begin{aligned}
				&\begin{pmatrix}
					0 & -ie^{i(\phi_k + C_k^f x + C_k^g t+\delta_{n-k})} \\
					-ie^{-i(\phi_k + C_k^f x + C_k^g t+\delta_{n-k})} & 0
				\end{pmatrix},\quad z\in(-E_{2k+1},-E_{2k}),\\
				&\begin{pmatrix}
					0 & -ie^{-i(\phi_k + C_k^f x + C_k^g t-\delta_{n+k})} \\
					-ie^{i(\phi_k + C_k^f x + C_k^g t-\delta_{n+k})} & 0
				\end{pmatrix},\quad z\in(E_{2k},E_{2k+1}).			
			\end{aligned}
			\right.
		\end{equation}
			Specifically, for genus $2n$,
		$$V^{(1)}(z)=\begin{pmatrix}0&-ie^{i\delta_{n}}\\-ie^{-i\delta_{n}}&0\end{pmatrix},\quad z\in(-E_{1},E_{1}). $$
		For $z\in \Sigma^{(1)}\backslash\Gamma$,
			\begin{align}\label{V1}
			V^{(1)}(z)=	\left\{ \begin{array}{ll}
				\begin{pmatrix}
					1 & 0\\
					-r_2(z)e^{-2it\theta(z)-2\delta(z)} & 1	
				\end{pmatrix},&  z\in\Sigma_1,\\[9pt]
				\begin{pmatrix}
					1 & r_1(z)e^{2it\theta(z)+2\delta(z)} \\
					0 & 1
				\end{pmatrix}, &z\in\Sigma_1^*,\\[9pt]
				\begin{pmatrix}
					1 &	\frac{ r_1(z)e^{2it\theta(z)+2\delta(z)}}{1-r_1(z)r_2(z)}\\
					0 & 1
				\end{pmatrix},&z\in\Sigma_2,\\[12pt]
				\begin{pmatrix}
					1 &	0\\
					-\frac{r_2(z)e^{-2it\theta(z)-2\delta(z)}}{1-r_1(z)r_2(z)} & 1
				\end{pmatrix},&z\in\Sigma_2^*,\\[12pt]
				\begin{pmatrix}
					1-r_1(z)r_2(z) & r_1(z)e^{2it\theta(z)+2\delta(z)}\\
					-r_2(z)e^{-2it\theta(z)-2\delta(z)} & 1
				\end{pmatrix},&z\in (-z_0,-E_{2k_0+1})\cup (E_{2k_0+1},z_0).\\[12pt]
			\end{array}\right.
		\end{align}	

			\item $M^{(1)}(z)$ satisfies the symmetry relations
		\begin{equation*}
			M^{(1)}(z)=\sigma_1\overline{M^{(1)}(\bar{z})}\sigma_1=e^{2\delta(\infty)\sigma_3}
\sigma_2M^{(1)}(-z)\sigma_2
e^{-2\delta(\infty)\sigma_3}.
		\end{equation*}
		\item  As $z\to\infty$, we have $$M^{(1)}(z)=I+\mathcal{O}(z^{-1}).$$
		\item  As  $ z\to p\in\boldsymbol{E}\setminus\{\pm E_{2k_0+1}\}$, we have	$	M^{(1)}(z)=\mathcal{O}\big((z-p)^{-\frac{1}{4}}\big).$
	\end{itemize}
\end{RHP}

\begin{figure}
	\centering
	\definecolor{mkdvAxisBlue}{RGB}{74,144,226}
	\definecolor{mkdvLensGreen}{RGB}{60,130,70}
	\resizebox{0.90\linewidth}{!}{%
		\begin{tikzpicture}[
			x=1cm,y=1cm,>=Latex,
			axis/.style={draw=mkdvAxisBlue,line width=0.72pt},
			band/.style={draw=black,line width=2.15pt},
			outer/.style={
				draw=mkdvLensGreen,line width=1.25pt,
				postaction={decorate},
				decoration={markings,mark=at position 0.60 with
					{\arrow{Latex[length=3.0mm,width=1.9mm]}}}
			},
			localcircle/.style={draw=mkdvLensGreen,line width=1.05pt},
			diskarrow/.style={
				draw=mkdvLensGreen,line width=1.25pt,
				-{Latex[length=3.0mm,width=1.9mm]}
			},
			inner/.style={draw=mkdvLensGreen,line width=1.15pt},
			innerarrow/.style={
				inner,postaction={decorate},
				decoration={markings,mark=at position 0.55 with
					{\arrow{Latex[length=3.0mm,width=1.9mm]}}}
			},
			point/.style={circle,fill=black,inner sep=1.45pt},
			every node/.style={font=\footnotesize}
			]
			\def\zouter{4.53}
			\def\Eodd{4.18}
			\def\zsaddle{2.70}
			\def\Eeven{1.65}
			\def\rho{1.68}
			\def\rayx{7.08}
			\def\rayy{1.58}
			\def\locR{0.84}
			
			% Real axis and spectral bands.
			\draw[axis,->] (-7.28,0)--(7.42,0) node[right] {$\mathbb R$};
			\draw[band] (-\Eodd,0)--(-\Eeven,0);
			\draw[band] ( \Eeven,0)--( \Eodd,0);
			\node[black!58] at (0,-0.19) {$\cdots$};
			
			% Outer contours.
			\draw[outer] (-\rayx, \rayy)--(-\zouter,0);
			\draw[outer] ( \zouter,0)--( \rayx, \rayy);
			\draw[outer] (-\rayx,-\rayy)--(-\zouter,0);
			\draw[outer] ( \zouter,0)--( \rayx,-\rayy);
			
			% Inner contours.
			\draw[innerarrow] (-\zsaddle,0)--(0, \rho);
			\draw[innerarrow] (0, \rho)--(\zsaddle,0);
			\draw[innerarrow] (-\zsaddle,0)--(0,-\rho);
			\draw[innerarrow] (0,-\rho)--(\zsaddle,0);

 % Clockwise local circles.
 \foreach \xc in {-\Eodd,\Eodd}{
 	\begin{scope}[shift={(\xc,0)}]
 		\draw[localcircle] (0,0) circle[radius=\locR];
 		
 		% Right-pointing arrowhead at the top of the circle.
 		\coordinate (arrowbase) at (100:\locR);
 		\fill[mkdvLensGreen]
 		(80:\locR)
 		-- ([yshift=1.1mm]arrowbase)
 		-- ([yshift=-1.1mm]arrowbase)
 		-- cycle;
 	\end{scope}
 }

			\node[mkdvLensGreen] at (-\Eodd,1.3) {$\partial U^{(l)}$};
			\node[mkdvLensGreen] at ( \Eodd,1.3) {$\partial U^{(r)}$};
			
			% Points.
			\foreach \x in {-\Eodd,-\Eeven,\Eeven,\Eodd}
			\node[point] at (\x,0) {};
			\node[point,fill=mkdvLensGreen] at (-\zouter,0) {};
			\node[point,fill=mkdvLensGreen] at ( \zouter,0) {};
			\node[point,fill=mkdvLensGreen] at (-\zsaddle,0) {};
			\node[point,fill=mkdvLensGreen] at ( \zsaddle,0) {};
			
			% Point labels.
			\node[above=5pt,xshift=-3pt] at (-\zouter,0) {$-z_0$};
			\node[above=5pt,xshift=3pt]  at ( \zouter,0) {$z_0$};
			\node[below=4pt] at (-\Eodd,0) {$-E_{2k_0+1}$};
			\node[below=4pt] at ( \Eodd,0) {$E_{2k_0+1}$};
			\node[above=4pt] at (-\zsaddle,0) {$-z_{k_0}$};
			\node[above=4pt] at ( \zsaddle,0) {$z_{k_0}$};
			\node[below=3.5pt] at (-\Eeven,0) {$-E_{2k_0}$};
			\node[below=3.5pt] at ( \Eeven,0) {$E_{2k_0}$};
			
			% Contour labels.
			\node[mkdvLensGreen] at (-6.42, 1.79) {$\Sigma_1^{(l)}$};
			\node[mkdvLensGreen] at ( 6.42, 1.79) {$\Sigma_1^{(r)}$};
			\node[mkdvLensGreen] at (-6.42,-1.79) {$\Sigma_1^{(l)*}$};
			\node[mkdvLensGreen] at ( 6.42,-1.79) {$\Sigma_1^{(r)*}$};
			\node[mkdvLensGreen,above=5pt] at (0, \rho) {$\Sigma_2$};
			\node[mkdvLensGreen,below=5pt] at (0,-\rho) {$\Sigma_2^*$};
			
			% Region labels.
			\node[black] at (-6.34, 0.49) {$\Omega_1^{(l)}$};
			\node[black] at ( 6.34, 0.49) {$\Omega_1^{(r)}$};
			\node[black] at (-6.34,-0.49) {$\Omega_1^{(l)*}$};
			\node[black] at ( 6.34,-0.49) {$\Omega_1^{(r)*}$};
			\node[black] at (0, 0.61) {$\Omega_2$};
			\node[black] at (0,-0.61) {$\Omega_2^*$};
		\end{tikzpicture}%
	}
	\caption{Lens contours and the boundaries of the local disks
		$U^{(l)}$ and $U^{(r)}$, centered at $-E_{2k_0+1}$ and
		$E_{2k_0+1}$, respectively, in transition region~$\mathcal{T}_{\mathrm{I}}$.}
	\label{fig:mkdv-region-I-local-disks}
\end{figure}
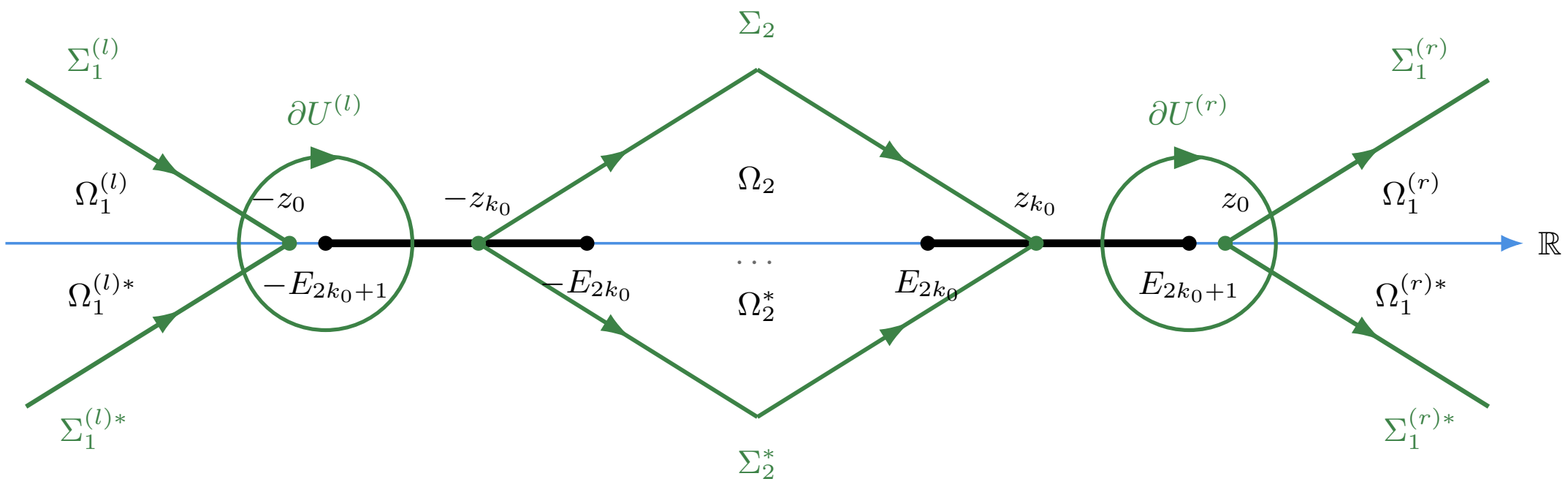

To analyze $M^{(1)}$, we introduce an approximation $M^{(2)}$ satisfying
\begin{equation}\label{a2e}
M^{(1)}(z)=E(z)M^{(2)}(z),
\end{equation}
where $E(z)$ is the error function.  The matrix
$M^{(2)}$ is constructed from a global parametrix  $M^{(glo)}$ and local parametrices $M^{(l)}$,  $M^{(r)}$ as follows:
\begin{equation}\label{m2}
M^{(2)}(z)=\left\{\begin{aligned}
&M^{(glo)}(z),\quad\ \, z\in\mathbb{C}\setminus(U^{(l)}\cup U^{(r)}), \\
&M^{(l)}(z),\quad\quad \, z\in U^{(l)},\\
&M^{(r)}(z),\quad\quad  z\in U^{(r)},
\end{aligned}\right.
\end{equation}
where
\begin{equation}\label{ulr}
	U^{(l)}=\{z\in\mathbb{C}: |z+E_{2k_{0}+1}|\leq c_{0} \}, \qquad
	U^{(r)}=\{z\in\mathbb{C}: |z-E_{2k_{0}+1}|\leq c_{0} \},
\end{equation}
and
\begin{align}\label{c0}
	c_0=\min\left\{\frac{1}{2}(E_{2k_0+1}-z_{k_0}),\ \frac{1}{2}(E_{2k_0+2}-E_{2k_0+1}),\ t^{-\frac{2}{3}+\epsilon}\right\},\quad\epsilon\in \Big(\frac{1}{3},\frac{2}{3}\Big).
\end{align}
Note that as $t\to \infty$, we have
\begin{align}
	z_0-E_{2k_0+1}=\mathcal{O}\big(t^{-\frac{2}{3}}\big).
\end{align}
Thus, $U^{(l)}$ and $U^{(r)}$ are disjoint shrinking disks,
with $-z_0$ and $z_0$ lying in their respective interiors. See Figure \ref{fig:mkdv-region-I-local-disks} for an illustration.
The construction of the global parametrix $M^{(glo)}$,
the local parametrices $M^{(l)}$, $M^{(r)}$, and the error
function $E(z)$ is carried out in Sections~\ref{global}--\ref{error} below.

\subsection{Global parametrix}\label{global}
We consider the following global parametrix $M^{(glo)}$.
\begin{RHP}\label{RHglo}
	\hfill
	\begin{itemize}
		\item $M^{(glo)}(z)$ is analytic in $\mathbb{C}\setminus\overline{\Gamma}$, where $\Gamma$ is given by \eqref{gamma}.
		\item  $M^{(glo)}(z)$  satisfies the jump condition
		\begin{equation}
			M^{(glo)}_{+}(z)=M^{(glo)}_{-}(z)V^{(glo)}(z),\quad z\in\Gamma,
		\end{equation}
		where
		\begin{equation}\label{Vglo}
			V^{(glo)}(z)=\left\{
			\begin{aligned}
				&\begin{pmatrix}
					0 & -ie^{i(\phi_k + C_k^f x + C_k^g t+\delta_{n-k})} \\
					-ie^{-i(\phi_k + C_k^f x + C_k^g t+\delta_{n-k})} & 0
				\end{pmatrix},\quad z\in(-E_{2k+1},-E_{2k}),\\
				&\begin{pmatrix}
					0 & -ie^{-i(\phi_k + C_k^f x + C_k^g t-\delta_{n+k})} \\
					-ie^{i(\phi_k + C_k^f x + C_k^g t-\delta_{n+k})} & 0
				\end{pmatrix},\quad z\in(E_{2k},E_{2k+1}),		
			\end{aligned}
			\right.
		\end{equation}
			with $k=1,\dots,n$ and $\delta_0=0$. 	Specifically, for genus $2n$,
			$$	V^{(glo)}(z)=\begin{pmatrix}0&-ie^{i\delta_{n}}\\-ie^{-i\delta_{n}}&0\end{pmatrix},\quad  z\in(-E_{1},E_{1}).$$
		\item $M^{(glo)}(z)$ satisfies the symmetry relations
		\begin{align}\label{sym:mglo}
		    M^{(glo)}(z)
		=\sigma_1\overline{M^{(glo)}(\bar z)}\sigma_1
		=e^{2\delta(\infty)\sigma_3}
		\sigma_2M^{(glo)}(-z)\sigma_2
		e^{-2\delta(\infty)\sigma_3}.
		\end{align}

		\item As $z\to\infty$, we have $$M^{(glo)}(z)=I+\mathcal{O}(z^{-1}).$$
		\item As $z\to p\in\boldsymbol{E}$, we have
		\begin{align}\label{gloas}
			M^{(glo)}(z)=\mathcal{O}\big((z-p)^{-\frac{1}{4}}\big).
		\end{align}
	\end{itemize}
\end{RHP}

Analogously to RH problem \ref{RHmu}, the solution of the RH problem \ref{RHglo} can be expressed as follows.

Invoking \eqref{equ:deletajdelta2n}-\eqref{equ:deltainfty}, we can define a shifted phase vector by
\begin{equation}\label{phi}
\begin{aligned}
    & \widetilde{\boldsymbol{\phi}}
 =(\widetilde{\phi}_1,\ldots,\widetilde{\phi}_n)^T,\\& \widetilde{\phi}_k
 =\phi_k+\delta_{n-k}+2i\delta(\infty)=\phi_k+\delta_{n-k}-\frac{\delta_{2n}}{2},
 \qquad k=1,\ldots,n.
\end{aligned}
 \end{equation}
 Since the constants \(\delta_j\) are real, \(\widetilde{\boldsymbol\phi}\in\mathbb R^n\).
Let \(\widetilde\mu^{(\mathrm{AG})}(z;x,t;\widetilde{\boldsymbol\phi})\) denote the solution of RH Problem \ref{RHmu} with \(\boldsymbol\phi\) replaced by \(\widetilde{\boldsymbol\phi}\). The explicit expression for this solution is given in  \eqref{muag}.
Comparing RH problem \ref{RHglo} with RH problem~\ref{RHmu}, we obtain
\begin{align}\label{def:Mglo}
    M^{(glo)}(z)
=e^{\delta(\infty)\sigma_3}
\widetilde{\mu}^{(AG)}(z;x,t)
e^{-\delta(\infty)\sigma_3}.
\end{align}
Consequently, if we write
\begin{equation}\label{asm2}
M^{(\mathrm{glo})}(z)
=
I+\frac{M_1^{(\mathrm{glo})}}{z}+\mathcal{O}(z^{-2}),\qquad z\to\infty,
\end{equation}
then we obtain
\begin{equation}\label{reco1}
		q^{(AG)}(x,t;\boldsymbol{E},\widetilde{\boldsymbol{\phi}})
	=e^{-2\delta(\infty)}
	\lim_{z\to\infty}2izM_{12}^{(glo)}(z),
\end{equation}
where $q^{(AG)}$ is given by \eqref{qAG} with \(\boldsymbol\phi\) replaced by \(\widetilde{\boldsymbol\phi}\).

\subsection{Local parametrices}\label{local}
In this section, we construct the local parametrices
$M^{(l)}$ and $M^{(r)}$ defined in \eqref{m2}. By the symmetry of the transformed RH problem $M^{(1)}$, the left local parametrix is obtained from the right one through
\begin{align}\label{sym:MrandMl}
    M^{(l)}(z)
=e^{2\delta(\infty)\sigma_3}
\sigma_2M^{(r)}(-z)\sigma_2
e^{-2\delta(\infty)\sigma_3},
\qquad z\in U^{(l)}\setminus\Sigma^{(1)}.
\end{align}
It therefore suffices to construct $M^{(r)}$
near $E_{2k_0+1}$.
\subsubsection{Local transformation from the \(z\)-plane to the \(\zeta\)-plane}
The right local parametrix  $M^{(r)}$ is constructed as follows.
\begin{RHP}\label{rhloc}
	\hfil
	\begin{itemize}
		\item  $M^{(r)}(z)$ is analytic in $U^{(r)}\setminus\Sigma^{(1)}$, where $U^{(r)}$ and $\Sigma^{(1)}$ are given by \eqref{ulr} and \eqref{sigma1}, respectively.
		\item  $M^{(r)}(z)$  satisfies the jump condition
		$$M^{(r)}_{+}(z)=M^{(r)}_{-}(z)V^{(1)}(z),\quad z\in\Sigma^{(1)}\cap U^{(r)},$$
		where $V^{(1)}$ is given by \eqref{V11}--\eqref{V1}.
		\item $M^{(r)}(z)$ has at most $-\frac{1}{4}$-singularity at $E_{2k_0+1}$.
		\item  As $t\to\infty$, $M^{(r)}(z)$ matches $M^{(glo)}(z)$ on $\partial U^{(r)}$.
	\end{itemize}
\end{RHP}
The phase function $\theta$ in the jump matrix $V^{(1)}$ plays a fundamental role, especially concerning its asymptotic behavior at $z=E_{2k_0+1}$. Accordingly, we first establish the following proposition, whose proof is similar to the one in \cite{NLSAG}.
	\begin{proposition}
	For any fixed $\xi$, as  $z\to E_{2k_0+1}$,  we have
	\begin{align}\label{alas}
		\theta(z;\xi)=\theta^{(0)}(\xi)+\theta^{(1)}(\xi)(\xi-\xi_{2k_0+1})(z-E_{2k_0+1})^{\frac{1}{2}}+\frac{2}{3}
		\theta^{(3)}(\xi)(z-E_{2k_0+1})^{\frac{3}{2}}+\mathcal{O}\left((z-E_{2k_0+1})^{\frac{5}{2}}\right),
	\end{align}
	where
\begin{align}
	\theta^{(0)}(\xi)
	&= \theta(E_{2k_0+1};\xi)=-\frac{1}{2}C_{k_0}^f \xi-\frac{1}{2}C_{k_0}^g,
	\label{theta0}
	\\[2mm]
	\theta^{(1)}(\xi)
	&=
	\begin{cases}
		\displaystyle
		-\frac{
			2E_{2k_0+1}
			\prod_{k=1}^{n}\bigl(E_{2k_0+1}^{2}-(z_k^f)^2\bigr)
		}{
			w^{(2k_0+1)}(E_{2k_0+1})
		},
		& \text{genus }2n,\\[3mm]
		\displaystyle
		-\frac{
			2
			\prod_{k=1}^{n}\bigl(E_{2k_0+1}^{2}-(z_k^f)^2\bigr)
		}{
			w^{(2k_0+1)}(E_{2k_0+1})
		},
		& \text{genus }2n-1,
	\end{cases}
	\label{theta1}
	\\[2mm]
	\theta^{(3)}(\xi)
	&=
	\begin{cases}
		\displaystyle
		-\left.
		\frac{d}{dz}
		\left(
		\frac{zF(z;\xi)}{w^{(2k_0+1)}(z)}
		\right)
		\right|_{z=E_{2k_0+1}},
		& \text{genus }2n,\\[3mm]
		\displaystyle
		-\left.
		\frac{d}{dz}
		\left(
		\frac{F(z;\xi)}{w^{(2k_0+1)}(z)}
		\right)
		\right|_{z=E_{2k_0+1}},
		& \text{genus }2n-1,
	\end{cases}
	\label{theta3}
\end{align}
	with
	\begin{equation}
	w^{(2k_0+1)}(z)=\left\{
		\begin{aligned}
			&\left[\prod_{j=1}^{2n+1}(z+E_{j})\prod_{\substack{1\le j\le 2n+1\\j\ne 2k_0+1}}(z-E_{j})\right]^{\frac{1}{2}},\quad \text{genus}\   2n,\\
			&\left[\prod_{j=2}^{2n+1}(z+E_{j})\prod_{\substack{2\le j\le 2n+1\\j\ne 2k_0+1}}(z-E_{j})\right]^{\frac{1}{2}},\quad \text{genus}\ 2n-1.
		\end{aligned}
		\right.
	\end{equation}
	The branch of $w^{(2k_0+1)}$ is chosen to be analytic
	near $E_{2k_0+1}$ and to satisfy
	\[
	w(z)=(z-E_{2k_0+1})^{\frac{1}{2}}w^{(2k_0+1)}(z).
	\]
	Here $\theta$ and $F$ are given by  \eqref{alpha} and \eqref{F}, respectively.  Furthermore, $\theta^{(3)}(\xi)<0$ holds in a sufficiently small neighborhood of $\xi_{2k_0+1}$.
\end{proposition}

Now, based on the expansion \eqref{alas}, we introduce a new variable $\zeta$ and formulate the RH problem on the $\zeta$-plane. Define
\begin{align}\label{zeta}
	\zeta(z):=\zeta(z;\xi)=
	\left[\frac{3it}{2}\left(\theta^{(0)}(\xi)+\theta^{(1)}(\xi)(\xi-\xi_{2k_0+1})(z-E_{2k_0+1})^{\frac{1}{2}}-
	\theta(z;\xi)\right)\right]^{\frac{2}{3}}, \quad  z\in U^{(r)}.
\end{align}
The branch is chosen so that $\zeta$ is analytic near
$E_{2k_0+1}$ and
\begin{align}\label{zeta1}
	\zeta(E_{2k_0+1})=0,\qquad \zeta'(E_{2k_0+1})=-|\theta^{(3)}(\xi)|^{\frac{2}{3}}t^{\frac{2}{3}}<0.
\end{align}
Taking the principal branches of
$(z-E_{2k_0+1})^{\frac{1}{2}}$ and $\zeta^{\frac{1}{2}}$, we have
\begin{align*}
	\zeta(z)^{\frac{3}{2}}=\left\{\begin{aligned}
		&-it\theta^{(3)}(\xi)(z-E_{2k_0+1})^{\frac{3}{2}}\big(1+\mathcal{O}(z-E_{2k_0+1})\big),\quad z\to E_{2k_0+1}~\ \textrm{from}~\ \mathbb{C}^+,\\
		&\ it\theta^{(3)}(\xi)(z-E_{2k_0+1})^{\frac{3}{2}}\big(1+\mathcal{O}(z-E_{2k_0+1})\big) ,\quad\ \ \, z\to E_{2k_0+1}\ ~\textrm{from}~\ \mathbb{C}^-,
	\end{aligned}\right.
\end{align*}
and
\begin{align*}
	\zeta(z)^{\frac{1}{2}}=\left\{\begin{aligned}
		&-it^{\frac{1}{3}}|\theta^{(3)}(\xi)|^{\frac{1}{3}}(z-E_{2k_0+1})^{\frac{1}{2}}\big(1+\mathcal{O}(z-E_{2k_0+1})\big),\quad z\to E_{2k_0+1}~\ \textrm{from}~\ \mathbb{C}^+,\\
		&\ it^{\frac{1}{3}}|\theta^{(3)}(\xi)|^{\frac{1}{3}}(z-E_{2k_0+1})^{\frac{1}{2}}\big(1+\mathcal{O}(z-E_{2k_0+1})\big),\quad\ \ \, z\to E_{2k_0+1}\ ~\textrm{from}~\ \mathbb{C}^-.
	\end{aligned}\right.
\end{align*}

Consequently, we are led to define
\begin{align}\label{L}
	S(z;\xi):=\left\{\begin{aligned}
		&-\frac{it\theta^{(1)}(\xi)(\xi-\xi_{2k_0+1})(z-E_{2k_0+1})^{\frac{1}{2}}}{\zeta(z)^{\frac{1}{2}}},\quad z\in \mathbb{C}^+\cap U^{(r)},\\
		&\frac{it\theta^{(1)}(\xi)(\xi-\xi_{2k_0+1})(z-E_{2k_0+1})^{\frac{1}{2}}}{\zeta(z)^{\frac{1}{2}}},\quad\quad \  z\in \mathbb{C}^-\cap U^{(r)}.
	\end{aligned}\right.
\end{align}
Observe that for sufficiently large positive $t$, the function $S$ extends analytically to $U^{(r)}$ and
\begin{align}\label{s1}
	s:=S(E_{2k_0+1};\xi)=\frac{\theta^{(1)}(\xi)(\xi-\xi_{2k_0+1})t^{\frac{2}{3}} }{|\theta^{(3)}(\xi)|^{\frac{1}{3}}}\in \mathbb{R}.
\end{align}
Based on the definitions of $\zeta$ and $S$ in \eqref{zeta}, \eqref{L} and expansion  \eqref{alas}, it follows directly that
\begin{align}\label{thetaze}
	it\theta(z;\xi)=\left\{ \begin{aligned}
		&it\theta^{(0)}(\xi)-S(z;\xi)\zeta(z)^{\frac{1}{2}}-\frac{2}{3}\zeta(z)^{\frac{3}{2}}, \quad \ z\in \mathbb{C}^+\cap U^{(r)},
		\\
		&it\theta^{(0)}(\xi)+S(z;\xi)\zeta(z)^{\frac{1}{2}}+\frac{2}{3}\zeta(z)^{\frac{3}{2}}, \quad \ z\in \mathbb{C}^-\cap U^{(r)}.
	\end{aligned}\right.
\end{align}

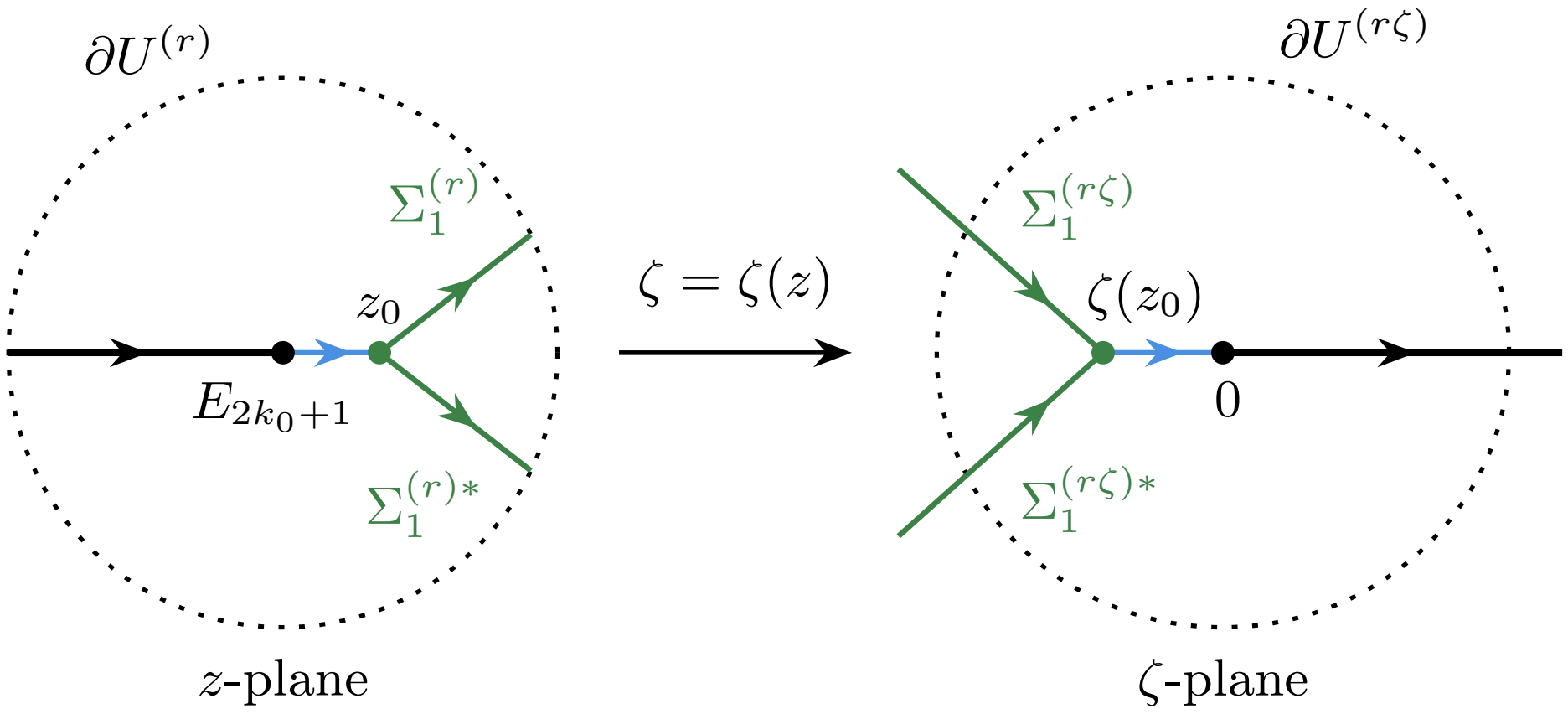
\begin{figure}
	\centering
	\definecolor{mkdvAxisBlue}{RGB}{74,144,226}
	\definecolor{mkdvLensGreen}{RGB}{60,130,70}
	\resizebox{0.78\linewidth}{!}{%
		\begin{tikzpicture}[
			x=1cm,y=1cm,
			oriented/.style={
				postaction={decorate},
				decoration={
					markings,
					mark=at position #1 with
					{\arrow{Stealth[length=2.3mm,width=1.7mm]}}
				}
			},
			band/.style={
				draw=black,line width=1.2pt
			},
			gap/.style={
				draw=mkdvAxisBlue,line width=0.9pt
			},
			lens/.style={
				draw=mkdvLensGreen,line width=0.95pt
			},
			boundary/.style={
				draw=black,
				line width=0.75pt,
				dash pattern=on 1pt off 2.5pt
			},
			every node/.style={
				font=\footnotesize,inner sep=2pt
			}
			]
			
			% Left panel: the z-plane.
			\begin{scope}
				\node at (0,-1.98) {$z$-plane};
				
				\draw[boundary] (0,0) circle (1.65);
				\node[black,anchor=south] at (-0.8,1.6)
				{$\partial U^{(r)}$};
				
				% Real contours, oriented to the right.
				\draw[band,oriented=0.51] (-1.65,0)--(0,0);
				\draw[gap,oriented=0.72] (0,0)--(0.58,0);
				
				% Nonreal contours, oriented away from z_0.
				\draw[lens,oriented=0.64] (0.58,0)--(1.49, 0.709);
				\draw[lens,oriented=0.64] (0.58,0)--(1.49,-0.709);
				
				% Distinguished points.
				\fill[black] (0,0) circle (2pt);
				\fill[mkdvLensGreen] (0.58,0) circle (2pt);
				
				\node[anchor=north] at (-0.06,-0.10) {$E_{2k_0+1}$};
				\node[anchor=south] at (0.58,0.10) {$z_0$};
				
				% Contour labels.
				\node[mkdvLensGreen,anchor=south east] at (1.29,0.63)
				{$\Sigma_1^{(r)}$};
				\node[mkdvLensGreen,anchor=north east] at (1.29,-0.63)
				{$\Sigma_1^{(r)*}$};
			\end{scope}
			
			% Coordinate transformation.
			\draw[-{Stealth[length=2.3mm]},line width=0.8pt]
			(2.02,0)--node[above=5pt] {$\zeta=\zeta(z)$} (3.42,0);
			
			% Right panel: the zeta-plane.
			\begin{scope}[shift={(5.65,0)}]
				\node at (0,-1.98) {$\zeta$-plane};
				
				% Image of the local-disk boundary.
				\draw[boundary] (0,0) ellipse (1.72 and 1.65);
				\node[black,anchor=south] at (0.8,1.7)
				{$\partial U^{(r\zeta)}$};
				
				% Real contours, oriented to the right.
				\draw[band,oriented=0.57] (0,0)--(2.04,0);
				\draw[gap,oriented=0.67] (-0.72,0)--(0,0);
				
				% Straight extensions outside the local region.
				\draw[lens] (-1.54, 0.735)--(-1.95, 1.1025);
				\draw[lens] (-1.54,-0.735)--(-1.95,-1.1025);
				
				% Local images, oriented towards zeta(z_0).
				\draw[lens,oriented=0.62] (-1.54,-0.735)--(-0.72,0);
				\draw[lens,oriented=0.62] (-1.54, 0.735)--(-0.72,0);
				
				% Distinguished points.
				\fill[black] (0,0) circle (2pt);
				\fill[mkdvLensGreen] (-0.72,0) circle (2pt);
				
				\node[anchor=north] at (0.03,-0.10) {$0$};
				\node[anchor=south west] at (-0.9,0.1) {$\zeta(z_0)$};
				
				% Contour labels.
				\node[mkdvLensGreen,anchor=south west] at (-1.3,0.6)
				{$\Sigma_1^{(r\zeta)}$};
				\node[mkdvLensGreen,anchor=north west] at (-1.3,-0.6)
				{$\Sigma_1^{(r\zeta)*}$};
			\end{scope}
			
		\end{tikzpicture}%
	}
	\caption{Local contours in the right disk $U^{(r)}$
		and the corresponding extended contours in the $\zeta$-plane.}
	\label{fig:mkdv-local-contours}
\end{figure}

Following the discussion above, we transfer the problem from the $z$-plane to the $\zeta$-plane. We begin by defining
\begin{align}
	U^{(r\zeta)}&:= \zeta(U^{(r)}), \label{ulzeta} \\
	\Sigma_1^{(r\zeta)}&:=
	\zeta\bigl(\Sigma_1^{(r)*}\cap U^{(r)}\bigr)
	\cup
	\left\{
	\zeta\bigl(\Sigma_1^{(r)*}\cap\partial U^{(r)}\bigr)
	+e^{i(\pi-\varphi_0)}\mathbb R^+
	\right\},
	\\
\Sigma^{(r\zeta)}&:=
(\zeta(z_0),+\infty)
\cup\Sigma_1^{(r\zeta)}
\cup\Sigma_1^{(r\zeta)*}.\label{lzeta}
\end{align}
The real contour in the $\zeta$-plane is oriented from left
to right, while the nonreal contours are oriented towards
$\zeta(z_0)$; see Figure~\ref{fig:mkdv-local-contours}.  Then, we define  $V^{(r)}(\zeta)$ by 
{\small\begin{align}\label{vl}
	V^{(r)}(\zeta) :=	\left\{ \begin{array}{ll}
		\begin{pmatrix}
			0 & ie^{-i(\phi_{k_{0}} + C_{k_{0}}^f x + C_{k_{0}}^g t-\delta_{n+k_{0}})} \\
			ie^{i(\phi_{k_{0}} + C_{k_{0}}^f x + C_{k_{0}}^g t-\delta_{n+k_{0}})} & 0
		\end{pmatrix} ,& \zeta\in \mathbb{R}^+,\\[12pt]
		\begin{pmatrix}
			1 & -r_1(z(\zeta))e^{2it\theta(z(\zeta))+2\delta(z(\zeta))} \\
			0 & 1
		\end{pmatrix} ,&  \zeta\in\Sigma_1^{(r\zeta)}\cap U^{(r\zeta)},\\[12pt]
		\begin{pmatrix}
			1 & 0\\
			r_2(z(\zeta))e^{-2it\theta(z(\zeta))-2\delta(z(\zeta))} & 1	
		\end{pmatrix}, &\zeta\in\Sigma_1^{(r\zeta)*}\cap U^{(r\zeta)},\\[12pt]
		\begin{pmatrix}
			1 & -r_1(z(\zeta))e^{2it\theta(z(\zeta))+2\delta(z(\zeta))}\\
			r_2(z(\zeta))e^{-2it\theta(z(\zeta))-2\delta(z(\zeta))} & 1	-r_1(z(\zeta))r_2(z(\zeta))
		\end{pmatrix}, & \zeta\in (\zeta(z_0),0),\\[12pt]
		I,&\zeta\in \big(\Sigma_1^{(r\zeta)}\cup\Sigma_1^{(r\zeta)*}\big)\setminus U^{(r\zeta)},
	\end{array}\right.
\end{align}}
which satisfies $V^{(r)}(\zeta(z))=[V^{(1)}(z)]^{-1}$ for $z\in\Sigma^{(1)}\cap U^{(r)}$, where $V^{(1)}(z)$ and $\Sigma^{(1)}$  are given by \eqref{V11}--\eqref{V1} and \eqref{sigma1}, respectively. Additionally, $V^{(r)}$ is readily seen to obey a factorization relation
\begin{equation}\label{facvl}
 V^{(r)}(\zeta)=\big[V_{-}^{(r)}(\zeta)\big]^{-1} V_{+}^{(r)}(\zeta),
\end{equation}
where
\begin{align}
	&V_{+}^{(r)}(\zeta)=\left\{ \begin{array}{ll}
		V^{(r)}(\zeta), & \zeta\in \mathbb{R}^+,\\
		\begin{pmatrix}
			1 & -r_1(z(\zeta))e^{2it\theta(z(\zeta))+2\delta(z(\zeta))} \\
			0 & 1
		\end{pmatrix},&  \zeta\in(\Sigma_1^{(r\zeta)}\cap U^{( r\zeta )})\cup(\zeta(z_0),0),\\
		I, & \text{elsewhere},
	\end{array}\right. \label{V+1}
	\\[4pt]
	&V_{-}^{(r)}(\zeta)=\left\{ \begin{array}{ll}
		I, & \zeta\in \mathbb{R}^+,\\
		\begin{pmatrix}
			1 & 0\\
			-r_2(z(\zeta))e^{-2it\theta(z(\zeta))-2\delta(z(\zeta))} & 1	
		\end{pmatrix},&  \zeta\in(\Sigma_1^{(r\zeta)*} \cap U^{( r\zeta )})\cup(\zeta(z_0),0),\\
		I, & \text{elsewhere}.
	\end{array}\right.
	\label{V-1}
\end{align}

\subsubsection{Construction of the model RH problem \(M^{(r,1)}\)}\label{sec332}
Next, we construct a new jump matrix  $V^{(r,1)}(\zeta)$ on  $\Sigma^{(r\zeta)}$, such that as  $t\to\infty$, $V^{(r)}(\zeta)$ is well-approximated by $V^{(r,1)}(\zeta)$ for $\zeta \in \Sigma_1^{(r\zeta)}\cap U^{(r\zeta)}$. Recalling the properties of  $r_{1}$ stated in  \eqref{eq:riloc}, we have, for large $t$,
 \begin{equation}\label{lin1}
 	r_1(z)=e^{-i\phi_{k_0}}+\mathcal{O}((z-E_{2k_0+1})^{\frac{1}{2}}),\qquad  z\in U^{(r)} \setminus (-\infty,E_{2k_0+1}].
 \end{equation}
Analogously, from the properties of   $\delta$ in \eqref{delp} it follows that
 \begin{equation}\label{lin2}
 	\delta(z)=\frac{i}{2}\delta_{n+k_0}+\mathcal{O}((z-E_{2k_0+1})^{\frac{1}{2}}),\qquad  z\in U^{(r)} \setminus (-\infty,E_{2k_0+1}].
 \end{equation}
From \eqref{ulr}--\eqref{c0} we find that, as $t\to\infty$,
 \begin{equation}\label{lin3}
 	z-E_{2k_0+1}=\mathcal{O}(t^{\epsilon-\frac{2}{3}}), \qquad z\in U^{(r)},
 \end{equation}
and then substituting \eqref{lin3} into \eqref{lin1}--\eqref{lin2} gives
 \begin{align}
 	r_1(z)&=e^{-i\phi_{k_0}}+\mathcal{O}(t^{\frac{\epsilon}{2}-\frac{1}{3}}),\label{rsda}\\
 	\delta(z)&=\frac{i}{2}\delta_{n+k_0}+\mathcal{O}(t^{\frac{\epsilon}{2}-\frac{1}{3}}).\label{rsdb}
 \end{align}
 Using the symmetry relation  between $r_1$ and $r_2$
 in Proposition \ref{propsca} (iv), we obtain
 \begin{equation}\label{rsdc}
 	r_2(z)=e^{i\phi_{k_0}}+\mathcal{O}(t^{\frac{\epsilon}{2}-\frac{1}{3}}).
 \end{equation}
From the expansion of $\theta$ given by \eqref{thetaze} together with \eqref{rsda}--\eqref{rsdc}, $V^{(r)}$ approaches  $V^{(r,1)}(\zeta)$
 as follows:
 {\small\begin{equation}\label{vl1}
 \begin{aligned}
 	V^{(r,1)}(\zeta):=
 	\begin{cases}
 		\begin{pmatrix}
 			0&
 			i e^{i(2t\theta^{(0)}(\xi)-\phi_{k_0}+\delta_{n+k_0})}\\
 			i e^{-i(2t\theta^{(0)}(\xi)-\phi_{k_0}+\delta_{n+k_0})}
 			&0
 		\end{pmatrix},
 		&\zeta\in\mathbb R^+,
 		\\[3mm]
 		\begin{pmatrix}
 			1&
 			-e^{2it\theta^{(0)}(\xi)+2\tilde\theta(\zeta)
 				-i\phi_{k_0}+i\delta_{n+k_0}}\\
 			0&1
 		\end{pmatrix},
 		&\zeta\in\Sigma_1^{(r\zeta)},
 		\\[3mm]
 		\begin{pmatrix}
 			1&0\\
 			e^{-2it\theta^{(0)}(\xi)+2\tilde\theta(\zeta)
 				+i\phi_{k_0}-i\delta_{n+k_0}}&1
 		\end{pmatrix},
 		&\zeta\in\Sigma_1^{(r\zeta)*},
 		\\[3mm]
 		\begin{pmatrix}
 			1&0\\
 			e^{-2it\theta^{(0)}(\xi)+2\tilde\theta_-(\zeta)
 				+i\phi_{k_0}-i\delta_{n+k_0}}&1
 		\end{pmatrix}
 		\begin{pmatrix}
 			1&
 			-e^{2it\theta^{(0)}(\xi)+2\tilde\theta_+(\zeta)
 				-i\phi_{k_0}+i\delta_{n+k_0}}\\
 			0&1
 		\end{pmatrix},
 		&\zeta\in(\zeta(z_0),0),
 	\end{cases}
 \end{aligned}
 \end{equation}}
 where
 \begin{align}
 	\tilde{\theta}(\zeta)=s\zeta^{\frac{1}{2}}+\frac{2}{3}\zeta^{\frac{3}{2}},
 \end{align}
and $s$ is given by \eqref{s1}.
On $(\zeta(z_0),0)$, $\tilde\theta_\pm$ denote
the upper and lower boundary values, respectively. Here we use the identity  $2t\theta^{(0)}(\xi)=-C_{k_0}^f x-C_{k_0}^g t$ as given in \eqref{theta0}, which is guaranteed by \eqref{gur}. Obviously, $V^{(r,1)}$ admits a factorization
 \begin{equation}\label{facvl1}
 	V^{(r,1)}(\zeta)=\big(V_{-}^{(r,1)}(\zeta)\big)^{-1} V_{+}^{(r,1)}(\zeta),
 \end{equation}
 where
  \begin{align}\label{Vl1+}
 V_+^{(r,1)}(\zeta)=
 \begin{cases}
 	V^{(r,1)}(\zeta),
 	&\zeta\in\mathbb R^+,
 	\\[2mm]
 	\begin{pmatrix}
 		1&
 		-e^{2it\theta^{(0)}(\xi)+2\tilde\theta(\zeta)
 			-i\phi_{k_0}+i\delta_{n+k_0}}\\
 		0&1
 	\end{pmatrix},
 	&\zeta\in\Sigma_1^{(r\zeta)}
 	\cup(\zeta(z_0),0),
 	\\[2mm]
 	I,&\text{elsewhere}.
 \end{cases}
  \end{align}
 \begin{align}	\label{Vl1-}
 	V_-^{(r,1)}(\zeta)=
 	\begin{cases}
 		\begin{pmatrix}
 			1&0\\
 			-e^{-2it\theta^{(0)}(\xi)+2\tilde\theta(\zeta)
 				+i\phi_{k_0}-i\delta_{n+k_0}}&1
 		\end{pmatrix},
 		&\zeta\in\Sigma_1^{(r\zeta)*}
 		\cup(\zeta(z_0),0),
 		\\[2mm]
 		I,&\text{elsewhere}.
 	\end{cases}
 \end{align}
On $(\zeta(z_0),0)$, $\tilde\theta$ is understood
as $\tilde\theta_+$ in $V_+^{(r,1)}$ and as
$\tilde\theta_-$ in $V_-^{(r,1)}$.

\begin{figure}
	\centering
	\definecolor{mkdvAxisBlue}{RGB}{74,144,226}
	\definecolor{mkdvLensGreen}{RGB}{60,130,70}
	\begin{tikzpicture}[
		x=1cm,y=1cm,
		every node/.style={font=\footnotesize},
		solidray/.style={
			draw=mkdvLensGreen,line width=0.95pt,
			postaction={decorate},
			decoration={markings,mark=at position 0.57 with
				{\arrow{Latex[length=2.1mm,width=1.4mm]}}}
		},
		oldray/.style={
			draw=black!55,densely dotted,line width=0.85pt,
			postaction={decorate},
			decoration={markings,mark=at position 0.57 with
				{\arrow{Latex[length=2.1mm,width=1.4mm]}}}
		},
		realray/.style={
			draw=black,line width=1.0pt,
			postaction={decorate},
			decoration={markings,mark=at position 0.72 with
				{\arrow{Latex[length=2.1mm,width=1.4mm]}}}
		},
		interval/.style={
			draw=mkdvAxisBlue,line width=1.0pt,
			postaction={decorate},
			decoration={markings,mark=at position 0.70 with
				{\arrow{Latex[length=2.0mm,width=1.3mm]}}}
		},
		point/.style={circle,fill=black,inner sep=1.45pt}
		]
		
		% Schematic positions.
		\coordinate (O)  at (0,0);
		\coordinate (Z)  at (-1.60,0);
		\coordinate (DU) at (-1.80, 1.75);
		\coordinate (DL) at (-1.80,-1.75);
		\coordinate (SU) at (-3.40, 1.75);
		\coordinate (SL) at (-3.40,-1.75);
		
		% Auxiliary rays, oriented towards 0.
		\draw[oldray] (DU)--(O);
		\draw[oldray] (DL)--(O);
		
		% Nonreal jump contours, oriented towards zeta(z_0).
		\draw[solidray] (SU)--(Z);
		\draw[solidray] (SL)--(Z);
		
		% Real jump contours, oriented from left to right.
		\draw[interval] (Z)--(O);
		\draw[realray] (O)--(1.35,0);
		
		% Points and their labels.
		\node[point,fill=mkdvLensGreen] at (Z) {};
		\node[point] at (O) {};
		\node at (-1.33,-0.32) {$\zeta(z_0)$};
		\node[below right=3pt] at (O) {$0$};
		
		% Contour labels.
		\node[mkdvLensGreen,above=5pt] at (-3.2,0.4)
		{$\Sigma_1^{(r\zeta)}$};
		\node[mkdvLensGreen] at (-3.1,-0.7)
		{$\Sigma_1^{(r\zeta)*}$};
		
		% Region labels.
		\node[black] at (-1.8,1.02)
		{$\Omega_1^{(\mathrm{loc})}$};
		\node[black] at (-1.8,-1.02)
		{$\Omega_1^{(\mathrm{loc})*}$};
		
	\end{tikzpicture}
	\caption{The regions $\Omega_1^{(\mathrm{loc})}$ and
		$\Omega_1^{(\mathrm{loc})*}$ in the definition of $G_2$.
		Solid lines show the jump contour of $M^{(r,1)}$ and
		dotted rays indicate the jumps removed by $G_2$.}
	\label{fig:mkdv-G2-regions}
\end{figure}
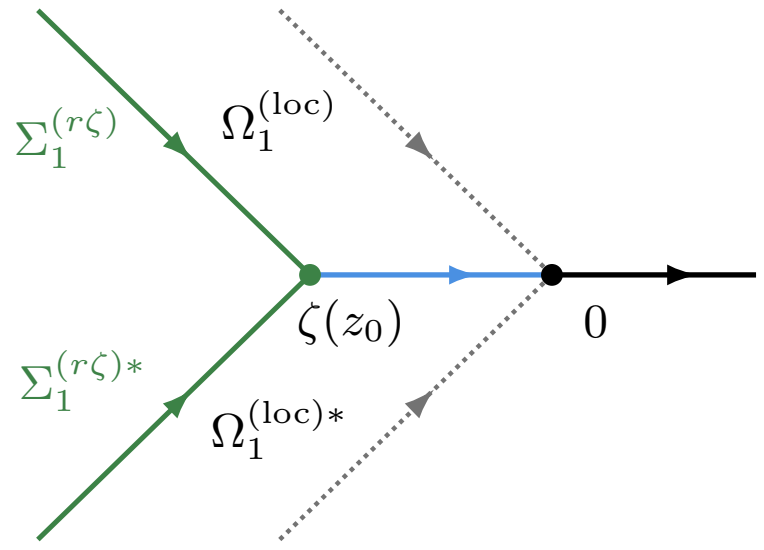

Based on the above analysis, we intend to consider the following RH problem.
\begin{RHP}\label{rhloc1}
	\hfill
	\begin{itemize}
		\item $M^{(r,1)}(\zeta)$ is analytic in $\mathbb{C}\setminus\Sigma^{(r\zeta)}$, where $\Sigma^{(r\zeta)}$ is defined in \eqref{lzeta}.
		\item
		 $M^{(r,1)}(\zeta)$ satisfies the jump condition $$M^{(r,1)}_+(\zeta)=M^{(r,1)}_{-}(\zeta)V^{(r,1)}(\zeta),\quad  \zeta\in\Sigma^{(r\zeta)},$$
		where $V^{(r,1)}$ is defined in \eqref{vl1}.
		\item $M^{(r,1)}(\zeta)$ has at most $-\frac{1}{4}$-singularity at $0$.
		\item As $\zeta\to\infty$, we have
		\begin{align}
			M^{(r,1)}(\zeta)=	\left (I
			+\mathcal{O}\left(  \zeta^{-1}\right) \right)
			\frac{\zeta^{-\frac{\sigma_3}{4}}}{\sqrt{2}}
			\begin{pmatrix}
				1 & i
				\\
				i & 1
			\end{pmatrix}G_1(\zeta),
		\end{align}
	   where
	   \begin{equation}\label{G1}
	   	G_1(\zeta):=
	   	\begin{cases}
	   		\begin{pmatrix}
	   			0&-1\\
	   			1&0
	   		\end{pmatrix}
	   		e^{\frac{\pi i}{4}\sigma_3}
	   		e^{-\frac{i}{2}
	   			(2t\theta^{(0)}(\xi)-\phi_{k_0}+\delta_{n+k_0})\sigma_3},
	   		&\zeta\in\mathbb C^+,
	   		\\[2mm]
	   		e^{\frac{\pi i}{4}\sigma_3}
	   		e^{-\frac{i}{2}
	   			(2t\theta^{(0)}(\xi)-\phi_{k_0}+\delta_{n+k_0})\sigma_3},
	   		&\zeta\in\mathbb C^-.
	   	\end{cases}
	   \end{equation}
	\end{itemize}
\end{RHP}
It can be verified that the solution to RH problem \ref{rhloc1} can be expressed in terms of the Painlev\'e XXXIV parametrix $M^{(P_{34})}(\zeta;s,\alpha,\omega)$ (see Appendix \ref{APP P34}) with  $\alpha=\frac{1}{4}$ and $\omega=0$:
\begin{align}\label{defl1}
	M^{(r,1)}(\zeta)=\begin{pmatrix}
		1 & 0\\
		ia(s) & 1
	\end{pmatrix}M^{(P_{34})}(\zeta;s,\tfrac{1}{4},0)e^{\tilde{\theta}(\zeta)\sigma_3}G_1(\zeta)G_2(\zeta),
\end{align}
where $a(s)=a(s;\tfrac14,0)$ is determined by \eqref{eq:Psi-infinity} and
\begin{equation}\label{def:G2}
	G_2(\zeta):=
	\begin{cases}
		\begin{pmatrix}
			1&
			-e^{2it\theta^{(0)}(\xi)+2\tilde\theta(\zeta)
				-i\phi_{k_0}+i\delta_{n+k_0}}\\
			0&1
		\end{pmatrix},
		&\zeta\in\Omega_1^{(\mathrm{loc})},
		\\[3mm]
		\begin{pmatrix}
			1&0\\
			-e^{-2it\theta^{(0)}(\xi)+2\tilde\theta(\zeta)
				+i\phi_{k_0}-i\delta_{n+k_0}}&1
		\end{pmatrix},
		&\zeta\in\Omega_1^{(\mathrm{loc})*},
		\\[2mm]
		I,&\text{elsewhere},
	\end{cases}
\end{equation}
with the region $\Omega_1^{(\mathrm{loc})}$ illustrated in  Figure~\ref{fig:mkdv-G2-regions}.

\subsubsection{Construction of \(M^{(r,2)}\)}
In order to deduce the solvability of the RH problem \ref{rhloc} by using the expression \eqref{defl1} of the solution to RH problem \ref{rhloc1}, we construct the following RH problem $M^{(r,2)}(\zeta)$, so that the jump matrix of $M^{(r,2)}(\zeta)$ is identical to that of $M^{(r)}(\zeta)$.
\begin{RHP}\label{rhloc2}
	\hfill
	\begin{itemize}
		\item $M^{(r,2)}(\zeta)$ is analytic in $\mathbb{C}\setminus\Sigma^{(r\zeta)}$, where $\Sigma^{(r\zeta)}$ is defined in \eqref{lzeta}.
		\item
		$M^{(r,2)}(\zeta)$ satisfies the jump condition $$M^{(r,2)}_+(\zeta)=M^{(r,2)}_{-}(\zeta)V^{(r)}(\zeta),\quad  \zeta\in\Sigma^{(r\zeta)},$$
		where $V^{(r)}$ is defined in \eqref{vl}.
		\item $M^{(r,2)}(\zeta)$ has at most $-\frac{1}{4}$-singularity at $0$.
		\item As $\zeta\to\infty$, we have
		\begin{align}
			M^{(r,2)}(\zeta)=	\left (I
			+\mathcal{O}\left(  \zeta^{-1}\right) \right)
			\frac{\zeta^{-\frac{\sigma_3}{4}}}{\sqrt{2}}
			\begin{pmatrix}
				1 & i
				\\
				i & 1
			\end{pmatrix}G_1(\zeta),
		\end{align}
		where $G_1$ is given by \eqref{G1}.
	\end{itemize}
\end{RHP}
The solvability of the RH problem \ref{rhloc2} is guaranteed by the following proposition.
\begin{proposition}
	The RH problem \ref{rhloc2} admits a unique solution for all sufficiently large $t>0$. Moreover, as $\zeta\to\infty$, the following asymptotic expansion holds:
\begin{align}\label{asy Mloc0I}
	M^{(r,2)}=\left (I+\frac{M^{(r,2)}_1}{\zeta}
	+\mathcal{O}\left(  \zeta^{-2}\right) \right)	\frac{\zeta^{-\frac{\sigma_3}{4}}}{\sqrt{2}}
	\begin{pmatrix}
		1 & i
		\\
		i & 1
	\end{pmatrix}G_1(\zeta),
\end{align}
where
\begin{equation}\label{asy loc1}
\begin{aligned}
	\big(M^{(r,2)}_1\big)_{12}=&\,\big(M^{(P_{34})}_1\big)_{12}+\mathcal{O}(t^{-\frac{1}{3}})\\ =&\,ia(s)+\mathcal{O}(t^{-\frac{1}{3}}),
\end{aligned}
\end{equation}
with $a(s)$ and $s$ given by \eqref{eq:Psi-infinity} and \eqref{s1}, respectively.
\end{proposition}
\begin{proof}
	Based on the RH problems \ref{rhloc1} and \ref{rhloc2} for $M^{(r,1)}$ and $M^{(r,2)}$, we define
	\begin{align}\label{varepsilon}
		E^{(\zeta)}(\zeta):=M^{(r,2)}(\zeta)\big(M^{(r,1)}(\zeta)\big)^{-1},
	\end{align}
	and set
	\[
	\Sigma^{(E,\zeta)}
	=\Sigma_1^{(r\zeta)}
	\cup\Sigma_1^{(r\zeta)*}
	\cup(\zeta(z_0),0).
	\]
	Then $E^{(\zeta)}$ is analytic in $\mathbb{C}\setminus\Sigma^{(E,\zeta)}$ and satisfies the jump condition
	$$E^{(\zeta)}_+(\zeta)=E^{(\zeta)}_-(\zeta)V^{(E,\zeta)}(\zeta),\qquad \zeta\in\Sigma^{(E,\zeta)},$$
	with
	\begin{align}\label{Vvarepsilon}
		V^{(E,\zeta)}=M^{(r,1)}_-V^{(r)}\big(V^{(r,1)}\big)^{-1}\big(M^{(r,1)}_-\big)^{-1}.
	\end{align}

	We now estimate $V^{(E,\zeta)}-I$. Using \eqref{Vvarepsilon} together with the factorizations \eqref{facvl} and \eqref{facvl1}, we obtain
	\begin{equation}\label{ak}
		\begin{aligned}
			&\|V^{(E,\zeta)}-I\|_{L^\infty\cap L^1\cap L^2(\Sigma_1^{(r\zeta)}\cup(\zeta(z_0),0))}\\
			=&\,\Big\|M^{(r,1)}_-\big(V^{(r)}(V^{(r,1)})^{-1}-I\big)(M^{(r,1)}_-)^{-1}\Big\|_{L^\infty\cap L^1\cap L^2(\Sigma_1^{(r\zeta)}\cup(\zeta(z_0),0))}\\
			\lesssim&\Big\|M^{(r,1)}_-\big(V^{(r)}(V^{(r,1)})^{-1}-I\big)(M^{(r,1)}_-)^{-1}\Big\|_{L^\infty\cap L^1\cap L^2((\Sigma_1^{(r\zeta)}\cup(\zeta(z_0),0))\cap\{\zeta:|\zeta|\leq C\})}\\
			&\quad+\Big\|M^{(r,1)}_-\big(V^{(r)}(V^{(r,1)})^{-1}-I\big)(M^{(r,1)}_-)^{-1}\Big\|_{L^\infty\cap L^1\cap L^2((\Sigma_1^{(r\zeta)}\cup(\zeta(z_0),0))\cap\{\zeta:|\zeta|\geq C\})}\\
			\lesssim&\Big\|\tfrac{1}{\sqrt{|\zeta|}}\big(V_{+}^{(r,1)}-V_{+}^{(r)}\big)\Big\|_{L^\infty\cap L^1\cap L^2((\Sigma_1^{(r\zeta)}\cup(\zeta(z_0),0))\cap\{\zeta:|\zeta|\leq C\})}\\
			&\quad+\Big\|\tfrac{1}{\sqrt{|\zeta|}}\big(V_{-}^{(r,1)}-V_{-}^{(r)}\big)\Big\|_{L^\infty\cap L^1\cap L^2((\Sigma_1^{(r\zeta)}\cup(\zeta(z_0),0))\cap\{\zeta:|\zeta|\leq C\})}\\
			&\quad+\Big\|\sqrt{|\zeta|}\big(V_{+}^{(r,1)}-V_{+}^{(r)}\big)\Big\|_{L^\infty\cap L^1\cap L^2((\Sigma_1^{(r\zeta)}\cup(\zeta(z_0),0))\cap\{\zeta:|\zeta|\geq C\})}\\
			&\quad+\Big\|\sqrt{|\zeta|}\big(V_{-}^{(r,1)}-V_{-}^{(r)}\big)\Big\|_{L^\infty\cap L^1\cap L^2((\Sigma_1^{(r\zeta)}\cup(\zeta(z_0),0))\cap\{\zeta:|\zeta|\geq C\})}.
		\end{aligned}
	\end{equation}
	In the above estimate we applied the bound
	$$
	|M^{(r,1)}|+|(M^{(r,1)}_-)^{-1}|\lesssim \begin{cases}
		|\zeta|^{-\frac14}, & |\zeta|\le C,\\[4pt]
		|\zeta|^{\frac14},  & |\zeta|\ge C,
	\end{cases}
	$$
	which is guaranteed by RH problem \ref{rhloc1}, and $C$ is a constant.
	
	For $\zeta\in\Sigma_1^{(r\zeta)}\cup(\zeta(z_0),0)$ with  $|\zeta|\leq C$, using the factorizations \eqref{V+1} and \eqref{Vl1+} together with \eqref{lin1}--\eqref{lin2} and \eqref{thetaze},  we obtain
	\begin{align*}
		\big|V_+^{(r,1)}(\zeta)-V_+^{(r)}(\zeta)\big|
		&=
		\big|
		r_1(E_{2k_0+1})
		e^{2it\theta^{(0)}(\xi)+2\tilde{\theta}(\zeta)
			+2\delta(E_{2k_0+1})}
		-r_1(z(\zeta))
		e^{2it\theta(z(\zeta))+2\delta(z(\zeta))}
		\big| \\
		&\le
		\big|
		r_1(E_{2k_0+1})e^{2\delta(E_{2k_0+1})}
		-r_1(z(\zeta))e^{2\delta(z(\zeta))}
		\big|
		\big|e^{2it\theta^{(0)}(\xi)+2\tilde{\theta}(\zeta)}\big| \\
		&\quad+
		\big|r_1(z(\zeta))
		e^{2it\theta(z(\zeta))+2\delta(z(\zeta))}\big|
		\big|1-e^{2(s-S(z(\zeta)))\zeta^{\frac{1}{2}}}\big| \\
		&\lesssim
		t^{-\frac{1}{3}}|\zeta^{\frac{1}{2}}|+t^{-\frac{2}{3}}|\zeta^{\frac{3}{2}}|.
	\end{align*}
	Here we used
	\[
	z(\zeta)-E_{2k_0+1}=\mathcal{O}(t^{-\frac{2}{3}}|\zeta|),
	\qquad
	S(z(\zeta))-s=\mathcal{O}(t^{-\frac{2}{3}}|\zeta|).
	\]
Consequently,
	\begin{equation}\label{askj}
		\frac{\big|V_{+}^{(r,1)}(\zeta)-V_{+}^{(r)}(\zeta)\big|}{\sqrt{|\zeta|}} \lesssim t^{-\frac{1}{3}}.
	\end{equation}

For $\zeta\in\Sigma_1^{(r\zeta)}\cup(\zeta(z_0),0)$ with $|\zeta|\geq C$, two cases arise depending on whether $\zeta$ belongs to $U^{(r\zeta)}$. If $\zeta\notin U^{(r\zeta)}$, then $V_+^{(r)}(\zeta)=I$
	and the model exponential gives
	\begin{equation}\label{notin}
		\big|V_+^{(r,1)}(\zeta)-V_+^{(r)}(\zeta)\big|
		\lesssim e^{-c|\zeta|^{\frac{3}{2}}},
		\qquad |\zeta|\gtrsim t^\epsilon.
	\end{equation}	
	If $\zeta\in U^{(r\zeta)}$, using the same decomposition
	as in the proof of \eqref{askj}, together with
	\[
	e^A-e^B=(A-B)\int_0^1e^{(1-\tau)B+\tau A}\,d\tau,
	\]
	we obtain
	\begin{equation}\label{in}
		\begin{aligned}
			\big|V_+^{(r,1)}(\zeta)-V_+^{(r)}(\zeta)\big|
			&\lesssim
			\left(
			|z(\zeta)-E_{2k_0+1}|^{\frac{1}{2}}
			+|S(z(\zeta))-s|\,|\zeta|^{\frac{1}{2}}
			\right)e^{-c|\zeta|^{\frac{3}{2}}}\\
			&\lesssim
			\left(t^{-\frac{1}{3}}|\zeta|^{\frac{1}{2}}
			+t^{-\frac{2}{3}}|\zeta|^{\frac{3}{2}}\right)e^{-c|\zeta|^{\frac{3}{2}}}.
		\end{aligned}
	\end{equation}
	Here we used \eqref{thetaze} and the exponential decay of both
	$e^{2it\theta(z(\zeta))}$ and
	$e^{2it\theta^{(0)}(\xi)+2\tilde{\theta}(\zeta)}$
	on the contour for $|\zeta|\ge C$. Combining \eqref{notin} and \eqref{in} yields
	\begin{equation}\label{askj1}
		\big\|
		|\zeta|^{\frac{1}{2}}
		\big(V_+^{(r,1)}(\zeta)-V_+^{(r)}(\zeta)\big)
		\big\|_{L^p(
			\Sigma_1^{(r\zeta)}\cap\{|\zeta|\ge C\})}
		\lesssim t^{-\frac{1}{3}},
		\qquad p=1,2,\infty.
\end{equation}

The bounds \eqref{askj} and \eqref{askj1} remain valid for  $\big|V_{-}^{(r,1)}(\zeta)-V_{-}^{(r)}(\zeta)\big|$. Substituting \eqref{askj} and \eqref{askj1}  into \eqref{ak}, we obtain
\[
\|V^{(E,\zeta)}-I\|_
{L^1\cap L^2\cap L^\infty(\Sigma^{(E,\zeta)})}
=\mathcal{O}(t^{-\frac{1}{3}}).
\]
The small-norm RH theory in \cite{PX3} therefore yields a unique solution
$E^{(\zeta)}$ for all sufficiently large $t$, satisfying
\begin{equation}\label{eq:estXi}
	E^{(\zeta)}(\zeta)=I+\mathcal{O}(t^{-\frac{1}{3}}),
\end{equation}
which guarantees that the RH problem \ref{rhloc2} is uniquely solvable for  sufficiently large $t>0$.

		The preceding estimates also give
		\[
		\|(1+|\zeta|)(V^{(E,\zeta)}-I)\|_
		{L^1\cap L^2(\Sigma^{(E,\zeta)})}
		=\mathcal{O}(t^{-\frac{1}{3}}).
		\]
		Hence, the Cauchy representation of the small-norm solution
		yields, as $\zeta\to\infty$ away from the contour,
	\[
	E^{(\zeta)}(\zeta)
	=I+\frac{E^{(\zeta)}_1}{\zeta}+\mathcal{O}(\zeta^{-2}),
	\]
	where
	\[
	E^{(\zeta)}_1
	=\mathcal{O}(t^{-\frac{1}{3}}).
	\]
		Combining this expansion with  \eqref{defl1}, \eqref{Psi-infinity} and \eqref{eq:12entry}
		gives \eqref{asy Mloc0I}, with
		\[
		\bigl(M_1^{(r,2)}\bigr)_{12}
		=\bigl(M_1^{(P_{34})}\bigr)_{12}
		+(E^{(\zeta)}_1)_{12}
		=ia(s)+\mathcal{O}(t^{-\frac{1}{3}}),
		\]
		which proves \eqref{asy loc1}.
\end{proof}

\subsubsection{Construction of the local parametrix $M^{(r)}$}
\begin{proposition}
	Let $M^{(r,2)}$ satisfy RH problem \ref{rhloc2}. Define
	\begin{align}\label{def Mloc}
		M^{(r)}(z)=H_1(z)t^{\frac{\sigma_3}{6}}M^{(r,2)}(\zeta(z)), \qquad z\in U^{(r)},
	\end{align}
	where
	\begin{align}\label{H1}
		H_1(z)=M^{(glo)}(z)G_1(\zeta(z))^{-1}\frac{I-i\sigma_1}{\sqrt{2}}
		\left(\zeta(z)t^{-\frac{2}{3}}\right)^{\frac{\sigma_3}{4}},
	\end{align}
	and $M^{(glo)}$  and  $G_1$ are given by \eqref{def:Mglo}, \eqref{G1}, respectively. Then $M^{(r)}(z)$ defined by \eqref{def Mloc} satisfies RH problem~\ref{rhloc}.
\end{proposition}

\begin{proof}
	First, we prove that \(H_1\) is analytic in \(U^{(r)}\). For \(z\in U^{(r)}\cap (\mathbb{R}\setminus\{E_{2k_0+1}\})\), using the jump matrix $V^{(glo)}$ of $M^{(glo)}$ in \eqref{Vglo} together with the boundary values of \(G_1\) in \eqref{G1}, a direct calculation yields
	\[
	H_{1,+}(z)=H_{1,-}(z).
	\]
	Since \(H_1\) is meromorphic in \(U^{(r)}\), and it follows from \eqref{gloas} that
	\[
	H_1(z)=\mathcal{O}\big((z-E_{2k_0+1})^{-\frac{1}{2}}\big),
	\]
	the singularity at \(E_{2k_0+1}\) is removable. Hence, \(H_1\) is analytic in \(U^{(r)}\).
	
	By the analyticity of \(H_1\) and the fact that $M^{(r,2)}$ satisfies RH problem~\ref{rhloc2}, it is straightforward to verify that $M^{(r)}(z)$ defined by \eqref{def Mloc} satisfies the first three conditions of RH problem~\ref{rhloc}. To verify the last condition of RH problem~\ref{rhloc}, it suffices to show that, for \(z\in\partial U^{(r)}\),
	\begin{equation}\label{lg}
		M^{(r)}(z)M^{(glo)}(z)^{-1}\to I,\qquad t\to\infty.
	\end{equation}
	For \(z\in\partial U^{(r)}\), it follows from \eqref{ulr}--\eqref{c0} and the definition of \(\zeta(z)\) in \eqref{zeta} that
	\begin{equation}\label{zetaas}
	|\zeta(z)|\sim t^\epsilon,\qquad t\to\infty,
	\end{equation}
	where \(\epsilon\in\left(\frac13,\frac23\right)\).
	
	Substituting \eqref{asy Mloc0I} into \eqref{def Mloc}, we obtain, for \(z\in\partial U^{(r)}\), as \(t\to\infty\),
	\begin{equation}\label{mlg}
		\begin{aligned}
			M^{(r)}(z)M^{(glo)}(z)^{-1}
			&=
			H_1(z)t^{\frac{\sigma_3}{6}}
			\left(
			I+\frac{M_1^{(r,2)}}{\zeta(z)}
			+\mathcal{O}(\zeta(z)^{-2})
			\right)
			t^{-\frac{\sigma_3}{6}}H_1(z)^{-1}\\
			&=	I+\frac{t^{\frac13}}{\zeta(z)}
			H_1(z)
			\begin{pmatrix}
				0 & ia(s)\\
				0 & 0
			\end{pmatrix}
			H_1(z)^{-1}
			+\mathcal{O}(t^{-\epsilon}).
		\end{aligned}
	\end{equation}
	Here we used \eqref{zetaas}, $M_1^{(r,2)}=\mathcal{O}(1)$ and
	$(M_1^{(r,2)})_{12}=ia(s)+\mathcal{O}(t^{-\frac13})$. Since $a(s)$ is bounded and
	$\frac{t^{\frac13}}{\zeta(z)}=\mathcal{O}(t^{\frac13-\epsilon})=o(1)$,
	the matching condition \eqref{lg}  follows.
\end{proof}

\subsection{Error estimate}\label{error}

\begin{figure}
	\centering
	\definecolor{mkdvLensGreen}{RGB}{60,130,70}
	\definecolor{mkdvAxisBlue}{RGB}{74,144,226}
	\resizebox{0.90\linewidth}{!}{%
		\begin{tikzpicture}[
			x=1cm,y=1cm,>=Latex,
			axis/.style={draw=mkdvAxisBlue,line width=0.72pt,densely dotted},
			band/.style={draw=black,line width=2.15pt},
			err/.style={
				draw=mkdvLensGreen,line width=1.25pt,
				postaction={decorate},
				decoration={markings,mark=at position 0.55 with
					{\arrow{Latex[length=3.0mm,width=1.9mm]}}}
			},
			disk/.style={draw=mkdvLensGreen,line width=1.45pt},
			diskarrow/.style={
				draw=mkdvLensGreen,line width=1.45pt,
				-{Latex[length=3.0mm,width=1.9mm]}
			},
			point/.style={circle,fill=black,inner sep=1.45pt},
			every node/.style={font=\footnotesize}
			]
			\def\zouter{4.53}
			\def\Eodd{4.18}
			\def\zsaddle{2.70}
			\def\Eeven{1.65}
			\def\rho{1.68}
			\def\rayx{7.08}
			\def\rayy{1.58}
			\def\locR{0.84}
			
			% Real axis and spectral bands.
			\draw[axis,->] (-7.28,0)--(7.42,0) node[right] {$\mathbb R$};
			\draw[band] (-\Eodd,0)--(-\Eeven,0);
			\draw[band] ( \Eeven,0)--( \Eodd,0);
			\node[black!58] at (0,-0.22) {$\cdots$};
			
			% Error contours outside the two local disks.
			\begin{scope}[even odd rule]
				\clip
				(-7.35,-2.12) rectangle (7.35,2.12)
				(-\Eodd,0) circle (\locR)
				( \Eodd,0) circle (\locR);
				
				\draw[err] (-\rayx, \rayy)--(-\zouter,0);
				\draw[err] ( \zouter,0)--( \rayx, \rayy);
				\draw[err] (-\rayx,-\rayy)--(-\zouter,0);
				\draw[err] ( \zouter,0)--( \rayx,-\rayy);
				
				\draw[err] (-\zsaddle,0)--(0, \rho);
				\draw[err] (0, \rho)--( \zsaddle,0);
				\draw[err] (-\zsaddle,0)--(0,-\rho);
				\draw[err] (0,-\rho)--( \zsaddle,0);
			\end{scope}
			
		% Clockwise circle boundaries.
		\foreach \xc in {-\Eodd,\Eodd}{
			\begin{scope}[shift={(\xc,0)}]
				\draw[disk] (0,0) circle[radius=\locR];
				\draw[diskarrow] (101:\locR) -- (79:\locR);
			\end{scope}
		}
			
			% Points.
			\foreach \x in {-\Eodd,-\Eeven,\Eeven,\Eodd}
			\node[point] at (\x,0) {};
			\node[point,fill=black] at (-\zouter,0) {};
			\node[point,fill=black] at ( \zouter,0) {};
			\node[point,fill=mkdvLensGreen] at (-\zsaddle,0) {};
			\node[point,fill=mkdvLensGreen] at ( \zsaddle,0) {};
			
			% Point labels.
			\node[above=3pt,xshift=-3pt] at (-\zouter,0) {$-z_0$};
			\node[above=3pt,xshift=3pt]  at ( \zouter,0) {$z_0$};
			\node[above=4pt] at (-\zsaddle,0) {$-z_{k_0}$};
			\node[above=4pt] at ( \zsaddle,0) {$z_{k_0}$};
			\node[below=4pt] at (-\Eodd,0) {$-E_{2k_0+1}$};
			\node[below=4pt] at ( \Eodd,0) {$E_{2k_0+1}$};
			\node[below=2.5pt] at (-\Eeven,0) {$-E_{2k_0}$};
			\node[below=2.5pt] at ( \Eeven,0) {$E_{2k_0}$};
			
			% Circle and contour labels.
			\node[mkdvLensGreen] at (-\Eodd,1.20) {$\partial U^{(l)}$};
			\node[mkdvLensGreen] at ( \Eodd,1.20) {$\partial U^{(r)}$};
			\node[mkdvLensGreen] at (-6.08, 1.49) {$\Sigma_1^{(l)}$};
			\node[mkdvLensGreen] at ( 6.08, 1.49) {$\Sigma_1^{(r)}$};
			\node[mkdvLensGreen] at (-6.08,-1.49) {$\Sigma_1^{(l)*}$};
			\node[mkdvLensGreen] at ( 6.08,-1.49) {$\Sigma_1^{(r)*}$};
			\node[mkdvLensGreen,above=5pt] at (0, \rho) {$\Sigma_2$};
			\node[mkdvLensGreen,below=5pt] at (0,-\rho) {$\Sigma_2^*$};
			
			% Region labels.
			\node[black] at (0, 0.61) {$\Omega_2$};
			\node[black] at (0,-0.61) {$\Omega_2^*$};
			\node[black] at ( 6.5, 0.49) {$\Omega_1^{(r)}$};
			\node[black] at ( 6.5,-0.49) {$\Omega_1^{(r)*}$};
			\node[black] at (-6.5, 0.49) {$\Omega_1^{(l)}$};
			\node[black] at (-6.5,-0.49) {$\Omega_1^{(l)*}$};
		\end{tikzpicture}%
	}
	\caption{The error jump contour $\Sigma^{(E)}$ (green) in transition region~$\mathcal{T}_{\mathrm{I}}$.}
	\label{fig:mkdv-region-I-error-contour}
\end{figure}
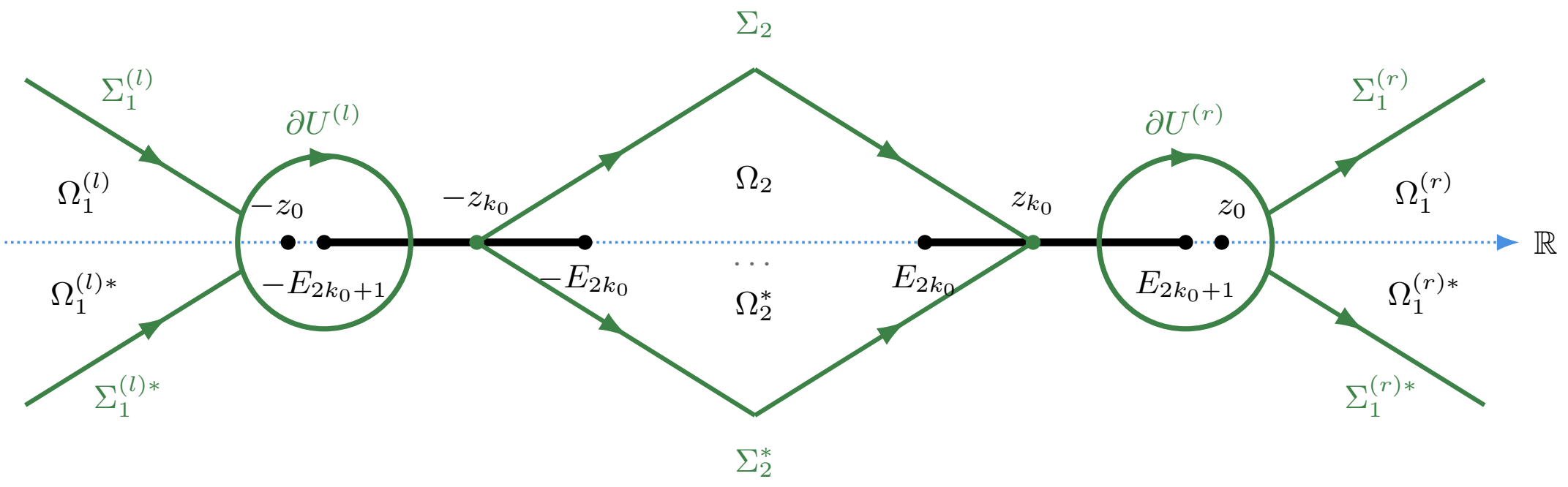

In this section, we analyze the error matrix $E$ introduced in \eqref{a2e}. It follows from \eqref{a2e}--\eqref{m2} that
\begin{equation}
	E(z)
	=M^{(1)}(z)\bigl(M^{(2)}(z)\bigr)^{-1}
	=
	\begin{cases}
		M^{(1)}(z)\bigl(M^{(glo)}(z)\bigr)^{-1},
		& z\in\mathbb{C}\setminus\bigl(U^{(l)}\cup U^{(r)}\bigr),\\
		M^{(1)}(z)\bigl(M^{(l)}(z)\bigr)^{-1},
		& z\in U^{(l)},\\
		M^{(1)}(z)\bigl(M^{(r)}(z)\bigr)^{-1},
		& z\in U^{(r)}.
	\end{cases}
\end{equation}
It is straightforward to verify that $E$ satisfies the following RH problem. The jumps on the original spectral cuts cancel, and the possible singularities of \(E\) at the branch points are removable.
\begin{RHP}
	\label{RHPE} \hfil
	\begin{itemize}
		\item $E(z)$ is analytic in  $\mathbb{C}\setminus\Sigma^{(E)}$, where
		\begin{align*}
			\Sigma^{(E)}
			:=
			\partial U^{(l)}\cup\partial U^{(r)}
			\cup
			\Bigl[
			\bigl(\Sigma_1\cup\Sigma_1^*
			\cup\Sigma_2\cup\Sigma_2^*\bigr)
			\setminus\bigl(U^{(l)}\cup U^{(r)}\bigr)
			\Bigr].
		\end{align*}
		See Figure~\ref{fig:mkdv-region-I-error-contour} for an illustration.
		\item  $E(z)$ satisfies the jump condition
		$$E_+(z)=E_-(z)V^{(E)}(z),\qquad z\in\Sigma^{(E)},$$ where
		\begin{align*}
			V^{(E)}(z)=	\left\{ \begin{array}{ll}
				M^{(glo)}(z)V^{(1)}(z)M^{(glo)}(z)^{-1}, & z\in \Sigma^{(E)} \setminus (\partial U^{(l)}\cup \partial U^{(r)}),\\
				M^{(l)}(z)M^{(glo)}(z)^{-1}, & z\in  \partial U^{(l)},\\
					M^{(r)}(z)M^{(glo)}(z)^{-1}, & z\in  \partial U^{(r)}.\\
			\end{array}\right.
		\end{align*}
		\item As $z\to\infty$, we have $$E(z)=I+\mathcal{O}(z^{-1}).$$
	\end{itemize}
\end{RHP}

\begin{proposition}
	For all sufficiently large $t>0$, RH problem \ref{RHPE} admits a unique solution. Moreover, as $z\to\infty$,
	\begin{equation}\label{ase}
		E(z)=I+\frac{E_1}{z}+\mathcal{O}(z^{-2}),
	\end{equation}
	where, as $t\to+\infty$,
	\begin{align}\label{eq:asyE1}
		E_1
		={}&
		-\frac{t^{-\frac{1}{3}}}
		{|\theta^{(3)}(\xi_{2k_0+1})|^{\frac{2}{3}}}
		H_1(E_{2k_0+1})
		\begin{pmatrix}
			0 & ia(s)\\
			0 & 0
		\end{pmatrix}
		H_1(E_{2k_0+1})^{-1}
		\nonumber\\
		&+
		\frac{t^{-\frac{1}{3}}}
		{|\theta^{(3)}(\xi_{2k_0+1})|^{\frac{2}{3}}}
			e^{2\delta(\infty)\sigma_3}
		\sigma_2 H_1(E_{2k_0+1})
		\begin{pmatrix}
			0&ia(s)\\
			0&0
		\end{pmatrix}
		H_1(E_{2k_0+1})^{-1}
		\sigma_2e^{-2\delta(\infty)\sigma_3}
		+\mathcal{O}(t^{-\epsilon}).
	\end{align}
	Here, $\theta^{(3)}$, $H_1$, and $a$ are defined in
	\eqref{theta3}, \eqref{H1}, and \eqref{eq:Psi-infinity}, respectively.
\end{proposition}

\begin{proof}
		By the definition of $V^{(1)}$ in \eqref{V11}--\eqref{V1}, together with \eqref{eq:inequality1}, \eqref{eq:inequality2} and \eqref{ulr}, we obtain
	\begin{equation}\label{small1}
		\big\|V^{(E)}-I\big\|_
		{L^p\left(\Sigma^{(E)}\setminus
		(\partial U^{(l)}\cup\partial U^{(r)})\right)}
		=
		\mathcal{O}\big(e^{-ct^{\frac{\epsilon}{2}}}\big),
	\end{equation}
	for $p=1,2,\infty$ and some $c>0$ independent of $t$,
	where $\epsilon\in\left(\tfrac{1}{3},\tfrac{2}{3}\right)$.
	Furthermore, it follows from \eqref{zetaas} and \eqref{mlg} that
	\begin{equation}\label{small2}
		\big\|V^{(E)}-I\big\|_
		{L^p\left(\partial U^{(l)}\cup\partial U^{(r)}\right)}
		=
		\mathcal{O}\big(t^{-\kappa_p}\big),\qquad p=2,\infty,
	\end{equation}
	where
	\[
	\kappa_\infty=\epsilon-\frac{1}{3},
	\qquad
	\kappa_2=\frac{\epsilon}{2}.
	\]
	Therefore, by \eqref{small1}--\eqref{small2} and the small-norm RH theory
	\cite{PX3}, RH problem \ref{RHPE} admits a unique solution for all
	sufficiently large $t$.
	
	Moreover, it follows from \cite{BC} that the solution of RH problem
	\ref{RHPE} can be represented as
	\begin{equation}\label{expression:E(z)}
		E(z)
		=
		I+\frac{1}{2\pi i}
		\int_{\Sigma^{(E)}}
		\frac{\upsilon(k)\bigl(V^{(E)}(k)-I\bigr)}
		{k-z}\,\dif k,
	\end{equation}
		where $\upsilon-I\in L^2(\Sigma^{(E)})$ and $\upsilon$
	satisfies the Beals--Coifman integral equation
	\begin{align*}
		\upsilon=I+\mathcal{C}_E\upsilon,
	\end{align*}
	and the operator $\mathcal{C}_E$ is defined by
	\[
	\mathcal{C}_E(f)(z)
	=
	\mathcal{C}_{\Sigma^{(E)}}^-
	\bigl[f(\,\cdot\,)\bigl(V^{(E)}(\,\cdot\,)-I\bigr)\bigr](z),
	\]
	with $\mathcal{C}_{\Sigma^{(E)}}^-$ denoting the Cauchy projection
	operator on $\Sigma^{(E)}$ defined in \eqref{def:opCpm}. By
	\eqref{small1}--\eqref{small2} and the uniform boundedness of
	$\mathcal{C}_{\Sigma^{(E)}}^-$ on $L^2(\Sigma^{(E)})$ as $t\to\infty$, we have
	\begin{align*}
		\|\mathcal{C}_E\|_
		{L^2(\Sigma^{(E)})\to L^2(\Sigma^{(E)})}
		&\leq
		\|\mathcal{C}_{\Sigma^{(E)}}^-\|_
		{L^2(\Sigma^{(E)})\to L^2(\Sigma^{(E)})}
		\|V^{(E)}-I\|_{L^\infty(\Sigma^{(E)})}\\
		&\lesssim
		t^{-\left(\epsilon-\frac{1}{3}\right)}.
	\end{align*}
	Consequently, $1-\mathcal{C}_E$ is invertible for all sufficiently
	large $t$, and
	\begin{equation}\label{normrho}
	\begin{aligned}
		\|\upsilon-I\|_{L^2(\Sigma^{(E)})}&=\|(1-\mathcal{C}_E)^{-1}(\mathcal{C}_EI)\|_{L^2(\Sigma^{(E)})}\\
		&\lesssim \|\mathcal{C}_EI\|_{L^2(\Sigma^{(E)})}\lesssim\|V^{(E)}-I\|_{L^2(\Sigma^{(E)})}\\
		&\lesssim  t^{-\frac{\epsilon}{2}}.
	\end{aligned}
	\end{equation}

	Expanding \eqref{expression:E(z)} as $z\to\infty$, we obtain
	\begin{align}\label{eq:Eneari}
		E(z)=I+\frac{E_1}{z}+\mathcal{O}(z^{-2}),
	\end{align}
	where
	\begin{align}
		E_1
		=
		-\frac{1}{2\pi i}
		\int_{\Sigma^{(E)}}
		\upsilon(k)\bigl(V^{(E)}(k)-I\bigr)\,\dif k.
		\label{def:E1}
	\end{align}
	By \eqref{small1}, the contribution from
	$\Sigma^{(E)}\setminus
	\bigl(\partial U^{(l)}\cup\partial U^{(r)}\bigr)$
	is exponentially small. Moreover, using \eqref{small2} and
	\eqref{normrho}, we find that
	\begin{align*}
		&\Big|
		\int_{\partial U^{(l)}\cup\partial U^{(r)}}
		(\upsilon-I)\bigl(V^{(E)}(k)-I\bigr)\,\dif k
		\Big|\\
		\lesssim&\
		\|\upsilon-I\|_
		{L^2(\partial U^{(l)}\cup\partial U^{(r)})}
		\|V^{(E)}-I\|_
		{L^2(\partial U^{(l)}\cup\partial U^{(r)})}\\
	\lesssim&\
		t^{-\epsilon}.
	\end{align*}
	It follows that
	\[
	E_1
	=
	-\frac{1}{2\pi i}
	\int_{\partial U^{(l)}\cup\partial U^{(r)}}
	\bigl(V^{(E)}(k)-I\bigr)\,\dif k
	+\mathcal{O}(t^{-\epsilon}).
	\]
	
	Finally, substituting  \eqref{mlg} into the
	above identity and using \eqref{zeta}--\eqref{zeta1}, together with the
	symmetry relation
		\[
	M^{(l)}(z)
	=e^{2\delta(\infty)\sigma_3}\sigma_2
	M^{(r)}(-z)\sigma_2
	e^{-2\delta(\infty)\sigma_3},
	\]
	\[
	M^{(glo)}(z)
	=
	e^{2\delta(\infty)\sigma_3}
	\sigma_2M^{(glo)}(-z)\sigma_2
	e^{-2\delta(\infty)\sigma_3},
	\]
	we obtain \eqref{eq:asyE1}. This completes the proof.
\end{proof}

\section{Asymptotic analysis of transition  region   $\mathcal{T}_{\mathrm{II}}$}\label{sec4}
In this section, we study the long-time asymptotics of $q$
in transition region  $\mathcal{T}_{\mathrm{II}}$. We fix $k_0\in\{1,\ldots,n\}$ and assume that
\[
0<(\xi-\xi_{2k_0})t^{\frac{2}{3}}\le C,
\qquad \xi=\frac{x}{t},
\]
since the analysis in the other half of the transition region is similar.
For sufficiently large $t$, we have
\[
\xi_{2k_0}<\xi<\xi_{2k_0-1},
\]
and hence, by the discussion in Section~\ref{sec25},
\[
z_0\in(E_{2k_0-1},E_{2k_0}),
\qquad
-z_0\in(-E_{2k_0},-E_{2k_0-1}).
\]
Thus, as $\xi\to\xi_{2k_0}$, $z_0$ approaches $E_{2k_0}$
from the left, while $-z_0$ approaches $-E_{2k_0}$
from the right;
see Figure~\ref{fig:mkdv-region-II-contours}.

To obtain jump factorizations suited to the sign distribution of
$\operatorname{Im}\theta$ in this region, we introduce an alternative
RH problem $N$ of the same scattering problem.
The phase function $\theta$ remains defined by \eqref{alpha}. We follow the analysis of Section~\ref{sec3} and detail only the necessary modifications.

\subsection{Construction of the RH problem for $N$} Starting from the scattering relation \eqref{Sz}, we define
\begin{align}	\label{def N}
	N(z)&=N(z;x,t) :=	\left\{ \begin{array}{ll}
		\left(  \mu^-_1  (z;x,t)  , \  \frac{\mu^+_2(z;x,t)}{S_{22}(z)}\right),   \quad  z\in \mathbb{C}^+,\\
		\left( \frac{\mu^+_1(z;x,t)}{S_{11}(z)}, \ \mu^-_2(z;x,t)\right), \quad z\in \mathbb{C}^-.\\
	\end{array}\right.
\end{align}
Then $N$ satisfies the following RH problem.
\begin{RHP} \label{RHN}
	\hfil
	\begin{itemize}
		\item $N(z)$ is analytic in $\mathbb{C}\setminus \mathbb{R}$.
		\item    $N(z)$  satisfies the jump condition $N_{+}(z)=N_{-}(z)\widetilde{V}(z)$ with
		\begin{equation}\label{tiledV}
			\widetilde{V}(z)=\left\{
			\begin{aligned}
				&
				\begin{pmatrix}
					0 & -ie^{i(\phi_k + C_k^f x + C_k^g t)} \\
					-ie^{-i(\phi_k + C_k^f x + C_k^g t)} & 0
				\end{pmatrix},\quad z\in(-E_{2k+1},-E_{2k}),\\
				&\begin{pmatrix}
					0 & -ie^{-i(\phi_k + C_k^f x + C_k^g t)} \\
					-ie^{i(\phi_k + C_k^f x + C_k^g t)} & 0
				\end{pmatrix},\quad z\in(E_{2k},E_{2k+1}),\\
				&\begin{pmatrix}
					1 & r_4(z)e^{2it\theta(z)}\\
					-r_3(z)e^{-2it\theta(z)} & 1-r_3(z)r_4(z)
				\end{pmatrix}, \quad z\in \mathbb{R}\setminus \overline{\Gamma},	
			\end{aligned}
			\right.
		\end{equation}
		where  $k=1,\cdots,n$, $\Gamma$ is given by \eqref{gamma} and $\theta$ is defined in \eqref{alpha}.  Specifically, for genus $2n$, $$\widetilde{V}(z)=\begin{pmatrix}
			0&-i\\-i&0
		\end{pmatrix},\qquad z\in(-E_{1},E_{1}).$$
		\item $N(z)$ satisfies the symmetry relations
		\begin{equation*}
			N(z)=\sigma_1\overline{N(\bar{z})}\sigma_1=\sigma_2N(-z)\sigma_2.
		\end{equation*}
		\item  As $z\to\infty$, we have $$N(z)=I+\mathcal{O}(z^{-1}).$$
		\item  For $p\in\boldsymbol{E}$, we have
		\begin{align*}
			N(z)=\left(\mathcal{O}\big((z-p)^{-\frac{1}{4}}\big),\mathcal{O}\big((z-p)^{\frac{1}{4}}\big)\right),\qquad z\to p, \quad  z\in\mathbb{C}^+,\\
			N(z)=\left(\mathcal{O}\big((z-p)^{\frac{1}{4}}\big),\mathcal{O}\big((z-p)^{-\frac{1}{4}}\big)\right),\qquad z\to p,\quad z\in\mathbb{C}^-.
		\end{align*}
	\end{itemize}
\end{RHP}
 RH problem \ref{RHN}  is uniquely solvable and the solution $q$ can be  recovered by
\begin{align}\label{reco2}
	q=2i\lim_{z\to\infty}zN_{12}(z).
\end{align}

Analogous to Lemma \ref{lem:jump-factorization}, $\widetilde{V}$ admits the following factorizations.  For  $z\in \bigcup_{k=1}^{n} \big( (-E_{2k+1}, -E_{2k}) \cup (E_{2k}, E_{2k+1}) \big) $,
\begin{equation}\label{frcN}
	\begin{aligned}
		\widetilde{V}(z) & =
		\begin{pmatrix}
			1&0\\
			-r_3(z)e^{-2it\theta_-(z)}&1
		\end{pmatrix}	\begin{pmatrix}
			0&-ie^{i\breve{\theta}(z)}\\
			-ie^{-i\breve{\theta}(z)}&0
		\end{pmatrix}  \begin{pmatrix}
			1&r_4(z)e^{2it\theta_+(z)}\\
			0&1
		\end{pmatrix}\\
		&=	\begin{pmatrix}
			1&\frac{r_4(z)e^{2it\theta_-(z)}}{1-r_3(z)r_4(z)}\\
			0&1
		\end{pmatrix}\begin{pmatrix}
			0&-ie^{i\breve{\theta}(z)}\\
			-ie^{-i\breve{\theta}(z)}&0
		\end{pmatrix} \begin{pmatrix}
			1&0\\
			\frac{-r_3(z)e^{-2it\theta_+(z)}}{1-r_3(z)r_4(z)}&1
		\end{pmatrix},
	\end{aligned}
\end{equation}
where $\breve{\theta}$ is given by \eqref{breal}.  For $z\in \mathbb{R}\setminus \overline{\Gamma}$,
\begin{equation}\label{fracN}
	\begin{aligned}
		\widetilde{V}(z)&=
		\begin{pmatrix}
			1&0\\
			-r_3(z)e^{-2it\theta(z)}&1
		\end{pmatrix}\begin{pmatrix}
			1&r_4(z)e^{2it\theta(z)}\\
			0&1	
		\end{pmatrix}\\
		&=	\begin{pmatrix}
			1&\frac{r_4(z)e^{2it\theta(z)}}{1-r_3(z)r_4(z)}\\
			0&1
		\end{pmatrix}\begin{pmatrix}
		\frac{1}{1-r_3(z)r_4(z)}&0\\
			0&	1-r_3(z)r_4(z)
		\end{pmatrix}	\begin{pmatrix}
			1&0\\
			-\frac{r_3(z)e^{-2it\theta(z)}}{1-r_3(z)r_4(z)}&1	
		\end{pmatrix}.
	\end{aligned}
\end{equation}

\subsection{Construction of $N^{(1)}$}
We define the first transformation of $N$ by
\begin{equation}\label{eq:II-N1}
	N^{(1)}(z)
	:=
	e^{\delta(\infty)\sigma_3}
	N(z)G(z)e^{-\delta(z)\sigma_3},
\end{equation}
where $\delta(\infty):=\lim\limits_{z\to\infty}\delta(z)$.
For genus $2n$, we define
\begin{equation}\label{eq:II-delta}
	\begin{aligned}
		\delta(z)
		:={}&
		\frac{w(z)}{2\pi i}
		\left(
		\sum_{j=1}^{n-1}\delta_j
		\int_{-E_{2n-2j+1}}^{-E_{2n-2j}}
		+\delta_n\int_{-E_1}^{E_1}
		+\sum_{j=n+1}^{2n}\delta_j
		\int_{E_{2j-2n}}^{E_{2j-2n+1}}
		\right)
		\frac{i\,ds}{w_+(s)(s-z)}
		\\
		&-
		\frac{w(z)}{2\pi i}
		\int_{I_0}
		\frac{\log(1-r_3(s)r_4(s))}{w(s)(s-z)}\,ds,
	\end{aligned}
\end{equation}
with
\begin{equation}\label{eq:II-I}
	I_0:=
	\bigl((-\infty,-E_{2k_0+1})
	\cup(E_{2k_0+1},\infty)\bigr)\setminus\Gamma.
\end{equation}
Here $w,r_3,r_4$ are defined in \eqref{w} and \eqref{r}.
The real constants $\delta_j$, $j=1,\ldots,2n$, are
determined by
\begin{equation}\label{eq:II-delta-moments}
	\begin{aligned}
		&
		\left(
		\sum_{j=1}^{n-1}\delta_j
		\int_{-E_{2n-2j+1}}^{-E_{2n-2j}}
		+\delta_n\int_{-E_1}^{E_1}
		+\sum_{j=n+1}^{2n}\delta_j
		\int_{E_{2j-2n}}^{E_{2j-2n+1}}
		\right)
		\frac{is^m}{w_+(s)}\,ds
		\\
		=&
		\int_{I_0}
		\frac{\log(1-r_3(s)r_4(s))s^m}{w(s)}\,ds,
		\qquad m=0,\ldots,2n-1.
	\end{aligned}
\end{equation}
For genus $2n-1$, we omit the central-cut terms
 $\delta_n\int_{-E_1}^{E_1}$ in
\eqref{eq:II-delta} and \eqref{eq:II-delta-moments},
and take $m=0,\ldots,2n-2$. The matrix $G$ in \eqref{eq:II-N1} is given by
\begin{equation}\label{eq:II-G}
	G(z)=
	\begin{cases}
		\begin{pmatrix}
			1 & -r_4(z)e^{2it\theta(z)}\\
			0 & 1
		\end{pmatrix},
		& z\in\Omega_1,
		\\[3mm]
		\begin{pmatrix}
			1 & 0\\
			-r_3(z)e^{-2it\theta(z)} & 1
		\end{pmatrix},
		& z\in\Omega_1^*,
		\\[3mm]
		\begin{pmatrix}
			1 & 0\\
			\dfrac{r_3(z)e^{-2it\theta(z)}}{1-r_3(z)r_4(z)} & 1
		\end{pmatrix},
		& z\in\Omega_2,
		\\[4mm]
		\begin{pmatrix}
			1 & \dfrac{r_4(z)e^{2it\theta(z)}}{1-r_3(z)r_4(z)}\\
			0 & 1
		\end{pmatrix},
		& z\in\Omega_2^*,
		\\[3mm]
		I, & \text{elsewhere}.
	\end{cases}
\end{equation}

\begin{figure}
	\centering
	\definecolor{mkdvAxisBlue}{RGB}{74,144,226}
	\definecolor{mkdvLensGreen}{RGB}{60,130,70}
	\resizebox{0.90\linewidth}{!}{%
		\begin{tikzpicture}[
			x=1cm,
			y=1cm,
			>=Latex,
			axis/.style={draw=mkdvAxisBlue,line width=0.72pt},
			band/.style={draw=black,line width=2.15pt},
			sigmaone/.style={
				draw=mkdvLensGreen,
				line width=1.25pt,
				postaction={decorate},
				decoration={markings,
					mark=at position 0.62 with
					{\arrow{Latex[length=3.0mm,width=1.9mm]}}
				}
			},
			sigmatwo/.style={
				draw=mkdvLensGreen,
				line width=1.15pt,
				postaction={decorate},
				decoration={markings,
					mark=at position 0.58 with
					{\arrow{Latex[length=3.0mm,width=1.9mm]}}
				}
			},
			point/.style={circle,fill=black,inner sep=1.45pt},
			every node/.style={font=\footnotesize}
			]
			% Schematic ordering: 0 < z_0 < E_{2k_0} < z_{k_0} < E_{2k_0+1}.
			\def\zinner{2.48}
			\def\Eeven{3.00}
			\def\zsaddle{3.72}
			\def\Eodd{5.35}
			\def\rho{1.55}
			\def\rayx{7.08}
			\def\rayy{1.90}
			
			% Real axis and spectral bands.
			\draw[axis,->] (-7.28,0)--(7.42,0) node[right] {$\mathbb R$};
			\draw[band] (-\Eodd,0)--(-\Eeven,0);
			\draw[band] ( \Eeven,0)--( \Eodd,0);
			\node[black!58] at (0,-0.19) {$\cdots$};
			
			% Sigma_1 and Sigma_1^*.
			\draw[sigmaone] (-\zinner,0)--(0, \rho);
			\draw[sigmaone] (0, \rho)--( \zinner,0);
			\draw[sigmaone] (-\zinner,0)--(0,-\rho);
			\draw[sigmaone] (0,-\rho)--( \zinner,0);
			
			% Sigma_2 and Sigma_2^*.
			\draw[sigmatwo] (-\rayx, \rayy)--(-\zsaddle,0);
			\draw[sigmatwo] ( \zsaddle,0)--( \rayx, \rayy);
			\draw[sigmatwo] (-\rayx,-\rayy)--(-\zsaddle,0);
			\draw[sigmatwo] ( \zsaddle,0)--( \rayx,-\rayy);
			
			% Points on the real axis.
			\foreach \x in {-\Eodd,-\Eeven,\Eeven,\Eodd}
			\node[point] at (\x,0) {};
			\foreach \x in {-\zinner,\zinner,-\zsaddle,\zsaddle}
			\node[point,fill=mkdvLensGreen] at (\x,0) {};
			
			% Point labels.
			\node[above=4pt] at (-\zinner,0)  {$-z_0$};
			\node[above=4pt] at ( \zinner,0)  {$z_0$};
			\node[below=4pt] at (-\Eeven,0)   {$-E_{2k_0}$};
			\node[below=4pt] at ( \Eeven,0)   {$E_{2k_0}$};
			\node[above=4pt] at (-\zsaddle,0) {$-z_{k_0}$};
			\node[above=4pt] at ( \zsaddle,0) {$z_{k_0}$};
			\node[below=4pt] at (-\Eodd,0)    {$-E_{2k_0+1}$};
			\node[below=4pt] at ( \Eodd,0)    {$E_{2k_0+1}$};
			
			% Contour labels.
			\node[mkdvLensGreen] at (0, 1.83)    {$\Sigma_1$};
			\node[mkdvLensGreen] at (0,-1.83)    {$\Sigma_1^*$};
			\node[mkdvLensGreen] at (-5.5, 1.5)  {$\Sigma_2^{(l)}$};
			\node[mkdvLensGreen] at ( 5.5, 1.5)  {$\Sigma_2^{(r)}$};
			\node[mkdvLensGreen] at (-5.5,-1.5)  {$\Sigma_2^{(l)*}$};
			\node[mkdvLensGreen] at ( 5.5,-1.5)  {$\Sigma_2^{(r)*}$};
			
			% Region labels.
			\node[black] at (0, 0.6)     {$\Omega_1$};
			\node[black] at (0,-0.7)     {$\Omega_1^*$};
			\node[black] at (-6.4, 0.7)  {$\Omega_2^{(l)}$};
			\node[black] at ( 6.4, 0.7)  {$\Omega_2^{(r)}$};
			\node[black] at (-6.4,-0.7)  {$\Omega_2^{(l)*}$};
			\node[black] at ( 6.4,-0.7)  {$\Omega_2^{(r)*}$};
		\end{tikzpicture}%
	}
	\caption{The opened contours $\Sigma_j,\Sigma_j^*$ and the lens regions
		$\Omega_j,\Omega_j^*$, $j=1,2$, in transition region~$\mathcal{T}_{\mathrm{II}}$.
		The positive saddle point $z_0$ approaches $E_{2k_0}$ from the left,
		while $-z_0$ approaches $-E_{2k_0}$ from the right.}
	\label{fig:mkdv-region-II-contours}
\end{figure}
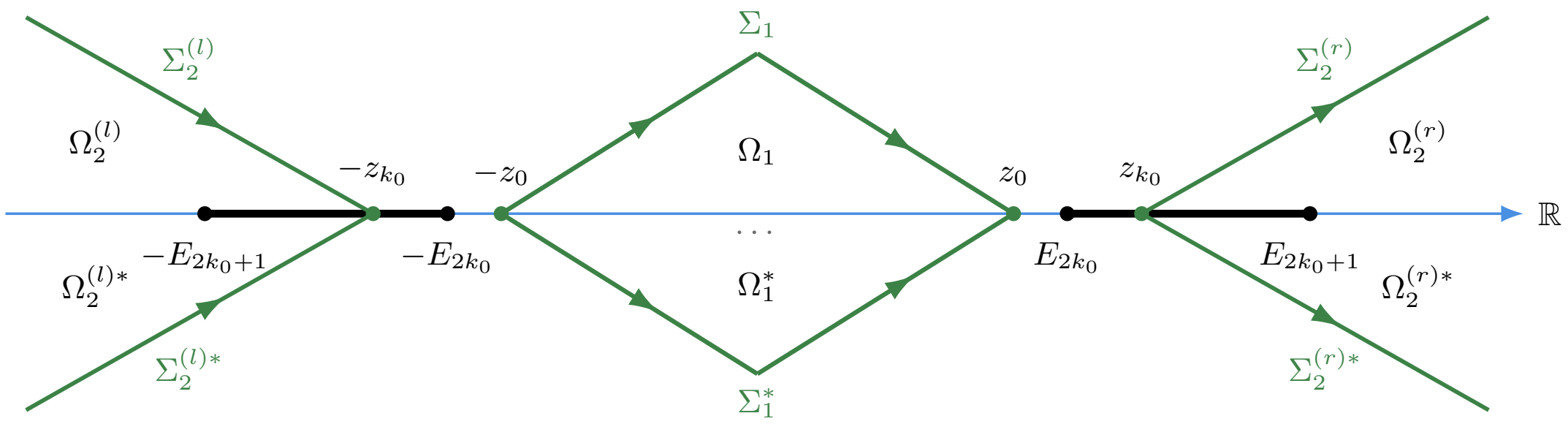

The contours and regions are defined as follows.
Choose a fixed point
$z_{k_0}\in(E_{2k_0},E_{2k_0+1})$.
For $\rho_0>0$, set
\begin{equation}\label{eq:II-Sigma1}
	\begin{aligned}
		\Sigma_1
		:=
		\bigl\{-(1-s)z_0+is\rho_0:0\le s\le1\bigr\}
	\cup
		\bigl\{sz_0+i(1-s)\rho_0:0\le s\le1\bigr\},
	\end{aligned}
\end{equation}
and
\begin{equation}\label{eq:II-Omega1}
	\Omega_1
	:=
	\left\{
	z\in\mathbb C:
	|\operatorname{Re}z|<z_0,\quad
	0<\operatorname{Im}z
	<\rho_0\left(1-\frac{|\operatorname{Re}z|}{z_0}\right)
	\right\}.
\end{equation}
The outer contours are
\begin{equation}\label{eq:II-Sigma2}
	\begin{aligned}
		\Sigma_2:=&\,\Sigma_2^{(l)}\cup\Sigma_2^{(r)}\\
		=&\,\{-z_{k_0}+se^{i(\pi-\varphi_0)}:s\ge0\}\cup \{z_{k_0}+se^{i\varphi_0}:s\ge0\}
	\end{aligned}
\end{equation}
with corresponding regions
\begin{equation}\label{eq:II-Omega2}
	\begin{aligned}
		\Omega_2:=&\, \Omega_2^{(l)}\cup\Omega_2^{(r)}\\
		=&\, \{z\in\mathbb C^+:
		\pi-\varphi_0<\arg(z+z_{k_0})<\pi\}\cup \{z\in\mathbb C^+:
		0<\arg(z-z_{k_0})<\varphi_0\}.
	\end{aligned}
\end{equation}
We choose $\rho_0$ and $\varphi_0$ so that
\begin{equation}\label{eq:II-phase-signs}
	\operatorname{Im}\theta(z)>0,
	\quad z\in\Sigma_1\setminus\{\pm z_0\},
	\qquad
	\operatorname{Im}\theta(z)<0,
	\quad z\in\Sigma_2\setminus\{\pm z_{k_0}\}.
\end{equation}
See Figure~\ref{fig:mkdv-region-II-contours} for an illustration.

According to the RH problem \ref{RHN} and  \eqref{frcN}--\eqref{fracN}, we consider the following RH problem.
\begin{RHP} \label{RHN1}
	\hfil
\begin{itemize}
	\item
	$N^{(1)}$ is analytic in $\mathbb C\setminus\overline{\Sigma^{(1)}}$,
	where
	\begin{equation}\label{eq:II-N1-contour}
		\Sigma^{(1)}
		:=
		\Gamma\cup\Sigma_1\cup\Sigma_1^*
		\cup\Sigma_2\cup\Sigma_2^*
		\cup(-E_{2k_0},-z_0)\cup(z_0,E_{2k_0}),
	\end{equation}
see Figure~\ref{fig:mkdv-region-II-contours}.
	\item
	$N^{(1)}$ satisfies the jump condition
	\[
	N^{(1)}_+(z)
	=
	N^{(1)}_-(z)\widetilde V^{(1)}(z),
	\qquad z\in\Sigma^{(1)}.
	\]
	For $z\in\Gamma$, with $k=1,\ldots,n$ and $\delta_0=0$,
	\begin{equation}\label{eq:II-N1-band-jumps}
		\widetilde V^{(1)}(z)
		=
		\begin{cases}
			\begin{pmatrix}
				0 & -ie^{i(\phi_k+C_k^f x+C_k^g t+\delta_{n-k})}\\
				-ie^{-i(\phi_k+C_k^f x+C_k^g t+\delta_{n-k})} & 0
			\end{pmatrix},
			&z\in(-E_{2k+1},-E_{2k}),
			\\[4mm]
			\begin{pmatrix}
				0 & -ie^{-i(\phi_k+C_k^f x+C_k^g t-\delta_{n+k})}\\
				-ie^{i(\phi_k+C_k^f x+C_k^g t-\delta_{n+k})} & 0
			\end{pmatrix},
			&z\in(E_{2k},E_{2k+1}).
		\end{cases}
	\end{equation}
	For genus $2n$, we also have
	\[
	\widetilde V^{(1)}(z)
	=
	\begin{pmatrix}
		0 & -ie^{i\delta_n}\\
		-ie^{-i\delta_n} & 0
	\end{pmatrix},
	\qquad z\in(-E_1,E_1).
	\]
	For $z\in\Sigma^{(1)}\setminus \Gamma$,
	\begin{equation}\label{eq:II-N1-lens-jumps}
		\widetilde V^{(1)}(z)
		=
		\begin{cases}
			\begin{pmatrix}
				1&r_4(z)e^{2it\theta(z)+2\delta(z)}\\
				0&1
			\end{pmatrix},
			&z\in\Sigma_1,
			\\[3mm]
			\begin{pmatrix}
				1&0\\
				-r_3(z)e^{-2it\theta(z)-2\delta(z)}&1
			\end{pmatrix},
			&z\in\Sigma_1^*,
			\\[3mm]
			\begin{pmatrix}
				1&0\\
				-\tfrac{r_3(z)e^{-2it\theta(z)-2\delta(z)}}
				{1-r_3(z)r_4(z)}&1
			\end{pmatrix},
			&z\in\Sigma_2,
			\\[5mm]
			\begin{pmatrix}
				1&\tfrac{r_4(z)e^{2it\theta(z)+2\delta(z)}}
				{1-r_3(z)r_4(z)}\\
				0&1
			\end{pmatrix},
			&z\in\Sigma_2^*,
			\\[4mm]
			\begin{pmatrix}
				1&r_4(z)e^{2it\theta(z)+2\delta(z)}\\
				-r_3(z)e^{-2it\theta(z)-2\delta(z)}
				&1-r_3(z)r_4(z)
			\end{pmatrix},
			&z\in(-E_{2k_0},-z_0)\cup(z_0,E_{2k_0}).
		\end{cases}
	\end{equation}
	
	\item
	As $z\to\infty$,
	\[
	N^{(1)}(z)=I+\mathcal O(z^{-1}).
	\]
	
	\item
	As $z\to p\in \boldsymbol{E} \setminus\{\pm E_{2k_0}\}$,
	\[
	N^{(1)}(z)=\mathcal O\bigl((z-p)^{-\frac{1}{4}}\bigr).
	\]
\end{itemize}
\end{RHP}

To approximate $N^{(1)}$, we introduce
\begin{equation}\label{eq:II-error-decomposition}
	N^{(1)}(z)=E(z)\times\begin{cases}
		N^{(glo)}(z),
		&z\in\mathbb C\setminus\bigl(U^{(l)}\cup U^{(r)}\bigr),\\
		N^{(l)}(z),&z\in U^{(l)},\\
		N^{(r)}(z),&z\in U^{(r)}.
	\end{cases}
\end{equation}
where $E$ is the error matrix.
Here $N^{(glo)}$ is the global parametrix,
and $N^{(l)}$, $N^{(r)}$ are the local parametrices in
\begin{equation}\label{eq:II-local-disks}
	U^{(l)}:=\{z\in\mathbb C:|z+E_{2k_0}|\le c_0\},
	\qquad
	U^{(r)}:=\{z\in\mathbb C:|z-E_{2k_0}|\le c_0\},
\end{equation}
with
\begin{equation}\label{eq:II-radius}
	c_0:=
	\min\left\{ \frac{1}{2}(z_{k_0}-E_{2k_0}),
	\frac{1}{2}(E_{2k_0}-E_{2k_0-1}),
	t^{-	\frac{2}{3}+\epsilon}
	\right\},
	\qquad \epsilon\in \left(\frac13,\frac23\right).
\end{equation}

\subsection{Parametrices and error estimates}
In this subsection,  we construct the global and local parametrices
$N^{(glo)}$, $N^{(l)}$, and $N^{(r)}$, and then estimate
the error matrix $E$.

The construction of $N^{(\mathrm{glo})}$ is analogous to
that of $M^{(\mathrm{glo})}$ in  Section~\ref{global}. In particular,
\begin{equation}\label{eq:II-global-reconstruction}
	q^{(\mathrm{AG})}(x,t;\mathbf E,\widetilde{\boldsymbol\phi})
	=
	e^{-2\delta(\infty)}
	\lim_{z\to\infty}2izN_{12}^{(\mathrm{glo})}(z),
\end{equation}
where the shifted phase vector is defined by
\begin{equation}\label{eq:II-shifted-phases}
	\widetilde{\boldsymbol\phi}
	:=(\widetilde\phi_1,\ldots,\widetilde\phi_n)^T,
	\qquad
	\widetilde\phi_k
	:=\phi_k+\delta_{n-k}+2i\delta(\infty),
	\quad k=1,\ldots,n,
\end{equation}
with $\delta_j$ and $\delta(\infty)$ now determined
by \eqref{eq:II-delta}--\eqref{eq:II-delta-moments} for transition region~$\mathcal{T}_{\mathrm{II}}$.

We next construct the local parametrices and focus on $N^{(r)}$.
The construction of $N^{(l)}$ is analogous.
We first consider the following RH problem.

\begin{RHP}\label{RH-II-local}
		\hfil
	\begin{itemize}
		\item $N^{(r)}$ is analytic in
		$U^{(r)}\setminus\Sigma^{(1)}$, where $U^{(r)}$ and
		$\Sigma^{(1)}$ are defined in \eqref{eq:II-local-disks} and \eqref{eq:II-N1-contour},
		respectively.
		
		\item $N^{(r)}$ satisfies the jump condition
		\[
		N^{(r)}_+(z)
		=N^{(r)}_-(z)\widetilde V^{(1)}(z),
		\qquad z\in\Sigma^{(1)}\cap U^{(r)},
		\]
		where $\widetilde V^{(1)}$ is given by \eqref{eq:II-N1-band-jumps}--\eqref{eq:II-N1-lens-jumps}.
		
		\item $N^{(r)}(z)$ has at most $-\frac{1}{4}$-singularity at $E_{2k_0}$.
		\item As $t\to+\infty$, $N^{(r)}(z)$ matches $N^{(\mathrm{glo})}(z)$ at $\partial U^{(r)}$.
	\end{itemize}
\end{RHP}
Following the construction in Section~\ref{local}, the solution to
RH problem~\ref{RH-II-local} is given by
\begin{align}\label{II-local-parametrix}
	N^{(r)}(z)
	={}&H_1(z)t^{\frac{\sigma_3}{6}}E^{(\zeta)}(\zeta(z))
	\begin{pmatrix}
		1&0\\
		ia(s)&1
	\end{pmatrix}
	M^{(P_{34})}\bigl(\zeta(z);s,\tfrac14,0\bigr)
	e^{\tilde\theta(\zeta(z))\sigma_3}
	G_1(\zeta(z))G_2(\zeta(z)).
\end{align}
Here $M^{(P_{34})}$ is the model in Appendix~\ref{APP P34} and
$a(s)=a(s;\tfrac14,0)$.

For transition
region $\mathcal{T}_{\mathrm{II}}$,
we now specify the local coordinate $\zeta$, the phase
$\tilde\theta$, the matrices $H_1,G_1,G_2$, and the
correction factor $E^{(\zeta)}$ appearing in
\eqref{II-local-parametrix}. The phase function $\theta$ retains its definition in \eqref{alpha}.
Near $E_{2k_0}$, it has the expansion
\begin{align}\label{II-phase-expansion}
\begin{aligned}
    	\theta(z;\xi)
	={}&\theta^{(0)}(\xi)
	+\theta^{(1)}(\xi)(\xi-\xi_{2k_0})(E_{2k_0}-z)^{\frac{1}{2}}\\&
	+\frac23\theta^{(3)}(\xi)(E_{2k_0}-z)^{\frac{3}{2}}+\mathcal{O}\bigl((E_{2k_0}-z)^{\frac{5}{2}}\bigr),
\end{aligned}
\end{align}
where
\begin{align}\label{II-phase-coefficients}
\begin{aligned}
  \theta^{(0)}(\xi)
&=\theta(E_{2k_0};\xi)
=-\frac12\bigl(C_{k_0}^f\xi+C_{k_0}^g\bigr),
\\[2mm]
\theta^{(1)}(\xi)
&=
\begin{cases}
\displaystyle
\frac{
2E_{2k_0}
\prod_{k=1}^{n}
\bigl(E_{2k_0}^{2}-(z_k^f)^2\bigr)
}{
w^{(2k_0)}(E_{2k_0})
},
& \text{genus }2n,\\[3mm]
\displaystyle
\frac{
2\prod_{k=1}^{n}
\bigl(E_{2k_0}^{2}-(z_k^f)^2\bigr)
}{
w^{(2k_0)}(E_{2k_0})
},
& \text{genus }2n-1,
\end{cases}
\\[3mm]
\theta^{(3)}(\xi)
&=
\begin{cases}
\displaystyle
-\left.
\frac{d}{dz}
\left(
\frac{zF(z;\xi)}{w^{(2k_0)}(z)}
\right)
\right|_{z=E_{2k_0}},
& \text{genus }2n,\\[3mm]
\displaystyle
-\left.
\frac{d}{dz}
\left(
\frac{F(z;\xi)}{w^{(2k_0)}(z)}
\right)
\right|_{z=E_{2k_0}},
& \text{genus }2n-1.
\end{cases}
\end{aligned}
\end{align}
Here $F$ is defined in  \eqref{F} and
\begin{equation}
w^{(2k_0)}(z)=
\begin{cases}
\displaystyle
\left[
-\prod_{j=1}^{2n+1}(z+E_j)
\prod_{\substack{1\le j\le 2n+1\\j\ne 2k_0}}
(z-E_j)
\right]^{\frac12},
& \text{genus }2n,\\[4mm]
\displaystyle
\left[
-\prod_{j=2}^{2n+1}(z+E_j)
\prod_{\substack{2\le j\le 2n+1\\j\ne 2k_0}}
(z-E_j)
\right]^{\frac12},
& \text{genus }2n-1.
\end{cases}
\end{equation}
The branch of $(E_{2k_0}-z)^{\frac{1}{2}}$ is chosen to be
positive for real $z<E_{2k_0}$ sufficiently close to
$E_{2k_0}$. The branch of $w^{(2k_0)}$ is chosen to be
analytic near $E_{2k_0}$ and to satisfy
\[
w(z)=(E_{2k_0}-z)^{\frac12}w^{(2k_0)}(z).
\]
 Furthermore,
$\theta^{(3)}(\xi)<0$ holds in a sufficiently small
neighborhood of $\xi_{2k_0}$.
Using \eqref{II-phase-expansion}, define
\begin{align}\label{II-local-coordinate}
	\zeta(z)
	:=
	\left[
	-\frac{3it}{2}
	\left(
	\theta^{(0)}(\xi)
	+\theta^{(1)}(\xi)(\xi-\xi_{2k_0})(E_{2k_0}-z)^{\frac{1}{2}}
	-\theta(z;\xi)
	\right)
	\right]^{\frac{2}{3}}.
\end{align}
The branch is chosen so that $\zeta$ is analytic in $U^{(r)}$ and
\[
\zeta(E_{2k_0})=0,\qquad
\zeta'(E_{2k_0})=|\theta^{(3)}(\xi)|^{\frac{2}{3}}t^{\frac{2}{3}}>0.
\]
The powers of $\zeta$ below are taken on the principal branch. Moreover,
\begin{align}\label{II-model-phase}
 s
	=\frac{\theta^{(1)}(\xi)(\xi-\xi_{2k_0})t^{\frac{2}{3}}}
	{|\theta^{(3)}(\xi)|^{\frac{1}{3}}},	\qquad \tilde\theta(\zeta)
	:=s\zeta^{\frac{1}{2}}+\frac23\zeta^{\frac{3}{2}}.
\end{align}
 We now define
\begin{align}\label{II-H1}
	H_1(z)
	:=
	N^{(\mathrm{glo})}(z)G_1(\zeta(z))^{-1}
	\frac{I-i\sigma_1}{\sqrt2}
	\left(\zeta(z)t^{-\frac23}\right)^{\frac{\sigma_3}{4}},
\end{align}
with
\begin{align}\label{II-G1}
	G_1(\zeta)
	:=
	\begin{cases}
		\begin{pmatrix}
			0&-1\\
			1&0
		\end{pmatrix}
		e^{-\frac{\pi i}{4}\sigma_3}
		e^{-\frac{i}{2}
			(2t\theta^{(0)}(\xi)-\phi_{k_0}
			+\delta_{n+k_0})\sigma_3},
		&\zeta\in\mathbb C^+,\\[2mm]
		e^{-\frac{\pi i}{4}\sigma_3}
		e^{-\frac{i}{2}
			(2t\theta^{(0)}(\xi)-\phi_{k_0}
			+\delta_{n+k_0})\sigma_3},
		&\zeta\in\mathbb C^-,
	\end{cases}
\end{align}
and
\begin{align}\label{II-G2}
	G_2(\zeta)
	:=
	\begin{cases}
		\begin{pmatrix}
			1&
			e^{2it\theta^{(0)}(\xi)
				+2\tilde\theta(\zeta)
				-i\phi_{k_0}+i\delta_{n+k_0}}\\
			0&1
		\end{pmatrix},
		&\zeta\in\Omega_1^{(\mathrm{loc})},\\[3mm]
		\begin{pmatrix}
			1&0\\
			e^{-2it\theta^{(0)}(\xi)
				+2\tilde\theta(\zeta)
				+i\phi_{k_0}-i\delta_{n+k_0}}&1
		\end{pmatrix},
		&\zeta\in\Omega_1^{(\mathrm{loc})*},\\[3mm]
		I,&\text{elsewhere},
	\end{cases}
\end{align}
where the region $\Omega_1^{(\mathrm{loc})}$ is defined as in Figure~\ref{fig:mkdv-G2-regions}, using the present local coordinate $\zeta$. Finally, $E^{(\zeta)}(\zeta)$ is defined analogously to \eqref{varepsilon}--\eqref{Vvarepsilon}, using the actual local jumps and the
corresponding endpoint model in the present region.

With the above parametrices, the same argument as in
Section~\ref{error} shows that the error matrix $E$ defined by
\eqref{eq:II-error-decomposition} has the expansion
\begin{align}\label{eq:II-error-expansion}
	E(z)=I+\frac{E_1}{z}+\mathcal O(z^{-2}),
	\qquad z\to\infty,
\end{align}
where, as $t\to+\infty$,
\begin{align}\label{eq:II-error-coefficient}
	E_1
	={}&
	\frac{t^{-\frac13}}
	{|\theta^{(3)}(\xi_{2k_0})|^{\frac23}}
	H_1(E_{2k_0})
	\begin{pmatrix}
		0&ia(s)\\
		0&0
	\end{pmatrix}
	H_1(E_{2k_0})^{-1}
	\notag\\
	&-
	\frac{t^{-\frac13}}
	{|\theta^{(3)}(\xi_{2k_0})|^{\frac23}}
	e^{2\delta(\infty)\sigma_3}\sigma_2
	H_1(E_{2k_0})
	\begin{pmatrix}
		0&ia(s)\\
		0&0
	\end{pmatrix}
	H_1(E_{2k_0})^{-1}
	\sigma_2e^{-2\delta(\infty)\sigma_3}
	+\mathcal O(t^{-\epsilon}).
\end{align}
Here $\theta^{(3)}$  and $H_1$ are given by \eqref{II-phase-coefficients}  and \eqref{II-H1}, respectively.

\section{Asymptotic analysis of the remaining regions}
\label{sec5}

\subsection{Zakharov-Manakov region  $\mathcal{T}_{\mathrm{III}}$}\label{sec51}
\begin{figure}
	\centering
	\definecolor{mkdvAxisBlue}{RGB}{74,144,226}
	\definecolor{mkdvLensGreen}{RGB}{60,130,70}
	\resizebox{0.90\linewidth}{!}{%
		\begin{tikzpicture}[
			x=1cm,
			y=1cm,
			>=Latex,
			axis/.style={draw=mkdvAxisBlue,line width=0.72pt},
			band/.style={draw=black,line width=2.15pt},
			outer/.style={
				draw=mkdvLensGreen,
				line width=1.25pt,
				postaction={decorate},
				decoration={markings,
					mark=at position 0.58 with
					{\arrow{Latex[length=3.0mm,width=1.9mm]}}
				}
			},
			inner/.style={draw=mkdvLensGreen,line width=1.15pt},
			innerarrow/.style={
				inner,
				postaction={decorate},
				decoration={markings,
					mark=at position 0.55 with
					{\arrow{Latex[length=3.0mm,width=1.9mm]}}
				}
			},
			point/.style={circle,fill=black,inner sep=1.45pt},
			every node/.style={font=\footnotesize}
			]
			
			% Representative component of region III:
			% E_{2k_0+1} < z_0 < E_{2k_0+2}.
			\def\Ez{1.20}
			\def\Eo{2.40}
			\def\zsaddle{4.02}
			\def\Et{5.45}
			\def\Eth{6.72}
			\def\rho{1.70}
			\def\rayx{7.12}
			\def\rayy{1.50}
			
			% Real axis and spectral bands.
			\draw[axis,->] (-7.48,0)--(7.62,0) node[right] {$\mathbb R$};
			\draw[band] (-\Eth,0)--(-\Et,0);
			\draw[band] (-\Eo,0)--(-\Ez,0);
			\draw[band] ( \Ez,0)--( \Eo,0);
			\draw[band] ( \Et,0)--( \Eth,0);
			\node[black!58] at (0,-0.23) {$\cdots$};
			
			% Sigma_1 and Sigma_1^*.
			\draw[outer] (-\rayx, \rayy)--(-\zsaddle,0);
			\draw[outer] ( \zsaddle,0)--( \rayx, \rayy);
			\draw[outer] (-\rayx,-\rayy)--(-\zsaddle,0);
			\draw[outer] ( \zsaddle,0)--( \rayx,-\rayy);
			
			% Sigma_2 and Sigma_2^*, oriented from -z_0 to z_0.
			\draw[innerarrow] (-\zsaddle,0)--(0, \rho);
			\draw[innerarrow] (0, \rho)--( \zsaddle,0);
			\draw[innerarrow] (-\zsaddle,0)--(0,-\rho);
			\draw[innerarrow] (0,-\rho)--( \zsaddle,0);
			
			% Branch points and saddle points.
			\foreach \x in {-\Eth,-\Et,-\Eo,-\Ez,\Ez,\Eo,\Et,\Eth}
			\node[point] at (\x,0) {};
			\node[point,fill=mkdvLensGreen] at (-\zsaddle,0) {};
			\node[point,fill=mkdvLensGreen] at ( \zsaddle,0) {};
			
			% Saddle-point labels.
			\node[above=5pt] at (-\zsaddle,0) {$-z_0$};
			\node[above=5pt] at ( \zsaddle,0) {$z_0$};
			
			% Branch-point labels.
			\node[below=3pt,font=\scriptsize] at (-\Eth,0) {$-E_{2k_0+3}$};
			\node[below=2pt,font=\scriptsize] at (-\Et,0)  {$-E_{2k_0+2}$};
			\node[below=2pt,font=\scriptsize] at (-\Eo,0)  {$-E_{2k_0+1}$};
			\node[below=2pt,font=\scriptsize,xshift=-3pt] at (-\Ez,0) {$-E_{2k_0}$};
			\node[below=2pt,font=\scriptsize,xshift=3pt]  at ( \Ez,0) {$E_{2k_0}$};
			\node[below=2pt,font=\scriptsize] at ( \Eo,0)  {$E_{2k_0+1}$};
			\node[below=2pt,font=\scriptsize] at ( \Et,0)  {$E_{2k_0+2}$};
			\node[below=3pt,font=\scriptsize] at ( \Eth,0) {$E_{2k_0+3}$};
			
			% Contour labels.
			\node[mkdvLensGreen] at (-5.7, 1.35) {$\Sigma_1^{(l)}$};
			\node[mkdvLensGreen] at ( 5.7, 1.35) {$\Sigma_1^{(r)}$};
			\node[mkdvLensGreen] at (-5.7,-1.35) {$\Sigma_1^{(l)*}$};
			\node[mkdvLensGreen] at ( 5.7,-1.35) {$\Sigma_1^{(r)*}$};
			\node[mkdvLensGreen,above=5pt] at (0, \rho) {$\Sigma_2$};
			\node[mkdvLensGreen,below=5pt] at (0,-\rho) {$\Sigma_2^*$};
			
			% Region labels.
			\node[black] at (-6.3, 0.6)    {$\Omega_1^{(l)}$};
			\node[black] at ( 6.3, 0.6)    {$\Omega_1^{(r)}$};
			\node[black] at (-6.68,-0.88)  {$\Omega_1^{(l)*}$};
			\node[black] at ( 6.7,-0.85)   {$\Omega_1^{(r)*}$};
			\node[black] at (0, 0.62)      {$\Omega_2$};
			\node[black] at (0,-0.62)      {$\Omega_2^*$};
			
		\end{tikzpicture}%
	}
	\caption{The opened contours $\Sigma_j,\Sigma_j^*$ and regions
		$\Omega_j,\Omega_j^*$, $j=1,2$, in Zakharov--Manakov region~$\mathcal{T}_{\mathrm{III}}$.}
	\label{fig:mkdv-region-III-contours}
\end{figure}
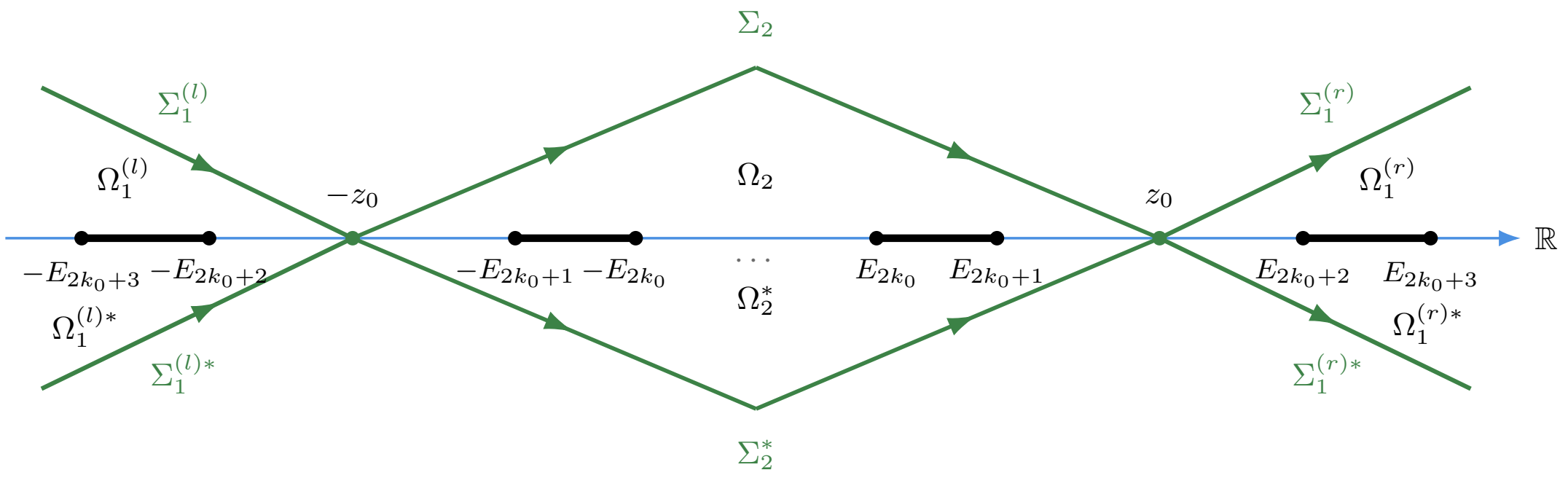

In this section, we study the long-time asymptotics of $q$ in
the Zakharov--Manakov region  $\mathcal{T}_{\mathrm{III}}$, where $\xi=\frac{x}{t}$ satisfies
\[
\xi\in
\begin{cases}
	\displaystyle
	(-\infty,\xi_{2n+1})
	\cup\bigcup_{k=1}^{n}(\xi_{2k},\xi_{2k-1}),
	& \text{ genus }2n,\\[2mm]
	\displaystyle
	(-\infty,\xi_{2n+1})
	\cup\bigcup_{k=2}^{n}(\xi_{2k},\xi_{2k-1})
	\cup(\xi_2,\xi_0),
	& \text{ genus }2n-1.
\end{cases}
\]
By the discussion in Section~\ref{sec25}, the phase $\theta$ defined in \eqref{alpha} has two
simple real saddle points $\pm z_0\in\mathbb R\setminus\Gamma$,
with $z_0>0$.
Figure~\ref{fig:mkdv-region-III-contours} shows the contours
and regions for
\[
\xi_{2k_0+2}<\xi<\xi_{2k_0+1},
\qquad
E_{2k_0+1}<z_0<E_{2k_0+2},
\qquad 1\leq k_0\leq n-1.
\]
The analysis follows the standard nonlinear steepest descent
method \cite{DZ93,PX3,MT}, and we only present the constructions and estimates
needed for the proof of Theorem~\ref{theorem}.

 For genus $2n$, we define
\begin{equation}\label{delta-region-III}
	\begin{aligned}
		\delta(z):=&\,\frac{w(z)}{2\pi i} \left(\sum_{j=1}^{n-1}\delta_j\int_{-E_{2n-2j+1}}^{-E_{2n-2j}}+\delta_{n}\int_{-E_1}^{E_1}+\sum_{j=n+1}^{2n}\delta_j\int_{E_{2j-2n}}^{E_{2j-2n+1}}\right)\frac{i}{w_+(s)(s-z)}\dif s\\
		&+\frac{w(z)}{2\pi i}\int_{I_0}\frac{\log (1-r_1(s)r_2(s))}{w(s)(s-z)}\dif s,
	\end{aligned}
\end{equation}
where, unlike in the preceding sections,
\begin{equation}
	I_0:=(-z_0,z_0)\setminus \Gamma,
\end{equation}
 $w$, $r_1$ and $r_2$ are defined in \eqref{w} and \eqref{r}, respectively, and the constants $\delta_j$, $j=1,\ldots,2n$, are determined by  the following linear systems:
\begin{equation}\label{deltaj-region-III}
	\begin{aligned}
	&\left(\sum_{j=1}^{n-1}\delta_j\int_{-E_{2n-2j+1}}^{-E_{2n-2j}}+\delta_{n}\int_{-E_1}^{E_1}+\sum_{j=n+1}^{2n}\delta_j\int_{E_{2j-2n}}^{E_{2j-2n+1}}\right)\frac{is^m}{w_+(s)}\dif s\\
	=&-\int_{I_0}\frac{\log (1-r_1(s)r_2(s))s^m}{w(s)}\dif s,
	\end{aligned}
\end{equation}
with $m=0,\cdots,2n-1$. For genus $2n-1$, the treatment is analogous: one simply deletes the integral term $\delta_{n}\int_{-E_1}^{E_1}$ in \eqref{delta-region-III} and \eqref{deltaj-region-III}, and takes $m=0,\cdots,2n-2$. And it also holds that
\begin{align}\label{deltainfty-region-III}
    \delta(\infty):=\lim_{z\to\infty} \delta(z)=\frac{i}{4}\delta_{2n}.
\end{align}

Starting from RH problem \ref{RHM},  we set
\begin{equation}\label{eq:III-first}
	M^{(1)}(z)=e^{\delta(\infty)\sigma_3}M(z)G(z)e^{-\delta(z)\sigma_3},
\end{equation}
where the matrix $G$ is given by \eqref{def:G}, with the
regions now chosen as in Figure~\ref{fig:mkdv-region-III-contours}. Then, we consider the following RH problem.

\begin{RHP}\label{RHPIII}
	\hfil
	\begin{itemize}
	\item $M^{(1)}(z)$ is analytic in
	$\mathbb C\setminus\overline{\Sigma^{(1)}}$, where
	\begin{equation}\label{eq:III-jump-contour}
		\Sigma^{(1)}
		:=\Gamma\cup\Sigma_1\cup\Sigma_1^*
		\cup\Sigma_2\cup\Sigma_2^*,
	\end{equation}
and
	\begin{equation*}
		\begin{aligned}
			\Sigma_1&
			:=\,\Sigma_1^{(l)}\cup\Sigma_1^{(r)}
			=\,\{-z_0+se^{i(\pi-\varphi_0)}:s\geq0\}
			\cup\{z_0+se^{i\varphi_0}:s\geq0\},\\
			\Sigma_2&
			:=\,\{-(1-s)z_0+is\rho_0:0\leq s\leq1\}
		\cup
			\{sz_0+i(1-s)\rho_0:0\leq s\leq1\}.
		\end{aligned}
	\end{equation*}
	We choose $\varphi_0$ and $\rho_0$ so that
	\begin{equation*}
		\operatorname{Im}\theta(z)<0,
		\quad z\in\Sigma_1\setminus\{\pm z_0\},
		\qquad
		\operatorname{Im}\theta(z)>0,
		\quad z\in\Sigma_2\setminus\{\pm z_0\},
	\end{equation*}
	see Figure~\ref{fig:mkdv-region-III-contours}.
	
	\item $M^{(1)}(z)$ satisfies the jump condition
	\[
	M_+^{(1)}(z)=M_-^{(1)}(z)V^{(1)}(z),
	\qquad z\in\Sigma^{(1)}.
	\]
	For $z\in\Gamma$, the jump matrix is given by \eqref{V11}.
	For  $z\in\Sigma^{(1)}\setminus\Gamma$, we have
	\begin{equation}\label{eq:III-lens-jumps}
		V^{(1)}(z)=
		\begin{cases}
			\begin{pmatrix}
				1&0\\
				-r_2(z)e^{-2it\theta(z)-2\delta(z)}&1
			\end{pmatrix},
			&z\in\Sigma_1,\\[4mm]
			\begin{pmatrix}
				1&r_1(z)e^{2it\theta(z)+2\delta(z)}\\
				0&1
			\end{pmatrix},
			&z\in\Sigma_1^*,\\[4mm]
			\begin{pmatrix}
				1&
				\frac{r_1(z)e^{2it\theta(z)+2\delta(z)}}
				{1-r_1(z)r_2(z)}\\
				0&1
			\end{pmatrix},
			&z\in\Sigma_2,\\[5mm]
			\begin{pmatrix}
				1&0\\
				-\frac{r_2(z)e^{-2it\theta(z)-2\delta(z)}}
				{1-r_1(z)r_2(z)}&1
			\end{pmatrix},
			&z\in\Sigma_2^*.
		\end{cases}
	\end{equation}
	\item $M^{(1)}(z)$ satisfies the symmetry relations
		\begin{equation*}
			M^{(1)}(z)=\sigma_1\overline{M^{(1)}(\bar{z})}\sigma_1=e^{2\delta(\infty)\sigma_3}
\sigma_2M^{(1)}(-z)\sigma_2
e^{-2\delta(\infty)\sigma_3}.
		\end{equation*}
	\item As $z\to\infty$,
	\(
	M^{(1)}(z)=I+\mathcal O(z^{-1}).
\)
	
	\item As $z\to p\in \mathbf E$,
	\(
	M^{(1)}(z)=\mathcal O\bigl((z-p)^{-\frac{1}{4}}\bigr).
	\)	
\end{itemize}
\end{RHP}

Next, we divide $M^{(1)}$ into the following parts
\begin{equation}\label{eq:III-error-transformation}
	M^{(1)}(z)=E(z)\times\begin{cases}
		M^{(glo)}(z),
		& z\in\mathbb C\setminus\bigl(U^{(l)}\cup U^{(r)}\bigr),\\
		M^{(l)}(z), & z\in U^{(l)},\\
		M^{(r)}(z), & z\in U^{(r)}.
	\end{cases}
\end{equation}
where $E$ is the error matrix.
Here $M^{(glo)}$ is the global parametrix, and
$M^{(l)}$, $M^{(r)}$ are the local parametrices in
\begin{equation}\label{urrr}
	U^{(l)}:=\{z\in\mathbb C:|z+z_0|\leq c_0\},
	\qquad
	U^{(r)}:=\{z\in\mathbb C:|z-z_0|\leq c_0\},
\end{equation}
with
\begin{equation*}
	c_0:=\frac12\min\left\{
	z_0,\ \operatorname{dist}(z_0,\Gamma)
	\right\}.
\end{equation*}

The construction of $M^{(glo)}$ is analogous to that
in Section~\ref{global}. In particular,
\begin{equation}\label{eq:III-global-reconstruction}
	q^{(AG)}(x,t; \mathbf E,\widetilde{\boldsymbol\phi})
	=e^{-2\delta(\infty)}
	\lim_{z\to\infty}2izM^{(glo)}_{12}(z),
\end{equation}
where \begin{equation}\label{phi-region-III}
    \widetilde{\boldsymbol{\phi}}
 =(\widetilde{\phi}_1,\ldots,\widetilde{\phi}_n)^T,\qquad \widetilde{\phi}_k
=\phi_k+\delta_{n-k}-\frac{\delta_{2n}}{2},
 \quad  k=1,\ldots,n,
 \end{equation} with $\delta_j$
now determined by the equations in \eqref{deltaj-region-III}.

We next construct the local parametrices and focus on $M^{(r)}$.
The construction of $M^{(l)}$ follows also by the symmetry as in \eqref{sym:MrandMl}. It therefore suffices to consider the following local
RH problem.

\begin{RHP}
		\hfil
\begin{itemize}
\item $M^{(r)}(z)$ is analytic in
$U^{(r)}\setminus\Sigma^{(1)}$,
where  $\Sigma^{(1)}$ and $U^{(r)}$ are defined by \eqref{eq:III-jump-contour} and \eqref{urrr}, respectively.

\item $M^{(r)}(z)$ satisfies the jump condition
\[
M^{(r)}_+(z)=M^{(r)}_-(z)V^{(1)}(z),
\qquad z\in\Sigma^{(1)}\cap U^{(r)},
\]
where $V^{(1)}$ is given by \eqref{eq:III-lens-jumps}.

\item As $t\to\infty$, $M^{(r)}(z)$ matches
$M^{(\mathrm{glo})}(z)$ on $\partial U^{(r)}$.
\end{itemize}
\end{RHP}

Following the analysis in
\cite{DZ93,DZ94case,DZ94higher,PX3,DM08,DMM19},
we obtain from the standard parabolic cylinder parametrix
in \cite{Its81}, as $t\to\infty$,
\begin{equation}\label{eq:III-local-asymptotics}
	M^{(r)}(z)
	=
	M^{(\mathrm{glo})}(z)
	\left[
	I+
	\frac{t^{-\frac{1}{2}}}
	{2(z-z_0)\sqrt{\theta^{(2)}(\xi)}}
	\begin{pmatrix}
		0&\gamma_{21}\\
		\gamma_{12}&0
	\end{pmatrix}
	+\mathcal{O}(t^{-1})
	\right].
\end{equation}
Here $\theta^{(2)}$ is determined by
\begin{equation}\label{eq:III-phase-expansion}
	\theta(z;\xi)
	=
	\theta(z_0)-\theta^{(2)}(\xi)(z-z_0)^2
	+\mathcal{O}((z-z_0)^3),
	\qquad z\to z_0,
\end{equation}
with
\begin{equation}\label{eq:III-phase-coefficient}
	\theta^{(2)}(\xi)
	=-\frac12\theta''(z_0)
	=
	\begin{cases}
		\displaystyle
		\frac{z_0\partial_zF(z_0;\xi)}{2w(z_0)},
		&g=2n,\\[6pt]
		\displaystyle
		\frac{\partial_zF(z_0;\xi)}{2w(z_0)},
		&g=2n-1,
	\end{cases}
\end{equation}
and  $\theta^{(2)}(\xi)>0.$
Set
\begin{equation*}
	r_0:=
	-r_2(z_0)e^{-2it\theta(z_0)-2\chi(z_0)}
	\left(2\sqrt{t\theta^{(2)}(\xi)}\right)^{2i\nu},
\end{equation*}
with
\begin{equation*}
	\nu:=-\frac{\log\bigl(1-r_1(z_0)r_2(z_0)\bigr)}{2\pi}
	,
	\qquad
	\chi(z_0):=
	\lim_{z\to z_0}
	\bigl[\delta(z)-i\nu\log(z-z_0)\bigr].
\end{equation*}
Here $\log(z-z_0)$ is taken on the principal branch,
with $-\pi<\arg(z-z_0)<\pi$.
Then $|r_0|=|r_2(z_0)|<1$, and the coefficients in
\eqref{eq:III-local-asymptotics} are
\begin{equation}\label{eq:III-beta}
	\gamma_{12}
	=
	\frac{\sqrt{2\pi}\,e^{-\frac{\pi\nu}{2}+\frac{i\pi}{4}}}
	{\overline{r}_0\,\Gamma(i\nu)},
	\qquad
	\gamma_{21}
	=
	\frac{\sqrt{2\pi}\,e^{-\frac{\pi\nu}{2}-\frac{i\pi}{4}}}
	{r_0\,\Gamma(-i\nu)}.
\end{equation}
Here $\Gamma$ denotes the Euler gamma function.

The error analysis is analogous to that in Section~\ref{error}.
For sufficiently large $t$, the error matrix $E$ has the expansion
\begin{equation}\label{eq:III-error-expansion}
	E(z)=I+\frac{E_1}{z}+\mathcal{O}(z^{-2}),
	\qquad z\to\infty,
\end{equation}
where, as $t\to\infty$,
\begin{equation}\label{eq:III-error-coefficient}
	E_1
	=
	t^{-\frac{1}{2}}
	\left[
	H_{z_0}
	-
	e^{2\delta(\infty)\sigma_3}
	\sigma_2H_{z_0}\sigma_2
	e^{-2\delta(\infty)\sigma_3}
	\right]
	+\mathcal{O}(t^{-1}),
\end{equation}
with
\begin{equation}\label{eq:III-Hz0}
	H_{z_0}
	:=
	\frac{1}{2\sqrt{\theta^{(2)}}}
	M^{(\mathrm{glo})}(z_0)
	\begin{pmatrix}
		0&\gamma_{21}\\
		\gamma_{12}&0
	\end{pmatrix}
	\bigl(M^{(\mathrm{glo})}(z_0)\bigr)^{-1}.
\end{equation}

\subsection{Fast decaying region  $\mathcal{T}_{\mathrm{IV}}$}\label{sec52}

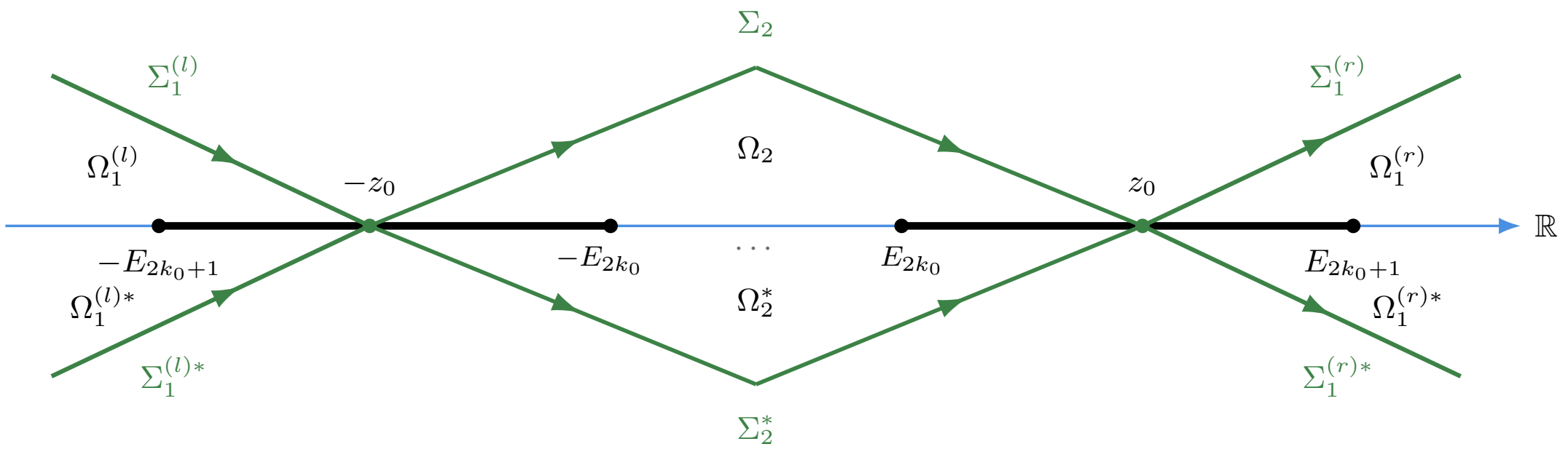
\begin{figure}
	\centering
	\definecolor{mkdvAxisBlue}{RGB}{74,144,226}
	\definecolor{mkdvLensGreen}{RGB}{60,130,70}
	\resizebox{0.90\linewidth}{!}{%
		\begin{tikzpicture}[
			x=1cm,
			y=1cm,
			>=Latex,
			axis/.style={draw=mkdvAxisBlue,line width=0.72pt},
			band/.style={draw=black,line width=2.15pt},
			outer/.style={
				draw=mkdvLensGreen,
				line width=1.25pt,
				postaction={decorate},
				decoration={markings,
					mark=at position 0.6 with
					{\arrow{Latex[length=3.0mm,width=1.9mm]}}
				}
			},
			inner/.style={draw=mkdvLensGreen,line width=1.15pt},
			innerarrow/.style={
				inner,
				postaction={decorate},
				decoration={markings,
					mark=at position 0.55 with
					{\arrow{Latex[length=3.0mm,width=1.9mm]}}
				}
			},
			point/.style={circle,fill=black,inner sep=1.45pt},
			every node/.style={font=\footnotesize}
			]
			
			% Representative component of region IV:
			% E_{2k_0} < z_0 < E_{2k_0+1}.
			\def\Eeven{1.45}
			\def\zsaddle{3.85}
			\def\Eodd{5.95}
			\def\rho{1.58}
			\def\rayx{7.02}
			\def\rayy{1.50}
			
			% Real axis and spectral bands.
			\draw[axis,->] (-7.48,0)--(7.62,0) node[right] {$\mathbb R$};
			\draw[band] (-\Eodd,0)--(-\Eeven,0);
			\draw[band] ( \Eeven,0)--( \Eodd,0);
			\node[black!58] at (0,-0.23) {$\cdots$};
			
			% Sigma_1 and Sigma_1^*.
			\draw[outer] (-\rayx, \rayy)--(-\zsaddle,0);
			\draw[outer] ( \zsaddle,0)--( \rayx, \rayy);
			\draw[outer] (-\rayx,-\rayy)--(-\zsaddle,0);
			\draw[outer] ( \zsaddle,0)--( \rayx,-\rayy);
			
			% Sigma_2 and Sigma_2^*, oriented from -z_0 to z_0.
			\draw[innerarrow] (-\zsaddle,0)--(0, \rho);
			\draw[innerarrow] (0, \rho)--( \zsaddle,0);
			\draw[innerarrow] (-\zsaddle,0)--(0,-\rho);
			\draw[innerarrow] (0,-\rho)--( \zsaddle,0);
			
			% Branch points and saddle points.
			\foreach \x in {-\Eodd,-\Eeven,\Eeven,\Eodd}
			\node[point] at (\x,0) {};
			\node[point,fill=mkdvLensGreen] at (-\zsaddle,0) {};
			\node[point,fill=mkdvLensGreen] at ( \zsaddle,0) {};
			
			% Point labels.
			\node[above=5pt] at (-\zsaddle,0) {$-z_0$};
			\node[above=5pt] at ( \zsaddle,0) {$z_0$};
			\node[below=3pt] at (-\Eodd,0) {$-E_{2k_0+1}$};
			\node[below=2pt,font=\scriptsize,xshift=-3pt] at (-\Eeven,0) {$-E_{2k_0}$};
			\node[below=2pt,font=\scriptsize,xshift=3pt]  at ( \Eeven,0) {$E_{2k_0}$};
			\node[below=3pt] at ( \Eodd,0) {$E_{2k_0+1}$};
			
			% Contour labels.
			\node[mkdvLensGreen] at (-5.8, 1.5) {$\Sigma_1^{(l)}$};
			\node[mkdvLensGreen] at ( 5.8, 1.5) {$\Sigma_1^{(r)}$};
			\node[mkdvLensGreen] at (-5.8,-1.5) {$\Sigma_1^{(l)*}$};
			\node[mkdvLensGreen] at ( 5.8,-1.5) {$\Sigma_1^{(r)*}$};
			\node[mkdvLensGreen,above=5pt] at (0, \rho) {$\Sigma_2$};
			\node[mkdvLensGreen,below=5pt] at (0,-\rho) {$\Sigma_2^*$};
			
			% Region labels.
			\pgfmathsetmacro{\innerlabely}{\zsaddle*\rho/
				(\zsaddle+sqrt(\zsaddle^2+\rho^2))}
			\node[black] at (-6.4, 0.6) {$\Omega_1^{(l)}$};
			\node[black] at ( 6.4, 0.6) {$\Omega_1^{(r)}$};
			\node[black] at (-6.5,-0.8) {$\Omega_1^{(l)*}$};
			\node[black] at ( 6.5,-0.8) {$\Omega_1^{(r)*}$};
			\node[black] at (0, \innerlabely) {$\Omega_2$};
			\node[black] at (0,-\innerlabely) {$\Omega_2^*$};
			
		\end{tikzpicture}%
	}
	\caption{The contours $\Sigma_j,\Sigma_j^*$ and regions
		$\Omega_j,\Omega_j^*$, $j=1,2$, in fast decaying region~$\mathcal{T}_{\mathrm{IV}}$.}
	\label{fig:mkdv-region-IV-contours}
\end{figure}

In this section, we consider  the long-time asymptotics of
$q$ in the  fast decaying region  $\mathcal{T}_{\mathrm{IV}}$,  where
\[
\xi \in \begin{cases}
	\displaystyle\bigcup_{k=1}^{n}(\xi_{2k+1},\xi_{2k})
	\cup(\xi_1,+\infty),
	& \text{genus }2n,\\[2mm]
	\displaystyle\bigcup_{k=1}^{n}(\xi_{2k+1},\xi_{2k})
	\cup(\xi_0,+\infty),
	& \text{genus }2n-1.
\end{cases}
\]
For $\xi_{2k_0+1}<\xi<\xi_{2k_0}$, we have
$E_{2k_0}<z_0<E_{2k_0+1}$, where $1\leq k_0\leq n$;
see Figure~\ref{fig:mkdv-region-IV-contours}.

The analysis is analogous to that in Section~\ref{sec51}.
 For genus $2n$, we define
\begin{equation}\label{delta-IV}
	\begin{aligned}
		\delta(z):=&\,\frac{w(z)}{2\pi i} \left(\sum_{j=1}^{n-1}\delta_j\int_{-E_{2n-2j+1}}^{-E_{2n-2j}}+\delta_{n}\int_{-E_1}^{E_1}+\sum_{j=n+1}^{2n}\delta_j\int_{E_{2j-2n}}^{E_{2j-2n+1}}\right)\frac{i}{w_+(s)(s-z)}\dif s\\
		&+\frac{w(z)}{2\pi i}\int_{I_0}\frac{\log (1-r_1(s)r_2(s))}{w(s)(s-z)}\dif s,
	\end{aligned}
\end{equation}
where
\[
I_0:=
\begin{cases}
	(-z_0,z_0)\setminus\Gamma,
	& \xi_{2k_0+1}<\xi<\xi_{2k_0},\quad k_0=1,\ldots,n,\\
	\varnothing,
	& \xi>\xi_1 \text{ for genus }2n
	\text{ or }\xi>\xi_0 \text{ for genus }2n-1.
\end{cases}
\]
Here $w$, $r_1$ and $r_2$ are defined in \eqref{w} and \eqref{r}, respectively, and the constants $\delta_j$, $j=1,\ldots,2n$, are determined by  the following linear systems:
\begin{equation}\label{deltaj-IV}
	\begin{aligned}
	&\left(\sum_{j=1}^{n-1}\delta_j\int_{-E_{2n-2j+1}}^{-E_{2n-2j}}+\delta_{n}\int_{-E_1}^{E_1}+\sum_{j=n+1}^{2n}\delta_j\int_{E_{2j-2n}}^{E_{2j-2n+1}}\right)\frac{is^m}{w_+(s)}\dif s\\
	=&-\int_{I_0}\frac{\log (1-r_1(s)r_2(s))s^m}{w(s)}\dif s,
	\end{aligned}
\end{equation}
with $m=0,\cdots,2n-1$. For genus $2n-1$, the treatment is analogous: one simply deletes the integral term $\delta_{n}\int_{-E_1}^{E_1}$ in \eqref{delta-IV} and \eqref{deltaj-IV}, and takes $m=0,\cdots,2n-2$.

We perform the same transformation \eqref{eq:III-first}, where
the matrix $G$ is given by \eqref{def:G}, with the regions in Figure \ref{fig:mkdv-region-IV-contours}. Then $M^{(1)}$ satisfies an RH problem of the same form as RH problem~\ref{RHPIII}, with the corresponding contours and scalar function $\delta$.

Since there are no real saddle points in
$\mathbb R\setminus\Gamma$, no local parabolic
cylinder parametrices are required. We therefore set
\begin{equation}\label{eq:IV-transformation}
	M^{(1)}(z)=E(z)M^{(glo)}(z).
\end{equation}
The construction of $M^{(glo)}$ is the same as in
Section~\ref{global}, with $\delta$ now defined for region~$\mathcal{T}_{\mathrm{IV}}$.
In particular,
\begin{equation}\label{eq:IV-global-reconstruction}
	q^{(AG)}(x,t;\mathbf E,\widetilde{\boldsymbol\phi})
	=e^{-2\delta(\infty)}
	\lim_{z\to\infty}2izM^{(glo)}_{12}(z),
\end{equation}
where $\widetilde{\boldsymbol\phi}$ is defined by \eqref{phi}, with $\delta_j$ and $\delta(\infty)$
now determined by the scalar function $\delta$ for
region~$\mathcal{T}_{\mathrm{IV}}$.

Using  an error analysis
analogous to that in Section~\ref{error}, we have
\begin{equation}\label{eq:IV-error}
	E(z)=I+\frac{E_1}{z}+\mathcal O(z^{-2}),
	\qquad z\to\infty,
\end{equation}
with
\begin{equation}\label{518}
E_1=\mathcal O(t^{-1}).
\end{equation}

\subsection{Central transition region $\mathcal{T}_{\mathrm{0}}$}
For genus $2n-1$, the origin lies in the central gap
$(-E_2,E_2)$, and the two real saddle points $\pm z_0$
will merge at $z=0$ as $\xi\to\xi_0$.
Therefore, in this section, we consider the long-time asymptotics of
$q(x,t)$ in the central transition region defined in
Definition~\ref{regions}, where
\[
|\xi-\xi_0|t^{\frac{2}{3}}\leq C,
\qquad \xi=\frac{x}{t},
\]
with $C>0$ fixed and $\xi_0$ given by \eqref{xi0}. For sufficiently large $t$, the real saddle points satisfy
$0\leq z_0<E_2$ when $\xi\leq\xi_0$.
The signature of $\operatorname{Im}\theta$ near the critical
value $\xi=\xi_0$ is illustrated in Figure~\ref{figim2}. It inspires us to consider a Painlev\'e II
parametrix, following the idea in \cite{DZ93,Lenellsmkdv,SA81}.

\begin{figure}
	\centering
	\definecolor{mkdvAxisBlue}{RGB}{74,144,226}
	\definecolor{mkdvLensGreen}{RGB}{60,130,70}
	\resizebox{0.84\linewidth}{!}{%
		\begin{tikzpicture}[
			x=1cm,
			y=0.50cm,
			>=Latex,
			axis/.style={draw=mkdvAxisBlue,line width=0.72pt},
			band/.style={draw=black,line width=2pt},
			outer/.style={
				draw=mkdvLensGreen,
				line width=1.15pt,
				postaction={decorate},
				decoration={markings,
					mark=at position 0.6 with
					{\arrow{Latex[length=3.0mm,width=1.9mm]}}
				}
			},
			point/.style={circle,fill=black,inner sep=1.45pt},
			every node/.style={font=\footnotesize}
			]
			% Spectral bands and lens boundaries.
			\def\Einner{2.30}
			\def\Einnerend{3.62}
			\def\Eouterstart{4.95}
			\def\Eouterend{6.27}
			\def\rayx{4.90}
			\pgfmathsetmacro{\rayy}{\rayx/sqrt(3)}
			
			% Real axis.
			\draw[axis,->] (-6.90,0)--(7.05,0) node[right] {$\mathbb R$};
			
			% Spectral bands.
			\draw[band] (-\Eouterend,0)--(-\Eouterstart,0);
			\draw[band] (-\Einnerend,0)--(-\Einner,0);
			\draw[band] ( \Einner,0)--( \Einnerend,0);
			\draw[band] ( \Eouterstart,0)--( \Eouterend,0);
			
			\node[black!58,below=4pt] at (-4.285,0) {$\cdots$};
			\node[black!58,below=4pt] at ( 4.285,0) {$\cdots$};
			
			% Lens boundaries, oriented from left to right.
			\draw[outer] (-\rayx, \rayy)--(0,0);
			\draw[outer] (0,0)--(\rayx, \rayy);
			\draw[outer] (-\rayx,-\rayy)--(0,0);
			\draw[outer] (0,0)--(\rayx,-\rayy);
			
			% Branch points and origin.
			\foreach \x in {
				-\Eouterend,-\Eouterstart,-\Einnerend,-\Einner,
				\Einner,\Einnerend,\Eouterstart,\Eouterend
			}{
				\node[point] at (\x,0) {};
			}
			\node[point,fill=mkdvLensGreen] at (0,0) {};
			\node[above=4pt] at (0,0) {$0$};
			
			% Endpoint labels.
			\node[below=3pt] at (-\Eouterend,0)   {$-E_{2n+1}$};
			\node[below=3pt] at (-\Eouterstart,0) {$-E_{2n}$};
			\node[below=3pt] at (-\Einnerend,0)   {$-E_3$};
			\node[below=3pt] at (-\Einner,0)      {$-E_2$};
			\node[below=3pt] at ( \Einner,0)      {$E_2$};
			\node[below=3pt] at ( \Einnerend,0)   {$E_3$};
			\node[below=3pt] at ( \Eouterstart,0) {$E_{2n}$};
			\node[below=3pt] at ( \Eouterend,0)   {$E_{2n+1}$};
			
			% Contour labels.
			\foreach \x in {-3.35,3.35}{
				\node[mkdvLensGreen,above=4pt] at (\x, 1.94) {$\Sigma_1$};
				\node[mkdvLensGreen,below=4pt] at (\x,-1.94) {$\Sigma_1^*$};
			}
			
			% Region labels.
			\foreach \x in {-4.28,4.28}{
				\node at (\x, 1.15) {$\Omega_1$};
				\node at (\x,-1.48) {$\Omega_1^*$};
			}
		\end{tikzpicture}%
	}
	\caption{The opened contours $\Sigma_1,\Sigma_1^*$ and the lens
		regions $\Omega_1,\Omega_1^*$ in the central transition
		region $\mathcal{T}_{\mathrm{0}}$.}
	\label{fig:central-lenses}
\end{figure}
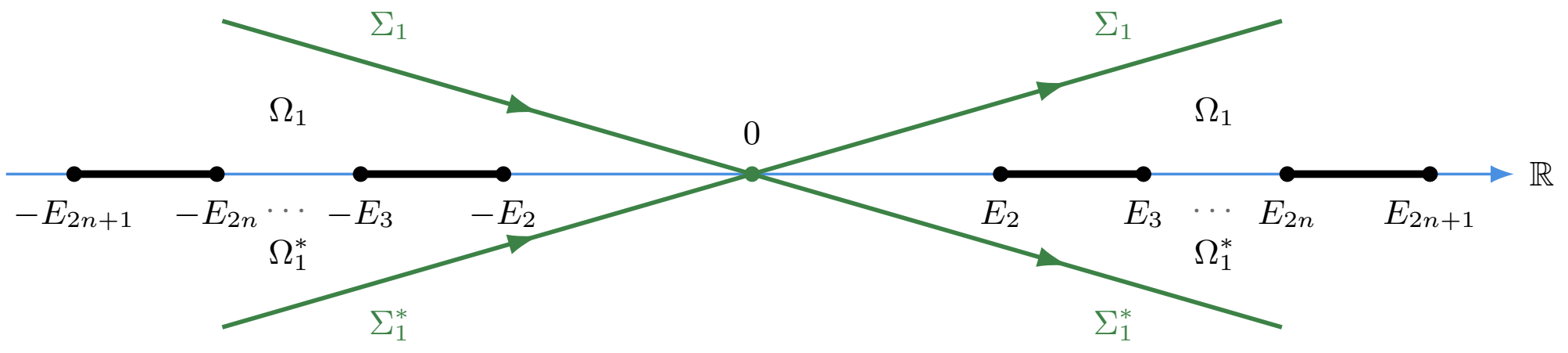

\medskip
\noindent\textit{The first transformation.}
We first open the lenses using the factorizations
\eqref{fac1} and \eqref{fac2}. Define
\begin{equation}\label{central-M1}
	M^{(1)}(z):=M(z)G(z),
\end{equation}
where
\begin{equation}\label{central-G}
	G(z):=
	\begin{cases}
		\begin{pmatrix}
			1 & 0\\
			r_2(z)e^{-2it\theta(z)} & 1
		\end{pmatrix},
		& z\in\Omega_1,\\[3mm]
		\begin{pmatrix}
			1 & r_1(z)e^{2it\theta(z)}\\
			0 & 1
		\end{pmatrix},
		& z\in\Omega_1^*,\\[3mm]
		I, & \text{elsewhere}.
	\end{cases}
\end{equation}
The contours and regions are defined by
\begin{equation*}
	\Sigma_1
	:=\{se^{\frac{\pi i}{6}}:s\geq0\}
	\cup\{se^{\frac{5\pi i}{6}}:s\geq0\},
	\qquad
	\Sigma_1^*:=\{\bar z:z\in\Sigma_1\},
\end{equation*}
and
\begin{equation*}
	\Omega_1
	:=\left\{z\in\mathbb C:
	0<\arg z<\frac{\pi}{6}
	\ \text{or}\
	\frac{5\pi}{6}<\arg z<\pi
	\right\},
	\qquad
	\Omega_1^*:=\{\bar z:z\in\Omega_1\}.
\end{equation*}
Both contours are oriented from left to right;
see Figure~\ref{fig:central-lenses}.

It follows from  RH problem \ref{RHM} that $M^{(1)}$ satisfies the following RH problem.
\begin{RHP}\label{RHcenter}
	\hfil
	\begin{itemize}
	\item
	$M^{(1)}$ is analytic in $\mathbb C\setminus\overline{\Sigma^{(1)}}$,
	where
	\begin{equation*}
		\Sigma^{(1)}:=\Gamma\cup\Sigma_1\cup\Sigma_1^*.
	\end{equation*}

	\item
	$M^{(1)}$ satisfies the jump condition
	\[
	M^{(1)}_+(z)=M^{(1)}_-(z)V^{(1)}(z),
	\qquad z\in\Sigma^{(1)},
	\]
	where
	\begin{equation}
		V^{(1)}(z)=
		\begin{cases}
			\begin{pmatrix}
				0 & -ie^{i(\phi_k+C_k^f x+C_k^g t)}\\
				-ie^{-i(\phi_k+C_k^f x+C_k^g t)} & 0
			\end{pmatrix},
			& z\in(-E_{2k+1},-E_{2k}),\\[4mm]
			\begin{pmatrix}
				0 & -ie^{-i(\phi_k+C_k^f x+C_k^g t)}\\
				-ie^{i(\phi_k+C_k^f x+C_k^g t)} & 0
			\end{pmatrix},
			& z\in(E_{2k},E_{2k+1}),\\[4mm]
			\begin{pmatrix}
				1 & 0\\
				-r_2(z)e^{-2it\theta(z)} & 1
			\end{pmatrix},
			& z\in\Sigma_1,\\[4mm]
			\begin{pmatrix}
				1 & r_1(z)e^{2it\theta(z)}\\
				0 & 1
			\end{pmatrix},
			& z\in\Sigma_1^*,
		\end{cases}
	\end{equation}
	with $k=1,\ldots,n$.
	
	\item
	As $z\to\infty$, we have
	\[
	M^{(1)}(z)=I+\mathcal O(z^{-1}).
\]
\end{itemize}
\end{RHP}

\medskip
\noindent\textit{Global and local parametrices.}
We write
\begin{equation}\label{eeerr}
M^{(1)}(z)=	E(z)\times\begin{cases}
		M^{(\mathrm{glo})}(z), & z\in\mathbb C\setminus U,\\
		M^{(\mathrm{loc})}(z), & z\in U,
	\end{cases}
\end{equation}
where
\[	U:=\{z\in\mathbb C:|z|<t^{-\frac{1}{3}+\epsilon}\},
\qquad0<\epsilon<\frac{1}{3}.\]
For sufficiently large $t$, we have
$\overline U\cap\Gamma=\varnothing $, and $U$ contains
both real saddle points $\pm z_0$  for $-C t^{-\frac{2}{3}}\leq\xi-\xi_0\leq0$ since $|z_0|\sim |\xi-\xi_0|^\frac{1}{2}$ as $t\to\infty$. Therefore, for sufficiently large $t$, $\Sigma_1\setminus U$ belongs to the region where Im$\theta<0.$ It thereby follows that the $V^{(1)}-I$ decays exponentially on $\Sigma_1\setminus U$.

The global parametrix $M^{(\mathrm{glo})}$ has jump $V^{(1)}$ on $\Gamma$. So it is constructed
as in Section~\ref{global}, with all $\delta_j=0$.
Thus,
\begin{equation}\label{IIQ}
	2i\lim_{z\to\infty}z
	\bigl(M^{(\mathrm{glo})}(z)\bigr)_{12}=q^{(\mathrm{AG})}(x,t;\mathbf E,\boldsymbol\phi),
\end{equation}
where $q^{(\mathrm{AG})}$ is given by \eqref{qAG}.

The construction of the  local parametrix follows the same step as in \cite{Lenellsmkdv}.
To construct the local parametrix, we first expand
the phase function $\theta$ defined in \eqref{alpha} near the origin. Since $\theta$ is odd and analytic near $z=0$, we have
\begin{equation}\label{central-phase-expansion}
	\theta(z;\xi)
	=\theta^{(1)}(0)(\xi-\xi_0)z
	+\frac{1}{6}\theta^{(3)}(0;\xi)z^3
	+\mathcal O(z^5),
\end{equation}
where %$\theta^{(j)}(0;\xi)(\xi-\xi_0)= \partial_z^j\theta(z;\xi)|_{z=0}$, and
\begin{align*}
	\theta^{(1)}(0)
	&=
	-\frac{
		\prod_{k=1}^{n}\bigl(-(z_k^f)^2\bigr)}
	{w(0)},
\\
	\theta^{(3)}(0;\xi)
	&=
	-\left.
	\frac{d^2}{dz^2}
	\left(\frac{F(z;\xi)}{w(z)}\right)
	\right|_{z=0}.
\end{align*}
Here $F$ and $w$ are given by \eqref{F} and \eqref{w},
respectively.
In particular,
$\theta^{(3)}(0;\xi_0)<0$. Next, we introduce the scaled variable $\zeta(z)$ such that
\begin{equation}\label{central-scaled-variables}
\frac43\zeta(z)^3
=-t\bigl(\theta(z;\xi)-\theta'(0;\xi)z\bigr).
\end{equation}
It follows from \eqref{central-phase-expansion} that $\zeta$ is analytic in $U$ with
\begin{align}
\zeta(0)=0,\qquad
\zeta'(0)=\frac{(-t\theta'''(0;\xi))^{\frac{1}{3}}}2>0.
\end{align}
Then we define
\begin{align}
    s(z)=-\frac{t\theta'(0;\xi)z}{\zeta(z)},
\end{align}
where
\[	
s:=s(0)=
-\frac{2\theta^{(1)}(0)(\xi-\xi_0)}
{|\theta^{(3)}(0;\xi)|^{\frac{1}{3}}}
t^{\frac{2}{3}}.\]
% Since
% $\theta^{(3)}(0;\xi)
% =\theta^{(3)}(0;\xi_0)+\mathcal O(|\xi-\xi_0|)$,
% we obtain
It thereby appears that
\begin{equation}\label{central-scaled-phase}
t\theta(z;\xi)
=-\left(\frac43\zeta(z)^3+s(z)\zeta(z)\right).
\end{equation}

The parameter $s$ is bounded in the central transition
region. Moreover, on $\partial U$,
\[
|\zeta(z)|
\sim
t^\epsilon\to\infty,
\qquad t\to\infty,
\]
and as $t\to\infty$,
\[\sup_{z\in U}|s(z)-s(0)|
=\mathcal O(t^{-\frac23+2\epsilon})\longrightarrow0,
\qquad 0<\epsilon<\frac13.\]
Now we are attempting to find $M^{(\mathrm{loc})}$ on the $\zeta$-plane admitting the following jump matrix on $\zeta\big((\Sigma_1\cup\Sigma_1^*)\cap U\big)$
	\begin{equation}
		V^{(\mathrm{loc})}(\zeta)=
		\begin{cases}
			\begin{pmatrix}
				1 & 0\\
				-r_2(z(\zeta))e^{2i\left(\frac{4}{3}\zeta^3+s(z(\zeta))\zeta\right)} & 1
			\end{pmatrix},
			& \zeta\in\zeta(\Sigma_1\cap U),\\[4mm]
			\begin{pmatrix}
				1 & r_1(z(\zeta))e^{-2i\left(\frac{4}{3}\zeta^3+s(z(\zeta))\zeta\right)}\\
				0 & 1
			\end{pmatrix},
			& \zeta\in\zeta(\Sigma_1^*\cap U).
		\end{cases}
	\end{equation}
By the symmetries of the reflection coefficients in
Proposition~\ref{propsca}, we have
\(
r_2(0)=\overline{r_1(0)}=-r_1(0).
\)
Define
\begin{equation}
\kappa:=ir_1(0)=-ir_2(0)\in(-1,1).
\end{equation}
Proposition~\ref{propsca} also implies that the above jump matrix $V^{(\mathrm{loc})}$ is well approximated by
	\begin{equation}\label{def:hat Vloc}
		\hat V^{(\mathrm{loc})}(\zeta)=
		\begin{cases}
			\begin{pmatrix}
				1 & 0\\
				-i\kappa e^{2i\left(\frac{4}{3}\zeta^3+s\zeta\right)} & 1
			\end{pmatrix},
			& \zeta\in\zeta(\Sigma_1\cap U),\\[4mm]
			\begin{pmatrix}
				1 & -i\kappa e^{-2i\left(\frac{4}{3}\zeta^3+s\zeta\right)}\\
				0 & 1
			\end{pmatrix},
			& \zeta\in\zeta(\Sigma_1^*\cap U),
		\end{cases}
	\end{equation}
    which yields the jump $V^{(P_2)}(\zeta;s,\kappa)$ in \eqref{PII-jumps} of $M^{(P_2)}(\zeta;s,\kappa)$ for $\zeta\in\zeta(U).$
%Replacing $r_j(z)$ by $r_j(0)$, $j=1,2$, and
%$t\theta(z;\xi)$ by
%$-\left(\frac{4}{3}\zeta^3+s\zeta\right)$
%in the local jumps of RH Problem~\ref{RHcenter}, and reversing
%the orientations of the two left rays, yields the jump matrices of RH Problem~\ref{RHII}.
% Since $M^{(\mathrm{glo})}$ is analytic in $U$, we define
% \begin{equation}
% 	\hat M^{(\mathrm{loc})}(z)
% 	:=
% 	M^{(\mathrm{glo})}(z)
% 	M^{(P_2)}(\zeta(z);s,\kappa)%,	\qquad z\in U\setminus\Sigma^{(1)}
%     .
% 	\label{central-local-parametrix}
% \end{equation}
% It thereby admits the jump $\hat V^{(\mathrm{loc})}(\zeta)$ for $\zeta\in \zeta(U).$
Let $E^{(0,\zeta)}$ be an error function  on the $\zeta$-plane such that
\[E^{(0,\zeta)}_+(\zeta)=E^{(0,\zeta)}_-(\zeta)V^{(E,0)}(\zeta),\qquad
V^{(E,0)}(\zeta)
=
I+
M^{(P_2)}_-(\zeta)
\bigl(V^{(\mathrm{loc})}-V^{(P_2)}\bigr)
\bigl(M^{(P_2)}_+\bigr)^{-1}.\]
Since
\begin{align*}
|V^{(E,0)}(\zeta)-I|
\le
C\left(t^{-\frac13}|\zeta|
+t^{-\frac23}|\zeta|^3\right)
e^{-c|\zeta|^3+C|\zeta|},
\end{align*}
we have
\begin{align*}
  \|V^{(E,0)}-I\|_{L^1\cap L^2\cap L^\infty}
=\mathcal O(t^{-\frac13}).
\end{align*}

The standard small-norm RH problem argument in \cite{PX3}
 gives that $E^{(0,\zeta)}$ exists and is unique. If we write,
\begin{align*}
    E^{(0,\zeta)}(\zeta)=I+\frac{E^{(0,\zeta)}_1}{\zeta}+\mathcal O(\zeta^{-2}),
\end{align*}
then \(E^{(0,\zeta)}_1=\mathcal O(t^{-\frac13}).\) Define
\begin{align}
    M^{(\mathrm{loc})}(\zeta)
:=
M^{(\mathrm{glo})}(z)\,
E^{(0,\zeta)}(\zeta)\,
M^{(P_2)}(\zeta;s,\kappa).
\end{align}
By the expansion in Appendix~\ref{app:PII}, for
$z\in\partial U$, as $t\to\infty,$
\begin{equation}\label{central-local-matching}
	\begin{aligned}
&M^{(\mathrm{loc})}(\zeta)
\bigl(M^{(\mathrm{glo})}(z)\bigr)^{-1}-I\\
={}&
\frac{
M^{(\mathrm{glo})}(z)
\bigl[M^{(P_2)}_1(s;\kappa)+E^{(0,\zeta)}_1\bigr]
\bigl(M^{(\mathrm{glo})}(z)\bigr)^{-1}
}{\zeta(z)}
+\mathcal O(\zeta(z)^{-2})\\
=&\frac{
M^{(\mathrm{glo})}(z)
M^{(P_2)}_1(s;\kappa)
\bigl(M^{(\mathrm{glo})}(z)\bigr)^{-1}
}{\zeta(z)}+\mathcal{O}(t^{-\epsilon-\frac{1}{3}}+t^{-2\epsilon}).
	\end{aligned}
\end{equation}
	Here $M_1^{(P_2)}(s;\kappa)$ is given in Appendix~\ref{app:PII}.
	In particular,
	\[
	M^{(\mathrm{loc})}(z)
	\bigl(M^{(\mathrm{glo})}(z)\bigr)^{-1}
	=I+\mathcal O(t^{-\epsilon}),
	\qquad z\in\partial U,\quad
	0<\epsilon<\frac{1}{3}.
	\]

\medskip
\noindent
\textit{Error estimate.}
It follows from \eqref{eeerr} that the error function satisfies
\begin{equation}
E_+(z)=E_-(z)V^{(E)}(z),
\qquad z\in\Sigma^{(E)},
\end{equation}
where
\begin{equation}
	V^{(E)}(z)=
	\begin{cases}
		M^{(\mathrm{loc})}(z)
		\bigl(M^{(\mathrm{glo})}(z)\bigr)^{-1},
		& z\in\partial U,\\[4pt]
		M^{(\mathrm{glo})}(z)V^{(1)}(z)
		\bigl(M^{(\mathrm{glo})}(z)\bigr)^{-1},
		& z\in(\Sigma_1\cup\Sigma_1^*)
		\setminus  U,\\[4pt]
		M^{(\mathrm{loc})}_-(z)V^{(1)}(z)
		\bigl(M^{(\mathrm{loc})}_+(z)\bigr)^{-1},
		& z\in(\Sigma_1\cup\Sigma_1^*)\cap U,
	\end{cases}
	\label{central-error-jump}
\end{equation}
and
\[
\Sigma^{(E)}
:=\partial U\cup\Sigma_1\cup\Sigma_1^*.
\]
The exterior contours are taken in the corresponding
decay regions, where
\[
\|V^{(E)}-I\|_{L^p(
	(\Sigma_1\cup\Sigma_1^*)\setminus U)}
=
\mathcal O\bigl(e^{-ct^{3\epsilon}}\bigr),
\qquad p=1,2,\infty,
\]
for some $c>0$.
Inside $U$, the analyticity of $r_j$ near the origin,
the phase expansion~\eqref{central-scaled-phase}, and the boundedness of the
Painlev\'e II model and its inverse yield
\begin{equation}
	\|V^{(E)}-I\|_{L^p(
		(\Sigma_1\cup\Sigma_1^*)\cap U)}
	=
	\mathcal O\left(t^{-\frac{1}{3}-\frac{1}{3p}}\right),
	\qquad p=1,2,\infty.
	\label{central-inner-error}
\end{equation}
Together with~\eqref{central-local-matching}, this gives
\begin{equation}\label{central-error-norms}
	\begin{aligned}
		\|V^{(E)}-I\|_{L^\infty(\Sigma^{(E)})}
		=\mathcal O(t^{-\epsilon}),\qquad
		\|V^{(E)}-I\|_{L^2(\Sigma^{(E)})}
		=\mathcal O\left(
		t^{-\frac{1}{6}-\frac{\epsilon}{2}}
		\right).
	\end{aligned}
\end{equation}

Applying the small-norm analysis of Section~\ref{error}, we obtain
\begin{equation}\label{central-error-expansion}
	E(z)=I+\frac{E_1}{z}+\mathcal O(z^{-2}),
	\qquad z\to\infty,
\end{equation}
with
\begin{equation}
	E_1
	=
	-\frac{1}{2\pi i}
	\int_{\partial U}
	\bigl(V^{(E)}(z)-I\bigr)\,dz
	+\mathcal O\left(
	t^{-\frac{1}{3}-\epsilon}
	\right).
	\label{central-error-integral}
\end{equation}
Using~\eqref{central-local-matching} and evaluating the integral by residues yields
\begin{equation}\label{central-error-coefficient}
	E_1
	=
	\frac{
		2M^{(\mathrm{glo})}(0)
		M_1^{(P_2)}(s;\kappa)
		\bigl(M^{(\mathrm{glo})}(0)\bigr)^{-1}
	}{
		|\theta^{(3)}(0;\xi_0)|^{\frac{1}{3}}
		t^{\frac{1}{3}}
	}
	+\mathcal O\left(
	t^{-\frac{1}{3}-\epsilon}
	\right),
	\qquad 0<\epsilon<\frac{1}{3}.
\end{equation}

\section{Proof of Theorem \ref{theorem}}\label{sec6}
	\paragraph{\textbf{Proof of \eqref{eq:main-I} in transition region $\mathcal{T}_{\mathrm{I}}$}}
	By  \eqref{M1} and \eqref{a2e}--\eqref{m2}, we have
	\[
	M(z)
	=
	e^{-\delta(\infty)\sigma_3}
	E(z)M^{(\mathrm{glo})}(z)e^{\delta(z)\sigma_3}G(z)^{-1},
	\qquad  z\in\mathbb{C}\setminus(U^{(l)}\cup U^{(r)}).
	\]
	Taking $z\to\infty$ outside the lens regions, where $G(z)=I$,
	and using \eqref{asde}, \eqref{asm2} and \eqref{ase}, we obtain
	\begin{align}\label{eq:proof-I-expansion}
		M(z)
		=
		I+\frac{1}{z}\left[
		e^{-\delta(\infty)\sigma_3}
		\bigl(E_1+M_1^{(\mathrm{glo})}\bigr)
		e^{\delta(\infty)\sigma_3}
		+\delta^{(1)}\sigma_3
		\right]+\mathcal{O}(z^{-2}).
	\end{align}
	Thus, by \eqref{reco} and \eqref{reco1},
	\begin{align}\label{eq:proof-I-reconstruction}
		q(x,t)
		&=
		2ie^{-2\delta(\infty)}
		\left((E_1)_{12}+(M_1^{(\mathrm{glo})})_{12}\right)
		\nonumber\\
		&=
		q^{(\mathrm{AG})}(x,t;\mathbf E,\widetilde{\boldsymbol\phi})
		+
		2ie^{-2\delta(\infty)}(E_1)_{12}.
	\end{align}
	Since $\det H_1=1$, we have
	\[
	\left(
	H_1
	\begin{pmatrix}0&ia(s)\\0&0\end{pmatrix}
	H_1^{-1}
	\right)_{12}
	=
	ia(s)(H_1)_{11}^{\,2},
	\qquad
	\left(
	H_1
	\begin{pmatrix}0&ia(s)\\0&0\end{pmatrix}
	H_1^{-1}
	\right)_{21}
	=
	-ia(s)(H_1)_{21}^{\,2}.
	\]
	Using these identities and
	$(\sigma_2 A\sigma_2)_{12}=-A_{21}$ in \eqref{eq:asyE1} gives
	\begin{align*}
		(E_1)_{12}
		=
		\frac{ia(s)t^{-\frac{1}{3}}}
		{|\theta^{(3)}(\xi_{2k_0+1})|^{\frac{2}{3}}}
		\left[
		-\bigl(H_1(E_{2k_0+1})\bigr)_{11}^{\,2}
		+
		e^{4\delta(\infty)}
		\bigl(H_1(E_{2k_0+1})\bigr)_{21}^{\,2}
		\right]
		+\mathcal{O}(t^{-\epsilon}).
	\end{align*}
	Substituting this into \eqref{eq:proof-I-reconstruction}
	proves \eqref{eq:main-I}, with
	\begin{align}\label{eq:coefficient-I}
		H_{E_{2k_0+1}}
		:=
		\frac{2}{|\theta^{(3)}(\xi_{2k_0+1})|^{\frac{2}{3}}}
		\left[
		e^{-2\delta(\infty)}
		\bigl(H_1(E_{2k_0+1})\bigr)_{11}^{\,2}
		-
		e^{2\delta(\infty)}
		\bigl(H_1(E_{2k_0+1})\bigr)_{21}^{\,2}
		\right],
	\end{align}
	where $\theta^{(3)}$ and $H_1$ are given by \eqref{theta3}
	and \eqref{H1}, respectively.
	\\[1pt]
	
	\paragraph{\textbf{Proof of \eqref{eq:main-II} in transition region $\mathcal{T}_{\mathrm{II}}$}} It follows from \eqref{eq:II-N1} and  \eqref{eq:II-error-decomposition} that
	\[
	N(z)
	=
	e^{-\delta(\infty)\sigma_3}
	E(z)N^{(\mathrm{glo})}(z)e^{\delta(z)\sigma_3}G(z)^{-1},
	\qquad z\in\mathbb{C}\setminus(U^{(l)}\cup U^{(r)}).
	\]
	Repeating the expansion at infinity used in region~$\mathcal{T}_{\mathrm{I}}$,
	and applying \eqref{reco2}, \eqref{eq:II-global-reconstruction} and \eqref{eq:II-error-expansion}, we obtain
	\[
	q(x,t)
	=
	q^{(\mathrm{AG})}(x,t;\mathbf E,\widetilde{\boldsymbol\phi})
	+
	2ie^{-2\delta(\infty)}(E_1)_{12}.
	\]
	By \eqref{eq:II-error-coefficient},
	\begin{align*}
		(E_1)_{12}
		=
		\frac{ia(s)t^{-\frac{1}{3}}}
		{|\theta^{(3)}(\xi_{2k_0})|^{\frac{2}{3}}}
		\left[
		\bigl(H_1(E_{2k_0})\bigr)_{11}^{\,2}
		-
		e^{4\delta(\infty)}
		\bigl(H_1(E_{2k_0})\bigr)_{21}^{\,2}
		\right]
		+\mathcal{O}(t^{-\epsilon}).
	\end{align*}
	Consequently, \eqref{eq:main-II} follows upon setting
	\begin{align}\label{eq:coefficient-II}
	 H_{E_{2k_0}}
		:=
		\frac{2}{|\theta^{(3)}(\xi_{2k_0})|^{\frac{2}{3}}}
		\left[
		e^{-2\delta(\infty)}
		\bigl(H_1(E_{2k_0})\bigr)_{11}^{\,2}
		-
		e^{2\delta(\infty)}
		\bigl(H_1(E_{2k_0})\bigr)_{21}^{\,2}
		\right],
	\end{align}
	where $\theta^{(3)}$ and $H_1$ are now given by
	\eqref{II-phase-coefficients} and \eqref{II-H1}, respectively.

\paragraph{\textbf{Proof of \eqref{eq:asymptotics-region-III} in Zakharov-Manakov region $\mathcal{T}_{\mathrm{III}}$}}
By \eqref{eq:III-first} and \eqref{eq:III-error-transformation}, we have
\[
M(z)=e^{-\delta(\infty)\sigma_3}
E(z)M^{(glo)}(z)e^{\delta(z)\sigma_3}G(z)^{-1},
\qquad
z\in\mathbb C\setminus\bigl(U^{(l)}\cup U^{(r)}\bigr).
\]
It follows from  \eqref{reco}, \eqref{eq:III-global-reconstruction} and \eqref{eq:III-error-expansion} that
\[
q(x,t)
=q^{(AG)}(x,t;\mathbf E,\widetilde{\boldsymbol\phi})
+2ie^{-2\delta(\infty)}(E_1)_{12}.
\]
Using $(\sigma_2 A\sigma_2)_{12}=-A_{21}$ in \eqref{eq:III-error-coefficient} gives
\[
(E_1)_{12}
=t^{-\frac{1}{2}}\left[
(H_{z_0})_{12}
+e^{4\delta(\infty)}(H_{z_0})_{21}
\right]
+\mathcal O(t^{-1}).
\]
Consequently,
\[
q(x,t)
=q^{(AG)}(x,t;\mathbf E,\widetilde{\boldsymbol\phi})
+2it^{-\frac{1}{2}}\left[
e^{-2\delta(\infty)}(H_{z_0})_{12}
+e^{2\delta(\infty)}(H_{z_0})_{21}
\right]
+\mathcal O(t^{-1}).
\]
%where $H_{z_0}$ is given by \eqref{eq:III-Hz0}.
From the exact expression of  $H_{z_0}$ in  \eqref{eq:III-Hz0} and \eqref{deltainfty-region-III}, we obtain
\begin{align*}
    	\begin{aligned}
				q(x,t)
				={}&q^{(AG)}(x,t;\mathbf E,\widetilde{\boldsymbol\phi})
				\\
                &+\frac{it^{-\frac12}e^{-\frac{i\delta_{2n}}{2}}}{\sqrt{\theta^{(2)}}}
    \left(
    \gamma_{21}
    \bigl(M^{(\mathrm{glo})}_{11}(z_0)^2
    -e^{i\delta_{2n}}
    M^{(\mathrm{glo})}_{21}(z_0)^2
    \bigr)\right.\\
     &\qquad\qquad\qquad\left.
    +\gamma_{12}\bigl(
    e^{i\delta_{2n}}
    M^{(\mathrm{glo})}_{22}(z_0)^2
    -M^{(\mathrm{glo})}_{12}(z_0)^2\bigr)
    \right)
				\\
                &+\mathcal O(t^{-1}).
			\end{aligned}
\end{align*}
From the symmetry \[M^{(\mathrm{glo})}_{21}(z_0)
=\overline{M^{(\mathrm{glo})}_{12}(z_0)},\quad
M^{(\mathrm{glo})}_{22}(z_0)
=\overline{M^{(\mathrm{glo})}_{11}(z_0)},
\quad
\gamma_{12}=\overline{\gamma_{21}},\] it follows that$$e^{-\frac{i\delta_{2n}}{2}}
    \gamma_{21}
    \bigl(M^{(\mathrm{glo})}_{11}(z_0)^2
    -e^{i\delta_{2n}}
    M^{(\mathrm{glo})}_{21}(z_0)^2
    \bigr)=\overline{e^{-\frac{i\delta_{2n}}{2}}\gamma_{12}\bigl(
    e^{i\delta_{2n}}
    M^{(\mathrm{glo})}_{22}(z_0)^2
    -M^{(\mathrm{glo})}_{12}(z_0)^2\bigr)}.$$
We then obtain \eqref{eq:asymptotics-region-III}.

	\paragraph{\textbf{Proof of \eqref{eq:asymptotics-region-IV} in fast decaying region $\mathcal{T}_{\mathrm{IV}}$}}
Applying  \eqref{reco}, \eqref{eq:IV-global-reconstruction} and \eqref{eq:IV-error}, we obtain
\[
q(x,t)
=
q^{(\mathrm{AG})}(x,t;\mathbf E,\widetilde{\boldsymbol\phi})
+2\mathrm{i}e^{-2\delta(\infty)}(E_1)_{12}.
\]
By \eqref{518}, this gives
\[
q(x,t)
=
q^{(\mathrm{AG})}(x,t;\mathbf E,\widetilde{\boldsymbol\phi})
+\mathcal O(t^{-1}),
\]
which proves the desired result.

	\paragraph{\textbf{Proof of \eqref{eq:central-asymptotics}	in the central transition region $\mathcal{T}_{\mathrm{0}}$}}
By \eqref{central-M1} and \eqref{eeerr}, we have
\[
M(z)
=
E(z)M^{(\mathrm{glo})}(z)G(z)^{-1},
\qquad z\in\mathbb C\setminus U.
\]
Taking $z\to\infty$ outside the lens regions, where $G(z)=I$,
and using \eqref{reco}, \eqref{IIQ} and \eqref{central-error-expansion}, we obtain
\[
q(x,t)
=
q^{(\mathrm{AG})}(x,t;\mathbf E,\boldsymbol\phi)
+2i(E_1)_{12}.
\]
By \eqref{central-error-coefficient},
\[
E_1
=
t^{-\frac{1}{3}}H_0
+\mathcal O\!\left(t^{-\frac{1}{3}-\epsilon}\right),
\qquad 0<\epsilon<\frac{1}{3},
\]
where
\begin{equation}\label{eq:central-H0}
	H_0
	:=
	\frac{
		2M^{(\mathrm{glo})}(0)
		M_1^{(P_2)}(s;\kappa)
		\bigl(M^{(\mathrm{glo})}(0)\bigr)^{-1}
	}{
		|\theta^{(3)}(0;\xi_0)|^{\frac{1}{3}}
	}.
\end{equation}
Here $M_1^{(P_2)}(s;\kappa)$ is given in Appendix~\ref{app:PII}. Then by \eqref{asy:MP2}--\eqref{asy:MP22}, we finally obtain
\begin{align*}
  \begin{aligned}
(H_0)_{12}
=\frac{1}{|\theta^{(3)}(0;\xi_0)|^{\frac13}}
\Bigg(
&v(s)
\bigl(M^{(\mathrm{glo})}_{11}(0)^2
-M^{(\mathrm{glo})}_{12}(0)^2\bigr)
\\
&\ \ +2iM^{(\mathrm{glo})}_{11}(0)
M^{(\mathrm{glo})}_{12}(0)
\int_s^\infty v(\eta)^2\,d\eta
\Bigg).
\end{aligned}
\end{align*}
Here, from the symmetry of $M^{(\mathrm{glo})}$ in \eqref{sym:mglo}, we obtain that
\begin{align*}
\begin{aligned}
M^{(\mathrm{glo})}_{11}(0)
&=M^{(\mathrm{glo})}_{22}(0)\in\mathbb R,\\
M^{(\mathrm{glo})}_{12}(0)
&=-M^{(\mathrm{glo})}_{21}(0)\in i\mathbb R.
\end{aligned}
\end{align*}
Thus $(H_0)_{12}\in\mathbb{R}$.
Substitution proves \eqref{eq:central-asymptotics}.

\appendix
\section{The Painlev\'e XXXIV  parametrix}
\label{APP P34}

We record the Painlev\'e XXXIV   model in the normalization used for the local
construction in Section~\ref{sec332}. The underlying RH problem and its
connection with Painlev\'e XXXIV are standard; see
\cite{DXZ20,ikj2008,ikj2009,XZ2011}. Throughout this appendix,
$s\in\mathbb R$, $\alpha>-\frac{1}{2}$, and $\omega\in\mathbb C\setminus(-\infty,0)$.
The specialization needed in Section~\ref{sec332} is
$(\alpha,\omega)=(\frac{1}{4},0)$.

\begin{figure}
	\centering
	\begin{tikzpicture}[
		x=1cm,y=1cm,
		xscale=1.18,yscale=0.56,
		every node/.style={font=\footnotesize},
		contour/.style={
			draw=black,line width=0.75pt,
			postaction={decorate},
			decoration={
				markings,
				mark=at position 0.58 with
				{\arrow{Latex[length=2.0mm,width=1.3mm]}}
			}
		},
		point/.style={circle,fill=black,inner sep=1.35pt}
		]
		
		% Oriented contours.
		\draw[contour] (0,0)--(2.25,0);
		\draw[contour] (120:3.48)--(0,0);
		\draw[contour] (-2.85,0)--(0,0);
		\draw[contour] (240:3.48)--(0,0);
		
		% Origin.
		\node[point] at (0,0) {};
		\node[above right=2pt] at (0,0) {$0$};
		
		% Contour labels.
		\node[above=2pt]       at ( 1.45,0)  {$\Sigma_1$};
		\node[above right=2pt] at (120:2.65) {$\Sigma_2$};
		\node[above=2pt]       at (-2.05,0)  {$\Sigma_3$};
		\node[below right=2pt] at (240:2.65) {$\Sigma_4$};
		
		% Sector labels.
		\node at ( 0.72, 1.32) {$\Omega_1$};
		\node at (-1.38, 1.3) {$\Omega_2$};
		\node at (-1.38,-1.3) {$\Omega_3$};
		\node at ( 0.72,-1.32) {$\Omega_4$};
		
	\end{tikzpicture}
	\caption{Oriented contours $\Sigma_j$ and the regions $\Omega_j$,
		$j=1,2,3,4$, for the $P_{34}$ model RH problem.}
	\label{fig:jumpsPsi}
\end{figure}
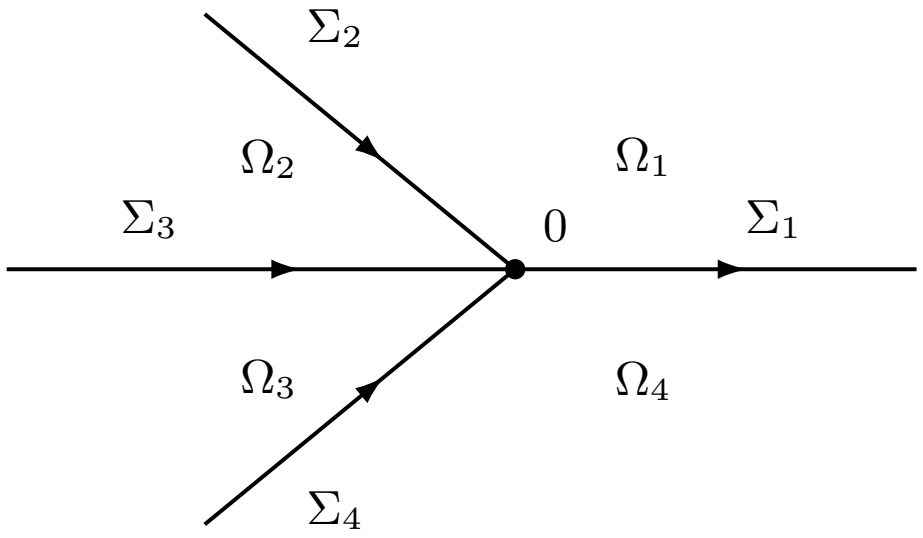

\begin{RHP}\label{RH P34}
	\hfill
		\begin{itemize}
		\item $M^{(P_{34})}$ is analytic  in
		$\mathbb{C} \setminus \big((\cup_{j=1}^4\Sigma_j)\cup\{0\}\big)$, where
		\begin{align}
			\Sigma_1=\mathbb{R}^+, \quad \Sigma_2=e^{\frac{2 \pi i}{3}}\mathbb{R}^+, \quad\Sigma_3=e^{   \pi i  }\mathbb{R}^+,\quad\Sigma_4=e^{-\frac{2 \pi i}{3}}\mathbb{R}^+.\label{def jumpcureve 34}
		\end{align}
The orientations of the contours $\Sigma_j$ and the regions $\Omega_j$,
$j=1,2,3,4$, are shown in
Figure~\ref{fig:jumpsPsi}.
	\item   $M^{(P_{34})}$  satisfies the jump condition $	M^{(P_{34})}_+ (\zeta)=M^{(P_{34})}_- (\zeta)J^{(P_{34})}(\zeta),$ with
\begin{equation}\label{eq:Psi-jump}
	J^{(P_{34})}(\zeta)=
	\begin{cases}
		\begin{pmatrix}1&\omega\\0&1\end{pmatrix},&\zeta\in\Sigma_1,\\[6pt]
		\begin{pmatrix}1&0\\e^{2\pi i\alpha}&1\end{pmatrix},&\zeta\in\Sigma_2,\\[6pt]
		\begin{pmatrix}0&1\\-1&0\end{pmatrix},&\zeta\in\Sigma_3,\\[6pt]
		\begin{pmatrix}1&0\\e^{-2\pi i\alpha}&1\end{pmatrix},&\zeta\in\Sigma_4.
	\end{cases}
\end{equation}
	\item As $\zeta \to \infty$,
\begin{equation} \label{Psi-infinity}
	\begin{aligned}
		M^{(P_{34})}(\zeta)
		&=\begin{pmatrix}1&0\\-ia(s)&1\end{pmatrix}
		\left(I+\frac{M_1^{(P_{34})}(s)}{\zeta}
		+\mathcal O(\zeta^{-2})\right)\\
		&\quad\times\frac{\zeta^{-\frac{\sigma_3}{4}}}{\sqrt2}
		\begin{pmatrix}1&i\\i&1\end{pmatrix}
		e^{-(s\zeta^{\frac{1}{2}}+\frac23\zeta^{\frac{3}{2}})\sigma_3},
	\end{aligned}
\end{equation}
where $a(s)=a(s;\alpha,\omega)$ and
\begin{equation}\label{eq:12entry}
	\bigl(M_1^{(P_{34})}(s)\bigr)_{12}=ia(s).
\end{equation}
	\item  For $-\frac{1}{2}<\alpha<0$, \begin{equation}\label{eq:resP34}
		M^{(P_{34})}(\zeta)=\mathcal{O}(|\zeta|^{\alpha}),
		\qquad   \zeta\to0.
	\end{equation}
For $\alpha>0$,
\begin{equation*}
	M^{(P_{34})}(\zeta)=
	\begin{cases}
		\begin{pmatrix}
			\mathcal O(|\zeta|^{\alpha})&\mathcal O(|\zeta|^{-\alpha})\\
			\mathcal O(|\zeta|^{\alpha})&\mathcal O(|\zeta|^{-\alpha})
		\end{pmatrix},&\zeta\in\Omega_1\cup\Omega_4,\\[8pt]
		\begin{pmatrix}
			\mathcal O(|\zeta|^{-\alpha})&\mathcal O(|\zeta|^{-\alpha})\\
			\mathcal O(|\zeta|^{-\alpha})&\mathcal O(|\zeta|^{-\alpha})
		\end{pmatrix},&\zeta\in\Omega_2\cup\Omega_3,
	\end{cases}\qquad \zeta\to0.
\end{equation*}
At $\alpha=0$, the corresponding condition is
\[
M^{(P_{34})}(\zeta)=
\begin{cases}
	\begin{pmatrix}
		\mathcal O(1)&\mathcal O(1+|\log|\zeta||)\\
		\mathcal O(1)&\mathcal O(1+|\log|\zeta||)
	\end{pmatrix},&\zeta\in\Omega_1\cup\Omega_4,\\[8pt]
	\mathcal O(1+|\log|\zeta||),&\zeta\in\Omega_2\cup\Omega_3,
\end{cases}
\qquad \zeta\to0.
\]
\end{itemize}
		\end{RHP}

For the stated parameter range, this RH problem has a unique solution
for every real $s$; see \cite{ikj2008,ikj2009,XZ2011}.
With $a(s)$ given in \eqref{Psi-infinity}, set
\[
u(s;\alpha,\omega)=a'(s;\alpha,\omega)-\frac{s}{2}.
\]
This function has no poles on the real axis and satisfies the Painlev\'e XXXIV equation
\begin{equation}\label{eq:app-p34-equation}
	u''(s)=4u(s)^2+2s\,u(s)
	+\frac{u'(s)^2-(2\alpha)^2}{2u(s)}.
\end{equation}
For $\omega=0$, the real-axis asymptotics are
\begin{equation}\label{eq:uasygen}
	u(s;\alpha,0)=
	\begin{cases}
		\displaystyle\frac{\alpha}{\sqrt{s}}+\mathcal O(s^{-2}),
		&s\to+\infty,\\[6pt]
		\displaystyle-\frac{s}{2}+\mathcal O(s^{-2}),
		&s\to-\infty.
	\end{cases}
\end{equation}
To fix the integration constant in $a$, note that our convention gives
$a(s;\alpha,\omega)=\frac{s^2}{4}-H(s;2\alpha,\omega)$,
where $H$ is the Hamiltonian used in \cite{DXZ20}.
Its asymptotics for $\omega=0$ yield
$H(s;2\alpha,0)=\frac{s^2}{4}+\mathcal O(|s|^{-1})$ as $s\to-\infty$.
Consequently,
\begin{equation}\label{eq:Psi-infinity}
	a(s;\alpha,0)=\int_{-\infty}^{s}
	\left(u(t;\alpha,0)+\frac{t}{2}\right)\,dt.
\end{equation}
In the analysis of transition regions $\mathcal{T}_{\mathrm{I}}$ and $\mathcal{T}_{\mathrm{II}}$, we use $a(s)=a(s;\frac14,0)$.

\section{The Painlev\'e II  parametrix}
\label{app:PII}

We recall the Painlev\'e II model RH problem, see \cite{FIKN06,DZ95}.
The corresponding solutions with Airy asymptotics were studied
in \cite{HM80}.

\begin{figure}
	\centering
	\begin{tikzpicture}[
		x=1cm,
		y=1cm,
		>=Latex,
		ray/.style={
			draw=black!80,
			line width=0.85pt,
			postaction={decorate},
			decoration={
				markings,
				mark=at position 0.60 with
				{\arrow{Latex[length=2.5mm,width=1.6mm]}}
			}
		},
		every node/.style={font=\small}
		]
		
		% All rays are oriented away from the origin.
		\draw[ray] (0,0)--(30:3);
		\draw[ray] (0,0)--(150:3);
		\draw[ray] (0,0)--(210:3);
		\draw[ray] (0,0)--(330:3);
		
		\node[below=5pt] at (0,0) {$0$};
		
		        % Contour labels.
		\node[above left=3pt]  at (30:2.15)  {$\Gamma_1$};
		\node[above right=3pt] at (150:2.15) {$\Gamma_3$};
		\node[below right=3pt] at (210:2.15) {$\Gamma_4$};
		\node[below left=3pt]  at (330:2.15) {$\Gamma_6$};
		
		% Stokes matrices in two symmetric columns.
		\def\matrixX{2.70}
		\def\matrixY{0.95}
		
		\node[anchor=west] at ( \matrixX, \matrixY)
		{$S_1=
			\begin{pmatrix}
				1 & 0 \\
				-i\kappa & 1
			\end{pmatrix}$};
		
		\node[anchor=east] at (-\matrixX, \matrixY)
		{$S_3=
			\begin{pmatrix}
				1 & 0 \\
				i\kappa & 1
			\end{pmatrix}$};
		
		\node[anchor=east] at (-\matrixX,-\matrixY)
		{$S_4=
			\begin{pmatrix}
				1 & i\kappa \\
				0 & 1
			\end{pmatrix}$};
		
		\node[anchor=west] at ( \matrixX,-\matrixY)
		{$S_6=
			\begin{pmatrix}
				1 & -i\kappa \\
				0 & 1
			\end{pmatrix}$};
		
	\end{tikzpicture}
	\caption{The contour $\Gamma_j$ and the corresponding matrices
		$S_j$, $j=1,3,4,6$, for the Painlev\'e II model.}
	\label{fig:PII-model}
\end{figure}
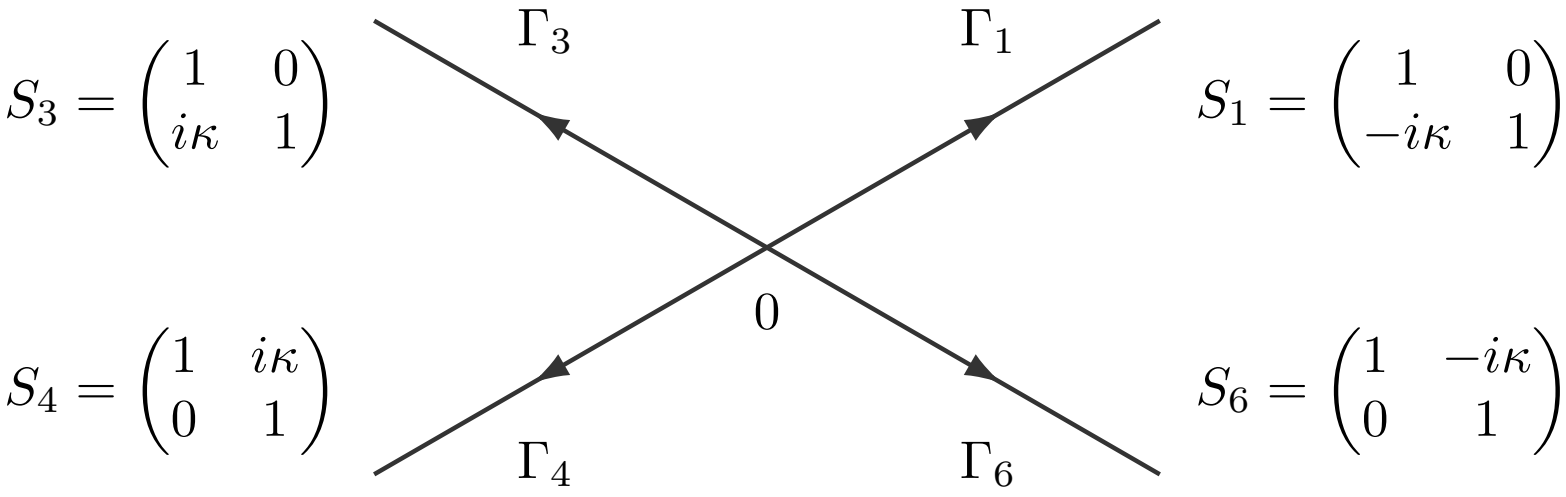

\begin{RHP}\label{RHII}
\hfill
\begin{itemize}
	\item
	$M^{(P_2)}$ is analytic in
	$\mathbb C\setminus\Gamma^{(P_2)}$, where $\Gamma^{(P_2)}
	:=\Gamma_1\cup\Gamma_3\cup\Gamma_4\cup\Gamma_6$ with
	\begin{equation*}
		\Gamma_j:=
		\left\{
		z\in\mathbb C:
		\arg z=\frac{\pi}{6}+\frac{\pi}{3}(j-1)
		\right\},
		\quad j=1,3,4,6,
	\end{equation*}
	see	Figure~\ref{fig:PII-model} for an illustration.

	\item
	$M^{(P_2)}$ satisfies the jump condition
	\begin{equation}
    \begin{aligned}
        &M^{(P_2)}_+(z;s,\kappa)
		=
		M^{(P_2)}_-(z;s,\kappa)V^{(P_2)}(z;s,\kappa),\\&
		V^{(P_2)}(z;s,\kappa)=e^{-i\left(\frac{4}{3}z^3+sz\right)\hat{\sigma}_3}
		S_j,
		\qquad z\in\Gamma_j,\quad j=1,3,4,6,
		\label{PII-jumps}
    \end{aligned}	
	\end{equation}
	where
	\begin{equation}
		\begin{aligned}
			S_1&=
			\begin{pmatrix}
				1&0\\
				-i\kappa&1
			\end{pmatrix},
			&
			S_3&=
			\begin{pmatrix}
				1&0\\
				i\kappa&1
			\end{pmatrix},
			\\[3pt]
			S_4&=
			\begin{pmatrix}
				1&i\kappa\\
				0&1
			\end{pmatrix},
			&
			S_6&=
			\begin{pmatrix}
				1&-i\kappa\\
				0&1
			\end{pmatrix}.
		\end{aligned}
		\label{PII-Stokes-matrices}
	\end{equation}
	
	\item
	As $z\to\infty$, we have
	\(
	M^{(P_2)}(z;s,\kappa)
	=I+\mathcal O(z^{-1}).
\)
	\item
		As $z\to0$, we have
	\(
	M^{(P_2)}(z;s,\kappa)
	=I+\mathcal O(1).
	\)
\end{itemize}
\end{RHP}

For $\kappa\in[-1,1]$, this RH problem has a unique solution
for every $s\in\mathbb R$; see \cite{FIKN06,DZ95}.
Moreover, for each integer $N\geq1$,
\begin{equation}
	M^{(P_2)}(z;s,\kappa)
	=
	I+\sum_{j=1}^{N}
	\frac{M_j^{(P_2)}(s;\kappa)}{z^j}
	+\mathcal O(z^{-N-1}),
	\qquad z\to\infty,
	\label{PII-expansion}
\end{equation}
where the first two coefficients are
	\begin{align}
		&M_1^{(P_2)}(s;\kappa)
		=
		\frac{1}{2}
		\begin{pmatrix}
			-i\int_s^\infty v(\eta)^2\,d\eta
			&
			v(s)
			\\[6pt]
			v(s)
			&
			i\int_s^\infty v(\eta)^2\,d\eta
		\end{pmatrix},\label{asy:MP2}\\
		&M_2^{(P_2)}(s;\kappa)
		=
		\frac{1}{8}
		\begin{pmatrix}
			-\left(\int_s^\infty
			v(\eta)^2\,d\eta\right)^2+v(s)^2
			&
			2i\left(
			v(s)\int_s^\infty v(\eta)^2\,d\eta
			+v'(s)
			\right)
			\\[8pt]
			-2i\left(
			v(s)\int_s^\infty v(\eta)^2\,d\eta
			+v'(s)
			\right)
			&
			-\left(\int_s^\infty
			v(\eta)^2\,d\eta\right)^2+v(s)^2\end{pmatrix}.\label{asy:MP22}
	\end{align}	
Here $v(s)=v(s;\kappa)$ is the  solution of
the Painlev\'e II equation
\begin{equation}
	v''(s)=s v(s)+2v(s)^3,
	\label{PII-equation}
\end{equation}
characterized by
\begin{equation}
	v(s;\kappa)
	=
	\kappa\operatorname{Ai}(s)
	+\mathcal O\big(
	s^{-\frac{1}{4}}e^{-2s^{\frac{3}{2}}}
	\big),
	\qquad s\to+\infty,
	\label{PII-Airy-condition}
\end{equation}
where $\operatorname{Ai}$ denotes the Airy function. We call RH problem \ref{RHII} the Painlev\'e  II parametrix.

\section*{Acknowledgements}
 The work of Fan is partially supported by NSFC under grant number 12671294. The work of Li is partially supported by NSFC under grant number 12647172.  The work of Yang is partially supported by
Natural Science Foundation of Chongqing under grant number CSTB2025NSCQ-GPX0727, the Fundamental Research Funds for the Central Universities under grant number 2025CDJ-IAISYB-011,  NSFC under grant number 12501325.

\hspace*{\parindent}
\\

\end{document}